\pdfoutput=1
\documentclass{redaction}%
\usepackage{not}%
\usepackage{meta}%
\date{October 27th, 2016}%
\def\MprefixNNNaaa{\Mprefixe{lin}{Polarization-free Quantization\\[2pt] of Linear Field Theories}}%
\def\seqBBBaao{\abbrevTheorem \hyperlink{PARaqf}{3.16}}%
\def\seqDDDaao{(\abbrevTheorem \hyperlink{PARaqf}{3.16})}%
\def\seqBBBaap{\abbrevSectionC \hyperlink{SECaju}{3.1} and \abbrevSectionD \hyperlink{SECbri}{B.1.1}}%
\def\seqBBBaaq{\abbrevSectionD \hyperlink{SECbri}{B.1.1}}%
\def\seqBBBaat{\bseqHHHabr{\abbrevSectionB 3}}%
\def\seqBBBaau{\abbrevSectionDs \hyperlink{SECbmy}{A.1.4} and \hyperlink{SECbyc}{B.1.4}}%
\def\seqBBBaav{the proof of \abbrevProposition \hyperlink{PARbbz}{4.5}}%
\def\seqBBBaaw{\abbrevSectionC \hyperlink{SECaam}{1.1}}%
\def\seqBBBaax{\abbrevSectionC \hyperlink{SECbfx}{5.3}}%
\def\seqBBBaba{\abbrevDef \hyperlink{PARbrz}{B.3}}%
\def\seqDDDaba{(\abbrevDef \hyperlink{PARbrz}{B.3})}%
\def\seqBBBabc{\abbrevSectionB \hyperlink{SECabt}{2}}%
\def\seqBBBabd{\abbrevProposition \hyperlink{PARaeq}{2.8}}%
\def\seqDDDabd{(\abbrevProposition \hyperlink{PARaeq}{2.8})}%
\def\seqBBBabe{\abbrevSectionB \hyperlink{SECajs}{3}}%
\def\seqCCCabe{\AbbrevSectionB \hyperlink{SECajs}{3}}%
\def\seqBBBabf{\abbrevTheorem \hyperlink{PARaif}{2.14}}%
\def\seqDDDabf{(\abbrevTheorem \hyperlink{PARaif}{2.14})}%
\def\seqDDDabh{(\abbrevTheorem \hyperlink{PARamj}{3.6})}%
\def\seqBBBabi{\abbrevTheorem \hyperlink{PARaot}{3.14}}%
\def\seqDDDabi{(\abbrevTheorem \hyperlink{PARaot}{3.14})}%
\def\seqBBBabj{\abbrevProposition \hyperlink{PARapc}{3.15}}%
\def\seqDDDabj{(\abbrevProposition \hyperlink{PARapc}{3.15})}%
\def\seqBBBabk{\abbrevSectionC \hyperlink{SECbgg}{A.1}}%
\def\seqBBBabl{\abbrevSectionC \hyperlink{SECbof}{A.2}}%
\def\seqDDDabl{(\abbrevSectionC \hyperlink{SECbof}{A.2})}%
\def\seqBBBabn{\abbrevSectionB \hyperlink{SECaws}{4}}%
\def\seqBBBabo{\abbrevTheorem \hyperlink{PARayj}{4.4}}%
\def\seqBBBabp{\abbrevProposition \hyperlink{PARbdh}{4.6}}%
\def\seqDDDabp{(\abbrevProposition \hyperlink{PARbdh}{4.6})}%
\def\seqBBBabr{\abbrevSectionD \hyperlink{SECcau}{B.1.5}}%
\def\seqDDDabr{(\abbrevSectionD \hyperlink{SECcau}{B.1.5})}%
\def\seqBBBabw{\abbrevSectionC \hyperlink{SECawu}{4.1}}%
\def\seqDDDaci{\bseqHHHabs{\abbrevDef 2.6}}%
\def\seqDDDacj{\bseqHHHabs{\abbrevDef 2.1 and \abbrevProposition 2.2}}%
\def\seqBBBack{\bseqHHHabs{\abbrevProposition 2.10}}%
\def\seqDDDack{\bseqHHHabs{\abbrevProposition 2.10}}%
\def\seqBBBacl{\abbrevFigure \hyperlink{PARacn}{2.1}}%
\def\seqBBBacm{\abbrevSectionC \hyperlink{SECaju}{3.1}}%
\def\seqDDDacm{(\abbrevSectionC \hyperlink{SECaju}{3.1})}%
\def\InputFigPaco{\includeStandaloneFig{1}{\figBBOXaco}}%
\def\seqBBBacq{\bseqHHHaas{\abbrevDef 2.1}}%
\def\seqBBBacz{\abbrevList \hyperlink{PARact}{\formatrefL{2.2}{3}}}%
\def\seqDDDacz{(\abbrevList \hyperlink{PARact}{\formatrefL{2.2}{3}})}%
\def\seqBBBada{\abbrevList \hyperlink{PARacw}{\formatrefL{2.2}{5}}}%
\def\seqCCCada{\AbbrevList \hyperlink{PARacw}{\formatrefL{2.2}{5}}}%
\def\seqDDDada{(\abbrevList \hyperlink{PARacw}{\formatrefL{2.2}{5}})}%
\def\seqBBBadb{\abbrevList \hyperlink{PARacv}{\formatrefL{2.2}{4}}}%
\def\seqDDDadb{(\abbrevList \hyperlink{PARacv}{\formatrefL{2.2}{4}})}%
\def\seqBBBadc{\bseqHHHabt{\abbrevSectionC 2.2}}%
\def\seqBBBadl{\abbrevList \hyperlink{PARadh}{\formatrefL{2.3}{1}}}%
\def\seqBBBadm{\abbrevList \hyperlink{PARadi}{\formatrefL{2.3}{2}}}%
\def\seqBBBadn{\abbrevList \hyperlink{PARadj}{\formatrefL{2.3}{3}}}%
\def\seqBBBaea{\abbrevList \hyperlink{PARadv}{\formatrefL{2.6}{1}}}%
\def\seqBBBaeb{\abbrevList \hyperlink{PARadw}{\formatrefL{2.6}{2}}}%
\def\seqBBBaec{\abbrevProposition \hyperlink{PARadd}{2.3}}%
\def\seqDDDaec{(\abbrevProposition \hyperlink{PARadd}{2.3})}%
\def\seqBBBaed{\abbrevList \hyperlink{PARadx}{\formatrefL{2.6}{3}}}%
\def\seqBBBaee{\abbrevList \hyperlink{PARady}{\formatrefL{2.6}{4}}}%
\def\seqBBBaei{\abbrevDef \hyperlink{PARacp}{2.2}}%
\def\seqDDDaei{(\abbrevDef \hyperlink{PARacp}{2.2})}%
\def\seqDDDaej{\bseqHHHaas{\abbrevEqua \formatrefE{2.1}{1} and \abbrevFigure 2.1}}%
\def\seqBBBaek{\bseqHHHabs{\abbrevProposition 2.8}}%
\def\seqBBBael{\abbrevLists \hyperlink{PARacv}{\formatrefL{2.2}{4}} and \hyperlink{PARacw}{\formatrefL{2.2}{5}}}%
\def\seqDDDael{(\abbrevLists \hyperlink{PARacv}{\formatrefL{2.2}{4}} and \hyperlink{PARacw}{\formatrefL{2.2}{5}})}%
\def\seqBBBaem{\abbrevFigure \hyperlink{PARaen}{2.2}}%
\def\InputFigPaeo{\includeStandaloneFig{2}{\figBBOXaeo}}%
\def\seqBBBaet{\abbrevDef \hyperlink{PARacp}{2.2} vs.\ \bseqHHHaas{\abbrevDef 2.1}}%
\def\seqBBBaew{\abbrevDef \hyperlink{PARado}{2.4}}%
\def\seqDDDaew{(\abbrevDef \hyperlink{PARado}{2.4})}%
\def\seqBBBaex{\abbrevDef \hyperlink{PARaef}{2.7}}%
\def\seqDDDaex{(\abbrevDef \hyperlink{PARaef}{2.7})}%
\def\seqBBBaey{\bseqHHHaas{\abbrevDef 2.2}}%
\def\seqBBBaez{\bseqHHHaas{\abbrevDef 2.3}}%
\def\seqBBBafa{\abbrevDefs \hyperlink{PARado}{2.4} and \hyperlink{PARaef}{2.7}}%
\def\seqBBBafb{\bseqHHHaas{\abbrevSectionC 2.1}}%
\def\seqBBBaff{\bseqHHHaas{\abbrevEqua \formatrefE{2.1}{1}}}%
\def\seqDDDaff{\bseqHHHaas{\abbrevEqua \formatrefE{2.1}{1}}}%
\def\seqBBBafh{\abbrevProposition \hyperlink{PARads}{2.6}}%
\def\seqBBBafk{\abbrevFigure \hyperlink{PARafl}{2.3}}%
\def\InputFigPafm{\includeStandaloneFig{3}{\figBBOXafm}}%
\def\seqBBBafo{\bseqHHHaas{\abbrevSectionC 2.2}}%
\def\seqBBBafp{the proof of \abbrevTheorem \hyperlink{PARaqf}{3.16}}%
\def\seqDDDafp{(proof of \abbrevTheorem \hyperlink{PARaqf}{3.16})}%
\def\seqBBBafq{\abbrevSectionC \hyperlink{SECahb}{2.3}}%
\def\seqBBBafy{\abbrevLists \hyperlink{PARafs}{\formatrefL{2.9}{1}} and \hyperlink{PARaft}{\formatrefL{2.9}{2}}}%
\def\seqCCCagk{\AbbrevLists \hyperlink{PARage}{\formatrefL{2.10}{1}} and \hyperlink{PARagf}{\formatrefL{2.10}{2}}}%
\def\seqCCCagl{\AbbrevLists \hyperlink{PARagh}{\formatrefL{2.10}{3}} and \hyperlink{PARagi}{\formatrefL{2.10}{4}}}%
\def\seqBBBagm{\abbrevList \hyperlink{PARafw}{\formatrefL{2.9}{4}}}%
\def\seqDDDagn{(\abbrevProposition \hyperlink{PARadd}{2.3} and \abbrevDef \hyperlink{PARado}{2.4})}%
\def\seqDDDago{(\abbrevProposition \hyperlink{PARads}{2.6} and \abbrevDef \hyperlink{PARaef}{2.7})}%
\def\seqBBBagv{\abbrevProposition \hyperlink{PARafz}{2.10}}%
\def\seqBBBaha{\abbrevDef \hyperlink{PARafr}{2.9}}%
\def\seqBBBahd{\abbrevSectionC \hyperlink{SECaay}{1.3}}%
\def\seqBBBahe{\abbrevSectionC \hyperlink{SECabu}{2.1}}%
\def\seqDDDahe{(\abbrevSectionC \hyperlink{SECabu}{2.1})}%
\def\seqBBBahf{\abbrevSectionBs \hyperlink{SECajs}{3} and \hyperlink{SECaws}{4}}%
\def\seqBBBahg{\bseqHHHaad{\abbrevSectionC 2.2}}%
\def\seqBBBahh{\bseqHHHaad{\abbrevTheorem 2.8}}%
\def\seqBBBahi{\abbrevFigure \hyperlink{PARahj}{2.4}}%
\def\InputFigPahk{\includeStandaloneFig{4}{\figBBOXahk}}%
\def\seqBBBahm{\abbrevProposition \hyperlink{PARagy}{2.12}}%
\def\seqDDDahm{(\abbrevProposition \hyperlink{PARagy}{2.12})}%
\def\seqBBBaht{\abbrevDef \hyperlink{PARahn}{2.13}}%
\def\seqDDDaht{(\abbrevDef \hyperlink{PARahn}{2.13})}%
\def\seqBBBahu{\abbrevSectionCs \hyperlink{SECayf}{4.2} and \hyperlink{SECbfv}{5.2}}%
\def\seqBBBahv{\abbrevLemma \hyperlink{PARaih}{2.15}}%
\def\seqCCCahv{\AbbrevLemma \hyperlink{PARaih}{2.15}}%
\def\seqBBBahw{\abbrevFigure \hyperlink{PARaia}{2.5}}%
\def\seqDDDahx{(\abbrevFigure \hyperlink{PARaid}{2.6})}%
\def\seqBBBahy{\abbrevList \hyperlink{PARahq}{\formatrefL{2.13}{3}}}%
\def\seqDDDahy{(\abbrevList \hyperlink{PARahq}{\formatrefL{2.13}{3}})}%
\def\seqBBBahz{the proof of \abbrevTheorem \hyperlink{PARaif}{2.14}}%
\def\InputFigPaib{\includeStandaloneFig{5}{\figBBOXaib}}%
\def\seqCCCaic{The proof of \abbrevLemma \hyperlink{PARaih}{2.15}}%
\def\InputFigPaie{\includeStandaloneFig{6}{\figBBOXaie}}%
\def\seqBBBaio{\abbrevList \hyperlink{PARaho}{\formatrefL{2.13}{1}}}%
\def\seqDDDaio{(\abbrevList \hyperlink{PARaho}{\formatrefL{2.13}{1}})}%
\def\seqCCCaiv{\AbbrevList \hyperlink{PARaij}{\formatrefL{2.15}{2}}}%
\def\seqBBBaix{\abbrevList \hyperlink{PARafv}{\formatrefL{2.9}{3}}}%
\def\seqDDDaix{(\abbrevList \hyperlink{PARafv}{\formatrefL{2.9}{3}})}%
\def\seqBBBaiy{\abbrevEqua \hyperlink{PARaip}{\formatrefE{2.15}{1}}}%
\def\seqCCCaiy{\AbbrevEqua \hyperlink{PARaip}{\formatrefE{2.15}{1}}}%
\def\seqBBBaiz{\abbrevEqua \hyperlink{PARait}{\formatrefE{2.15}{2}}}%
\def\seqCCCaiz{\AbbrevEqua \hyperlink{PARait}{\formatrefE{2.15}{2}}}%
\def\seqBBBaja{\abbrevList \hyperlink{PARaii}{\formatrefL{2.15}{1}}}%
\def\seqBBBajb{\abbrevList \hyperlink{PARahp}{\formatrefL{2.13}{2}}}%
\def\seqDDDajb{(\abbrevList \hyperlink{PARahp}{\formatrefL{2.13}{2}})}%
\def\seqBBBajc{\abbrevList \hyperlink{PARaik}{\formatrefL{2.15}{3}}}%
\def\seqBBBajj{\abbrevLists \hyperlink{PARajf}{\formatrefL{2.14}{2}} and \hyperlink{PARajh}{\formatrefL{2.14}{4}}}%
\def\seqBBBajk{\abbrevLists \hyperlink{PARafv}{\formatrefL{2.9}{3}} and \hyperlink{PARafw}{\formatrefL{2.9}{4}}}%
\def\seqBBBajl{\abbrevEqua \hyperlink{PARaig}{\formatrefE{2.14}{1}}}%
\def\seqBBBajq{\abbrevEqua \hyperlink{PARajo}{\formatrefE{2.16}{1}}}%
\def\seqBBBajw{\abbrevSectionC \hyperlink{SECbrh}{B.1}}%
\def\seqBBBajx{\abbrevSectionC \hyperlink{SECceu}{B.2}}%
\def\seqBBBajy{\abbrevDef \hyperlink{PARbgn}{A.2}}%
\def\seqDDDajy{(\abbrevDef \hyperlink{PARbgn}{A.2})}%
\def\seqBBBajz{\abbrevFigure \hyperlink{PARakc}{3.1}}%
\def\seqDDDajz{(\abbrevFigure \hyperlink{PARakc}{3.1})}%
\def\seqBBBaka{\abbrevProposition \hyperlink{PARbia}{A.6}}%
\def\seqDDDaka{(\abbrevProposition \hyperlink{PARbia}{A.6})}%
\def\seqBBBakb{\abbrevProposition \hyperlink{PARbsq}{B.5}}%
\def\InputFigPakd{\includeStandaloneFig{7}{\figBBOXakd}}%
\def\seqDDDakf{(\abbrevPropositions \hyperlink{PARbia}{A.6} and \hyperlink{PARbsq}{B.5})}%
\def\seqBBBakg{\abbrevDef \hyperlink{PARbgk}{A.1}}%
\def\seqDDDakg{(\abbrevDef \hyperlink{PARbgk}{A.1})}%
\def\seqBBBakh{\abbrevProposition \hyperlink{PARbgp}{A.3}}%
\def\seqDDDakh{(\abbrevProposition \hyperlink{PARbgp}{A.3})}%
\def\seqBBBaki{\abbrevDef \hyperlink{PARbwh}{B.9}}%
\def\seqDDDaki{(\abbrevDef \hyperlink{PARbwh}{B.9})}%
\def\seqDDDakj{(\abbrevFigure \hyperlink{PARakk}{3.2})}%
\def\InputFigPakl{\includeStandaloneFig{8}{\figBBOXakl}}%
\def\seqBBBakn{\bseqHHHaas{\abbrevSectionC 3.2}}%
\def\seqBBBako{\abbrevSectionC \hyperlink{SECaar}{1.2}}%
\def\seqDDDako{(\abbrevSectionC \hyperlink{SECaar}{1.2})}%
\def\seqBBBaks{the proof of \abbrevProposition \hyperlink{PARbgp}{A.3}}%
\def\seqBBBaky{\abbrevProposition \hyperlink{PARbjl}{A.9}}%
\def\seqDDDaky{(\abbrevProposition \hyperlink{PARbjl}{A.9})}%
\def\seqBBBalc{\abbrevEqua \hyperlink{PARala}{\formatrefE{3.2}{1}}}%
\def\seqCCCalc{\AbbrevEqua \hyperlink{PARala}{\formatrefE{3.2}{1}}}%
\def\seqBBBald{\abbrevProposition \hyperlink{PARbjf}{A.7}}%
\def\seqDDDalf{(\abbrevSectionCs \hyperlink{SECbof}{A.2} and \hyperlink{SECceu}{B.2})}%
\def\seqBBBalg{\abbrevDef \hyperlink{PARamc}{3.5}}%
\def\seqDDDalg{(\abbrevDef \hyperlink{PARamc}{3.5})}%
\def\seqBBBalo{\abbrevProposition \hyperlink{PARbpf}{A.22}}%
\def\seqDDDalo{(\abbrevProposition \hyperlink{PARbpf}{A.22})}%
\def\seqBBBalq{\abbrevProposition \hyperlink{PARbpy}{A.24}}%
\def\seqBBBalu{\abbrevEqua \hyperlink{PARals}{\formatrefE{3.4}{1}}}%
\def\seqBBBalv{\abbrevLists \hyperlink{PARalk}{\formatrefL{3.3}{3}} and \hyperlink{PARall}{\formatrefL{3.3}{4}}}%
\def\seqBBBalw{\abbrevList \hyperlink{PARall}{\formatrefL{3.3}{4}}}%
\def\seqBBBalx{\abbrevProposition \hyperlink{PARbpw}{A.23}}%
\def\seqBBBaly{\abbrevProposition \hyperlink{PARbok}{A.19}}%
\def\seqBBBama{\abbrevSectionD \hyperlink{SECbwe}{B.1.3}}%
\def\seqDDDama{(\abbrevSectionD \hyperlink{SECbwe}{B.1.3})}%
\def\seqBBBamb{\abbrevSectionD \hyperlink{SECchh}{B.2.3}}%
\def\seqBBBame{\abbrevDef \hyperlink{PARbrq}{B.2}}%
\def\seqDDDame{(\abbrevDef \hyperlink{PARbrq}{B.2})}%
\def\seqBBBamf{\abbrevDef \hyperlink{PARcex}{B.19}}%
\def\seqBBBamh{\abbrevProposition \hyperlink{PARcgd}{B.22}}%
\def\seqBBBami{\abbrevDef \hyperlink{PARcho}{B.24}}%
\def\seqDDDami{(\abbrevDef \hyperlink{PARcho}{B.24})}%
\def\seqBBBaml{\abbrevEqua \hyperlink{PARbwo}{\formatrefE{B.10}{2}}}%
\def\seqBBBamm{\abbrevEqua \hyperlink{PARchv}{\formatrefE{B.25}{2}}}%
\def\seqBBBamn{\abbrevProposition \hyperlink{PARbxb}{B.11}}%
\def\seqBBBamo{\abbrevProposition \hyperlink{PARchz}{B.26}}%
\def\seqBBBamq{\abbrevSectionD \hyperlink{SECbgh}{A.1.1}}%
\def\seqBBBana{\abbrevEquas \hyperlink{PARbwk}{\formatrefE{B.10}{1}} and \hyperlink{PARbsw}{\formatrefE{B.5}{3}}}%
\def\seqBBBand{\abbrevEqua \hyperlink{PARbif}{\formatrefE{A.6}{1}}}%
\def\seqBBBane{\abbrevDef \hyperlink{PARbgu}{A.4}}%
\def\seqDDDane{(\abbrevDef \hyperlink{PARbgu}{A.4})}%
\def\seqBBBang{\abbrevEqua \hyperlink{PARamt}{\formatrefE{3.8}{1}}}%
\def\seqBBBanh{\abbrevEqua \hyperlink{PARbsr}{\formatrefE{B.5}{1}}}%
\def\seqBBBanj{\abbrevEqua \hyperlink{PARcfh}{\formatrefE{B.19}{1}}}%
\def\seqDDDanj{(\abbrevEqua \hyperlink{PARcfh}{\formatrefE{B.19}{1}})}%
\def\seqBBBanl{\abbrevDef \hyperlink{PARcfl}{B.20}}%
\def\seqDDDanl{(\abbrevDef \hyperlink{PARcfl}{B.20})}%
\def\seqBBBanq{\abbrevProposition \hyperlink{PARams}{3.8}}%
\def\seqBBBanr{\abbrevEquas \hyperlink{PARchr}{\formatrefE{B.25}{1}} and \hyperlink{PARcgg}{\formatrefE{B.22}{2}}}%
\def\seqBBBans{\abbrevDef \hyperlink{PARbop}{A.20}}%
\def\seqBBBant{\abbrevProposition \hyperlink{PARcfr}{B.21}}%
\def\seqDDDanw{(\abbrevDef \hyperlink{PARamc}{3.5} and \abbrevTheorem \hyperlink{PARamj}{3.6})}%
\def\seqBBBany{\abbrevDef \hyperlink{PARbkc}{A.11}}%
\def\seqDDDany{(\abbrevDef \hyperlink{PARbkc}{A.11})}%
\def\seqDDDanz{(\abbrevDef \hyperlink{PARbqe}{A.25})}%
\def\seqBBBaof{\abbrevSectionC \hyperlink{SECafi}{2.2}}%
\def\seqBBBaog{\abbrevDef \hyperlink{PARanx}{3.11}}%
\def\seqDDDaog{(\abbrevDef \hyperlink{PARanx}{3.11})}%
\def\seqBBBaoi{\abbrevDef \hyperlink{PARagp}{2.11}}%
\def\seqBBBaok{\abbrevDef \hyperlink{PARamr}{3.7}}%
\def\seqBBBaol{\abbrevDef \hyperlink{PARank}{3.9}}%
\def\seqBBBaon{\abbrevProposition \hyperlink{PARajn}{2.16}}%
\def\seqDDDaoo{(\abbrevSectionD \hyperlink{SECbqc}{A.2.3})}%
\def\seqBBBaop{\abbrevSectionD \hyperlink{SECbju}{A.1.3}}%
\def\seqDDDaop{(\abbrevSectionD \hyperlink{SECbju}{A.1.3})}%
\def\seqBBBaoq{\abbrevProposition \hyperlink{PARbnm}{A.17}}%
\def\seqDDDaoq{(\abbrevProposition \hyperlink{PARbnm}{A.17})}%
\def\seqBBBaos{\abbrevDef \hyperlink{PARbjx}{A.10}}%
\def\seqDDDaos{(\abbrevDef \hyperlink{PARbjx}{A.10})}%
\def\seqBBBaou{\abbrevDef \hyperlink{PARbkc}{A.11} and \abbrevProposition \hyperlink{PARbkh}{A.12}}%
\def\seqDDDaou{(\abbrevDef \hyperlink{PARbkc}{A.11} and \abbrevProposition \hyperlink{PARbkh}{A.12})}%
\def\seqBBBaow{the proof of \abbrevProposition \hyperlink{PARbkr}{A.13}}%
\def\seqBBBaox{\abbrevDef \hyperlink{PARbjx}{A.10} and \abbrevProposition \hyperlink{PARbkh}{A.12}}%
\def\seqBBBaoy{\abbrevProposition \hyperlink{PARbkr}{A.13}}%
\def\seqBBBaoz{\abbrevProposition \hyperlink{PARbqh}{A.26}}%
\def\seqBBBapb{\abbrevTheorem \hyperlink{PARaif}{2.14} and \abbrevProposition \hyperlink{PARajn}{2.16}}%
\def\seqBBBapm{\abbrevDefs \hyperlink{PARamr}{3.7} and \hyperlink{PARbrz}{B.3}}%
\def\seqBBBapn{\abbrevDefs \hyperlink{PARank}{3.9} and \hyperlink{PARcfl}{B.20}}%
\def\seqDDDapo{(\abbrevList \hyperlink{PARafs}{\formatrefL{2.9}{1}})}%
\def\seqBBBapq{\abbrevList \hyperlink{PARbjy}{\formatrefL{A.10}{1}}}%
\def\seqCCCapq{\AbbrevList \hyperlink{PARbjy}{\formatrefL{A.10}{1}}}%
\def\seqDDDapq{(\abbrevList \hyperlink{PARbjy}{\formatrefL{A.10}{1}})}%
\def\seqBBBapu{\abbrevSectionDs \hyperlink{SECbmy}{A.1.4} and \hyperlink{SECbqr}{A.2.4}}%
\def\seqDDDapv{(\abbrevSectionDs \hyperlink{SECbyc}{B.1.4} and \hyperlink{SECcje}{B.2.4})}%
\def\seqBBBapw{\abbrevSectionCs \hyperlink{SECahb}{2.3} and \hyperlink{SECanu}{3.2}}%
\def\seqDDDapw{(\abbrevSectionCs \hyperlink{SECahb}{2.3} and \hyperlink{SECanu}{3.2})}%
\def\seqBBBapy{\abbrevEqua \hyperlink{PARala}{\formatrefE{3.2}{1}}, resp.\ \hyperlink{PARals}{\formatrefE{3.4}{1}}}%
\def\seqBBBapz{\abbrevFigure \hyperlink{PARaqc}{3.3}}%
\def\seqDDDapz{(\abbrevFigure \hyperlink{PARaqc}{3.3})}%
\def\seqBBBaqb{\bseqHHHaas{\abbrevTheorem 2.9}}%
\def\seqDDDaqb{\bseqHHHaas{\abbrevTheorem 2.9}}%
\def\InputFigPaqd{\includeStandaloneFig{9}{\figBBOXaqd}}%
\def\seqBBBaqg{\abbrevDef \hyperlink{PARbnb}{A.16}}%
\def\seqDDDaqg{(\abbrevDef \hyperlink{PARbnb}{A.16})}%
\def\seqBBBaqh{\abbrevDef \hyperlink{PARbqu}{A.28}}%
\def\seqDDDaqh{(\abbrevDef \hyperlink{PARbqu}{A.28})}%
\def\seqBBBaqt{\bseqHHHaas{\abbrevProposition 2.6}}%
\def\seqDDDaqx{(\abbrevEqua \hyperlink{PARbzb}{\formatrefE{B.14}{2}}, resp.\ \hyperlink{PARcjs}{\formatrefE{B.30}{2}})}%
\def\seqBBBaqz{\abbrevEqua \hyperlink{PARaqv}{\formatrefE{3.16}{1}}}%
\def\seqDDDara{(\abbrevList \hyperlink{PARamg}{\formatrefL{3.5}{2}})}%
\def\seqBBBarb{\abbrevEqua \hyperlink{PARbyz}{\formatrefE{B.14}{1}}, resp.\ \hyperlink{PARcjq}{\formatrefE{B.30}{1}}}%
\def\seqBBBare{\abbrevEqua \hyperlink{PARarc}{\formatrefE{3.16}{2}}}%
\def\seqBBBarj{\abbrevEqua \hyperlink{PARarf}{\formatrefE{3.16}{3}}}%
\def\seqBBBarn{\abbrevEqua \hyperlink{PARarh}{\formatrefE{3.16}{4}} and \bseqHHHaas{\abbrevEqua \formatrefE{2.1}{1}}}%
\def\seqBBBarz{\abbrevDef \hyperlink{PARbye}{B.12}, resp.\ \hyperlink{PARcjg}{B.28}}%
\def\seqBBBase{\abbrevDef \hyperlink{PARbrk}{B.1}, resp.\ \hyperlink{PARcex}{B.19}}%
\def\seqBBBasf{\abbrevProposition \hyperlink{PARbgp}{A.3}, resp.\ \hyperlink{PARbok}{A.19}}%
\def\seqBBBasn{\abbrevLists \hyperlink{PARaqi}{\formatrefL{3.16}{1}}, \hyperlink{PARaqk}{\formatrefL{3.16}{3}} and \hyperlink{PARaql}{\formatrefL{3.16}{4}}}%
\def\seqBBBasp{\bseqHHHaas{proof of \abbrevTheorem 2.9}}%
\def\seqDDData{(\abbrevDef \hyperlink{PARamr}{3.7}, resp.\ \hyperlink{PARank}{3.9})}%
\def\seqDDDatb{(\abbrevEqua \hyperlink{PARcag}{\formatrefE{B.15}{1}}, resp.\ \hyperlink{PARcjy}{\formatrefE{B.31}{1}})}%
\def\seqDDDatg{(\abbrevProposition \hyperlink{PARams}{3.8}, resp.\ \hyperlink{PARanm}{3.10})}%
\def\seqBBBath{\abbrevEqua \hyperlink{PARasy}{\formatrefE{3.16}{6}}}%
\def\seqBBBatk{\abbrevList \hyperlink{PARaqj}{\formatrefL{3.16}{2}}}%
\def\seqBBBatm{\abbrevProposition \hyperlink{PARatu}{3.17}}%
\def\seqCCCatm{\AbbrevProposition \hyperlink{PARatu}{3.17}}%
\def\seqBBBatn{\abbrevSectionC \hyperlink{SECayf}{4.2}}%
\def\seqBBBato{\abbrevProposition \hyperlink{PARbbz}{4.5}}%
\def\seqBBBatp{\abbrevSectionC \hyperlink{SECanu}{3.2}}%
\def\seqBBBatr{\abbrevProposition \hyperlink{PARbvp}{B.7}, resp.\ \hyperlink{PARcgt}{B.23}}%
\def\seqBBBatt{\abbrevList \hyperlink{PARaql}{\formatrefL{3.16}{4}}}%
\def\seqDDDatt{(\abbrevList \hyperlink{PARaql}{\formatrefL{3.16}{4}})}%
\def\seqBBBaty{the proof of \abbrevTheorem \hyperlink{PARaot}{3.14}}%
\def\seqBBBaub{\abbrevEqua \hyperlink{PARatz}{\formatrefE{3.17}{1}}}%
\def\seqBBBauc{\abbrevLists \hyperlink{PARaqj}{\formatrefL{3.16}{2}} and \hyperlink{PARaqk}{\formatrefL{3.16}{3}}}%
\def\seqBBBauf{\bseqHHHaas{\abbrevList \formatrefL{2.10}{1}}}%
\def\seqBBBauj{\abbrevEqua \hyperlink{PARcag}{\formatrefE{B.15}{1}}}%
\def\seqCCCauj{\AbbrevEqua \hyperlink{PARcag}{\formatrefE{B.15}{1}}}%
\def\seqBBBauq{\abbrevEqua \hyperlink{PARbwk}{\formatrefE{B.10}{1}}}%
\def\seqBBBaut{\abbrevEqua \hyperlink{PARbxv}{\formatrefE{B.11}{1}}}%
\def\seqBBBauw{\abbrevEqua \hyperlink{PARaum}{\formatrefE{3.17}{4}}}%
\def\seqBBBaux{\abbrevEqua \hyperlink{PARbyz}{\formatrefE{B.14}{1}}}%
\def\seqCCCaux{\AbbrevEqua \hyperlink{PARbyz}{\formatrefE{B.14}{1}}}%
\def\seqBBBavc{\abbrevEqua \hyperlink{PARauy}{\formatrefE{3.17}{5}}}%
\def\seqDDDave{(\abbrevEqua \hyperlink{PARaru}{\formatrefE{3.16}{5}})}%
\def\seqDDDavj{(\abbrevProposition \hyperlink{PARbur}{B.6})}%
\def\seqBBBavo{\abbrevEqua \hyperlink{PARaug}{\formatrefE{3.17}{3}}}%
\def\seqBBBavt{\abbrevProposition \hyperlink{PARbvp}{B.7}}%
\def\seqDDDavt{(\abbrevProposition \hyperlink{PARbvp}{B.7})}%
\def\seqBBBavw{\abbrevEqua \hyperlink{PARavr}{\formatrefE{3.17}{6}}}%
\def\seqBBBavz{\abbrevEqua \hyperlink{PARaud}{\formatrefE{3.17}{2}}}%
\def\seqBBBawf{\abbrevLists \hyperlink{PARaqi}{\formatrefL{3.16}{1}}, \hyperlink{PARaqj}{\formatrefL{3.16}{2}} and \hyperlink{PARaqk}{\formatrefL{3.16}{3}}}%
\def\seqBBBawi{\abbrevEqua \hyperlink{PARcjy}{\formatrefE{B.31}{1}}}%
\def\seqBBBawn{\abbrevEqua \hyperlink{PARawg}{\formatrefE{3.17}{7}}}%
\def\seqBBBawq{\abbrevProposition \hyperlink{PARcgt}{B.23}}%
\def\seqBBBaww{\abbrevFigure \hyperlink{PARawx}{4.1}}%
\def\InputFigPawy{\includeStandaloneFig{10}{\figBBOXawy}}%
\def\seqBBBaxg{\abbrevDef \hyperlink{PARaxi}{4.2}}%
\def\seqBBBaxh{the proof of \abbrevTheorem \hyperlink{PARayj}{4.4}}%
\def\seqBBBaxx{\abbrevPropositions \hyperlink{PARajn}{2.16} and \hyperlink{PARapc}{3.15}}%
\def\seqBBBaxz{\abbrevDef \hyperlink{PARaxa}{4.1}}%
\def\seqBBBaye{\abbrevList \hyperlink{PARaya}{\formatrefL{4.3}{1}}}%
\def\seqBBBayh{\abbrevProposition \hyperlink{PARbkh}{A.12}}%
\def\seqDDDayi{(\abbrevDef \hyperlink{PARakp}{3.1})}%
\def\seqBBBbap{\abbrevEqua \hyperlink{PARbah}{\formatrefE{4.4}{1}}}%
\def\seqBBBbaz{\abbrevEquas \hyperlink{PARbaj}{\formatrefE{4.4}{2}} and \hyperlink{PARbaq}{\formatrefE{4.4}{3}}}%
\def\seqCCCbbh{\AbbrevEqua \hyperlink{PARbaj}{\formatrefE{4.4}{2}}}%
\def\seqBBBbbi{\abbrevEqua \hyperlink{PARbba}{\formatrefE{4.4}{6}}}%
\def\seqBBBbbx{\abbrevList \hyperlink{PARayd}{\formatrefL{4.3}{2}}}%
\def\seqBBBbcb{\abbrevEqua \hyperlink{PARbat}{\formatrefE{4.4}{4}}}%
\def\seqBBBbcg{\abbrevEqua \hyperlink{PARbbt}{\formatrefE{4.4}{7}}}%
\def\seqDDDbcq{(\abbrevList \hyperlink{PARazp}{\formatrefL{4.4}{1}})}%
\def\seqDDDbcw{(\abbrevEqua \hyperlink{PARbax}{\formatrefE{4.4}{5}})}%
\def\seqBBBbdg{\abbrevProposition \hyperlink{PARbyy}{B.14}}%
\def\seqBBBber{\abbrevEqua \hyperlink{PARbee}{\formatrefE{4.6}{2}}}%
\def\seqBBBbfi{\abbrevEqua \hyperlink{PARbfe}{\formatrefE{4.6}{3}}}%
\def\seqBBBbfl{\abbrevEqua \hyperlink{PARbdk}{\formatrefE{4.6}{1}}}%
\def\seqCCCbfo{The proof of \abbrevProposition \hyperlink{PARbdh}{4.6}}%
\def\seqBBBbfp{\abbrevDef \hyperlink{PARaxy}{4.3}}%
\def\seqBBBbfu{\abbrevSectionC \hyperlink{SECaps}{3.3}}%
\def\seqBBBbfz{\bseqHHHabs{\abbrevSectionB A}}%
\def\seqBBBbgf{\abbrevSectionB \hyperlink{SECbrf}{B}}%
\def\seqBBBbgj{\abbrevProposition \hyperlink{PARaoh}{3.13}}%
\def\seqBBBbgr{\abbrevEqua \hyperlink{PARbgl}{\formatrefE{A.1}{1}}}%
\def\seqCCCbgr{\AbbrevEqua \hyperlink{PARbgl}{\formatrefE{A.1}{1}}}%
\def\seqDDDbgr{(\abbrevEqua \hyperlink{PARbgl}{\formatrefE{A.1}{1}})}%
\def\seqBBBbhc{\abbrevSectionD \hyperlink{SECbmy}{A.1.4}}%
\def\seqDDDbhc{(\abbrevSectionD \hyperlink{SECbmy}{A.1.4})}%
\def\seqBBBbhd{the proof of \abbrevProposition \hyperlink{PARbia}{A.6}}%
\def\seqDDDbhm{(\abbrevEqua \hyperlink{PARbgz}{\formatrefE{A.4}{1}})}%
\def\seqBBBbhp{\abbrevList \hyperlink{PARbhh}{\formatrefL{A.5}{1}}}%
\def\seqBBBbhx{\abbrevList \hyperlink{PARbhi}{\formatrefL{A.5}{2}}}%
\def\seqBBBbhz{the proof of \abbrevProposition \hyperlink{PARbsq}{B.5}}%
\def\seqDDDbji{(\abbrevLists \hyperlink{PARaca}{\formatrefL{2.1}{3}}, \hyperlink{PARacv}{\formatrefL{2.2}{4}} and \hyperlink{PARakx}{\formatrefL{3.1}{5}})}%
\def\seqBBBbjw{\abbrevList \hyperlink{PARbjz}{\formatrefL{A.10}{2}}}%
\def\seqBBBbkn{\abbrevLists \hyperlink{PARbjy}{\formatrefL{A.10}{1}} and \hyperlink{PARbjz}{\formatrefL{A.10}{2}}}%
\def\seqBBBbmc{\abbrevLemma \hyperlink{PARblo}{A.15}}%
\def\seqBBBbmd{\abbrevLists \hyperlink{PARbks}{\formatrefL{A.13}{1}} to \hyperlink{PARbku}{\formatrefL{A.13}{3}}}%
\def\seqBBBbmg{\abbrevLemma \hyperlink{PARbkv}{A.14}}%
\def\seqBBBbmt{\abbrevList \hyperlink{PARbku}{\formatrefL{A.13}{3}}}%
\def\seqBBBbmw{\abbrevList \hyperlink{PARbks}{\formatrefL{A.13}{1}}}%
\def\seqBBBbmx{\abbrevList \hyperlink{PARbkt}{\formatrefL{A.13}{2}}}%
\def\seqBBBbna{\abbrevSectionD \hyperlink{SECbyc}{B.1.4}}%
\def\seqDDDbna{(\abbrevSectionD \hyperlink{SECbyc}{B.1.4})}%
\def\seqBBBbnl{\abbrevPropositions \hyperlink{PARbbz}{4.5} and \hyperlink{PARbdh}{4.6}}%
\def\seqDDDboo{(\abbrevProposition \hyperlink{PARalr}{3.4})}%
\def\seqBBBbow{\abbrevDef \hyperlink{PARbjj}{A.8}}%
\def\seqBBBboy{\abbrevProposition \hyperlink{PARbhe}{A.5}}%
\def\seqDDDboy{(\abbrevProposition \hyperlink{PARbhe}{A.5})}%
\def\seqBBBboz{\abbrevLists \hyperlink{PARbhh}{\formatrefL{A.5}{1}} vs.\ \hyperlink{PARbpb}{\formatrefL{A.21}{1}}}%
\def\seqBBBbpt{\abbrevDefs \hyperlink{PARbjj}{A.8} and \hyperlink{PARbop}{A.20}}%
\def\seqBBBbpu{the proof of \abbrevProposition \hyperlink{PARbjl}{A.9}}%
\def\seqDDDbpu{(proof of \abbrevProposition \hyperlink{PARbjl}{A.9})}%
\def\seqDDDbqt{(\abbrevSectionD \hyperlink{SECcje}{B.2.4})}%
\def\seqBBBbso{the proof of \abbrevProposition \hyperlink{PARbyy}{B.14}}%
\def\seqBBBbsp{the proof of \abbrevProposition \hyperlink{PARbse}{B.4}}%
\def\seqBBBbtd{\abbrevProposition \hyperlink{PARbse}{B.4}}%
\def\seqBBBbty{\abbrevEqua \hyperlink{PARbiw}{\formatrefE{A.6}{2}}}%
\def\seqBBBbub{\abbrevEqua \hyperlink{PARbtn}{\formatrefE{B.5}{4}}}%
\def\seqBBBbug{\abbrevEqua \hyperlink{PARbue}{\formatrefE{B.5}{6}}}%
\def\seqBBBbuh{\abbrevEqua \hyperlink{PARbsu}{\formatrefE{B.5}{2}}}%
\def\seqBBBbuj{\abbrevEqua \hyperlink{PARbtn}{\formatrefE{B.5}{4}} and \abbrevProposition \hyperlink{PARbse}{B.4}}%
\def\seqBBBbuo{\abbrevEqua \hyperlink{PARbsw}{\formatrefE{B.5}{3}}}%
\def\seqBBBbuz{\abbrevList \hyperlink{PARbus}{\formatrefL{B.6}{1}}}%
\def\seqBBBbvc{\bseqHHHabu{proof of \abbrevProposition A.5}}%
\def\seqBBBbvl{\abbrevList \hyperlink{PARbuw}{\formatrefL{B.6}{2}}}%
\def\seqCCCbvn{The proof of \abbrevProposition \hyperlink{PARatu}{3.17}}%
\def\seqBBBbvo{\bseqHHHabu{\abbrevSectionB A}}%
\def\seqBBBbvu{\bseqHHHabu{\abbrevProposition A.10}}%
\def\seqBBBbvv{\bseqHHHabu{\abbrevPropositions A.5 and A.7}}%
\def\seqBBBbwg{the proof of \abbrevTheorem \hyperlink{PARamj}{3.6}}%
\def\seqBBBbxa{\abbrevEqua \hyperlink{PARbuc}{\formatrefE{B.5}{5}}}%
\def\seqBBBbxo{\abbrevProposition \hyperlink{PARbwj}{B.10}}%
\def\seqBBBbxz{\abbrevProposition \hyperlink{PARbvx}{B.8}}%
\def\seqBBBbyf{\abbrevEqua \hyperlink{PARbzb}{\formatrefE{B.14}{2}}}%
\def\seqBBBbym{\abbrevProposition \hyperlink{PARboa}{A.18}}%
\def\seqBBBbyq{\abbrevEqua \hyperlink{PARbyn}{\formatrefE{B.13}{1}}}%
\def\seqDDDbyq{(\abbrevEqua \hyperlink{PARbyn}{\formatrefE{B.13}{1}})}%
\def\seqBBBbyw{\abbrevProposition \hyperlink{PARbyl}{B.13}}%
\def\seqBBBbyx{\abbrevDef \hyperlink{PARbnb}{A.16} and \abbrevProposition \hyperlink{PARboa}{A.18}}%
\def\seqDDDbze{(\abbrevEqua \hyperlink{PARbwi}{\formatrefE{B.9}{1}})}%
\def\seqBBBbzf{the proof of \abbrevProposition \hyperlink{PARbyl}{B.13}}%
\def\seqBBBbzk{\abbrevEqua \hyperlink{PARbzg}{\formatrefE{B.14}{3}}}%
\def\seqBBBcap{\abbrevEquas \hyperlink{PARbyz}{\formatrefE{B.14}{1}} and \hyperlink{PARbwk}{\formatrefE{B.10}{1}}}%
\def\seqBBBcar{\abbrevEqua \hyperlink{PARcaj}{\formatrefE{B.15}{2}}}%
\def\seqBBBcca{\abbrevEqua \hyperlink{PARcbo}{\formatrefE{B.18}{1}}}%
\def\seqDDDccf{\bseqHHHabu{proof of \abbrevProposition A.10}}%
\def\seqBBBccr{\abbrevEquas \hyperlink{PARbsw}{\formatrefE{B.5}{3}} and \hyperlink{PARbwk}{\formatrefE{B.10}{1}}}%
\def\seqBBBcdd{\abbrevDef \hyperlink{PARbye}{B.12}}%
\def\seqBBBcdg{\abbrevEqua \hyperlink{PARcax}{\formatrefE{B.16}{1}}}%
\def\seqBBBcdj{\abbrevEqua \hyperlink{PARccy}{\formatrefE{B.16}{2}}}%
\def\seqBBBcdq{\abbrevLemma \hyperlink{PARcbn}{B.18}}%
\def\seqBBBcdv{\abbrevLemma \hyperlink{PARcaz}{B.17}}%
\def\seqBBBcec{\abbrevLemma \hyperlink{PARbql}{A.27}}%
\def\seqBBBcfa{\abbrevDef \hyperlink{PARbrk}{B.1}}%
\def\seqBBBcfj{\abbrevProposition \hyperlink{PARbpa}{A.21}}%
\def\seqDDDcfj{(\abbrevProposition \hyperlink{PARbpa}{A.21})}%
\def\seqBBBcfk{\abbrevDef \hyperlink{PARbrz}{B.3} and \abbrevProposition \hyperlink{PARbsq}{B.5}}%
\def\seqBBBcfq{\abbrevPropositions \hyperlink{PARbhe}{A.5} and \hyperlink{PARbpa}{A.21}}%
\def\seqBBBcfv{\abbrevList \hyperlink{PARbpb}{\formatrefL{A.21}{1}}}%
\def\seqBBBcgn{the proof of \abbrevProposition \hyperlink{PARbpf}{A.22}}%
\def\seqBBBcgo{\abbrevEqua \hyperlink{PARcge}{\formatrefE{B.22}{1}}}%
\def\seqBBBcgq{\abbrevEqua \hyperlink{PARcgg}{\formatrefE{B.22}{2}}}%
\def\seqDDDchj{(\abbrevEqua \hyperlink{PARchp}{\formatrefE{B.24}{1}})}%
\def\seqBBBchk{\abbrevDef \hyperlink{PARank}{3.9} and \abbrevProposition \hyperlink{PARanm}{3.10}}%
\def\seqBBBchl{\abbrevDef \hyperlink{PARbop}{A.20} and \abbrevProposition \hyperlink{PARbpf}{A.22}}%
\def\seqBBBchm{\abbrevLemma \hyperlink{PARcid}{B.27}}%
\def\seqBBBchn{\abbrevPropositions \hyperlink{PARchq}{B.25} and \hyperlink{PARchz}{B.26}}%
\def\seqBBBchy{\abbrevEqua \hyperlink{PARbwr}{\formatrefE{B.10}{3}}}%
\def\seqBBBcii{\abbrevEqua \hyperlink{PARcig}{\formatrefE{B.27}{1}}}%
\def\seqBBBcik{the proof of \abbrevProposition \hyperlink{PARbxb}{B.11}}%
\def\seqBBBcir{\abbrevDefs \hyperlink{PARcfl}{B.20} vs.\ \hyperlink{PARbrz}{B.3}}%
\def\seqBBBckd{\abbrevProposition \hyperlink{PARcaf}{B.15}}%
\def\seqBBBcke{\abbrevEquas \hyperlink{PARchr}{\formatrefE{B.25}{1}}, \hyperlink{PARcgg}{\formatrefE{B.22}{2}} and \hyperlink{PARcfs}{\formatrefE{B.21}{1}}}%
\def\seqBBBckf{\abbrevEquas \hyperlink{PARbwk}{\formatrefE{B.10}{1}}, \hyperlink{PARbsw}{\formatrefE{B.5}{3}} and \hyperlink{PARbsr}{\formatrefE{B.5}{1}}}%
\def\MsectionBiblio#1{\MsectionA{ckh}{C}{#1}}%
\def\figBBOXaco{0.0pt 0.0pt 512.14154pt 227.61847pt}%
\def\figBBOXaeo{0.0pt 0.0pt 312.97539pt 227.61847pt}%
\def\figBBOXafm{0.0pt 0.0pt 284.52307pt 227.61847pt}%
\def\figBBOXahk{0.0pt 0.0pt 284.52307pt 125.19016pt}%
\def\figBBOXaib{0.0pt 0.0pt 512.14154pt 227.61847pt}%
\def\figBBOXaie{0.0pt 0.0pt 512.14154pt 227.61847pt}%
\def\figBBOXakd{0.0pt 0.0pt 512.14154pt 284.52307pt}%
\def\figBBOXakl{6.06876pt 1.5648pt 495.44847pt 143.82634pt}%
\def\figBBOXaqd{0.0pt 0.0pt 500.76064pt 256.07077pt}%
\def\figBBOXawy{0.0pt 0.0pt 415.4037pt 227.61847pt}%
\def\bseqJJJaaa{\cite{abz}}%
\def\bseqHHHaaa#1{\cite[#1]{abz}}%
\def\bseqJJJaab{\cite{aba}}%
\def\bseqHHHaab#1{\cite[#1]{aba}}%
\def\bseqJJJaac{\cite{aaq,*aap,*aao}}%
\def\bseqJJJaad{\cite{aac}}%
\def\bseqHHHaad#1{\cite[#1]{aac}}%
\def\bseqJJJaae{\cite{aag}}%
\def\bseqHHHaae#1{\cite[#1]{aag}}%
\def\bseqHHHaaf#1{\cite[#1]{aan}}%
\def\bseqJJJaag{\cite{aas}}%
\def\bseqHHHaag#1{\cite[#1]{aas}}%
\def\bseqJJJaah{\cite{abi}}%
\def\bseqHHHaai#1{\cite[#1]{abg}}%
\def\bseqJJJaaj{\cite{abv}}%
\def\bseqJJJaak{\cite{aau}}%
\def\bseqJJJaal{\cite{abo}}%
\def\bseqJJJaam{\cite{abh}}%
\def\bseqHHHaam#1{\cite[#1]{abh}}%
\def\bseqHHHaan#1{\cite[#1]{aaa}}%
\def\bseqHHHaao#1{\cite[#1]{abc}}%
\def\bseqHHHaap#1{\cite[#1]{abn}}%
\def\bseqHHHaaq#1{\cite[#1]{aaj}}%
\def\bseqJJJaar{\cite{aba,aau,abh,aad}}%
\def\bseqJJJaas{\cite{aad}}%
\def\bseqHHHaas#1{\cite[#1]{aad}}%
\def\bseqJJJaat{\cite{aby}}%
\def\bseqJJJaau{\cite{aay}}%
\def\bseqJJJaav{\cite{abx}}%
\def\bseqJJJaaw{\cite{abe,*aaw,*aai,*abl}}%
\def\bseqHHHaax#1{\cite[#1]{abr}}%
\def\bseqHHHaay#1{\cite[#1]{aat}}%
\def\bseqHHHaaz#1{\cite[#1]{aar}}%
\def\bseqHHHaba#1{\cite[#1]{abm}}%
\def\bseqHHHabb#1{\cite[#1]{abu}}%
\def\bseqJJJabc{\cite{aam}}%
\def\bseqJJJabd{\cite{aca,*aak}}%
\def\bseqJJJabe{\cite{abe,*aaw,*aai,*abl}}%
\def\bseqJJJabf{\cite{aaz}}%
\def\bseqHHHabg#1{\cite[#1]{abk}}%
\def\bseqHHHabh#1{\cite[#1]{abf}}%
\def\bseqHHHabi#1{\cite[#1]{abb}}%
\def\bseqHHHabj#1{\cite[#1]{abd}}%
\def\bseqHHHabk#1{\cite[#1]{aal}}%
\def\bseqHHHabl#1{\cite[#1]{aav}}%
\def\bseqHHHabm#1{\cite[#1]{aab}}%
\def\bseqJJJabn{\cite{abq,*abt,*abs,*abw}}%
\def\bseqJJJabo{\cite{abq,*abt,*abs,*abw}}%
\def\bseqHHHabp#1{\cite[#1]{abj}}%
\def\bseqHHHabq#1{\cite[#1]{aah}}%
\def\bseqHHHabr#1{\cite[#1]{aae}}%
\def\bseqHHHabs#1{\cite[#1]{aaf}}%
\def\bseqHHHabt#1{\cite[#1]{aax}}%
\def\bseqHHHabu#1{\cite[#1]{abp}}%
\begin{document}%
\MprefixNNNaaa%
\AuthorsAffil%
\begin{abstract}\Mpar \vspace{\Saut}\hspace{\Alinea}It is well-known that there exist infinitely-many \emph{inequivalent} representations of the canonical (anti)-commutation relations of Quantum Field Theory (QFT).
A way out, suggested by Algebraic QFT {\bseqJJJaaa}, is to instead \emph{define} the quantum theory as encompassing \emph{all} possible (abstract) states.
In the present paper, we describe a quantization scheme for general \emph{linear} (aka.~free) field theories that can be seen as intermediate between traditional Fock quantization and full Algebraic QFT, in the sense that:%
\Mpar \MPhyList it provides a \emph{constructive}, explicit description of the resulting space of quantum states;%
\MStopList \Mpar \MPhyList it does \emph{not} require the choice of a polarization, aka.~the splitting of classical solutions into positive vs.~negative-frequency modes: in fact, any Fock representation corresponding to a "reasonable" choice of polarization is naturally \emph{embedded};%
\MStopList \Mpar \MPhyList it supports the implementation of a "large enough" class of \emph{linear} symplectomorphisms of the classical, infinite-dimensional phase space.%
\MStopList \Mpar %
\Mpar \vspace{\Saut}\hspace{\Alinea}This approach relies on three main ingredients:%
\Mpar \MPhyList Kijowski's projective formalism for QFT {\bseqJJJaab}, which constructs quantum states of the full theory as projective families of \emph{partial} states on finite-dimensional \emph{truncations} (a short standalone introduction to this formalism is included);%
\MStopList \Mpar \MPhyList the Stone–von Neumann theorem {\bseqJJJaac} stating the equivalence of representations in the \emph{finite-dimensional} (aka.~quantum mechanical) case, which the projective formalism allows to \emph{lift} to the infinite dimensional case;%
\MStopList \Mpar \MPhyList a prescription {\bseqJJJaad} to select \emph{countable} collections of truncations in a way that does not compromise the universality of the resulting quantum state space: in the present context, admissible collections are shown to be in one-to-one correspondence with \emph{dense}, countably-generated vector subspaces of the (dual of the) classical phase space.
\MStopList \Mpar \vspace{\Saut}\hspace{\Alinea}The proposed quantization (like Algebraic QFT) is notably meant for use on curved spacetimes, where the lack of a preferred choice of polarization makes the introduction of a Fock representation problematic.
Accordingly, we illustrate it in the simple case of a \emph{free} Klein–Gordon field on a spatially-compact, cosmological spacetime.
Specifically, we exhibit a quantum state space that supports \emph{arbitrarily good} approximations of the time evolution.
For comparison, we examine how well the same time evolution could be approximated by unitary transformations on a suitable Fock representation: in the case of a flat spacial geometry (on a torus), the minimal error is proved to be \emph{bounded below}.

\end{abstract}%
\tableofcontents%
\Mnomdefichier{lin01}%
\hypertarget{SECaal}{}\MsectionA{aal}{1}{Introduction}%
\hypertarget{SECaam}{}\MsectionB{aam}{1.1}{Geometric Quantization and the Choice of a Polarization}%
\vspace{\PsectionB}\Mpar \MstartPhyMode\hspace{\Alinea}Geometric quantization {\bseqJJJaae} is meant to provide a systematic procedure to turn any (finite-dimensional) classical phase space (aka.~symplectic manifold, \bseqHHHaae{section 1.1}) into a quantum theory, namely an Hilbert space carrying a representation of a suitable algebra of classical \emph{elementary} observables.
The latter are functions on the phase space, with commutation relations determined by their Poisson brackets.%
\Mpar \vspace{\Saut}\hspace{\Alinea}The first step of this procedure, called \emph{pre-quantization} \bseqHHHaae{chap.~8}, yields a Hilbert space on which the \emph{full} classical Poisson algebra (consisting of \emph{all} functions on the classical phase space) can be represented \emph{exactly}.
To get a physically admissible \emph{quantum} theory, it is however necessary to impose a \emph{polarization} \bseqHHHaae{chap.~9}, which selects a \emph{subspace} out of the very large pre-quantum Hilbert space. At the same time, it restricts the algebra of observables to a sub-algebra of \emph{elementary} observables, which are the ones that can be directly quantized (for example linear observables, assuming the classical phase space from which we started carries a linear structure).
Further observables are then quantized as \emph{composite} operators in terms of the elementary ones (modulo a choice of operator ordering), and their commutators receive quantum corrections of order {$\hbar$}.%
\Mpar \vspace{\Saut}\hspace{\Alinea}In the present article, we will mostly be concerned with so-called \emph{complex} polarizations.\footnote{\emph{Real} polarizations, such as the one underlying the position representation of quantum mechanics, can be seen as a \emph{limit case} \bseqHHHaae{section 9.3}, however this limit is not well-defined for an infinite-dimensional theory: the equivalent of Fock spaces does not exist for real polarizations (due to the lack of Lebesgue measures on infinite dimensional spaces). On the other hand, projective state spaces can be constructed without problems in real polarizations {\bseqJJJaab}, and, in fact, the state space we will construct not only contains all holomorphic polarizations at same time, but it is easy to see that it even encompasses all real ones (see the note \ref{RealPolas} before {\seqBBBaao}).}
Geometric quantization along a complex polarization is also known as \emph{holomorphic} quantization \bseqHHHaae{section 9.2}, the prime example of which is the quantization of the harmonic oscillator.
Considering a one-dimensional harmonic oscillator, its two-dimensional phase space can be parametrized by a complex variable \MMath{z \mathrel{\mathop:}=  \alpha \,  q + i\, \beta \,  p} (with \MMath{q,p} the position and impulsion, respectively). This complex variable, resp.~its complex conjugate, is then quantized as the annihilation, resp.~creation, operator acting on a (single mode) Fock space, which provides the Hilbert space of the quantum theory.
Coherent states \MMath{\left|z\right\rangle } picked at a certain value of the variable \MMath{z} can be formed by modeling them on the vacuum state (which is a coherent state picked at \MMath{z=0}) and, for any quantum state \MMath{\left| \psi  \right\rangle }, the function \MMath{z \mapsto  \left\langle z\middlewithspace|\psi \right\rangle } is holomorphic (up to a Gaussian measure factor), hence the name of this quantization scheme (although, from the point of view of geometric quantization sketched above, one would arrive at it differently, starting from a pre-quantum Hilbert space which would consists of arbitrary functions on the phase space, and recovering the usual Fock space by imposing holomorphic dependence in the complex variable \MMath{z}).%
\Mpar \vspace{\Saut}\hspace{\Alinea}The natural choice for the constants \MMath{\alpha ,\beta } entering the complex parametrization of the phase space is \MMath{\alpha  \mathrel{\mathop:}=  \sqrt{{\nicefrac{m\omega }{2\hbar }}} \& \beta  \mathrel{\mathop:}=  \sqrt{{\nicefrac{1}{2m\hbar \omega }}}}. This choice \emph{adjusts} the polarization to the \emph{dynamics}: it ensures that the time evolution preserves the particle number (and, in particular, the vacuum state), so that the particle states are precisely the energy eigenstates.
In the case of \emph{finitely} many degrees of freedom (\dofs), such a fine-tuning of the polarization to the dynamics is however not strictly necessary: the Stone–von Neumann theorem {\bseqJJJaac} ensures that the representations obtained using different \emph{linear} complex polarizations are \emph{unitarily equivalent}.
In other words, these representations describe the \emph{same} quantum theory: they yield identical \emph{physical predictions} (provided, of course, that the Hamiltonian is properly quantized in the particular representation used: the annihilation and creation operators represent different quantized variables in different polarizations, so the expression of the Hamiltonian in term of those operators will also have be different, if it is to keep representing the same classical function).
Alternatively, this unitary equivalence relating different linear complex polarizations can be stated as the existence of a (projective, see {\seqBBBaap}, as well as \bseqHHHaaf{section 2.7 and appendix 2.B}) unitary representation of the group of Poisson-brackets-preserving linear transformations of the classical phase space (aka.~symplectomorphisms), since those are the transformations that relate different polarizations at the classical level. This so-called metaplectic representation (see \bseqHHHaae{chap.~10} and/or {\seqBBBaaq}) will play an important role in the construction we are going to develop.%
\Mpar \vspace{\Saut}\hspace{\Alinea}Note that the Stone–von Neumann theorem also guarantees the unitary equivalence of the finite-dimensional\footnote{By \emph{finite-dimensional} Fock representation, we mean a representation built on finitely many \dofs, although the Hilbert space on which the representation acts may be infinite-dimensional (it \emph{is} in the bosonic case).} Fock and Schrödinger representations (in the terminology of geometric quantization, the latter arises from a \emph{real} linear polarization).
By contrast, even when only dealing with finitely many \dofs, polarizations that select \emph{different} sub-algebras of elementary observables (eg.~linear vs.~non-linear polarizations) yields \emph{physically distinct} quantum theories: in fact, the notion of equivalence of representations would not even make sense in this case, since we would be trying to compare representations of different algebras.%
\MleavePhyMode \hypertarget{SECaar}{}\MsectionB{aar}{1.2}{Polarizations in Quantum Field Theory}%
\vspace{\PsectionB}\Mpar \MstartPhyMode\hspace{\Alinea}Unfortunately, the Stone–von Neumann theorem breaks down in the case of theories with \emph{infinitely} many \dofs, aka.~\emph{field theories}.
While Fock spaces can still be constructed for such theories, different linear complex polarizations often (see eg.~{\bseqJJJaag}) give rise to physically \emph{inequivalent} quantum field theories (QFTs).
One may then wonder what is the "right" choice of polarization.%
\Mpar \vspace{\Saut}\hspace{\Alinea}If we consider the Schrödinger equation of quantum mechanics as a classical field theory (aka.~\emph{first-quantized} theory), the "classical" phase space for this theory (ie.~the space of solutions of the field equations) is just the Hilbert space of the quantum mechanical theory (see eg.~{\bseqJJJaah} and/or {\seqBBBaat}).
As a \emph{complex} Hilbert space, it is naturally equipped with a complex polarization, and the corresponding (\emph{second-quantized}) Fock space is precisely the Fock space describing an arbitrary number of particles obeying the first-quantized theory.%
\Mpar \vspace{\Saut}\hspace{\Alinea}However, if we go over to \emph{relativistic} field theories, we no longer get the particle structure of the QFT for free. For example, in the case of a scalar field obeying the Klein–Gordon equation, the phase space is parametrized by the field and its time derivative as conjugate variables (since the Klein–Gordon equation is of \emph{2nd order} in \MMath{\partial _{t}}).
Even if the scalar field happens to be complex, this does \emph{not} provide us with a usable complex polarization. This is because, in order to build a Fock representation, the complex structure must \emph{combine} with the Poisson bracket structure into the inner product of a complex Hilbert space (see {\seqBBBaau}), while if we use the naive complex structure on the phase space of the complex scalar field, the resulting inner product (known as Klein–Gordon inner product, see eg.~\bseqHHHaai{eq.~1.3}) fails to be \emph{positive-definite}.%
\Mpar \vspace{\Saut}\hspace{\Alinea}On the other hand, when working on a Minkowski background, there exists a canonical splitting of the phase space into positive- vs.~negative-frequency modes, and, by \emph{reversing} the complex structure over the negative-frequency modes, one gets an admissible complex polarization\footnote{More generally, a preferred complex polarization can always be deduced from a \emph{static}, linear time-evolution operator (as illustrated for example in {\seqBBBaav}). In the case of relativistic field theory, the time evolution operator is schematically given by \MMath{\partial _{t} = \pm i\, \sqrt{{\vec{p}^{2} + m^{2}}}} and the \MMath{\pm i} pre-factor appearing in this expression is precisely the desired complex structure.}. The resulting Fock space then consists of 2 particle species (aka.~a particle and its anti-particle), in agreement with the phase space (which gets identified with the 1-particle Hilbert space) having twice the dimension of the field space (since, as mentioned above, the Klein–Gordon equation is of 2nd order).%
\Mpar \vspace{\Saut}\hspace{\Alinea}By proceeding along these lines to quantize relativistic field theories, the symmetries of the Minkowski background enter \emph{critically} the construction of the quantum state space of QFTs.
Unsurprisingly, this approach breaks down in the case of QFT on \emph{curved} spacetime, where there is in general no preferred polarization, ie.~no preferred particle structure.
Worse, if we simply elect some arbitrary Fock space for the theory, the time evolution may actually kick the quantum states \emph{out} of this space (ie.~time translations may not be implementable as \emph{unitary} transformations on the chosen Fock space): heuristically, the spacetime curvature tends to create particles, which may void the normalizability of the quantum state we started from (see below).%
\Mpar \vspace{\Saut}\hspace{\Alinea}A similar issue affects \emph{interacting} field theories: even on flat spacetime, the Fock representation used to describe the free theory will no longer support the time evolution if we add interactions (this statement can be made precise as the Haag's \emph{no-go theorem}, {\bseqJJJaaj}).
In other words, and in contrast to the quantum mechanics of finite-dimensional systems, the quantization of the (kinematical) classical phase space needs to be \emph{fine-tuned} to the \emph{dynamics} of the theory, requiring a precise understanding of the latter (which we typically don't have...).
In the present article, we will focus on the quantization of free (aka.~linear) field theories: indeed, we will mainly be concerned with the extension to the infinite-dimensional case of the Stone–von Neumann theorem, which, as mentioned at the end of {\seqBBBaaw}, does not make away with the linear structure of the classical phase space. Nevertheless, we will briefly comment on perspectives for interacting theories in {\seqBBBaax}.%
\MleavePhyMode \hypertarget{SECaay}{}\MsectionB{aay}{1.3}{Universal Quantum State Spaces}%
\vspace{\PsectionB}\Mpar \MstartPhyMode\hspace{\Alinea}To understand the \emph{polarization dependence} of the infinite-dimensional Fock representation, mentioned at the beginning of the previous subsection, it is useful to realize the special role played by the \emph{vacuum state} in the definition of this representation: the entire Fock space is spanned by states describing only \emph{discrete} excitations on top of the vacuum.
In systems with finitely many \dofs, any change of (linear) polarization can be implemented as a unitary transformation (as discussed in {\seqBBBaaw}), but this transformation does not preserve the number of particles, and, in particular, does \emph{not} preserve the vacuum state (see {\seqBBBaba}).
In the case of infinitely many \dofs, a change of polarization may thus map a Fock state to state excited over infinitely many modes, which is no longer \emph{normalizable} in the original Fock space.
In other words, the issue is that a given Fock representation only explores a small neighborhood around the chosen vacuum state: the so-called \emph{vacuum sector}.%
\Mpar \vspace{\Saut}\hspace{\Alinea}The state space would be more \emph{robust} if it would extend beyond the boundaries of the individual Fock representations, bundling together the various vacuum sectors, and we would start with better chances to support arbitrary field dynamics.
A way to achieve this is provided by Algebraic QFT: instead of restricting states to be vectors in a particular Hilbert space, one \emph{defines} the quantum state space as encompassing \emph{all} possible quantum states.
Note that quantum states can be discussed independently of any supporting Hilbert space, since a state can be completely specified by the expectation values it associates to all products of elementary observables (aka.~by its moments, or \MMath{n}-point functions).
However, such abstract states are hard to construct in practice, because the various expectations values cannot simply be chosen independently: the state need to obey non-trivial \emph{positivity conditions} to ensure a valid probability interpretation for all possible measurements.%
\Mpar \vspace{\Saut}\hspace{\Alinea}In the present article, we want to propose an approach, that, while preserving most of the \emph{universality} of the algebraic approach, restores the \emph{computational} convenience of the traditional Fock quantization. Indeed, it yields a quantum state space that spans a large class of vacuum sectors, and still admits an explicit, constructive parametrization.
This is achieved with the help of a formalism that was introduced by Jerzy Kijowski {\bseqJJJaab} and further developed by Andrzej Okołów {\bseqJJJaak}. The idea is to describe quantum states as projective families of \emph{partial} density matrices (aka.~mixed states), which capture the properties of the state over \emph{finitely} many \dofs. The partial theories, extracting finite selections of \dofs from the continuum field theory, will be realized on finite-dimensional Fock representations, which will allow us to \emph{leverage} the Stone–von Neumann theorem, while \emph{coarse-graining} projections will be given by \emph{partial traces} over suitable tensor product factorizations.%
\Mpar \vspace{\Saut}\hspace{\Alinea}The physical justification for this approach comes from an operational perspective, where one focuses on information that can be both \emph{experimentally measured} and \emph{algorithmically computed}, at an arbitrary precision (at least in principle).
Then, one realizes that, while a field theory may hold infinitely many \dofs, any \emph{given} experiment only measures finitely many observables: hence, when computing predictions for its outcome, it is sufficient to work in a partial theory, extracting just the \dofs we need.%
\Mpar \vspace{\Saut}\hspace{\Alinea}The article is organized as follows:%
\Mpar \MPhyList We start in {\seqBBBabc} by reviewing Kijowski's projective formalism. The formulation we will be using differs in two respects from the original one.
The first difference is that we will equip the collection of partial theories with the structure of a (small) category. While the resulting formalism is shown in {\seqBBBabd} to be equivalent to the original one (where one relies instead on a structure of partially ordered set), it will allow us to avoid introducing spurious input into the quantization procedure on which the final quantum state space does not actually depend. This will make the construction presented in {\seqBBBabe} \emph{manifestly} universal.
Second, we will present a prescription to extract from the collection of partial theories a \emph{countably-generated} sub-collection: this is necessary to ensure that the projective limit state space can be described \emph{constructively}, which, as explained above, is one of our main goals.
This prescription can be understood as restricting the observable algebra to a \emph{"dense"} sub-algebra, and, crucially, it can be performed \emph{without} voiding the universality of the quantum theory {\seqDDDabf}.%
\MStopList \Mpar \MPhyList {\seqCCCabe} forms the core of the article. We set up projective quantum state spaces for \emph{general} infinite-dimensional \emph{linear} phase spaces {\seqDDDabh}.
Admissible sub-collections of partial theories according to the just mentioned restriction prescription are shown to be in 1-to-1 correspondence with dense, countably-generated, linear subspaces in the phase space {\seqDDDabi}. In this context, the universality result from {\seqBBBabf} is expressed as the existence of an \emph{arbitrarily small} phase space automorphism mapping any two such dense subspaces into each other, which ensures that the choice to work on a particular dense subspace has no physically measurable consequences. Moreover, the linear phase space transformations which preserve a given dense subspace form a \emph{dense sub-group} in the group of all phase space automorphisms {\seqDDDabj}, so that eg.~time evolution can be implemented at an arbitrary precision on the restricted quantum theory.
Finally, there is a natural embedding of the space of density matrices over an infinite-dimensional Fock space into the thus constructed projective state space {\seqDDDaao}.
All results in this section are derived both for the \emph{bosonic} case (where the classical phase space is a symplectic vector space, see {\seqBBBabk}), as well as for the \emph{fermionic} case (where the phase space is an Euclidean space, see {\seqBBBabl}).%
\MStopList \Mpar \MPhyList For illustrative purposes, and as a confirmation that the input structure assumed in {\seqBBBabe} is indeed physically reasonable, we work out a concrete example in {\seqBBBabn}, namely the description of a free scalar field (obeying the Klein–Gordon equation) on an arbitrary spatially-compact cosmological spacetime. The applicability of the framework from {\seqBBBabe} is established in {\seqBBBabo}, and the resulting quantum theory is shown to improve on Fock quantization, as the latter in general would not support a good enough approximation of the time evolution {\seqDDDabp}.%
\MStopList \Mpar \MPhyList Two extensive appendices have been added, covering the mathematical tools to study classical and quantum \emph{linear} field theory (both in the bosonic and fermionic cases). This includes fairly standard material, with the aim of fixing the notations and conventions, as well as proofs of various technical results used in the main text.
In particular, we present a new proof of Shale's characterization {\bseqJJJaag} of those symplectomorphisms that are unitarily implementable on an infinite-dimensional bosonic Fock representation {\seqDDDabr}.
\MStopList \MleavePhyMode %
\Mnomdefichier{lin10}%
\hypertarget{SECabt}{}\MsectionA{abt}{2}{Projective Quantum State Spaces}%
\hypertarget{SECabu}{}\MsectionB{abu}{2.1}{General Formalism}%
\vspace{\PsectionB}\Mpar \MstartPhyMode\hspace{\Alinea}We consider a classical (field) theory, characterized by its (infinite-dimensional) phase space, which we would like to quantize (see {\seqBBBabw} for more details what this phase space concretely may be).%
\Mpar \vspace{\Saut}\hspace{\Alinea}The first input we need is a collection of \emph{partial theories}, labeled by selections of \emph{finitely many} \dofs.
Each such selection (which we will call a \emph{label}) is meant as the arena to describe some concrete experiment, measuring finitely many observables.
Next, we need to ensure \emph{consistency} of the partial descriptions of a given system: namely, any partial theory \emph{detailed enough} to support the observables we are measuring should lead to the \emph{same physical predictions} for our experiment.
To achieve this, we need \emph{coarse-graining} prescriptions, specifying how to \emph{extract} from a large collection of \dofs (aka.~a \emph{finer} label) a smaller sub-collection (aka.~a \emph{coarser} label).
Such prescriptions will be labeled by \emph{arrows} going from finer to coarser labels.
Thus, the collection of partial theories gets naturally equipped with the structure of a (small) category.%
\MleavePhyMode \hypertarget{PARabx}{}\Mpar \Mdefinition{2.1}A (small) category {$\mathcal{L}$} consists of {\bseqJJJaal}:%
\Mpar \MList{1}a label set \MMath{\mathcal{L} _{}} (aka.~objects);%
\MStopList \Mpar \MList{2}for any \MMath{\lambda _{1}, \lambda _{2} \in  \mathcal{L} _{}}, a set of arrows \MMath{\mathcal{L} _{{\lambda _{2} \rightarrow  \lambda _{1}}}} (aka.~morphisms);%
\MStopList \hypertarget{PARaca}{}\Mpar \MList{3}for any \MMath{\lambda _{1}, \lambda _{2}, \lambda _{3} \in  \mathcal{L} _{}}, a composition operation \MMath{\cdot : \mathcal{L} _{{\lambda _{2} \rightarrow  \lambda _{1}}} \times  \mathcal{L} _{{\lambda _{3} \rightarrow  \lambda _{2}}} \rightarrow  \mathcal{L} _{{\lambda _{3} \rightarrow  \lambda _{1}}}};%
\MStopList \Mpar such that:%
\Mpar \MList{4}\italique{(identity)} for any \MMath{\lambda  \in  \mathcal{L} _{}} there exists an identity arrow \MMath{1 _{\lambda } \in  \mathcal{L} _{{\lambda  \rightarrow \lambda }}} such that, for any \MMath{\lambda ' \in  \mathcal{L} _{}}, \MMath{\forall  \mu  \in  \mathcal{L} _{{\lambda ' \rightarrow  \lambda }}, 1 _{\lambda }  \cdot  \mu  = \mu } and \MMath{\forall  \mu  \in  \mathcal{L} _{{\lambda  \rightarrow  \lambda '}}, \mu  \cdot  1 _{\lambda }  = \mu } (this implies that \MMath{1 _{\lambda }} is unique);%
\MStopList \Mpar \MList{5}\italique{(associativity)} for any \MMath{\lambda _{1}, \lambda _{2}, \lambda _{3}, \lambda _{4} \in  \mathcal{L} _{}} and any \MMath{\mu _{12} \in  \mathcal{L} _{{\lambda  _{2} \rightarrow  \lambda _{1}}}, \mu _{23} \in  \mathcal{L} _{{\lambda  _{3} \rightarrow  \lambda _{2}}}, \mu _{34} \in  \mathcal{L} _{{\lambda  _{4} \rightarrow  \lambda _{3}}}}, \MMath{\mu _{12} \cdot  (\mu _{23} \cdot  \mu _{34}) = (\mu _{12} \cdot  \mu _{23}) \cdot  \mu _{34}}.%
\MStopList \Mpar \vspace{\Sssaut}\hspace{\Alinea}We define a binary relation \MMath{\leqslant } on \MMath{\mathcal{L} _{}} by:%
\Mpar \MStartEqua \MMath{\forall  \lambda _{1}, \lambda _{2} \in  \mathcal{L} _{}, {\lambda _{1} \leqslant  \lambda _{2} \Leftrightarrow  \mathcal{L} _{{\lambda _{2} \rightarrow  \lambda _{1}}} \neq  \varnothing }}.%
\MStopEqua \Mpar \MMath{\leqslant } is a pre-order, ie.\ a reflexive and transitive binary relation (the reflexivity follows from the identity property above, and the transitivity from the existence of the composition operation).%
\Mpar \MstartPhyMode\hspace{\Alinea}A quantization scheme yielding a projective quantum state space will associate to each label {$\lambda$} a \emph{Hilbert space} \MMath{\mathcal{H} _{\lambda }} (understood as the quantization of the partial theory labeled by {$\lambda$}), and to each arrow \MMath{\mu : \lambda _{2} \rightarrow  \lambda _{1}} a \emph{tensor product decomposition} of \MMath{\mathcal{H} _{{\lambda _{2}}}} as \MMath{\mathcal{H} _{{\lambda _{1}}} \otimes  \mathcal{H} _{\mu }} (which identifies the \dofs retained by \MMath{\mathcal{H} _{{\lambda _{1}}}} within \MMath{\mathcal{H} _{{\lambda _{2}}}}).%
\Mpar \vspace{\Saut}\hspace{\Alinea}To understand where such a structure would come from, it is useful to consider its classical precursor.
On the classical side, we can represent the infinite-dimensional phase space of the continuum theory as (a dense subset in) a \emph{projective limit} of finite-dimensional phase spaces \MMath{\left( \mathcal{M}_{\lambda } \right)_{\lambda }} {\seqDDDaci}. Each of these "small" phase spaces accommodates a partial theory, and the coarse-graining relations between them are expressed as \emph{projections} going from finer (higher dimensional) phase spaces into coarser (lower dimensional) ones.
These projections allow to partially forget information about a physical system (described by a point in the phase space), and, in a dual way, to lift observables (described by functions on the phase space): this is how the coarser algebra of observables can be identified as a \emph{sub-algebra} of the finer one.
As this lifting of observables has to respect the Poisson-brackets {\seqDDDacj}, one can show {\seqDDDack} that it selects a preferred Cartesian product factorization\footnote{More precisely, such a factorization always exists locally. There might be obstructions for it to hold globally; see the discussion preceding {\seqBBBack}.} of the finer phase space \MMath{\mathcal{M}_{{\lambda _{2}}}} as \MMath{\mathcal{M}_{{\lambda _{1}}} \times  \mathcal{M}_{\mu }}, with \MMath{\mathcal{M}_{\mu }} holding the \dofs discarded by the coarse-graining (ie.~those \dofs that Poisson-commute with the ones in \MMath{\mathcal{M}_{{\lambda _{1}}}}).%
\Mpar \vspace{\Saut}\hspace{\Alinea}These Cartesian product decompositions of the small phase spaces \MMath{\mathcal{M}_{\lambda }} are then naturally quantized as tensor product decompositions of the Hilbert spaces \MMath{\mathcal{H} _{\lambda }}.
The consistency conditions satisfied by the latter reflect the ones satisfied by their classical counterparts.
The first such condition, represented on the left part of {\seqBBBacl}, determines the \emph{composition} of arrows: it states that, instead of directly trimming a very fine label \MMath{\lambda _{3}} to a very coarse one \MMath{\lambda _{1}}, it is always possible to perform the coarse-graining in two steps, through an intermediate label \MMath{\lambda _{2}}.
The second condition (right part of {\seqBBBacl}) encapsulates the \emph{unambiguity} of all coarse-graining relations: while there may be \emph{multiple} distinct arrows going from a label \MMath{\lambda _{2}} to a label \MMath{\lambda _{1}}, the tensor product factorizations they define should be \emph{equivalent}, ie.~should prescribe identical embeddings of the \MMath{\lambda _{1}}-\dofs within \MMath{\mathcal{H} _{{\lambda _{2}}}}.
The corresponding condition at the classical level tells us that the Cartesian product factorization \MMath{\mathcal{M}_{{\lambda _{2}}} \approx  \mathcal{M}_{{\lambda _{1}}} \times  \mathcal{M}_{\mu }} derived from a Poisson-bracket-preserving projection \MMath{\mathcal{M}_{{\lambda _{2}}} \rightarrow  \mathcal{M}_{{\lambda _{1}}}} is \emph{unique} \emph{up to} reparametrization of the complementary \MMath{\mathcal{M}_{\mu }}: this is precisely the uniqueness part of {\seqBBBack}.%
\Mpar \vspace{\Saut}\hspace{\Alinea}Such multiple, redundant arrows are useful when there exists many equivalent ways of writing down the coarse-graining relations, and no reason to prefer one over the others (indeed, we will take advantage of this possibility in {\seqBBBacm}).
In the same spirit, it is worth noting that the formalism also allows for \emph{redundant} labels: distinct labels \MMath{\lambda ,\lambda '} with arrows in both directions between them (so that both \MMath{\lambda  \leqslant  \lambda '} and \MMath{\lambda ' \leqslant  \lambda } hold: that's why {$\leqslant$} is only a \emph{pre}-order). Thus, we can use the labels to encode more that just particular selections of \dofs: for example, we can decorate the partial theories with some extra structure (such as preferred coordinates, or a choice of polarization, allowing us to make the \MMath{\mathcal{H} _{\lambda }} completely explicit), while at the same time retaining universality (since all possible choices of the extra structure can \emph{coexist} in the label set, as long as suitable arrows can be defined to relate them; again, see {\seqBBBacm}).%
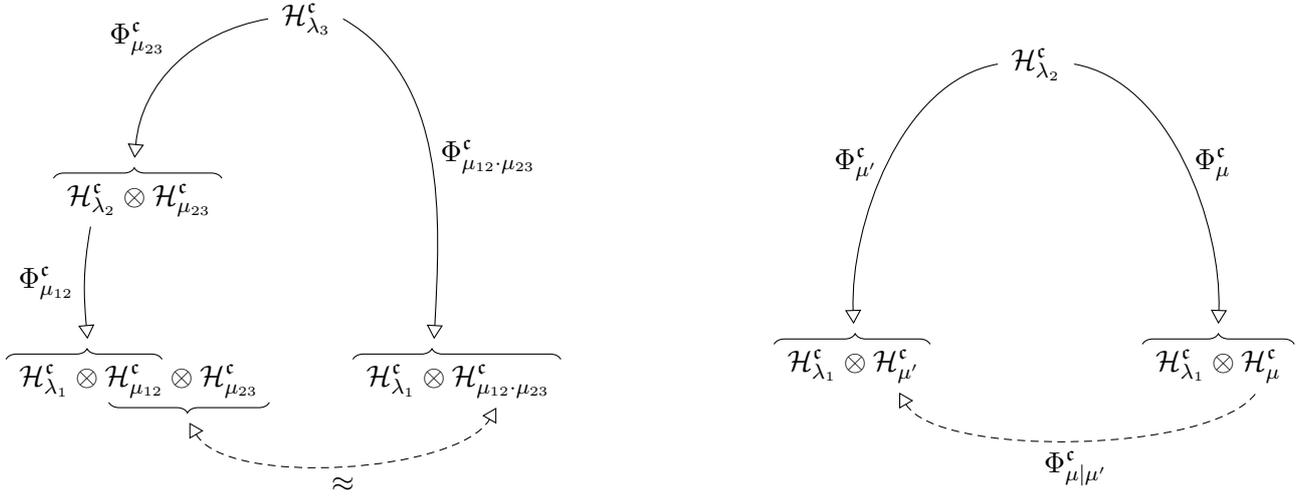
\Myfigure{2.1}{\hypertarget{PARacn}{}}{Consistency conditions in a system of factorized Hilbert spaces: composition (left) and unambiguity (right) of arrows}{{\InputFigPaco}}%
\MleavePhyMode \hypertarget{PARacp}{}\Mpar \Mdefinition{2.2}A system of factorized Hilbert spaces {$\mathfrak{c}$} consists of (see also {\seqBBBacq} and \bseqHHHaam{def.~2.4}):%
\Mpar \MList{1}a (small) category \MMath{\mathcal{L} ^{\mathfrak{c} }};%
\MStopList \Mpar \MList{2}for any \MMath{\lambda  \in  \mathcal{L} ^{\mathfrak{c} }_{}}, a complex Hilbert space \MMath{\mathcal{H} ^{\mathfrak{c} }_{\lambda }} (with \MMath{\mathcal{H} ^{\mathfrak{c} }_{\lambda } \neq \left\{0\right\}});%
\MStopList \hypertarget{PARact}{}\Mpar \MList{3}for any \MMath{\lambda _{1}, \lambda _{2} \in  \mathcal{L} ^{\mathfrak{c} }_{}} and any \MMath{\mu  \in  \mathcal{L} ^{\mathfrak{c} }_{{\lambda _{2} \rightarrow  \lambda _{1}}}}, a complex Hilbert space \MMath{\mathcal{H} ^{\mathfrak{c} }_{\mu }} and a Hilbert space isomorphism \MMath{\Phi ^{\mathfrak{c} }_{\mu } : \mathcal{H} ^{\mathfrak{c} }_{{\lambda _{2}}} \rightarrow  \mathcal{H} ^{\mathfrak{c} }_{{\lambda _{1}}} \otimes  \mathcal{H} ^{\mathfrak{c} }_{\mu }};%
\MStopList \Mpar such that:%
\hypertarget{PARacv}{}\Mpar \MList{4}\italique{(composition)} for any \MMath{\lambda _{1}, \lambda _{2}, \lambda _{3} \in  \mathcal{L} ^{\mathfrak{c} }_{}} and any \MMath{\mu _{12} \in  \mathcal{L} ^{\mathfrak{c} }_{{\lambda _{2} \rightarrow  \lambda _{1}}}, \mu _{23} \in  \mathcal{L} ^{\mathfrak{c} }_{{\lambda _{3} \rightarrow  \lambda _{2}}}}, \MMath{\mathcal{H} ^{\mathfrak{c} }_{{\mu _{12} \cdot  \mu _{23}}} \approx  \mathcal{H} ^{\mathfrak{c} }_{{\mu _{12}}} \otimes  \mathcal{H} ^{\mathfrak{c} }_{{\mu _{23}}}} and \MMath{\Phi ^{\mathfrak{c} }_{{\mu _{12} \cdot  \mu _{23}}} \approx  \left( \Phi ^{\mathfrak{c} }_{{\mu _{12}}} \otimes  \mathds{1} ^{\mathfrak{c} }_{{\mu _{23}}} \right) \circ  \Phi ^{\mathfrak{c} }_{{\mu _{23}}}} (where {$\approx$} denotes, respectively, natural isomorphic identification and equality up to this identification, and \MMath{\mathds{1} ^{\mathfrak{c} }_{{\mu _{23}}}} denotes the identity operator on \MMath{\mathcal{H} ^{\mathfrak{c} }_{{\mu _{23}}}});%
\MStopList \hypertarget{PARacw}{}\Mpar \MList{5}\italique{(unambiguity)} for any \MMath{\lambda _{1}, \lambda _{2} \in  \mathcal{L} ^{\mathfrak{c} }_{}} and any \MMath{\mu ,\mu ' \in  \mathcal{L} ^{\mathfrak{c} }_{{\lambda _{2} \rightarrow  \lambda _{1}}}}, there exists a Hilbert space isomorphism \MMath{\Phi ^{\mathfrak{c} }_{{\mu  |  \mu '}}: \mathcal{H} ^{\mathfrak{c} }_{\mu } \rightarrow  \mathcal{H} ^{\mathfrak{c} }_{\mu '}} such that \MMath{\Phi ^{\mathfrak{c} }_{\mu '} = \left( \mathds{1} ^{\mathfrak{c} }_{{\lambda _{1}}} \otimes  \Phi ^{\mathfrak{c} }_{{\mu  |  \mu '}} \right) \circ  \Phi ^{\mathfrak{c} }_{\mu }} (this implies that \MMath{\Phi ^{\mathfrak{c} }_{{\mu  |  \mu '}}} is unique).%
\MStopList \Mpar \vspace{\Sssaut}\hspace{\Alinea}For any \MMath{\lambda  \in  \mathcal{L} ^{\mathfrak{c} }_{}}, we will denote by \MMath{\mathcal{S} ^{\mathfrak{c} }_{\lambda }} the ordered vector space of self-adjoint trace-class operators on \MMath{\mathcal{H} ^{\mathfrak{c} }_{\lambda }} (the order structure being determined by the subset of non-negative operators in \MMath{\mathcal{S} ^{\mathfrak{c} }_{\lambda }}, which forms a proper convex cone; see \bseqHHHaan{section II.2.5}) and by \MMath{\mathcal{B} ^{\mathfrak{c} }_{\lambda }} the \MMath{C^{*}}-algebra of bounded operators on \MMath{\mathcal{H} ^{\mathfrak{c} }_{\lambda }} \bseqHHHaaa{section III.2}.%
\Mpar \MstartPhyMode\hspace{\Alinea}Representing \emph{partial} information about a quantum state as a density matrix on \MMath{\mathcal{H} _{\lambda }}, the Hilbert space factorizations from {\seqBBBacz} provide a natural way to selectively forget about \emph{some} of this information, by \emph{tracing over} the complementary tensor product factor \MMath{\mathcal{H} _{\mu }}.
Importantly, the thus-defined partial trace provides a \emph{well-defined} coarse-graining mapping going from the space of density matrices over \MMath{\mathcal{H} _{{\lambda _{2}}}} to the space of density matrices over \MMath{\mathcal{H} _{{\lambda _{1}}}}: indeed, the unambiguity property ({\seqBBBada} and right part of {\seqBBBacl}) is precisely what we need to ensure that different arrows between the same two labels \MMath{\lambda _{2} \rightarrow  \lambda _{1}} all define the \emph{same} partial trace operation.%
\Mpar \vspace{\Saut}\hspace{\Alinea}Moreover, the composition property ({\seqBBBadb} and left part of {\seqBBBacl}) is \emph{inherited} by the partial traces, so that the complete quantum state space can simply be defined as a projective limit (aka.~inverse limit): a particular quantum state can be specified by a \emph{consistent} family of partial states, where consistent means that the partial states on coarser labels match what they should be as deduced from partial states on finer labels.
In order for the overall consistency of a state to be fully enforced simply by relations between \emph{pairs} of comparable labels, it is however necessary that the label set \MMath{\mathcal{L} ^{\mathfrak{c} }_{}, \leqslant }  be \emph{directed}: ie.~for any two labels \MMath{\lambda _{1},\lambda _{2}}, there should exist a label \MMath{\lambda _{3} \geqslant  \lambda _{1},\lambda _{2}}. Then, the appropriate consistency between the partial states over \MMath{\lambda _{1}} and \MMath{\lambda _{2}} will be imposed \emph{indirectly}, through the existence a \emph{common} refining state over \MMath{\lambda _{3}}.%
\Mpar \vspace{\Saut}\hspace{\Alinea}Note that it is \emph{mandatory} to work with \emph{density matrices} (aka.~"mixed" states) rather than vectors in the Hilbert space (aka.~"pure" states) because the partial trace is the right tool to forget about some \dofs in a quantum theory, and it can only be defined at the level of density matrices (see {\seqBBBadc} for further discussion as to why working with density matrices may anyway be physically preferable in the context of quantum \emph{field} theories ).%
\MleavePhyMode \hypertarget{PARadd}{}\Mpar \Mproposition{2.3}Let \MMath{\mathfrak{c} } be a system of factorized Hilbert spaces. For any \MMath{\lambda _{1},\lambda _{2} \in  \mathcal{L} ^{\mathfrak{c} }_{}} and any \MMath{\mu  \in  \mathcal{L} ^{\mathfrak{c} }_{{\lambda _{2} \rightarrow  \lambda _{1}}}}, we define:%
\Mpar \MStartEqua \MMath{\definitionFonction{\mkop{Tr} _{\mu }}{\mathcal{S} ^{\mathfrak{c} }_{{\lambda _{2}}}}{\mathcal{S} ^{\mathfrak{c} }_{{\lambda _{1}}}}{\rho }{\mkop{Tr} _{{\otimes \mu }} \left( \Phi ^{\mathfrak{c} }_{\mu } \,  \rho  \,  \Phi ^{{\mathfrak{c} ,-1}}_{\mu } \right)}},%
\MStopEqua \Mpar where \MMath{\mkop{Tr} _{{\otimes \mu }}} denotes the partial trace \bseqHHHaao{section 2.3} over the second tensor product factor in \MMath{\mathcal{H} ^{\mathfrak{c} }_{{\lambda _{1}}} \otimes  \mathcal{H} ^{\mathfrak{c} }_{\mu }}.%
\Mpar \vspace{\Sssaut}\hspace{\Alinea}Let \MMath{\lambda _{1},\lambda _{2},\lambda _{3} \in  \mathcal{L} ^{\mathfrak{c} }_{}}. We have:%
\hypertarget{PARadh}{}\Mpar \MList{1}for any \MMath{\mu  \in  \mathcal{L} ^{\mathfrak{c} }_{{\lambda _{2} \rightarrow  \lambda _{1}}}}, \MMath{\mkop{Tr} _{\mu }} is surjective (aka.\ onto), linear and order-preserving;%
\MStopList \hypertarget{PARadi}{}\Mpar \MList{2}for any \MMath{\mu ,\mu ' \in  \mathcal{L} ^{\mathfrak{c} }_{{\lambda _{2} \rightarrow  \lambda _{1}}}}, \MMath{\mkop{Tr} _{\mu } = \mkop{Tr} _{\mu '}};%
\MStopList \hypertarget{PARadj}{}\Mpar \MList{3}for any \MMath{\mu _{12} \in  \mathcal{L} ^{\mathfrak{c} }_{{\lambda _{2} \rightarrow  \lambda _{1}}}, \mu _{23} \in  \mathcal{L} ^{\mathfrak{c} }_{{\lambda _{3} \rightarrow  \lambda _{2}}}}, \MMath{\mkop{Tr} _{{\mu _{12} \cdot  \mu _{23}}} = \mkop{Tr} _{{\mu _{12}}} \circ  \mkop{Tr} _{{\mu _{23}}}}.%
\MStopList \Mpar \Mproof Respectively, {\seqBBBadl} follows from the properties of the partial trace, {\seqBBBadm} from \abbrevDef{\seqBBBada}, and {\seqBBBadn} from \abbrevDef{\seqBBBadb}.%
\MendOfProof \hypertarget{PARado}{}\Mpar \Mdefinition{2.4}Let \MMath{\mathfrak{c} } be a system of factorized Hilbert spaces. For any \MMath{\lambda _{1} \leqslant  \lambda _{2} \in  \mathcal{L} ^{\mathfrak{c} }_{}}, we define \MMath{\mkop{Tr} _{{\lambda _{2} \rightarrow  \lambda _{1}}} \mathrel{\mathop:}=  \mkop{Tr} _{\mu }} for some \MMath{\mu  \in  \mathcal{L} ^{\mathfrak{c} }_{{\lambda _{2} \rightarrow  \lambda _{1}}}}.%
\Mpar \vspace{\Sssaut}\hspace{\Alinea}If \MMath{\mathcal{L} ^{\mathfrak{c} }_{}, \leqslant } is \emph{directed} (ie.\ \MMath{\forall  \lambda _{1}, \lambda _{2} \in  \mathcal{L} ^{\mathfrak{c} }_{}, \exists  \lambda _{3} \in  \mathcal{L} ^{\mathfrak{c} }_{} \mathrel{\big/}  \lambda _{1}, \lambda _{2} \leqslant  \lambda _{3}}), we define \MMath{\mathcal{S} ^{\mathfrak{c} }} as the projective (aka.\ inverse) limit of the projective system \MMath{\left( \left(\mathcal{S} ^{\mathfrak{c} }_{\lambda }\right)_{{\lambda  \in  \mathcal{L} ^{\mathfrak{c} }_{}}}, \left(\mkop{Tr} _{{\lambda _{2} \rightarrow  \lambda _{1}}}\right)_{{\lambda _{1} \leqslant  \lambda _{2}}} \right)}.%
\hypertarget{PARadp}{}\Mpar \Mproposition{2.5}\MMath{\mathcal{S} ^{\mathfrak{c} }} is an ordered vector space with positive cone \MMath{\mathcal{S} ^{\mathfrak{c} }_{+} \mathrel{\mathop:}=  \left\{ \left( \rho _{\lambda } \right)_{{\lambda  \in  \mathcal{L} ^{\mathfrak{c} }_{}}} \in  \mathcal{S} ^{\mathfrak{c} } \middlewithspace| \forall  \lambda  \in  \mathcal{L} ^{\mathfrak{c} }_{}, \rho _{\lambda }  \geqslant  0 \right\}}.%
\Mpar \Mproof The infinite Cartesian product \MMath{{\textstyle \prod_{{\lambda  \in  \mathcal{L} ^{\mathfrak{c} }_{}}}} \mathcal{S} ^{\mathfrak{c} }_{\lambda }} equipped with componentwise operations and product order is an ordered vector space. Since the projections are linear (\abbrevProposition{\seqBBBadl}), \MMath{\mathcal{S} ^{\mathfrak{c} }} is a vector subspace of \MMath{{\textstyle \prod_{{\lambda  \in  \mathcal{L} ^{\mathfrak{c} }_{}}}} \mathcal{S} ^{\mathfrak{c} }_{\lambda }}, hence an ordered vector space.%
\MendOfProof \Mpar \MstartPhyMode\hspace{\Alinea}The complete algebra of quantum observables is defined in a dual way, and, accordingly, builds an \emph{inductive} limit (aka.~direct limit): observables are \emph{lifted} from coarse theories into finer ones (see \bseqHHHaab{section 6} as well as \bseqHHHaap{section 1}).
Thanks to the properties of the partial trace, the lifting of observables is compatible with the coarse-graining of states, in the sense that the evaluation of expectation values is suitably \emph{intertwined}.%
\Mpar \vspace{\Saut}\hspace{\Alinea}For any projective state \MMath{\rho  = \left( \rho _{\lambda } \right)_{{\lambda }}}, the evaluation \MMath{A \mapsto  \mkop{Tr}  (\rho \,  A )} defines a linear functional on the \MMath{C^{*}}-algebra of quantum observables and it satisfies the positivity constraints needed to ensure a well-defined probabilistic interpretation of quantum measurements (namely, for any observable \MMath{A}, \MMath{\mkop{Tr}  (\rho \,  A^{\dag} \, A ) \geqslant  0}): in other words, it defines a \emph{state} in the \emph{algebraic} sense \bseqHHHaaa{part III, def.~2.2.8}.
Thus, the projective state space is naturally embedded in the space of \emph{all} possible quantum states considered in algebraic QFT.%
\MleavePhyMode \hypertarget{PARads}{}\Mpar \Mproposition{2.6}Let \MMath{\mathfrak{c} } be a system of factorized Hilbert spaces. For any \MMath{\lambda _{1},\lambda _{2} \in  \mathcal{L} ^{\mathfrak{c} }_{}} and any \MMath{\mu  \in  \mathcal{L} ^{\mathfrak{c} }_{{\lambda _{2} \rightarrow  \lambda _{1}}}}, we define:%
\Mpar \MStartEqua \MMath{\definitionFonction{\iota _{\mu }}{\mathcal{B} ^{\mathfrak{c} }_{{\lambda _{1}}}}{\mathcal{B} ^{\mathfrak{c} }_{{\lambda _{2}}}}{A}{\Phi ^{{\mathfrak{c} ,-1}}_{\mu } \,  \left( A \otimes  \mathds{1} ^{\mathfrak{c} }_{\mu } \right) \,  \Phi ^{\mathfrak{c} }_{\mu }}}.%
\MStopEqua \Mpar \vspace{\Sssaut}\hspace{\Alinea}Let \MMath{\lambda _{1},\lambda _{2},\lambda _{3} \in  \mathcal{L} ^{\mathfrak{c} }_{}}. We have:%
\hypertarget{PARadv}{}\Mpar \MList{1}for any \MMath{\mu  \in  \mathcal{L} ^{\mathfrak{c} }_{{\lambda _{2} \rightarrow  \lambda _{1}}}}, \MMath{\iota _{\mu }} is an injective (aka.\ one-to-one), isometric morphism of \MMath{C^{*}}-algebras;%
\MStopList \hypertarget{PARadw}{}\Mpar \MList{2}for any \MMath{\mu  \in  \mathcal{L} ^{\mathfrak{c} }_{{\lambda _{2} \rightarrow  \lambda _{1}}}}, any \MMath{\rho  \in  \mathcal{S} ^{\mathfrak{c} }_{{\lambda _{2}}}} and any \MMath{A \in  \mathcal{B} ^{\mathfrak{c} }_{{\lambda _{1}}}}, \MMath{\mkop{Tr}  \big( (\mkop{Tr} _{\mu } \rho ) A \big) = \mkop{Tr}  \big( \rho  (\iota _{\mu } A) \big)};%
\MStopList \hypertarget{PARadx}{}\Mpar \MList{3}for any \MMath{\mu ,\mu ' \in  \mathcal{L} ^{\mathfrak{c} }_{{\lambda _{2} \rightarrow  \lambda _{1}}}}, \MMath{\iota _{\mu } = \iota _{\mu '}};%
\MStopList \hypertarget{PARady}{}\Mpar \MList{4}for any \MMath{\mu _{12} \in  \mathcal{L} ^{\mathfrak{c} }_{{\lambda _{2} \rightarrow  \lambda _{1}}}, \mu _{23} \in  \mathcal{L} ^{\mathfrak{c} }_{{\lambda _{3} \rightarrow  \lambda _{2}}}}, \MMath{\iota _{{\mu _{12} \cdot  \mu _{23}}} = \iota _{{\mu _{23}}} \circ  \iota _{{\mu _{12}}}}.%
\MStopList \Mpar \Mproof Respectively, {\seqBBBaea} follows from the properties of the tensor product of operators, {\seqBBBaeb} from the definition of \MMath{\mkop{Tr} _{\mu }} {\seqDDDaec} and the properties of the partial trace, {\seqBBBaed} from \abbrevDef{\seqBBBada}, and {\seqBBBaee} from \abbrevDef{\seqBBBadb}.%
\MendOfProof \hypertarget{PARaef}{}\Mpar \Mdefinition{2.7}Let \MMath{\mathfrak{c} } be a system of factorized Hilbert spaces. For any \MMath{\lambda _{1} \leqslant  \lambda _{2} \in  \mathcal{L} ^{\mathfrak{c} }_{}}, we define \MMath{\iota _{{\lambda _{2} \leftarrow  \lambda _{1}}} \mathrel{\mathop:}=  \iota _{\mu }} for some \MMath{\mu  \in  \mathcal{L} ^{\mathfrak{c} }_{{\lambda _{2} \rightarrow  \lambda _{1}}}}.%
\Mpar \vspace{\Sssaut}\hspace{\Alinea}If \MMath{\mathcal{L} ^{\mathfrak{c} }_{}, \leqslant } is \emph{directed}, we define \MMath{\mathcal{B} ^{\mathfrak{c} }} as the inductive (aka.\ direct) limit of the inductive system \MMath{\left( \left(\mathcal{B} ^{\mathfrak{c} }_{\lambda }\right)_{{\lambda  \in  \mathcal{L} ^{\mathfrak{c} }_{}}}, \left(\iota _{{\lambda _{2} \leftarrow  \lambda _{1}}}\right)_{{\lambda _{1} \leqslant  \lambda _{2}}} \right)}. We define \MMath{\overline{\mathcal{B} ^{\mathfrak{c} }}} as the completion of \MMath{\mathcal{B} ^{\mathfrak{c} }} with respect to the operator norm. \MMath{\overline{\mathcal{B} ^{\mathfrak{c} }}} is a \MMath{C^{*}}-algebra, with a unit that we will denote by \MMath{\mathds{1} ^{\mathfrak{c} }}.
Thanks to \abbrevProposition{\seqBBBaeb}, \MMath{\mkop{Tr} } is well-defined as a bilinear map \MMath{\mkop{Tr}  \big( (\, \cdot \, ) (\, \cdot \, ) \big) : \mathcal{S} ^{\mathfrak{c} } \times  \mathcal{B} ^{\mathfrak{c} } \rightarrow  \mathds{C}}, which can be extended (by continuity in the second argument, \bseqHHHaaq{theorem VI.26}) as a map \MMath{\mathcal{S} ^{\mathfrak{c} } \times  \overline{\mathcal{B} ^{\mathfrak{c} }} \rightarrow  \mathds{C}}. We define the quantum state space on {$\mathfrak{c}$} as:%
\Mpar \MStartEqua \MMath{\mathcal{S} ^{\mathfrak{c} }_{{+,1 }} \mathrel{\mathop:}=  \left\{ \rho  \in  \mathcal{S} ^{\mathfrak{c} } \middlewithspace| \rho  \geqslant  0 \& \mkop{Tr}  \rho  \mathrel{\mathop:}=  \mkop{Tr}  \rho  \,  \mathds{1} ^{\mathfrak{c} } = 1 \right\}}.%
\MStopEqua \Mpar \MstartPhyMode\hspace{\Alinea}The original formulation of projective quantum state spaces simply used a structure of directed \emph{(pre-)ordered set} on the collection of partial theories {\bseqJJJaar}.
In other words, the tensor product factorization connecting the partial Hilbert spaces \MMath{\mathcal{H} _{{\lambda _{2}}}} and \MMath{\mathcal{H} _{{\lambda _{1}}}} over two labels \MMath{\lambda _{2} \geqslant  \lambda _{1}} was supposed to be completely \emph{fixed} once the involved labels were known, excluding the option of "redundant" arrows\footnote{Redundant labels have been also excluded in all formulations except the one from {\bseqJJJaas}, ie.~the order on \MMath{\mathcal{L} ^{\mathfrak{c} }_{}} has been required to be a partial order, rather than a pre-order (the latter drops the \emph{antisymmetry} property satisfied by the former).} which we discussed in the comments before {\seqBBBaei}.
The compatibility between the various coarse-grainings was then formulated as a single "3-spaces" consistency condition, involving 3 increasingly fine labels \MMath{\lambda _{1} \leqslant  \lambda _{2} \leqslant  \lambda _{3}} {\seqDDDaej}.%
\Mpar \vspace{\Saut}\hspace{\Alinea}The formulation used in the present article reduces to the original one by \emph{choosing} a \emph{preferred} arrow\footnote{And, defining on \MMath{\mathcal{L} ^{\mathfrak{c} }_{}} the equivalence relation \MMath{\lambda  \sim \lambda ' \, \Leftrightarrow \,  (\lambda  \leqslant  \lambda ' \& \lambda ' \leqslant  \lambda )}, the pre-order {$\leqslant$} on \MMath{\mathcal{L} ^{\mathfrak{c} }_{}} can be turned into a partial order on the quotient \MMath{\mathcal{L} ^{\mathfrak{c} }_{} / \sim}, see {\seqBBBaek}.} for every pair of labels \MMath{\lambda _{1} \leqslant  \lambda _{2}}.
The 3-spaces consistency condition is then recovered by combining the composition and unambiguity properties of arrows {\seqDDDael} as shown on {\seqBBBaem}.%
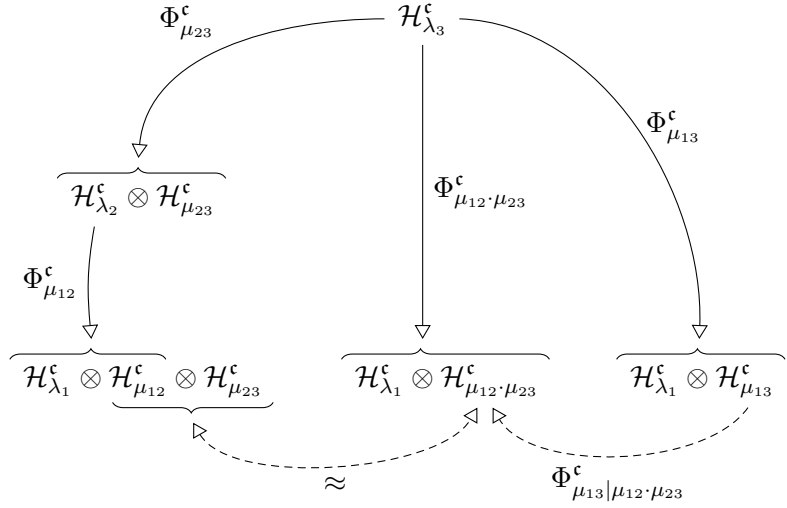
\Myfigure{2.2}{\hypertarget{PARaen}{}}{Recovering the "3-spaces" consistency condition of the original formulation from the composition and unambiguity properties of arrows}{{\InputFigPaeo}}%
\Mpar \vspace{\Saut}\hspace{\Alinea}Since both formulations turn out to be equivalent, enriching the label set with a category structure may at first seems a gratuitous complication.
As mentioned above, the motivation for allowing redundant arrows (and labels) comes from the pursuit of a \emph{manifestly} universal quantization (in particular in view of the construction that we will present in {\seqBBBacm}): at the end of the day, the quantum state space does \emph{not} depend on the assignment of preferred arrows (provided, of course, that the unambiguity property from {\seqBBBada} holds), hence we may as well dispense from choosing them in the first place.
A further advantage of the categorical reformulation is conceptual: by splitting the original 3-space consistency condition into two separate conditions, its physical meaning, and in particular its role in the well-definiteness of coarse-grainings, can be made more transparent.%
\MleavePhyMode \MleavePhyMode \hypertarget{PARaeq}{}\Mpar \Mproposition{2.8}Let {$\mathfrak{c}$} be a system of factorized Hilbert spaces and assume that \MMath{\mathcal{L} ^{\mathfrak{c} }_{}, \leqslant } is \emph{directed}. For any \MMath{\lambda _{1} \leqslant  \lambda _{2} \in  \mathcal{L} ^{\mathfrak{c} }_{}}, choose \MMath{\mu _{{\lambda _{2} \rightarrow  \lambda _{1}}} \in  \mathcal{L} ^{\mathfrak{c} }_{{\lambda _{2} \rightarrow  \lambda _{1}}}} (choosing \MMath{\mu _{{\lambda _{2} \rightarrow  \lambda _{1}}} = 1 _{{\lambda _{1}}}} if \MMath{\lambda _{1} = \lambda _{2}}), and define:%
\Mpar \MList{1}for any \MMath{\lambda  \in  \mathcal{L} ^{\mathfrak{c} }_{}}, \MMath{\mathcal{H} _{\lambda } \mathrel{\mathop:}=  \mathcal{H} ^{\mathfrak{c} }_{\lambda }};%
\MStopList \Mpar \MList{2}for any \MMath{\lambda _{1} \leqslant  \lambda _{2} \in  \mathcal{L} ^{\mathfrak{c} }_{}}, \MMath{\mathcal{H} _{{\lambda _{2} \rightarrow  \lambda _{1}}} \mathrel{\mathop:}=  \mathcal{H} ^{\mathfrak{c} }_{{\mu _{{\lambda _{2} \rightarrow  \lambda _{1}}}}} \& \Phi _{{\lambda _{2} \rightarrow  \lambda _{1}}} \mathrel{\mathop:}=  \text{Swap}_{{\lambda _{2} \rightarrow  \lambda _{1}}} \circ  \Phi ^{\mathfrak{c} }_{{\mu _{{\lambda _{2} \rightarrow  \lambda _{1}}}}}}, where \MMath{\text{Swap}_{{\lambda _{2} \rightarrow  \lambda _{1}}}} is the isomorphism \MMath{\mathcal{H} _{{\lambda _{1}}} \otimes  \mathcal{H} _{{\lambda _{2} \rightarrow  \lambda _{1}}} \rightarrow  \mathcal{H} _{{\lambda _{2} \rightarrow  \lambda _{1}}} \otimes  \mathcal{H} _{{\lambda _{1}}}} swapping the tensor product factors\footnote{This is necessary due to opposed conventions in {\seqBBBaet}.};%
\MStopList \Mpar \MList{3}for any \MMath{\lambda _{1} \leqslant  \lambda _{2} \leqslant  \lambda _{3}}, \MMath{\Phi _{{\lambda _{3} \rightarrow  \lambda _{2} \rightarrow  \lambda _{1}}} \mathrel{\mathop:}=  \text{Swap}_{{\lambda _{3} \rightarrow  \lambda _{2} \rightarrow  \lambda _{1}}} \circ  \Phi ^{\mathfrak{c} }_{{\mu  |  \mu '}}} where \MMath{\mu  \mathrel{\mathop:}=  \mu _{{\lambda _{3} \rightarrow  \lambda _{1}}}}, \MMath{\mu ' \mathrel{\mathop:}=  \mu _{{\lambda _{2} \rightarrow  \lambda _{1}}} \cdot  \mu _{{\lambda _{3} \rightarrow  \lambda _{2}}}}, and \MMath{\text{Swap}_{{\lambda _{3} \rightarrow  \lambda _{2} \rightarrow  \lambda _{1}}}} is the isomorphism \MMath{\mathcal{H} _{{\lambda _{2} \rightarrow  \lambda _{1}}} \otimes  \mathcal{H} _{{\lambda _{3} \rightarrow  \lambda _{2}}} \rightarrow  \mathcal{H} _{{\lambda _{3} \rightarrow  \lambda _{2}}} \otimes  \mathcal{H} _{{\lambda _{2} \rightarrow  \lambda _{1}}}} swapping the tensor product factors.%
\MStopList \Mpar Then, \MMath{\left( \mathcal{L} ^{\mathfrak{c} }_{},\,  \left(\mathcal{H} _{\lambda }\right)_{{\lambda  \in  \mathcal{L} ^{\mathfrak{c} }}},\,  \left(\mathcal{H} _{{\lambda _{2} \rightarrow  \lambda _{1}}}\right)_{{\lambda _{1} \leqslant  \lambda _{2}}},\,  \left(\Phi _{{\lambda _{2} \rightarrow  \lambda _{1}}}\right)_{{\lambda _{1} \leqslant  \lambda _{2}}},\,  \left(\Phi _{{\lambda _{3} \rightarrow  \lambda _{2} \rightarrow  \lambda _{1}}}\right)_{{\lambda _{1} \leqslant  \lambda _{2} \leqslant  \lambda _{3}}} \right)} fulfills {\seqBBBacq}. Moreover, for any \MMath{\lambda _{1} \leqslant  \lambda _{2} \in  \mathcal{L} ^{\mathfrak{c} }}, the definition of \MMath{\mkop{Tr} _{{\lambda _{2} \rightarrow  \lambda _{1}}}}, resp.~\MMath{\iota _{{\lambda _{2} \leftarrow  \lambda _{1}}}}, given in {\seqBBBaew}, resp.~in {\seqBBBaex}, coincides with the one given in {\seqBBBaey}, resp.~{\seqBBBaez}. Thus, the definitions for the state space and for the algebra of operators given in {\seqBBBafa} coincide with the ones from {\seqBBBafb}.%
\Mpar \Mproof Let \MMath{\lambda _{1} \leqslant  \lambda _{2} \in  \mathcal{L} ^{\mathfrak{c} }} and let \MMath{\mu  \mathrel{\mathop:}=  \mu _{{\lambda _{2} \rightarrow  \lambda _{1}}}}. By definition of \MMath{\Phi ^{\mathfrak{c} }_{\mu }} {\seqDDDacz}, \MMath{\Phi _{{\lambda _{2} \rightarrow  \lambda _{1}}}} is a Hilbert space isomorphism \MMath{\mathcal{H} _{{\lambda _{2}}} \rightarrow  \mathcal{H} _{{\lambda _{2} \rightarrow  \lambda _{1}}} \otimes  \mathcal{H} _{{\lambda _{1}}}}. Moreover, if \MMath{\lambda _{1} = \lambda _{2} = \lambda }, \MMath{\mu  =  1 _{\lambda }} by assumption, and, applying {\seqBBBadb} to \MMath{1 _{\lambda } = 1 _{\lambda } \cdot  1 _{\lambda }}, we get \MMath{\mathcal{H} ^{\mathfrak{c} }_{{1 _{\lambda }}} \approx  \mathcal{H} ^{\mathfrak{c} }_{{1 _{\lambda }}} \otimes  \mathcal{H} ^{\mathfrak{c} }_{{1 _{\lambda }}}} (in a natural identification), hence \MMath{\mathcal{H} ^{\mathfrak{c} }_{{1 _{\lambda }}} \approx  \mathds{C}}, as well as \MMath{\Phi ^{\mathfrak{c} }_{{1 _{\lambda }}} \approx  \left( \Phi ^{\mathfrak{c} }_{{1 _{\lambda }}} \otimes  \mathds{1} ^{\mathfrak{c} }_{{1 _{\lambda }}} \right) \circ  \Phi ^{\mathfrak{c} }_{{1 _{\lambda }}}}, hence \MMath{\Phi ^{\mathfrak{c} }_{{1 _{\lambda }}} \approx  \mathds{1} ^{\mathfrak{c} }_{\lambda } : \mathcal{H} _{\lambda } \rightarrow  \mathcal{H} _{\lambda } \otimes  \mathds{C}}.%
\Mpar \vspace{\Sssaut}\hspace{\Alinea}Let \MMath{\lambda _{1} \leqslant  \lambda _{2} \leqslant  \lambda _{3} \in  \mathcal{L} ^{\mathfrak{c} }} and let \MMath{\mu  \mathrel{\mathop:}=  \mu _{{\lambda _{3} \rightarrow  \lambda _{1}}}}, \MMath{\mu ' \mathrel{\mathop:}=  \mu _{{\lambda _{2} \rightarrow  \lambda _{1}}} \cdot  \mu _{{\lambda _{3} \rightarrow  \lambda _{2}}}}. \MMath{\Phi ^{\mathfrak{c} }_{{\mu  |  \mu '}}} is an isomorphism \MMath{\mathcal{H} ^{\mathfrak{c} }_{{\mu _{{\lambda _{3} \rightarrow  \lambda _{1}}}}} \rightarrow  \mathcal{H} ^{\mathfrak{c} }_{\mu '} \approx  \mathcal{H} ^{\mathfrak{c} }_{{\mu _{{\lambda _{2} \rightarrow  \lambda _{1}}}}} \otimes  \mathcal{H} ^{\mathfrak{c} }_{{\mu _{{\mu _{{\lambda _{3} \rightarrow  \lambda _{2}}}}}}}}, and, combining {\seqBBBadb} with {\seqBBBada}, we get that \MMath{\Phi _{{\lambda _{3} \rightarrow  \lambda _{2} \rightarrow  \lambda _{1}}}} is a Hilbert space isomorphism \MMath{\mathcal{H} _{{\lambda _{3} \rightarrow  \lambda _{1}}} \rightarrow  \mathcal{H} _{{\lambda _{3} \rightarrow  \lambda _{2}}} \otimes  \mathcal{H} _{{\lambda _{2} \rightarrow  \lambda _{1}}}}, satisfying:%
\Mpar \MStartEqua \MMath{\left( \Phi _{{\lambda _{3} \rightarrow  \lambda _{2} \rightarrow  \lambda _{1}}} \otimes  \mathds{1} ^{\mathfrak{c} }_{{\lambda _{1}}} \right) \circ  \Phi _{{\lambda _{3} \rightarrow  \lambda _{1}}} = \left( \mathds{1} ^{\mathfrak{c} }_{{\mu _{{\lambda _{3} \rightarrow  \lambda _{2}}}}} \otimes  \Phi _{{\lambda _{2} \rightarrow  \lambda _{1}}} \right) \circ  \Phi _{{\lambda _{3} \rightarrow  \lambda _{2}}}}.%
\MStopEqua \Mpar Thus, {\seqBBBaff} is fulfilled. In particular, this implies that \MMath{\Phi _{{\lambda _{3} \rightarrow  \lambda _{2} \rightarrow  \lambda _{1}}}} is trivial whenever \MMath{\lambda _{1} = \lambda _{2}} and/or \MMath{\lambda _{2} = \lambda _{3}}.%
\Mpar \vspace{\Sssaut}\hspace{\Alinea}For any \MMath{\lambda _{1} \leqslant  \lambda _{2} \in  \mathcal{L} ^{\mathfrak{c} }_{}}, \MMath{\mkop{Tr} _{{\lambda _{2} \rightarrow  \lambda _{1}}} \mathrel{\mathop:}=  \mkop{Tr} _{{\mu _{{\lambda _{2} \rightarrow  \lambda _{1}}}}}} and the definition of the latter in {\seqBBBaec} coincides with {\seqBBBaey}. Similarly, \MMath{\iota _{{\lambda _{2} \leftarrow  \lambda _{1}}} \mathrel{\mathop:}=  \iota _{{\mu _{{\lambda _{2} \rightarrow  \lambda _{1}}}}}} and the definition of the latter in {\seqBBBafh} coincides with {\seqBBBaez}.%
\MendOfProof \hypertarget{SECafi}{}\MsectionB{afi}{2.2}{Isomorphisms of Projective State Spaces}%
\vspace{\PsectionB}\Mpar \MstartPhyMode\hspace{\Alinea}We now want to discuss the isomorphisms of the structures introduced in the previous subsection, and in particular their automorphisms, which will play an important role in the next subsection.
To build an isomorphism between two systems of factorized Hilbert spaces {\seqDDDaei} {$\mathfrak{c}$} and {$\mathfrak{d}$}, we need a one-to-one mapping between the labels in each system, as well as \emph{unitary identifications} of the corresponding Hilbert spaces. We do \emph{not} require a one-to-one mapping between the arrows, but we do require that the orders on each side match, and, for any \MMath{\lambda _{1} \leqslant  \lambda _{2}}, we demand that, at least for \emph{one} pair of arrows \MMath{(\mu ,\mu ')} (with \MMath{\mu : \lambda _{2} \rightarrow  \lambda _{1}} in {$\mathfrak{c}$} and \MMath{\mu '} between the corresponding labels in {$\mathfrak{d}$}), the unitary identifications be compatible with the coarse-graining decompositions, as represented in {\seqBBBafk} (thanks to {\seqBBBada}, this is enough to ensure that {\seqBBBafk} holds for \emph{any} pair of arrows).%
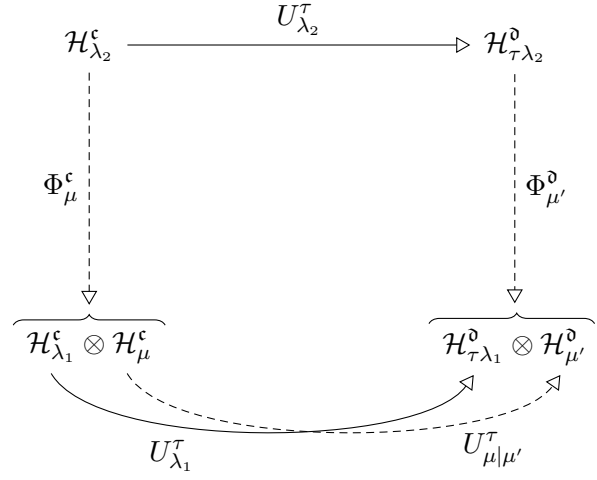
\Myfigure{2.3}{\hypertarget{PARafl}{}}{Compatibility with the coarse-graining decompositions of an isomorphism between two systems of factorized Hilbert spaces}{{\InputFigPafm}}%
\Mpar \vspace{\Saut}\hspace{\Alinea}An isomorphism {$\tau$} defined in this way provides a bijective mapping from the algebra of observables over {$\mathfrak{c}$} into the one over {$\mathfrak{d}$}, and, by duality, a bijective mapping in the other direction, ie.~from {$\mathfrak{d}$} to {$\mathfrak{c}$}, between the quantum state spaces.
The reason why {$\tau$} acts in the right direction on the \emph{observables} is because it is first defined as a mapping between \emph{labels}, and the latter represent sub-algebra of observables (what we called above selections of \dofs).%
\Mpar \vspace{\Saut}\hspace{\Alinea}Of course, since these are bijective mappings, the direction in which they are more naturally defined is irrelevant. But it would become significant if we were to consider \emph{morphisms}, rather than isomorphisms: such morphisms could be defined by relaxing the requirement for the map between the label sets to be \emph{bijective}.
For example, it is easy to see that, if the image of this map were only a \emph{subset} {$\mathcal{K}$} of the target label set, observables could still be push-forwarded and states could still be pull-backed: indeed observables, being elements in an inductive limit, are represented as equivalence classes (see {\seqBBBaez}), which can be \emph{extended} beyond {$\mathcal{K}$}, while states are represented as projective families, which can be \emph{restricted} to {$\mathcal{K}$}.
Such label set restrictions are extensively discussed in {\seqBBBafo} and will play a role in  {\seqBBBafp} (leading to a standard pattern in the statement of {\seqBBBaao}, namely mappings in opposite directions between states and observables, suitably intertwining the trace evaluation).
Interestingly, if {$\mathcal{K}$} happens to be \emph{cofinal} in \MMath{\mathcal{L} ^{\mathfrak{d} }_{}} (ie.~such that any label in \MMath{\mathcal{L} ^{\mathfrak{d} }_{}} is \emph{dominated} by a label in {$\mathcal{K}$}), both the mapping between the observable algebras and the one between the state spaces would \emph{still} be \emph{bijective} (see the explanations at the beginning of {\seqBBBafq}): arguably, such a morphism could then merit the qualification of \emph{iso}morphism, although we choose here to adopt a more straightforward definition.%
\MleavePhyMode \MleavePhyMode \hypertarget{PARafr}{}\Mpar \Mdefinition{2.9}Let \MMath{\mathfrak{c} , \mathfrak{d} } be two systems of factorized Hilbert spaces. An isomorphism {$\tau$} from {$\mathfrak{c}$} to {$\mathfrak{d}$} consists of:%
\hypertarget{PARafs}{}\Mpar \MList{1}a bijective map \MMath{\mathcal{L} ^{\mathfrak{c} }_{} \rightarrow  \mathcal{L} ^{\mathfrak{d} }_{}, \lambda  \mapsto  \tau  \lambda };%
\MStopList \hypertarget{PARaft}{}\Mpar \MList{2}for any \MMath{\lambda  \in  \mathcal{L} ^{\mathfrak{c} }_{}}, a Hilbert space isomorphism \MMath{U ^{\tau }_{\lambda } : \mathcal{H} ^{\mathfrak{c} }_{\lambda } \rightarrow  \mathcal{H} ^{\mathfrak{d} }_{{\tau  \lambda }}};%
\MStopList \Mpar such that:%
\hypertarget{PARafv}{}\Mpar \MList{3}for any \MMath{\lambda _{1}, \lambda _{2} \in  \mathcal{L} ^{\mathfrak{c} }_{}}, \MMath{\lambda _{1} \leqslant  \lambda _{2} \Leftrightarrow  \tau  \lambda _{1} \leqslant  \tau  \lambda _{2}};%
\MStopList \hypertarget{PARafw}{}\Mpar \MList{4}for any \MMath{\lambda _{1} \leqslant  \lambda _{2} \in  \mathcal{L} ^{\mathfrak{c} }_{}}, there exist \MMath{\mu  \in  \mathcal{L} ^{\mathfrak{c} }_{{\lambda _{2} \rightarrow  \lambda _{1}}}}, \MMath{\mu ' \in  \mathcal{L} ^{\mathfrak{d} }_{{\tau  \lambda _{2} \rightarrow  \tau  \lambda _{1}}}} and a Hilbert space isomorphism \MMath{U ^{\tau }_{\mu | \mu '} : \mathcal{H} ^{\mathfrak{c} }_{\mu } \rightarrow  \mathcal{H} ^{\mathfrak{d} }_{\mu '}} such that \MMath{\Phi ^{\mathfrak{d} }_{\mu '} = \left( U ^{\tau }_{{\lambda _{1}}} \otimes  U ^{\tau }_{\mu | \mu '} \right) \circ  \Phi ^{\mathfrak{c} }_{\mu } \circ  U ^{{\tau ,-1}}_{{\lambda _{2}}}}.%
\MStopList \Mpar \vspace{\Sssaut}\hspace{\Alinea}The composition of two such isomorphisms is obtained by taking the composition of the individual maps {\seqBBBafy} and is associative. The identity automorphism \MMath{\text{id}^{\mathfrak{c} }} on {$\mathfrak{c}$} is obtained by choosing these maps to be identity maps. The inverse \MMath{\tau ^{-1}} of \MMath{\tau } is obtained by taking the inverse of each map (in particular, for any \MMath{\lambda ' \in  \mathcal{L} ^{\mathfrak{d} }_{}}, \MMath{U ^{{\tau ^{-1}}}_{\lambda '} \mathrel{\mathop:}=  U ^{{\tau ,-1}}_{{\tau ^{-1} \lambda '}}}).%
\hypertarget{PARafz}{}\Mpar \Mproposition{2.10}Let \MMath{\mathfrak{c} , \mathfrak{d} } be two systems of factorized Hilbert spaces and let {$\tau$} be an isomorphism from {$\mathfrak{c}$} to {$\mathfrak{d}$}. For any \MMath{\lambda  \in  \mathcal{L} ^{\mathfrak{c} }_{}}, we define the maps \MMath{\Sigma ^{\tau }_{\lambda }} and \MMath{\Lambda ^{\tau }_{\lambda }} by:%
\Mpar \MStartEqua \MMath{\definitionFonction{\Sigma ^{\tau }_{\lambda }}{\mathcal{S} ^{\mathfrak{d} }_{{\tau  \lambda }}}{\mathcal{S} ^{\mathfrak{c} }_{\lambda }}{\rho }{U ^{{\tau ,-1}}_{\lambda } \,  \rho  \,  U ^{\tau }_{\lambda }}},%
\MStopEqua \Mpar and:%
\Mpar \MStartEqua \MMath{\definitionFonction{\Lambda ^{\tau }_{\lambda }}{\mathcal{B} ^{\mathfrak{c} }_{\lambda }}{\mathcal{B} ^{\mathfrak{d} }_{{\tau  \lambda }}}{A}{U ^{\tau }_{\lambda } \,  A \,  U ^{{\tau ,-1}}_{\lambda }}}.%
\MStopEqua \Mpar \vspace{\Sssaut}\hspace{\Alinea}For any \MMath{\lambda  \in  \mathcal{L} ^{\mathfrak{c} }_{}}, we have:%
\hypertarget{PARage}{}\Mpar \MList{1}\MMath{\Sigma ^{\tau }_{\lambda }} is a bijective, linear, order-preserving map \MMath{\mathcal{S} ^{\mathfrak{d} }_{{\tau  \lambda }} \rightarrow  \mathcal{S} ^{\mathfrak{c} }_{\lambda }} and \MMath{\Lambda ^{\tau }_{\lambda }} is an isometric \MMath{*}-algebra isomorphism \MMath{\mathcal{B} ^{\mathfrak{c} }_{\lambda } \rightarrow  \mathcal{B} ^{\mathfrak{d} }_{{\tau  \lambda }}};%
\MStopList \hypertarget{PARagf}{}\Mpar \MList{2}\MMath{\forall  \rho  \in  \mathcal{S} ^{\mathfrak{d} }_{{\tau  \lambda }}, \forall  A \in  \mathcal{B} ^{\mathfrak{c} }_{\lambda }, {\mkop{Tr}  \big( \Sigma ^{\tau }_{\lambda }(\rho ) \,  A \big) = \mkop{Tr}  \big( \rho  \,  \Lambda ^{\tau }_{\lambda }(A) \big)}}.%
\MStopList \Mpar and for any \MMath{\lambda _{1} \leqslant  \lambda _{2} \in  \mathcal{L} ^{\mathfrak{c} }_{}}:%
\hypertarget{PARagh}{}\Mpar \MList{3}\MMath{\mkop{Tr} _{{\lambda _{2} \rightarrow  \lambda _{1}}} \circ  \Sigma ^{\tau }_{{\lambda _{2}}} = \Sigma ^{\tau }_{{\lambda _{1}}} \circ  \mkop{Tr} _{{\tau  \lambda _{2} \rightarrow  \tau  \lambda _{1}}}};%
\MStopList \hypertarget{PARagi}{}\Mpar \MList{4}\MMath{\Lambda ^{\tau }_{{\lambda _{2}}} \circ  \iota _{{\lambda _{2} \leftarrow  \lambda _{1}}} = \iota _{{\tau  \lambda _{2} \leftarrow  \tau  \lambda _{1}}} \circ  \Lambda ^{\tau }_{{\lambda _{1}}}};%
\MStopList \Mpar \Mproof {\seqCCCagk} can be checked from the definitions of \MMath{\Sigma ^{\tau }_{\lambda }} and \MMath{\Lambda ^{\tau }_{\lambda }}. {\seqCCCagl} follow from \abbrevDef{\seqBBBagm} together with the definitions of \MMath{\mkop{Tr} _{{\lambda _{2} \rightarrow  \lambda _{1}}}} {\seqDDDagn} and \MMath{\iota _{{\lambda _{2} \leftarrow  \lambda _{1}}}} {\seqDDDago}.%
\MendOfProof \hypertarget{PARagp}{}\Mpar \Mdefinition{2.11}Let \MMath{\mathfrak{c} , \mathfrak{d} } be two systems of factorized Hilbert spaces with \MMath{\mathcal{L} ^{\mathfrak{c} }_{}, \leqslant } and \MMath{\mathcal{L} ^{\mathfrak{d} }_{}, \leqslant } directed, and let {$\tau$} be an isomorphism from {$\mathfrak{c}$} to {$\mathfrak{d}$}. We define the maps \MMath{\Sigma ^{\tau }} \italique{(inverse active transformation)} and \MMath{\Lambda ^{\tau }} \italique{(passive transformation)} by:%
\Mpar \MStartEqua \MMath{\definitionFonction{\Sigma ^{\tau }}{\mathcal{S} ^{\mathfrak{d} }}{\mathcal{S} ^{\mathfrak{c} }}{\left( \rho _{\lambda '} \right)_{{\lambda ' \in  \mathcal{L} ^{\mathfrak{d} }_{}}}}{\left( \Sigma ^{\tau }_{\lambda }(\rho _{{\tau  \lambda }}) \right)_{{\lambda  \in  \mathcal{L} ^{\mathfrak{c} }_{}}}}},%
\MStopEqua \Mpar and:%
\Mpar \MStartEqua \MMath{\definitionFonction{\Lambda ^{\tau }}{\mathcal{B} ^{\mathfrak{c} }}{\mathcal{B} ^{\mathfrak{d} }}{\left[ A_{\lambda } \right]^{\mathfrak{c} }}{\left[ \Lambda ^{\tau }_{\lambda }(A_{\lambda }) \right]^{\mathfrak{d} }}},%
\MStopEqua \Mpar where \MMath{\left[ A_{\lambda } \right]^{\mathfrak{c} }} denotes the equivalence class of \MMath{A_{\lambda } \in  \mathcal{B} ^{\mathfrak{c} }_{\lambda }} in \MMath{\mathcal{B} ^{\mathfrak{c} }} (representing the inductive limit \MMath{\mathcal{B} ^{\mathfrak{c} }} as a quotient set, see {\seqBBBaez}).%
\Mpar \vspace{\Sssaut}\hspace{\Alinea}From {\seqBBBagv}, \MMath{\Sigma ^{\tau }} is well-defined as a bijective, linear, order-preserving map \MMath{\mathcal{S} ^{\mathfrak{d} } \rightarrow  \mathcal{S} ^{\mathfrak{c} }}, \MMath{\Lambda ^{\tau }} can be extended to a \MMath{C^{*}}-algebra isomorphism \MMath{\overline{\mathcal{B} ^{\mathfrak{c} }} \rightarrow  \overline{\mathcal{B} ^{\mathfrak{d} }}}, and they satisfy:%
\Mpar \MStartEqua \MMath{\forall  \rho  \in  \mathcal{S} ^{\mathfrak{d} }, \forall  A \in  \overline{\mathcal{B} ^{\mathfrak{c} }}, {\mkop{Tr}  \big( \Sigma ^{\tau }(\rho ) \,  A \big) = \mkop{Tr}  \big( \rho  \,  \Lambda ^{\tau }(A) \big)}}.%
\MStopEqua \Mpar Moreover, \MMath{\Sigma ^{\tau }} and \MMath{\Lambda ^{\tau }} are bijective maps (respectively with inverse \MMath{\Sigma ^{{\tau ^{-1}}}} and \MMath{\Lambda ^{{\tau ^{-1}}}}).%
\hypertarget{PARagy}{}\Mpar \Mproposition{2.12}Let {$\mathfrak{c}$} be a system of factorized Hilbert spaces. The space \MMath{\mathcal{A} ^{\mathfrak{c} }} of automorphisms of {$\mathfrak{c}$} is a group.%
\Mpar \vspace{\Sssaut}\hspace{\Alinea}Assuming that \MMath{\mathcal{L} ^{\mathfrak{c} }_{}, \leqslant } is directed, \MMath{\mathcal{A} ^{\mathfrak{c} }} acts on the right (resp.\ on the left) on \MMath{\mathcal{S} ^{\mathfrak{c} }} (resp.\ \MMath{\overline{\mathcal{B} ^{\mathfrak{c} }}}).%
\Mpar \Mproof As underlined in {\seqBBBaha} the composition of isomorphisms is associative, there exists an identity automorphism on {$\mathfrak{c}$}, and every isomorphism admits an inverse, hence \MMath{\mathcal{A} ^{\mathfrak{c} }} is a group.%
\Mpar \vspace{\Sssaut}\hspace{\Alinea}We now assume that \MMath{\mathcal{L} ^{\mathfrak{c} }_{}, \leqslant } is directed. By definition of the identity automorphism \MMath{\text{id}^{\mathfrak{c} }} on {$\mathfrak{c}$}, \MMath{\Sigma ^{{\text{id}^{\mathfrak{c} }}} = \text{id}_{{\mathcal{S} ^{\mathfrak{c} }}}} and \MMath{\Lambda ^{{\text{id}^{\mathfrak{c} }}} = \text{id}_{{\mathcal{B} ^{\mathfrak{c} }}}}. Let \MMath{\tau ,\tau '} be two automorphisms of {$\mathfrak{c}$}. For any \MMath{\lambda  \in  \mathcal{L} ^{\mathfrak{c} }_{}}, \MMath{U^{\tau '\tau }_{\lambda } \mathrel{\mathop:}=  U^{\tau '}_{\tau \lambda } \circ  U^{\tau }_{\lambda }} (where \MMath{\tau '\tau } denotes the composition). Thus, \MMath{\Sigma ^{\tau '\tau } = \Sigma ^{\tau } \circ  \Sigma ^{\tau '}} and \MMath{\Lambda ^{\tau '\tau } = \Lambda ^{\tau '} \circ  \Lambda ^{\tau }}. Therefore, \MMath{\tau  \mapsto  \Sigma ^{\tau }} (resp.\ \MMath{\tau  \mapsto  \Lambda ^{\tau }}) is a right (resp.\ left) action of \MMath{\mathcal{A} ^{\mathfrak{c} }} on \MMath{\mathcal{S} ^{\mathfrak{c} }} (resp.\ \MMath{\overline{\mathcal{B} ^{\mathfrak{c} }}}).%
\MendOfProof \hypertarget{SECahb}{}\MsectionB{ahb}{2.3}{Universal Label Subsets}%
\vspace{\PsectionB}\Mpar \MstartPhyMode\hspace{\Alinea}We mentioned in {\seqBBBahd} that our motivation to use the projective formalism set up in {\seqBBBahe} was to make the \emph{construction} of states easier (eg.~compared to an approach using general algebraic states), and ideally to have a our disposal a complete, explicit \emph{parametrization} of the whole quantum state space.
But writing the state space as a projective limit is not enough to achieve this goal. Indeed projective families cannot be arbitrary: they have to fulfill the consistency relations coming from coarse-graining (namely, for any \MMath{\lambda _{1} \leqslant  \lambda _{2}}, we must have \MMath{\rho _{{\lambda _{1}}} \stackrel{!}{=} \mkop{Tr} _{{\lambda _{2} \rightarrow  \lambda _{1}}} \rho _{{\lambda _{2}}}}). It turns out that, when the label set is \emph{uncountably infinite}, constructing such families can be extremely hard: worse, a seemingly valid projective system can even yields an \emph{empty} projective limit {\bseqJJJaat}.\footnote{Note that for the construction we will develop in {\seqBBBabe}, the emptiness issue is easily averted: as long as there exists \emph{some} admissible complex structure on the classical phase space (which is typically the case for QFT on curved space-time, even though there may be no way to \emph{single out} a \emph{specific} one), the embedding result from {\seqBBBaao} guarantees that the projective state space is non-empty.
Furthermore, if we were only concerned about possible emptiness of the quantum state space, there exists a prescription to extend \emph{any} projective state space so that it will contain \emph{"enough"} states {\bseqJJJaam}. Unfortunately, this prescription does not tell us how to actually construct said states (existence is proved via the axiom of choice).}%
\Mpar \vspace{\Saut}\hspace{\Alinea}There \emph{is} a case in which constructing (all possible) projective families is unproblematic: if the label set happens to be isomorphic (as a pre-ordered set) to \MMath{\mathds{N}, \leqslant }, states can be constructed \emph{recursively}, by choosing each \MMath{\rho _{n+1}} in the \emph{pre-image} of \MMath{\mkop{Tr} _{{n+1 \rightarrow  n}} \left\langle  \rho _{n} \right\rangle } (the partial trace is always surjective, aka.~onto).
From there, we can identify a larger class of (directed) labels sets that will support a constructive description of the state space.
Indeed, we know that restricting the label set \MMath{\mathcal{L} ^{\mathfrak{c} }_{}} to a \emph{cofinal} part {$\mathcal{K}$} (ie.~some \MMath{\mathcal{K}  \subseteq \mathcal{L} ^{\mathfrak{c} }_{}} such that \MMath{\forall  \lambda  \in  \mathcal{L} ^{\mathfrak{c} }_{}, \exists  \kappa  \in  \mathcal{K} } with \MMath{\kappa  \geqslant  \lambda }) does \emph{not} change the label set: this is because knowing the projective family \MMath{\left(\rho _{\lambda }\right)_{\lambda }} over {$\mathcal{K}$} is enough to \emph{reconstruct} it over the whole label set \MMath{\mathcal{L} ^{\mathfrak{c} }_{}} (defining \MMath{\rho _{\lambda } \mathrel{\mathop:}=  \mkop{Tr} _{{\kappa  \rightarrow  \lambda }} \rho _{\kappa }}; the composition property of traces from {\seqBBBadb} ensures that the thus reconstructed family will be well-defined and projective as long as the one over {$\mathcal{K}$} is).%
\Mpar \vspace{\Saut}\hspace{\Alinea}To summarize, we can achieve our goal provided the label set admits a \emph{cofinal, increasing sequence} (or, equivalently, provided it is directed and of \emph{countable cofinality}).
However, typical classical field theories do not naively fulfill this requirement: they possess a \emph{continuum} of \dofs, which \emph{cannot} be captured by a \emph{countably-generated} collection of labels (recall that labels are \emph{finite} selections of \dofs\footnote{More precisely, the number of \dofs can be thought as (half) the \emph{dimension} of the phase space: labels correspond to finite-dimensional phase spaces, while countably-cofinal label sets can span phase spaces of \emph{at most} countable dimension. See {\seqBBBahf} for concrete examples.}).
On the other hand, this continuum of \dofs is really a mathematical abstraction rather than a physical reality: from an experimental point of view, observables can only be resolved with limited precision (the experimental protocol, specifying which observable(s) a particular experiment will be measuring, can only convey a \emph{finite} amount of information).%
\Mpar \vspace{\Saut}\hspace{\Alinea}This realization motivates the approach proposed in {\bseqJJJaad}\footnote{The present subsection is mostly taken from {\seqBBBahg}, albeit starting from a slightly different definition, which allows us to streamline the proof of the universality property ({\seqBBBabf} below, corresponding to {\seqBBBahh} in the original article).}, which \emph{restricts} the original theory to a countably-generated one, by extracting from the "naive" label set \MMath{\mathcal{L} ^{\mathfrak{c} }_{}} a countably-cofinal subset {$\mathcal{K}$}.
To ensure that {$\mathcal{K}$} is large enough, ie.~that it supports \emph{arbitrarily good approximations} of all observables from the original theory, we demand that it be \emph{dense} in \MMath{\mathcal{L} ^{\mathfrak{c} }_{}}: namely, that any label {$\lambda$} can be made an element of {$\mathcal{K}$} with the help of an arbitrarily small \emph{deformation}. Furthermore, we demand that, whenever {$\lambda$} happens to include sublabels which are \emph{already} in {$\mathcal{K}$}, the deformation can be chosen such that it leaves these sublabels \emph{invariant}, as schematized in {\seqBBBahi}. This extra condition turns out to be important to guarantee that restricting ourselves to {$\mathcal{K}$} is physically innocuous.%
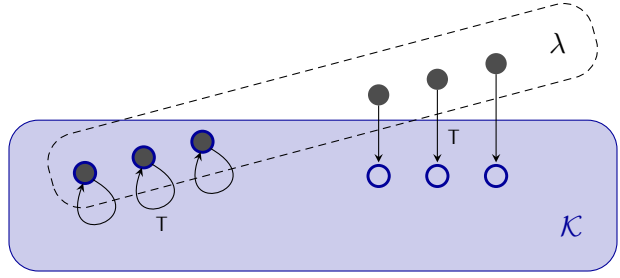
\Myfigure{2.4}{\hypertarget{PARahj}{}}{Density property for universal label subsets (in all figures of the present subsection, \dofs are symbolically represented by discs and circles, labels by groups of such discs, and label subsets by shaded regions)}{{\InputFigPahk}}%
\Mpar \vspace{\Saut}\hspace{\Alinea}Since our notion of \emph{density} in the label set will rely on "small deformations" of the labels, we need to start by telling what those deformations should be: this is achieved by giving ourselves the action of a \emph{topological} group {$\mathcal{T}$} on the system of factorized Hilbert spaces, ie.~a homomorphism from {$\mathcal{T}$} into the group of automorphisms introduced in {\seqBBBahm}.
This approach notably has the advantage that only labels whose observable algebras are isomorphic can be deemed "close".
Note that the concrete choice of {$\mathcal{T}$} will \emph{depend} on the particular physical system at hand, and this choice will determine \emph{in which sense} the restriction to {$\mathcal{K}$} respects the universality of the quantization and the symmetries of the theory (see below).%
\MleavePhyMode \MleavePhyMode \hypertarget{PARahn}{}\Mpar \Mdefinition{2.13}Let {$\mathfrak{c}$} be a system of factorized Hilbert spaces. Let {$\mathcal{T}$} be a \emph{topological} group and let \MMath{\mathcal{T}  \rightarrow  \mathcal{A} ^{\mathfrak{c} }, \text{{\sc t}}  \mapsto  \grpAct{\text{{\sc t}}}} be a group homomorphism (aka.\ an action of {$\mathcal{T}$} on {$\mathfrak{c}$}). A {$\mathcal{T}$}-universal label subset for {$\mathfrak{c}$} is a subset {$\mathcal{K}$} of \MMath{\mathcal{L} ^{\mathfrak{c} }_{}} such that:%
\hypertarget{PARaho}{}\Mpar \MList{1}{$\mathcal{K}$} is directed and of countable cofinality (ie.\ it admits a sequence \MMath{\left(\kappa _{n}\right)_{n\in \mathds{N}}} such that \MMath{\forall  \kappa  \in  \mathcal{K} , \exists  n \in  \mathds{N} \mathrel{\big/}  \kappa  \leqslant  \kappa _{n}});%
\MStopList \hypertarget{PARahp}{}\Mpar \MList{2}{$\mathcal{K}$} is a lower set of \MMath{\mathcal{L} ^{\mathfrak{c} }_{}} (ie.\ \MMath{\forall  \kappa  \in  \mathcal{K} , \forall  \lambda  \in  \mathcal{L} ^{\mathfrak{c} }_{}, {\lambda  \leqslant  \kappa  \Rightarrow  \lambda  \in  \mathcal{K} }});%
\MStopList \hypertarget{PARahq}{}\Mpar \MList{3}for any neighborhood {$\mathcal{V}$} of {$\mathbf{1}$} in {$\mathcal{T}$} (where {$\mathbf{1}$} denotes the identity element of {$\mathcal{T}$}) and any \MMath{\lambda  \in  \mathcal{L} ^{\mathfrak{c} }_{}}, there exists \MMath{\text{{\sc t}} \in  \mathcal{V} } such that:%
\Mpar \MStartEqua \MMath{\grpAct{\text{{\sc t}}} \lambda  \in  \mathcal{K}  \& \forall  \kappa  \in  \mathcal{K}  \mathrel{\big/}  \kappa  \leqslant  \lambda , \left( \grpAct{\text{{\sc t}}} \kappa  = \kappa  \& U ^{\grpAct{\text{{\sc t}}}} _{\kappa } = \mathds{1} ^{\mathfrak{c} }_{\kappa } \right)}.%
\MStopEqua \MStopList \Mpar \MstartPhyMode\hspace{\Alinea}While it is true that the formulation of classical field theories in terms of a continuum of \dofs is a matter of mathematical convenience, it \emph{does} accomplish an important objective, to which any attempt at a more discrete formulation should carefully pay heed: it \emph{minimizes} the amount of \emph{arbitrariness} entering the definition of the theory. By restricting the quantum theory to some label subset {$\mathcal{K}$}, are we making the quantization critically \emph{dependent} on the choice of this {$\mathcal{K}$}? In which sense can different choices be said to be equivalent? %
\Mpar \vspace{\Saut}\hspace{\Alinea}Fortunately, we have the following \emph{universality} result: for any two label subsets {$\mathcal{K}$} and \MMath{\mathcal{K} '} \emph{obeying} {\seqBBBaht} above, the quantum theories built on {$\mathcal{K}$} vs.~\MMath{\mathcal{K} '} are \emph{isomorphic}. The isomorphism between them can \emph{almost} be thought of as an equivalence of representations.
Since the algebras of observables on both sides are only dense sub-algebras of the continuous one, which a priori do \emph{not} \emph{coincide}, it is not possible to map an observable on one side to the \emph{exact} same observable on the other side, but we can arrange for the correspondence between observables to be an \emph{arbitrarily good approximation}. Namely, a given observable \MMath{A} in the sub-algebra selected by {$\mathcal{K}$} will be mapped to an observable \MMath{A'} lying in the sub-algebra selected by \MMath{\mathcal{K} '}, with \MMath{A} and \MMath{A'} related by an arbitrarily small deformation (again, closeness is defined with respect to the topological group {$\mathcal{T}$}).
While the small deformations involved are not necessarily the same for \emph{all} pairs of observables (in the statement of {\seqBBBabf} below, they are label-dependent; see however {\seqBBBabj}), the closeness criterion they satisfy is \emph{uniform}\footnote{This alone distinguishes the present universality result from a similar sounding result, known as Fell's theorem {\bseqJJJaau}, which states the approximate equivalence of different Hilbert spaces representations of a given algebra. The comparison between the two will be further analyzed in {\seqBBBahu}.} across the observable algebra (it is parametrized by a \emph{single} small open in {$\mathcal{T}$}).%
\Mpar \vspace{\Saut}\hspace{\Alinea}Keeping in mind the limited \emph{experimental resolution} of observables mentioned above, this result ensures that the choice to work on a particular {$\mathcal{K}$} has no physically measurable consequences, as long as {$\mathcal{K}$} fulfills {\seqBBBaht}, which justifies the terminology "\emph{universal} label subset".%
\Mpar \vspace{\Saut}\hspace{\Alinea}The proof of {\seqBBBabf} is a recursive one, with {\seqBBBahv} capturing one recursion step, as illustrated in {\seqBBBahw}. The idea is to construct the isomorphism {$\tau$} mapping {$\mathcal{K}$} into \MMath{\mathcal{K} '} by defining it first on a small label of {$\mathcal{K}$}, and by progressively \emph{extending} it to incorporate more and more labels from each side.
{\seqCCCahv} itself is proved by passing an auxiliary label back and forth between {$\mathcal{K}$} and \MMath{\mathcal{K} '}, aggregating additional \dofs as we go: one step consists in first using the universal label subset property for \MMath{\mathcal{K} '}, to build a deformation from a certain label of {$\mathcal{K}$} into \MMath{\mathcal{K} '}, before using the property for {$\mathcal{K}$}, to deform a slightly enlarged label back into {$\mathcal{K}$} {\seqDDDahx}.
It is to for this reason that we needed the extra stabilizing condition in {\seqBBBahy}: it ensures that, as we extend {$\tau$}, its definition on earlier labels will no longer change: otherwise, we would have no guarantee that the procedure would ultimately converge to a well-defined isomorphism.%
\Myfigure{2.5}{\hypertarget{PARaia}{}}{The recursion step for {\seqBBBahz} as captured by {\seqBBBahv}}{{\InputFigPaib}}%
\Myfigure{2.6}{\hypertarget{PARaid}{}}{{\seqCCCaic}, using the density property represented on {\seqBBBahi} successively in both directions}{{\InputFigPaie}}%
\MleavePhyMode \hypertarget{PARaif}{}\Mpar \Mtheorem{2.14}Let {$\mathfrak{c}$} be a system of factorized Hilbert spaces and let {$\mathcal{T}$} be a topological group acting on {$\mathfrak{c}$}. Let \MMath{\mathcal{K} , \mathcal{K} '} be two {$\mathcal{T}$}-universal label subsets for {$\mathfrak{c}$} and let {$\mathfrak{d}$} (resp.\ \MMath{\mathfrak{d} '}) denote the restriction of {$\mathfrak{c}$} on {$\mathcal{K}$} (resp.\ \MMath{\mathcal{K} '}).
For any neighborhood {$\mathcal{V}$} of {$\mathbf{1}$} in {$\mathcal{T}$}, there exists an isomorphism {$\tau$} from {$\mathfrak{d}$} to \MMath{\mathfrak{d} '} such that:%
\hypertarget{PARaig}{}\Mpar \MStartEqua \MMath{\forall  \kappa  \in  \mathcal{K} , \exists  \text{{\sc t}} \in  \mathcal{V}  \mathrel{\big/}  \big( \tau \kappa  = \grpAct{\text{{\sc t}}} \kappa  \& U ^{\tau }_{\kappa } = U ^{\grpAct{\text{{\sc t}}}} _{\kappa } \big)}.%
\NumeroteEqua{2.14}{1}\MStopEqua \hypertarget{PARaih}{}\Mpar \Mlemma{2.15}Let {$\mathfrak{c}$} be a system of factorized Hilbert spaces, {$\mathcal{T}$} be a topological group acting on {$\mathfrak{c}$} and \MMath{\mathcal{K} , \mathcal{K} '} be two {$\mathcal{T}$}-universal label subsets for {$\mathfrak{c}$}. Let {$\mathcal{V}$} be an \emph{open} neighborhood of {$\mathbf{1}$} in {$\mathcal{T}$} and suppose that there exist \MMath{\kappa _{1} \in  \mathcal{K} , \kappa '_{1} \in  \mathcal{K} '} and \MMath{\text{{\sc t}}_{1} \in  \mathcal{V} } such that \MMath{\grpAct{{\text{{\sc t}}_{1}}} \kappa _{1} = \kappa '_{1}}. Then, for any \MMath{\kappa _{o} \in  \mathcal{K} , \kappa '_{o} \in  \mathcal{K} '}, there exist \MMath{\kappa _{2} \in  \mathcal{K} , \kappa '_{2} \in  \mathcal{K} '} and \MMath{\text{{\sc t}}_{2} \in  \mathcal{V} } such that:%
\hypertarget{PARaii}{}\Mpar \MList{1}\MMath{\kappa _{2} \geqslant  \kappa _{1}, \kappa _{o}}, resp.\ \MMath{\kappa '_{2} \geqslant  \kappa '_{1}, \kappa '_{o}};%
\MStopList \hypertarget{PARaij}{}\Mpar \MList{2}\MMath{\grpAct{{\text{{\sc t}}_{2}}} \kappa _{2} = \kappa '_{2}};%
\MStopList \hypertarget{PARaik}{}\Mpar \MList{3}\MMath{\forall  \kappa  \in  \mathcal{K}  \mathrel{\big/}  \kappa  \leqslant  \kappa _{1}, \big( \grpAct{{\text{{\sc t}}_{2}}} \kappa  = \grpAct{{\text{{\sc t}}_{1}}} \kappa  \& U ^{\grpAct{{\text{{\sc t}}_{2}}}}_{\kappa } = U ^{\grpAct{{\text{{\sc t}}_{1}}}}_{\kappa } \big)}.%
\MStopList \Mpar \Mproof Let \MMath{\kappa _{o} \in  \mathcal{K} , \kappa '_{o} \in  \mathcal{K} '}. We define:%
\Mpar \MStartEqua \MMath{\mathcal{V} _{\alpha } \mathrel{\mathop:}=  \left\{ \text{{\sc t}} \in  \mathcal{T}  \middlewithspace| \text{{\sc t}} \mathrel{.}  \text{{\sc t}}_{1} \in  \mathcal{V}  \right\}}.%
\MStopEqua \Mpar \MMath{\mathcal{V} _{\alpha }} is a neighborhood of {$\mathbf{1}$} in {$\mathcal{T}$}. Since {$\mathcal{K}$} is directed (\abbrevDef{\seqBBBaio}), there exists \MMath{\kappa _{\alpha } \in  \mathcal{K} } such that \MMath{\kappa _{\alpha } \geqslant  \kappa _{1}, \kappa _{o}}. We define \MMath{\lambda _{\alpha } \mathrel{\mathop:}=  \grpAct{{\text{{\sc t}}_{1}}} \kappa _{\alpha } \in  \mathcal{L} ^{\mathfrak{c} }_{}}. Applying \abbrevDef{\seqBBBahy} (for the {$\mathcal{T}$}-universal label subset \MMath{\mathcal{K} '}) to \MMath{\mathcal{V} _{\alpha }}, \MMath{\lambda _{\alpha }}, there exists \MMath{\text{{\sc t}}_{\alpha } \in  \mathcal{V} _{\alpha }} such that \MMath{\kappa '_{\alpha } \mathrel{\mathop:}=  \grpAct{{\text{{\sc t}}_{\alpha }}} \lambda _{\alpha } \in  \mathcal{K} '} and:%
\hypertarget{PARaip}{}\Mpar \MStartEqua \MMath{\forall  \kappa ' \in  \mathcal{K} ' \mathrel{\big/}  \kappa ' \leqslant  \lambda _{\alpha }, \left( \grpAct{{\text{{\sc t}}_{\alpha }}} \kappa ' = \kappa ' \& U ^{\grpAct{{\text{{\sc t}}_{\alpha }}}}_{\kappa '} = \mathds{1} ^{\mathfrak{c} }_{\kappa '} \right)}.%
\NumeroteEqua{2.15}{1}\MStopEqua \Mpar We define:%
\Mpar \MStartEqua \MMath{\mathcal{V} _{\beta } \mathrel{\mathop:}=  \left\{ \text{{\sc t}} \in  \mathcal{T}  \middlewithspace| \text{{\sc t}}_{\alpha } \mathrel{.}  \text{{\sc t}}_{1} \mathrel{.}  \text{{\sc t}}^{-1} \in  \mathcal{V}  \right\}}.%
\MStopEqua \Mpar \MMath{\mathcal{V} _{\beta }} is a neighborhood of {$\mathbf{1}$} in {$\mathcal{T}$}. Since {$\mathcal{K}$}' is directed, there exists \MMath{\kappa '_{2} \in  \mathcal{K} '} such that \MMath{\kappa '_{2} \geqslant  \kappa '_{\alpha }, \kappa '_{o}}. We define \MMath{\lambda _{\beta } \mathrel{\mathop:}=  \grpAct{{(\text{{\sc t}}_{\alpha } \mathrel{.}  \text{{\sc t}}_{1})^{-1}}} \kappa '_{2} \in  \mathcal{L} ^{\mathfrak{c} }_{}}. Applying \abbrevDef{\seqBBBahy} (for the {$\mathcal{T}$}-universal label subset \MMath{\mathcal{K} }) to \MMath{\mathcal{V} _{\beta }}, \MMath{\lambda _{\beta }}, there exists \MMath{\text{{\sc t}}_{\beta } \in  \mathcal{V} _{\beta }} such that \MMath{\kappa _{2} \mathrel{\mathop:}=  \grpAct{{\text{{\sc t}}_{\beta }}} \lambda _{\beta } \in  \mathcal{K} } and:%
\hypertarget{PARait}{}\Mpar \MStartEqua \MMath{\forall  \kappa  \in  \mathcal{K}  \mathrel{\big/}  \kappa  \leqslant  \lambda _{\beta }, \left( \grpAct{{\text{{\sc t}}_{\beta }}} \kappa  = \kappa  \& U ^{\grpAct{{\text{{\sc t}}_{\beta }}}}_{\kappa } = \mathds{1} ^{\mathfrak{c} }_{\kappa } \right)}.%
\NumeroteEqua{2.15}{2}\MStopEqua \Mpar We define \MMath{\text{{\sc t}}_{2} \mathrel{\mathop:}=  \text{{\sc t}}_{\alpha } \mathrel{.}  \text{{\sc t}}_{1} \mathrel{.}  \text{{\sc t}}_{\beta }^{-1} \in  \mathcal{V} }.
Then, {\seqCCCaiv} holds by construction.%
\Mpar \vspace{\Sssaut}\hspace{\Alinea}From \abbrevDef{\seqBBBaix}, we have \MMath{\kappa '_{1} \leqslant  \lambda _{\alpha }} and \MMath{\kappa _{\alpha } \leqslant  \lambda _{\beta }}. Applying {\seqBBBaiy} (resp.\ {\seqBBBaiz}), we get \MMath{\grpAct{{\text{{\sc t}}_{\alpha }}} \kappa '_{1} = \kappa '_{1}} (resp.\ \MMath{\grpAct{{\text{{\sc t}}_{\beta }}} \kappa _{\alpha } = \kappa _{\alpha }}). Therefore, \MMath{\kappa _{2} \geqslant  \kappa _{\alpha }} and \MMath{\kappa '_{\alpha } \geqslant  \kappa '_{1}}, so {\seqBBBaja} holds.%
\Mpar \vspace{\Sssaut}\hspace{\Alinea}Let \MMath{\kappa  \in  \mathcal{K} } such that \MMath{\kappa  \leqslant  \kappa _{1}}. We have \MMath{\grpAct{{\text{{\sc t}}_{1}}} \kappa  \leqslant  \kappa '_{1}}, hence \MMath{\grpAct{{\text{{\sc t}}_{1}}} \kappa  \in  \mathcal{K} '} (from \abbrevDef{\seqBBBajb}) and \MMath{\grpAct{{\text{{\sc t}}_{1}}} \kappa  \leqslant  \lambda _{\alpha }}. {\seqCCCaiy} then yields \MMath{\grpAct{(\text{{\sc t}}_{\alpha } \mathrel{.}  \text{{\sc t}}_{1})} \kappa  = \grpAct{{\text{{\sc t}}_{1}}} \kappa } and \MMath{U ^{\grpAct{{\text{{\sc t}}_{\alpha } \mathrel{.}  \text{{\sc t}}_{1}}}}_{\kappa } = U ^{\grpAct{{\text{{\sc t}}_{1}}}}_{\kappa }}. Moreover, \MMath{\grpAct{(\text{{\sc t}}_{\alpha } \mathrel{.}  \text{{\sc t}}_{1})} \kappa  \leqslant  \kappa '_{\alpha }}, hence \MMath{\kappa  \leqslant  \lambda _{\beta }}. {\seqCCCaiz} then yields \MMath{\grpAct{{\text{{\sc t}}_{\beta }^{-1}}} \kappa  = \kappa } and \MMath{U ^{\grpAct{{\text{{\sc t}}_{\beta }^{-1}}}}_{\kappa } = \mathds{1} ^{\mathfrak{c} }_{\kappa }}, so {\seqBBBajc} holds.%
\MendOfProof \Mpar \MproofTheorem{2.14}Let {$\mathcal{V}$} be a neighborhood of {$\mathbf{1}$} in {$\mathcal{T}$}. Let \MMath{\mathcal{V} _{o}} be an \emph{open} neighborhood of {$\mathbf{1}$} in {$\mathcal{V}$}. Let \MMath{\left( \kappa _{{o,n}} \right)_{n\geqslant 0}} (resp.\ \MMath{\left( \kappa '_{{o,n}} \right)_{n\geqslant 1}}) be a cofinal sequence (\abbrevDef{\seqBBBaio}) in {$\mathcal{K}$} (resp.\ \MMath{\mathcal{K} '}). Applying \abbrevDef{\seqBBBahy} (for \MMath{\mathcal{K} '}) to \MMath{\mathcal{V} _{o}} and \MMath{\kappa _{o} \mathrel{\mathop:}=  \kappa _{{o,o}} \in  \mathcal{K} }, let \MMath{\text{{\sc t}}_{o} \in  \mathcal{V} _{o}} such that \MMath{\kappa '_{o} \mathrel{\mathop:}=  \grpAct{{\text{{\sc t}}_{o}}} \kappa _{o} \in  \mathcal{K} '}. Using {\seqBBBahv}, we construct recursively three sequences \MMath{\left( \kappa _{n} \right)_{n\geqslant 0}} in {$\mathcal{K}$}, \MMath{\left( \kappa '_{n} \right)_{n\geqslant 0}} in \MMath{\mathcal{K} '} and \MMath{\left( \text{{\sc t}}_{n} \right)_{n\geqslant 0}} in \MMath{\mathcal{V} _{o}}, such that:%
\hypertarget{PARaje}{}\Mpar \MList{1}\MMath{\left( \kappa _{n} \right)_{n\geqslant 0}} and \MMath{\left( \kappa '_{n} \right)_{n\geqslant 0}} are \emph{increasing};%
\MStopList \hypertarget{PARajf}{}\Mpar \MList{2}for any \MMath{n\geqslant 0}, \MMath{\kappa _{n} \geqslant  \kappa _{{o,n}}}, so \MMath{\left( \kappa _{n} \right)_{n\geqslant 0}} is cofinal in {$\mathcal{K}$}, and, similarly, \MMath{\left( \kappa '_{n} \right)_{n\geqslant 0}} is cofinal in \MMath{\mathcal{K} '};%
\MStopList \hypertarget{PARajg}{}\Mpar \MList{3}for any \MMath{n\geqslant 0}, \MMath{\grpAct{{\text{{\sc t}}_{n}}} \kappa _{n} = \kappa '_{n}};%
\MStopList \hypertarget{PARajh}{}\Mpar \MList{4}for any \MMath{m \geqslant  n \geqslant  0} and any \MMath{\kappa  \in  \mathcal{K} } such that \MMath{\kappa  \leqslant  \kappa _{n}}, \MMath{\grpAct{{\text{{\sc t}}_{m}}} \kappa  = \grpAct{{\text{{\sc t}}_{n}}} \kappa } and \MMath{U ^{\grpAct{{\text{{\sc t}}_{m}}}}_{\kappa } = U ^{\grpAct{{\text{{\sc t}}_{n}}}}_{\kappa }}.%
\MStopList \Mpar \vspace{\Sssaut}\hspace{\Alinea}For any \MMath{\kappa  \in  \mathcal{K} }, we define \MMath{\tau  \kappa  \mathrel{\mathop:}=  \grpAct{{\text{{\sc t}}_{n}}} \kappa } and \MMath{U ^{\tau }_{\kappa } \mathrel{\mathop:}=  U ^{\grpAct{{\text{{\sc t}}_{n}}}}_{\kappa }}, for some \MMath{n\geqslant 0} such that \MMath{\kappa  \leqslant  \kappa _{n}} (this is well-defined thanks to {\seqBBBajj}). Since \MMath{\tau  \kappa  \leqslant  \kappa '_{n}}, we have \MMath{\tau  \kappa  \in  \mathcal{K} '} (using \abbrevDef{\seqBBBajb} for \MMath{\mathcal{K} '}). Next, for any \MMath{\kappa ' \in  \mathcal{K} '}, there exists \MMath{n\geqslant 0} such that \MMath{\kappa ' \leqslant  \kappa '_{n}}, so defining \MMath{\kappa  \mathrel{\mathop:}=  \grpAct{{\text{{\sc t}}_{n}^{-1}}} \kappa ' \leqslant  \kappa _{n}}, we get \MMath{\kappa  \in  \mathcal{K} } (using \abbrevDef{\seqBBBajb} for {$\mathcal{K}$}) and \MMath{\kappa ' = \tau  \kappa }, therefore \MMath{\kappa  \mapsto  \tau  \kappa } is a surjective map \MMath{\mathcal{K}  \rightarrow  \mathcal{K} '}. Finally, for any \MMath{\kappa _{1},\kappa _{2} \in  \mathcal{K} }, there exists \MMath{n\geqslant 0} such that \MMath{\kappa _{1},\kappa _{2} \leqslant  \kappa _{n}}, so the injectivity of \MMath{\kappa  \mapsto  \tau  \kappa }, as well as \abbrevDef{\seqBBBajk}, follow from the corresponding properties of \MMath{\grpAct{{\text{{\sc t}}_{n}}}}. Therefore, {$\tau$} is an isomorphism from {$\mathfrak{d}$} to \MMath{\mathfrak{d} '}, and it satisfies {\seqBBBajl} by construction.%
\MendOfProof \Mpar \MstartPhyMode\hspace{\Alinea}An immediate corollary of the previous result is that any element {\sc t}{} of {$\mathcal{T}$} (and, more generally, any automorphism of {$\mathfrak{c}$} whose adjoint action induces a bi-continuous homomorphism on {$\mathcal{T}$}) can be \emph{approximately realized} as an automorphism of the restricted theory: this works because the image of {$\mathcal{K}$} under {\sc t}{} is \emph{again} a {$\mathcal{T}$}-universal label subset, which can be \emph{deformed back} into {$\mathcal{K}$} by virtue of {\seqBBBabf}, yielding an automorphism that \emph{stabilizes} {$\mathcal{K}$}.
This property can be used to ensure that the symmetries of the classical theory will be (approximately) implemented in the quantum theory constructed over {$\mathcal{K}$} (see eg.~{\seqBBBabj}).%
\MleavePhyMode \hypertarget{PARajn}{}\Mpar \Mproposition{2.16}Let {$\mathfrak{c}$} be a system of factorized Hilbert spaces and let {$\mathcal{T}$} be a topological group acting on {$\mathfrak{c}$}. Let \MMath{\mathcal{K} } be a {$\mathcal{T}$}-universal label subset for {$\mathfrak{c}$} and let {$\mathfrak{d}$} denote the restriction of {$\mathfrak{c}$} on {$\mathcal{K}$}.
For any \MMath{\text{{\sc t}}_{o} \in  \mathcal{T} } and any neighborhood \MMath{\mathcal{V} _{o}} of \MMath{\text{{\sc t}}_{o}} in {$\mathcal{T}$}, there exists an automorphism {$\tau$} of {$\mathfrak{d}$} such that:%
\hypertarget{PARajo}{}\Mpar \MStartEqua \MMath{\forall  \kappa  \in  \mathcal{K} , \exists  \text{{\sc t}} \in  \mathcal{V} _{o} \mathrel{\big/}  \big( \tau \kappa  = \grpAct{\text{{\sc t}}} \kappa  \& U ^{\tau }_{\kappa } = U ^{\grpAct{\text{{\sc t}}}} _{\kappa } \big)}.%
\NumeroteEqua{2.16}{1}\MStopEqua \Mpar \Mproof Let \MMath{\text{{\sc t}}_{o} \in  \mathcal{T} } and let \MMath{\mathcal{V} _{o}} be a neighborhood of \MMath{\text{{\sc t}}_{o}} in {$\mathcal{T}$}. Let \MMath{\mathcal{K} ' \mathrel{\mathop:}=  \left\{ \grpAct{{\text{{\sc t}}_{o}}}  \kappa  \middlewithspace| \kappa  \in  \mathcal{K}  \right\}} and \MMath{\mathcal{V}  \mathrel{\mathop:}=  \left\{ \text{{\sc t}} \in  \mathcal{T}  \middlewithspace| \text{{\sc t}}^{-1} \mathrel{.}  \text{{\sc t}}_{o}  \in  \mathcal{V} _{o}\right\}}. \MMath{\mathcal{K} '} is a {$\mathcal{T}$}-universal label subset for {$\mathfrak{c}$}  (thanks to \MMath{\grpAct{{\text{{\sc t}}_{o}}}} being an automorphism of {$\mathfrak{c}$} and \MMath{\text{{\sc t}} \mapsto  \text{{\sc t}}_{o} \mathrel{.}  \text{{\sc t}} \mathrel{.}  \text{{\sc t}}_{o}^{-1}} being an automorphism of {$\mathcal{T}$}) and {$\mathcal{V}$} is a neighborhood of {$\mathbf{1}$} in {$\mathcal{T}$}. Let \MMath{\mathfrak{d} '} denote the restriction of {$\mathfrak{c}$} on \MMath{\mathcal{K} '}. Applying {\seqBBBabf}, there exists an isomorphism \MMath{\tilde{\tau }} from {$\mathfrak{d}$}  to \MMath{\mathfrak{d} '} fulfilling {\seqBBBajl}. Then \MMath{\tau  \mathrel{\mathop:}=  \tilde{\tau }^{-1}  \grpAct{{\text{{\sc t}}_{o}}}} is an automorphism of {$\mathfrak{d}$} fulfilling {\seqBBBajq}.
\MendOfProof %
\Mnomdefichier{lin20}%
\hypertarget{SECajs}{}\MsectionA{ajs}{3}{Quantization of Linear Field Theories}%
\vspace{\PsectionA}\Mpar \MstartPhyMode\hspace{\Alinea}We want to develop the quantization of a general \emph{linear} classical field theory along the lines of the previous section.
Linear here means that the classical phase space is taken to be a (infinite-dimensional) \emph{symplectic} ({\seqBBBabk}, in the \emph{bosonic} case) or \emph{inner-product} ({\seqBBBabl}, in the \emph{fermionic} case) \emph{vector space}, ideally with its linear structure being \emph{preserved} by the time evolution (ie.~the Hamiltonian function is quadratic, or the field theory is "free"; see however {\seqBBBaax}).%
\MleavePhyMode \hypertarget{SECaju}{}\MsectionB{aju}{3.1}{State Space and Observables}%
\vspace{\PsectionB}\Mpar \MstartPhyMode\hspace{\Alinea}We begin by recalling some properties of the (Fock) quantization of a \emph{finite-dimensional} vector space {$F$} (of dimension \MMath{2d }), as this will form the basic \emph{building bloc} of our construction (for concreteness, let us focus on the symplectic, or bosonic, case, a detailed overview of which is given in {\seqBBBajw}; the fermionic case, covered in {\seqBBBajx}, is very similar).
Given a symplectic frame \MMath{\lambda  = \left(\mathbf{e} _{1},\dots ,\mathbf{e} _{2d }\right)} in {$F$} (aka.~Darboux linear coordinates, see {\seqBBBajy}), we can build a corresponding Fock representation \MMath{\mathcal{H} ^{\mathfrak{lin} }_{\lambda }}: each pair \MMath{\left(\mathbf{e} _{2k+1}, \mathbf{e} _{2k+2}\right)} of \emph{conjugate} frame vectors constitutes a \emph{mode}, and the \emph{canonical} orthonormal basis for \MMath{\mathcal{H} ^{\mathfrak{lin} }_{\lambda }} will be indexed by the number of particles in each of these modes.%
\Mpar \vspace{\Saut}\hspace{\Alinea}Of course, we know from the Stone–von Neumann theorem that the choice of the frame {$\lambda$} is irrelevant: if we consider a different frame \MMath{\lambda '}, there should exist a \emph{unitary identification} relating \MMath{\mathcal{H} ^{\mathfrak{lin} }_{\lambda }} to \MMath{\mathcal{H} ^{\mathfrak{lin} }_{\lambda '}}.
We can think of all possible Fock quantizations, resulting from all possible choices of frames, as forming a \emph{vector bundle}: the \emph{base space} being the space of all symplectic frames, and the \emph{fiber} over a frame {$\lambda$} being the Fock space \MMath{\mathcal{H} ^{\mathfrak{lin} }_{\lambda }}. Then, the natural identification relating the Fock representations built over different frames can be thought of as a \emph{connection} in this bundle (right part of {\seqBBBajz}).
We have some freedom in the definition of this connection, as the unitary identifications provided by the Stone–von Neumann theorem are only fixed \emph{up to a phase}. Exploiting this freedom, it is possible to arrange for the connection to be \emph{flat} (aka.~curvature-free), which is the choice we will adopt.
There is however a catch: the space of frames is \emph{not} simply-connected, it contains a topologically non-trivial loop {\seqDDDaka}, and the price we pay for making the connection flat is that, when parallel-transporting around this loop, we pick-up an extra sign factor (see the explanations before {\seqBBBakb}).%
\Myfigure{3.1}{\hypertarget{PARakc}{}}{Relating finite-dimensional Fock representations build over different frames}{{\InputFigPakd}}%
\Mpar \vspace{\Saut}\hspace{\Alinea}A different way of formulating the same mechanism is to match the canonical basis in each \MMath{\mathcal{H} ^{\mathfrak{lin} }_{\lambda }} to define a \emph{global trivialization} of the bundle, with an "anonymous" Fock space \MMath{\mathcal{F} ^{(d )}_{\mathfrak{lin} }} as model fiber.
This "naive" global trivialization should \emph{not} be confused with the \emph{proper} identification of Fock representations, which matches the \emph{physical interpretations} of the observables (in fact, due to the above-mentioned topological non-triviality, the latter does not provide a global trivialization).
Instead, we want to look at the \emph{deviation} between the two: by \emph{casting} over \MMath{\mathcal{F} ^{(d )}_{\mathfrak{lin} }} the proper parallel transport, we obtain on \MMath{\mathcal{F} ^{(d )}_{\mathfrak{lin} }} a \emph{unitary representation} of the symplectic group, the group of all transformations mapping a symplectic frame into another {\seqDDDajz}.
From this perspective, the topological non-triviality of the connection manifests itself in the \emph{projective} nature of the representation (meaning that we may pick up additional phases when composing transformations, see \bseqHHHaaf{section 2.7 and appendix 2.B}): it is actually a representation of the \emph{double cover} of the symplectic group, the so-called \emph{metaplectic group} {\seqDDDakf}.%
\Mpar \vspace{\Saut}\hspace{\Alinea}With this tool at our disposal, we can now turn to the quantization of an \emph{infinite-dimensional} symplectic vector space {$V$}. The basic observables to be implemented in the quantum theory will be the \emph{linear forms} on {$V$} (sufficiently regular ones, ie.~those that can written as \MMath{\Omega (\mathbf{v} ,\, \cdot \, )}; see {\seqBBBakg} and the comments preceding it).
Thus, a selection of \dofs, adequate for an experiment measuring finitely many of those linear observables, will correspond to a \emph{linear subspace} in the \emph{dual} of {$V$}.
Since the symplectic structure of {$V$} gives us a natural identification between {$V$} and (the sufficiently regular part of) its dual, we can equivalently think of selections of \dofs as subspaces in {$V$} \emph{itself}.
Taking advantage of the admissibility of \emph{redundant labels}, which was highlighted in the discussion preceding {\seqBBBaei}, our labels will be the \emph{finite symplectic families} in {$V$} (ie.~\emph{incomplete} symplectic frames, {\seqBBBajy}).
Not every finite-dimensional subspace of {$V$} is the span of a symplectic family, but those who are, are the ones that correspond to valid partial theories (keeping in mind that \dofs are \emph{pairs} of \emph{conjugate} variables), and those who are not, can always be \emph{embedded} into a larger one, who is {\seqDDDakh}.%
\Mpar \vspace{\Saut}\hspace{\Alinea}Next, we need to establish the coarse-graining relations between these labels. Clearly a label \MMath{\lambda _{1}} should be considered \emph{coarser} than a label \MMath{\lambda _{2}} if, and only if, the subspace spanned by \MMath{\lambda _{1}} is \emph{included} in the subspace spanned by \MMath{\lambda _{2}} (in order for their respective counterparts in the \emph{dual} to be included in the same way: see the above explanation of how labels select \dofs).
If the symplectic family \MMath{\lambda _{2}} simply \emph{extends} \MMath{\lambda _{1}} (ie.~\MMath{\lambda _{1}} consists of the first \MMath{2n} vectors of \MMath{\lambda _{2}}), it is straightforward to write a suitable tensor product decomposition {\seqDDDacz} of the Fock space built on \MMath{\lambda _{2}}: each vector of the particle basis can be {$\otimes$}-factorized by putting on one side all particles in the \MMath{\lambda _{1}}-modes, and on the other all particles in the remaining \MMath{\lambda _{2}}-modes {\seqDDDaki}.
If the coarser modes cannot so easily be split out of \MMath{\lambda _{2}}, the solution is to \emph{first} transform \MMath{\lambda _{2}} into a label \MMath{\lambda '_{2}} that \emph{does} extend \MMath{\lambda _{1}}, using the metaplectic representation to relate \MMath{\mathcal{H} ^{\mathfrak{lin} }_{{\lambda _{2}}}} and \MMath{\mathcal{H} ^{\mathfrak{lin} }_{{\lambda '_{2}}}}, and \emph{then} factor \MMath{\mathcal{H} ^{\mathfrak{lin} }_{{\lambda _{1}}}} out of \MMath{\mathcal{H} ^{\mathfrak{lin} }_{{\lambda '_{2}}}} {\seqDDDakj}.
Accordingly, any element of the metaplectic group that can transform \MMath{\lambda _{2}} into an extension of \MMath{\lambda _{1}} will define an \emph{arrow} from \MMath{\lambda _{2}} to \MMath{\lambda _{1}}.%
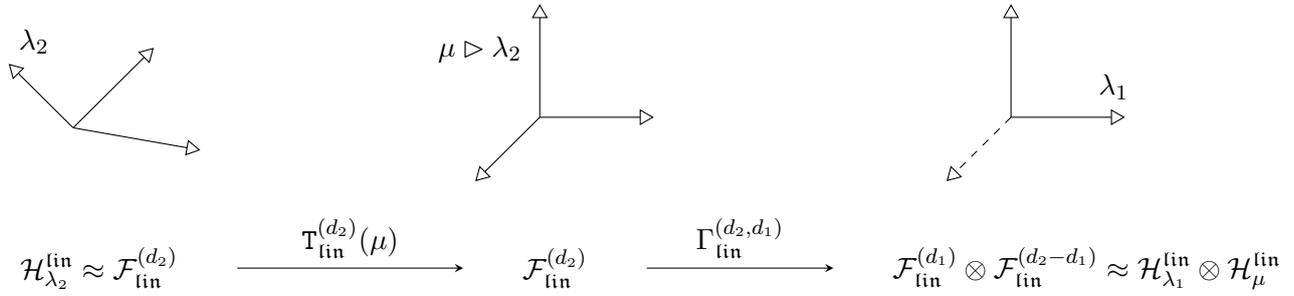
\Myfigure{3.2}{\hypertarget{PARakk}{}}{Coarse-graining labels (this schematic visualization should not be taken too literally though: symplectic frames are always even-dimensional)}{{\InputFigPakl}}%
\Mpar \vspace{\Saut}\hspace{\Alinea}The ability to connect labels which are not \emph{obviously} compatible is crucial.
If we would only have arrows between labels which extend each other, the directedness requirement for the label set {\seqDDDaew} would enforce that \emph{all} labels be extracted from a \emph{common}, global, symplectic frame: depending on the choice of such a complete basis in {$V$} would be even worse\footnote{And if we could only connect labels when the \emph{partial polarizations} from which they are built extend each other, we would end up depending precisely on a global choice of polarization: this is in fact what was done in {\seqBBBakn}.} than depending on the choice of a polarization for {$V$} (recall {\seqBBBako}).
In this sense, the \emph{polarization-independence} of the construction we are going to present can really be understood as a direct consequence of the finite-dimensional Stone–von Neumann theorem.
By contrast, the availability of a \emph{flat} connexion in the bundle of finite-dimensional quantizations is not indispensable: if we were working with a \emph{curved} connection, we could simply associate arrows to \emph{paths} in the space of symplectic frames.
In other words, our use of the metaplectic representation is only a cosmetic preference (it makes the construction tighter, by not keeping track of the precise paths but merely of their \emph{winding parities}; cf.~the definition of the metaplectic group in {\seqBBBaka}).%
\Mpar \vspace{\Saut}\hspace{\Alinea}In line with the discussion preceding {\seqBBBaei}, we have many different labels describing the \emph{same} selection of \dofs (namely symplectic families spanning the same subspace of {$V$}), as well as many different arrows between two given labels (there is some freedom in extending \MMath{\lambda _{1}} into a family \MMath{\lambda '_{2}} with the same span as \MMath{\lambda _{2}}, as well as the additional freedom of picking one of the two elements of the metaplectic group which map \MMath{\lambda _{2}} into \MMath{\lambda '_{2}}).
This redundancy allows for a very \emph{explicit} description of the Fock spaces \MMath{\mathcal{H} ^{\mathfrak{lin} }_{\lambda }} and \MMath{\mathcal{H} ^{\mathfrak{lin} }_{\mu }}, without threatening the \emph{overall consistency} (which will automatically follows from {\seqBBBael}; see below). We could try to dispense from relying on explicit frames by introducing more "abstract" Hilbert spaces: the idea would be to consider "Fock spaces \emph{modulo} change of frame", in the same way as an \MMath{n}-dimensional vector space can be though as "\MMath{\mathds{R}^{n}} modulo change of basis". However, the topological non-triviality of the connection discussed at the beginning of the present subsection stands in the way of such an approach: to make the construction consistent, we would need to fix various choices of phases throughout the label set. This would not only be awkward, but in fact an \emph{artificial} complication, since global phases end up \emph{canceling out}, when representing quantum states as density matrices.%
\MleavePhyMode \MleavePhyMode \MleavePhyMode \hypertarget{PARakp}{}\Mpar \Mdefinition{3.1}Let \MMath{V , \Omega } be a symplectic vector space {\seqDDDakg}. We define:%
\Mpar \MList{1}\MMath{\mathcal{L} ^{\mathfrak{bos} }_{} \mathrel{\mathop:}=  \left\{ \left(\mathbf{e} _{1},\dots ,\mathbf{e} _{2n}\right) \in  V ^{2n} \middlewithspace| n\geqslant 0 \& \forall  i,j \leqslant  2n, {\Omega (\mathbf{e} _{i}, \mathbf{e} _{j}) = \Omega ^{(n)}_{ij}} \right\}} (where \MMath{\Omega ^{(n)}} denotes the canonical \MMath{2n\times 2n} symplectic matrix, see {\seqBBBajy});%
\MStopList \Mpar \MList{2}for any \MMath{\lambda  = \left(\mathbf{e} _{1},\dots ,\mathbf{e} _{2n}\right) \in  \mathcal{L} ^{\mathfrak{bos} }_{}}, \MMath{d _{\lambda } \mathrel{\mathop:}=  n \geqslant  0} and \MMath{V _{\lambda } \mathrel{\mathop:}=  \mkop{Span}  \left\{\mathbf{e} _{1},\dots ,\mathbf{e} _{2n}\right\} \subset  V } (note that \MMath{\mkop{dim}  V _{\lambda }  = 2 d _{\lambda }}, see {\seqBBBaks});%
\MStopList \Mpar \MList{3}for any \MMath{\lambda _{1},\lambda _{2} \in  \mathcal{L} ^{\mathfrak{bos} }_{}} such that \MMath{V _{{\lambda _{1}}} \nsubseteq V _{{\lambda _{2}}}}, \MMath{\mathcal{L} ^{\mathfrak{bos} }_{{\lambda _{2} \rightarrow  \lambda _{1}}} \mathrel{\mathop:}=  \varnothing };%
\MStopList \Mpar \MList{4}for any \MMath{\lambda _{1} = \left(\mathbf{e} _{1},\dots ,\mathbf{e} _{2n}\right), \lambda _{2} = \left(\mathbf{f} _{1},\dots ,\mathbf{f} _{2m}\right) \in  \mathcal{L} ^{\mathfrak{bos} }_{}} such that \MMath{V _{{\lambda _{1}}} \subseteq V _{{\lambda _{2}}}}:%
\Mpar \MStartEqua \MMath{\mathcal{L} ^{\mathfrak{bos} }_{{\lambda _{2} \rightarrow  \lambda _{1}}} \mathrel{\mathop:}=  \left\{ \mu  \in  \text{Mp}^{(m)} \middlewithspace| \exists  \left(\mathbf{e} _{2n+1},\dots ,\mathbf{e} _{2m}\right) \in  V ^{{2(m-n)}} \mathrel{\big/}  \mu  \rhd  \left(\mathbf{f} _{1},\dots ,\mathbf{f} _{2m}\right) = \left(\mathbf{e} _{1},\dots ,\mathbf{e} _{2m}\right) \right\}},%
\MStopEqua \Mpar where \MMath{\text{Mp}^{(m)}} denotes the metaplectic group over \MMath{\mathds{R}^{2m}} and {$\rhd$} its action on a symplectic \MMath{(2m)}-family {\seqDDDaka};%
\MStopList \hypertarget{PARakx}{}\Mpar \MList{5}for any \MMath{\lambda _{1}, \lambda _{2}, \lambda _{3} \in  \mathcal{L} ^{\mathfrak{bos} }_{}} and any \MMath{\mu _{12} \in  \mathcal{L} ^{\mathfrak{bos} }_{{\lambda _{2} \rightarrow  \lambda _{1}}}, \mu _{23} \in  \mathcal{L} ^{\mathfrak{bos} }_{{\lambda _{3} \rightarrow  \lambda _{2}}}}, \MMath{\mu _{12} \cdot  \mu _{23} \mathrel{\mathop:}=  \ell _{{d _{{\lambda _{3}}} \leftarrow  d _{{\lambda _{2}}}}} \big( \mu _{12} \big) \,  \mu _{23}} (ie.\ the composition as group elements, identifying \MMath{\text{Mp}^{(d _{{\lambda _{2}}})}} as a subgroup of \MMath{\text{Mp}^{(d _{{\lambda _{3}}})}}, see {\seqBBBaky}).%
\MStopList \hypertarget{PARakz}{}\Mpar \Mproposition{3.2}\MMath{\mathcal{L} ^{\mathfrak{bos} }} is a (small) category. Moreover, \MMath{\mathcal{L} ^{\mathfrak{bos} }_{}, \leqslant } is directed, with:%
\hypertarget{PARala}{}\Mpar \MStartEqua \MMath{\forall  \lambda ,\lambda ' \in  \mathcal{L} ^{\mathfrak{bos} }_{}, {\lambda  \leqslant  \lambda ' \Leftrightarrow  V _{\lambda } \subseteq V _{\lambda '}}}.%
\NumeroteEqua{3.2}{1}\MStopEqua \Mpar \Mproof Let \MMath{\lambda _{1} = \left(\mathbf{e} _{1},\dots ,\mathbf{e} _{2n}\right), \lambda _{2} = \left(\mathbf{f} _{1},\dots ,\mathbf{f} _{2m}\right), \lambda _{3} = \left(\mathbf{g} _{1},\dots ,\mathbf{g} _{2l}\right) \in  \mathcal{L} ^{\mathfrak{bos} }_{}} and \MMath{\mu _{12} \in  \mathcal{L} ^{\mathfrak{bos} }_{{\lambda _{2} \rightarrow  \lambda _{1}}}, \mu _{23} \in  \mathcal{L} ^{\mathfrak{bos} }_{{\lambda _{3} \rightarrow  \lambda _{2}}}}. Let \MMath{\left(\mathbf{e} _{2n+1},\dots ,\mathbf{e} _{2m}\right) \in  V ^{{2(m-n)}}} such that \MMath{\mu _{12} \rhd  \left(\mathbf{f} _{1},\dots ,\mathbf{f} _{2m}\right) = \left(\mathbf{e} _{1},\dots ,\mathbf{e} _{2m}\right)} and \MMath{\left(\mathbf{e} _{2m+1},\dots ,\mathbf{e} _{2l}\right) \in  V ^{{2(l-m)}}} such that \MMath{\mu _{23} \rhd  \left(\mathbf{g} _{1},\dots ,\mathbf{g} _{2l}\right) = \left(\mathbf{f} _{1},\dots ,\mathbf{f} _{2m},\mathbf{e} _{2m+1},\dots ,\mathbf{e} _{2l}\right)}. Then, \MMath{(\mu _{12} \cdot  \mu _{23}) \rhd  \left(\mathbf{g} _{1},\dots ,\mathbf{g} _{2l}\right) = \left(\mathbf{e} _{1},\dots ,\mathbf{e} _{2l}\right)} {\seqDDDaky}, so the composition is well defined as a map \MMath{\mathcal{L} ^{\mathfrak{bos} }_{{\lambda _{2} \rightarrow  \lambda _{1}}} \times  \mathcal{L} ^{\mathfrak{bos} }_{{\lambda _{3} \rightarrow  \lambda _{2}}} \rightarrow  \mathcal{L} ^{\mathfrak{bos} }_{{\lambda _{3} \rightarrow  \lambda _{1}}}}.%
\Mpar \vspace{\Sssaut}\hspace{\Alinea}For any \MMath{\lambda  \in  \mathcal{L} ^{\mathfrak{bos} }_{}}, \MMath{1 _{\lambda }} is the identity element of \MMath{\text{Mp}^{(d _{\lambda })}}. The associativity of the composition operation follows from the properties of \MMath{\ell _{m\leftarrow n}} {\seqDDDaky}.%
\Mpar \vspace{\Sssaut}\hspace{\Alinea}{\seqCCCalc} follows from {\seqBBBald}. Let \MMath{\lambda ,\lambda ' \in  \mathcal{L} ^{\mathfrak{bos} }_{}}. Using {\seqBBBakh}, there exists \MMath{\lambda '' \in  \mathcal{L} ^{\mathfrak{bos} }_{}} such that \MMath{V_{\lambda } + V_{\lambda '} \subseteq V_{\lambda ''}}, hence \MMath{\mathcal{L} ^{\mathfrak{bos} }_{}, \leqslant } is directed.%
\MendOfProof \Mpar \MstartPhyMode\hspace{\Alinea}The classical precursor of a \emph{fermionic} QFT is a "phase space" with an \emph{Euclidean} rather than symplectic structure {\seqDDDalf}. This leads to a subtlety regarding \emph{orientation}: while symplectomorphisms are always orientation-preserving, ie.~all symplectic frames share the same orientation, this is \emph{not} true for orthonormal frames. As detailed below (before {\seqBBBalg}), we cannot allow arrows to be associated to \emph{orientation-reversing} transformations. Therefore, two orthonormal families \MMath{\lambda _{1}} and \MMath{\lambda '_{1}} spanning the \emph{same} vector subspace, but with \emph{reverse orientation}, will be \emph{incomparable} in the ordering of labels, although they naively correspond to the same selection of \dofs.
Fortunately, this does not spoil the \emph{directedness} of the label set {\seqDDDaew}, since the two families can still be embedded into a strictly larger subspace (assuming {$V$} to be infinite-dimensional), and completed there into families \MMath{\lambda _{2}} and \MMath{\lambda '_{2}}, which \emph{do} share the same orientation.%
\MleavePhyMode \hypertarget{PARalh}{}\Mpar \Mdefinition{3.3}Let \MMath{V , \left(\, \cdot \, \middlewithspace|\, \cdot \, \right)} be a real, \emph{infinite-dimensional} pre-Hilbert space {\seqDDDabl}. We define:%
\Mpar \MList{1}\MMath{\mathcal{L} ^{\mathfrak{ferm} }_{} \mathrel{\mathop:}=  \left\{ \left(\mathbf{e} _{1},\dots ,\mathbf{e} _{2n}\right) \in  V ^{2n} \middlewithspace| n\geqslant 0 \& \forall  i,j \leqslant  2n, {\left(\mathbf{e} _{i} \middlewithspace| \mathbf{e} _{j}\right) = \delta _{ij}} \right\}};%
\MStopList \Mpar \MList{2}for any \MMath{\lambda  = \left(\mathbf{e} _{1},\dots ,\mathbf{e} _{2n}\right) \in  \mathcal{L} ^{\mathfrak{ferm} }_{}}, \MMath{d _{\lambda } \mathrel{\mathop:}=  n \geqslant  0} and \MMath{V _{\lambda } \mathrel{\mathop:}=  \mkop{Span}  \left\{\mathbf{e} _{1},\dots ,\mathbf{e} _{2n}\right\} \subset  V } (\MMath{\mkop{dim}  V _{\lambda }  = 2 d _{\lambda }});%
\MStopList \hypertarget{PARalk}{}\Mpar \MList{3}for any \MMath{\lambda _{1},\lambda _{2} \in  \mathcal{L} ^{\mathfrak{ferm} }_{}} such that \MMath{V _{{\lambda _{1}}} \nsubseteq V _{{\lambda _{2}}}}, \MMath{\mathcal{L} ^{\mathfrak{ferm} }_{{\lambda _{2} \rightarrow  \lambda _{1}}} \mathrel{\mathop:}=  \varnothing };%
\MStopList \hypertarget{PARall}{}\Mpar \MList{4}for any \MMath{\lambda _{1} = \left(\mathbf{e} _{1},\dots ,\mathbf{e} _{2n}\right), \lambda _{2} = \left(\mathbf{f} _{1},\dots ,\mathbf{f} _{2m}\right) \in  \mathcal{L} ^{\mathfrak{ferm} }_{}} such that \MMath{V _{{\lambda _{1}}} \subseteq V _{{\lambda _{2}}}}:%
\Mpar \MStartEqua \MMath{\mathcal{L} ^{\mathfrak{ferm} }_{{\lambda _{2} \rightarrow  \lambda _{1}}} \mathrel{\mathop:}=  \left\{ \mu  \in  \text{Spin}^{(m)} \middlewithspace| \exists  \left(\mathbf{e} _{2n+1},\dots ,\mathbf{e} _{2m}\right) \in  V ^{{2(m-n)}} \mathrel{\big/}  \mu  \rhd  \left(\mathbf{f} _{1},\dots ,\mathbf{f} _{2m}\right) = \left(\mathbf{e} _{1},\dots ,\mathbf{e} _{2m}\right) \right\}},%
\MStopEqua \Mpar where \MMath{\text{Spin}^{(m)}} denotes the spin group over \MMath{\mathds{R}^{2m}} and {$\rhd$} its action on an orthonormal \MMath{(2m)}-family {\seqDDDalo};%
\MStopList \hypertarget{PARalp}{}\Mpar \MList{5}for any \MMath{\lambda _{1}, \lambda _{2}, \lambda _{3} \in  \mathcal{L} ^{\mathfrak{ferm} }_{}} and any \MMath{\mu _{12} \in  \mathcal{L} ^{\mathfrak{ferm} }_{{\lambda _{2} \rightarrow  \lambda _{1}}}, \mu _{23} \in  \mathcal{L} ^{\mathfrak{ferm} }_{{\lambda _{3} \rightarrow  \lambda _{2}}}}, \MMath{\mu _{12} \cdot  \mu _{23} \mathrel{\mathop:}=  \ell _{{d _{{\lambda _{3}}} \leftarrow  d _{{\lambda _{2}}}}} \big( \mu _{12} \big) \,  \mu _{23}} (ie.\ the composition as group elements, identifying \MMath{\text{Spin}^{(d _{{\lambda _{2}}})}} as a subgroup of \MMath{\text{Spin}^{(d _{{\lambda _{3}}})}}, see {\seqBBBalq}).%
\MStopList \hypertarget{PARalr}{}\Mpar \Mproposition{3.4}\MMath{\mathcal{L} ^{\mathfrak{ferm} }} is a (small) category. Moreover, \MMath{\mathcal{L} ^{\mathfrak{ferm} }_{}, \leqslant } is directed, with:%
\hypertarget{PARals}{}\Mpar \MStartEqua \MMath{\forall  \lambda ,\lambda ' \in  \mathcal{L} ^{\mathfrak{ferm} }_{}, {\lambda  \leqslant  \lambda ' \Leftrightarrow  \Big[ V _{\lambda } \subseteq V _{\lambda '} \& \left( V _{\lambda } \neq  V _{\lambda '} \text{{ or }} \lambda  \text{{ has the same orientation as }} \lambda '\right) \Big]}}.%
\NumeroteEqua{3.4}{1}\MStopEqua \Mpar \Mproof The proof that \MMath{\mathcal{L} ^{\mathfrak{ferm} }}  is a category works as in the bosonic case, using {\seqBBBalq} in place of {\seqBBBaky}.%
\Mpar \vspace{\Sssaut}\hspace{\Alinea}The direction `\hspace{0.2cm}$\Rightarrow$\hspace{0.2cm}{}' of {\seqBBBalu} follows from {\seqBBBalv} together with the definition of the spin group {\seqDDDalo}.
The direction `\hspace{0.2cm}$\Leftarrow$\hspace{0.2cm}{}' follows from {\seqBBBalw} together with {\seqBBBalx} or {\seqBBBalo}.
Since {$V$} is infinite-dimensional, {\seqBBBalu} is sufficient to ensure the directedness of \MMath{\mathcal{L} ^{\mathfrak{ferm} }_{}} using {\seqBBBaly}.%
\MendOfProof \Mpar \MstartPhyMode\hspace{\Alinea}We can now define the system of factorized Hilbert spaces along the lines sketched at the beginning of the present subsection.
The \emph{composition} property {\seqDDDadb}, resp.~the \emph{unambiguity} property {\seqDDDada}, is secured thanks to the compatibility of the metaplectic representation with the natural tensor-product decomposition of Fock spaces: a symplectomorphism which only acts on the first $2n$, resp.~the last $2m-2n$ vectors of a symplectic frame, will be represented by a unitary transformation which only acts on the first, resp.~the second, tensor-product factor {\seqDDDama}.%
\Mpar \vspace{\Saut}\hspace{\Alinea}The same is true for \emph{orientation-preserving} orthogonal transformations (aka.~rotations) in the fermionic case. But even when an orientation-reversing transformation only acts on one group of modes, its unitary representation may pick-up an extra sign-prefactor, which \emph{depends} on the particular state over the \emph{other} group of modes: such an operator then cannot be written as a tensor product with only the identity operator acting on one side (depending of our choice of conventions, the problem may affect either transformations acting on the first vectors or those acting on the last ones, see the comment at the beginning of {\seqBBBamb}). This is the reason why we had to exclude such transformations above: it would otherwise not be possible to arrange for both the composition and the unambiguity properties to hold.%
\MleavePhyMode \hypertarget{PARamc}{}\Mpar \Mdefinition{3.5}Let \MMath{\mathfrak{lin}  \in  \left\{ \mathfrak{bos} , \mathfrak{ferm}  \right\}}. We define:%
\hypertarget{PARamd}{}\Mpar \MList{1}for any \MMath{\lambda  \in  \mathcal{L} ^{\mathfrak{lin} }_{}}, \MMath{\mathcal{H} ^{\mathfrak{lin} }_{\lambda } \mathrel{\mathop:}=  \mathcal{F} ^{(d _{\lambda })}_{\mathfrak{lin} }} (where \MMath{\mathcal{F} ^{(n)}_{\mathfrak{bos} /\mathfrak{ferm} }} denotes the bosonic/fermionic Fock space over \MMath{n} states, see {\seqBBBame} and {\seqBBBamf});%
\MStopList \hypertarget{PARamg}{}\Mpar \MList{2}for any \MMath{\lambda _{1}, \lambda _{2} \in  \mathcal{L} ^{\mathfrak{lin} }_{}} and any \MMath{\mu  \in  \mathcal{L} ^{\mathfrak{lin} }_{{\lambda _{2} \rightarrow  \lambda _{1}}}}, \MMath{\mathcal{H} ^{\mathfrak{lin} }_{\mu } \mathrel{\mathop:}=  \mathcal{F} ^{(d _{{\lambda _{2}}} - d _{{\lambda _{1}}})}_{\mathfrak{lin} }} and \MMath{\Phi ^{\mathfrak{lin} }_{\mu } \mathrel{\mathop:}=  \Gamma ^{(d _{{\lambda _{2}}},d _{{\lambda _{1}}})}_{\mathfrak{lin} } \circ  \mathtt{T} ^{(d _{{\lambda _{2}}})}_{\mathfrak{lin} }(\mu )} (where \MMath{\mathtt{T} ^{(m)}_{\mathfrak{bos} /\mathfrak{ferm} }} denotes the representation of the metaplectic/spin group on \MMath{\mathcal{F} ^{(m)}_{\mathfrak{bos} /\mathfrak{ferm} }}, see {\seqBBBakb}, resp.~{\seqBBBamh}, and \MMath{\Gamma ^{(m,n)}_{\mathfrak{bos} /\mathfrak{ferm} }} is the factorization introduced in {\seqBBBaki}, resp.~{\seqBBBami}).%
\MStopList \hypertarget{PARamj}{}\Mpar \Mtheorem{3.6}The objects from {\seqBBBalg} constitute a system of factorized Hilbert spaces {\seqDDDaei}.%
\Mpar \Mproof Let \MMath{\mathfrak{lin}  \in  \left\{ \mathfrak{bos} , \mathfrak{ferm}  \right\}}. For any \MMath{\lambda _{1}, \lambda _{2} \in  \mathcal{L} ^{\mathfrak{lin} }_{}} and any \MMath{\mu  \in  \mathcal{L} ^{\mathfrak{lin} }_{{\lambda _{2} \rightarrow  \lambda _{1}}}}, \MMath{\mathtt{T} ^{(d _{{\lambda _{2}}})}_{\mathfrak{lin} }(\mu )} is a unitary isomorphism \MMath{\mathcal{H} ^{\mathfrak{lin} }_{{\lambda _{2}}} \rightarrow  \mathcal{H} ^{\mathfrak{lin} }_{{\lambda _{2}}}} ({\seqBBBakb}, resp.~{\seqBBBamh}), and \MMath{\Gamma ^{(d _{{\lambda _{2}}},d _{{\lambda _{1}}})}_{\mathfrak{lin} }} is a unitary isomorphism \MMath{\mathcal{H} ^{\mathfrak{lin} }_{{\lambda _{2}}} \rightarrow  \mathcal{H} ^{\mathfrak{lin} }_{{\lambda _{1}}} \otimes  \mathcal{H} ^{\mathfrak{lin} }_{\mu }} ({\seqBBBaki}, resp.~{\seqBBBami}).%
\Mpar \vspace{\Sssaut}\hspace{\Alinea}Let \MMath{\lambda _{1}, \lambda _{2}, \lambda _{3} \in  \mathcal{L} ^{\mathfrak{lin} }_{}}, \MMath{\mu _{12} \in  \mathcal{L} ^{\mathfrak{lin} }_{{\lambda _{2} \rightarrow  \lambda _{1}}}}, and \MMath{\mu _{23} \in  \mathcal{L} ^{\mathfrak{lin} }_{{\lambda _{3} \rightarrow  \lambda _{2}}}}. \MMath{\Gamma ^{(d _{{\lambda _{3}}} - d _{{\lambda _{1}}}, d _{{\lambda _{2}}} - d _{{\lambda _{1}}})}_{\mathfrak{lin} }} provides a natural identification \MMath{\mathcal{H} ^{\mathfrak{lin} }_{{\mu _{12} \cdot  \mu _{23}}} \rightarrow  \mathcal{H} ^{\mathfrak{lin} }_{{\mu _{12}}} \otimes  \mathcal{H} ^{\mathfrak{lin} }_{{\mu _{23}}}}, and, modulo this identification \MMath{\left( \Gamma ^{(d _{{\lambda _{2}}},d _{{\lambda _{1}}})}_{\mathfrak{lin} } \otimes  \mathds{1} ^{\mathfrak{lin} }_{{\mu _{23}}} \right) \circ  \Gamma ^{(d _{{\lambda _{3}}},d _{{\lambda _{2}}})}_{\mathfrak{lin} } \approx  \Gamma ^{(d _{{\lambda _{3}}},d _{{\lambda _{1}}})}_{\mathfrak{lin} }}, hence {\seqBBBadb} follows from {\seqBBBaml}, resp.~{\seqBBBamm}.%
\Mpar \vspace{\Sssaut}\hspace{\Alinea}Let \MMath{\lambda _{1}, \lambda _{2} \in  \mathcal{L} ^{\mathfrak{lin} }_{}} and \MMath{\mu ,\mu ' \in  \mathcal{L} ^{\mathfrak{lin} }_{{\lambda _{2} \rightarrow  \lambda _{1}}}}. {\seqCCCada} is obtained by applying {\seqBBBamn}, resp.~{\seqBBBamo}, to \MMath{\mu ' \,  \mu ^{-1}}.%
\MendOfProof \Mpar \MstartPhyMode\hspace{\Alinea}To confirm that we are indeed quantizing the classical symplectic space \MMath{V , \Omega }, we identify the \emph{quantized observables} corresponding to linear forms on {$V$} (exponentiated to avoid issues with unbounded operators), and check their \emph{commutator relations}.
These observables are labeled by \emph{vectors} {$\mathbf{v}$} in {$V$}, because, as mentioned in the introduction of the present subsection, we only consider linear forms which can be written as \MMath{\Omega (\mathbf{v} , \, \cdot \, )} (these are the ones among which Poisson-brackets can be defined, see {\seqBBBamq}).%
\MleavePhyMode \hypertarget{PARamr}{}\Mpar \Mdefinition{3.7}Let \MMath{\mathbf{v}  \in  V } and let \MMath{\lambda  = \left(\mathbf{e} _{1},\dots ,\mathbf{e} _{2n}\right) \in  \mathcal{L} ^{\mathfrak{bos} }_{}} such that \MMath{\mathbf{v}  \in  V _{\lambda }}. Let \MMath{\mathbf{x}  \in  \mathds{R}^{2n}} such that \MMath{\mathbf{v}  = \sum_{i=1}^{2n} \mathbf{x} _{i} \mathbf{e} _{i}}. We define a bounded operator \MMath{\mathcal{O} ^{\mathfrak{bos} }_{\lambda }(\mathbf{v} ) \in  \mathcal{B} ^{\mathfrak{bos} }_{\lambda }} by \MMath{\mathcal{O} ^{\mathfrak{bos} }_{\lambda }(\mathbf{v} ) \mathrel{\mathop:}=  \exp(i \,  \hat{\mathbf{x} })} (where \MMath{\hat{\mathbf{x} }} is the essentially self-adjoint operator on \MMath{\mathcal{H} ^{\mathfrak{bos} }_{\lambda }} defined in {\seqBBBaba}).%
\hypertarget{PARams}{}\Mpar \Mproposition{3.8}Let \MMath{\mathbf{v}  \in  V }, \MMath{\lambda _{1} \in  \mathcal{L} ^{\mathfrak{bos} }_{}} such that \MMath{\mathbf{v}  \in  V _{{\lambda _{1}}}}, and \MMath{\lambda _{2} \geqslant  \lambda _{1}} (hence \MMath{V _{{\lambda _{1}}} \subseteq V _{{\lambda _{2}}}}, so \MMath{\mathbf{v}  \in  V _{{\lambda _{2}}}}). Then, \MMath{\mathcal{O} ^{\mathfrak{bos} }_{{\lambda _{2}}}(\mathbf{v} ) = \iota _{{\lambda _{2} \leftarrow  \lambda _{1}}} \big( \mathcal{O} ^{\mathfrak{bos} }_{{\lambda _{1}}}(\mathbf{v} ) \big)}. Thus, we can associate to any \MMath{\mathbf{v}  \in  V } an observable \MMath{\mathcal{O} ^{\mathfrak{bos} }(\mathbf{v} ) \in  \mathcal{B} ^{\mathfrak{bos} }} {\seqDDDaex}.%
\Mpar \vspace{\Sssaut}\hspace{\Alinea}Let \MMath{\mathbf{v} ,\mathbf{w}  \in  V } and \MMath{\lambda  \in  \mathcal{L} ^{\mathfrak{bos} }_{}} such that \MMath{\mathbf{v} ,\mathbf{w}  \in  V _{\lambda }}. We have:%
\hypertarget{PARamt}{}\Mpar \MStartEqua \MMath{\mathcal{O} ^{\mathfrak{bos} }_{\lambda }(\mathbf{v} ) \,  \mathcal{O} ^{\mathfrak{bos} }_{\lambda }(\mathbf{w} ) = e^{{\nicefrac{i}{2} \,  \Omega (\mathbf{v} ,\mathbf{w} )}} \,  \mathcal{O} ^{\mathfrak{bos} }_{\lambda }(\mathbf{v}  + \mathbf{w} )}.%
\NumeroteEqua{3.8}{1}\MStopEqua \Mpar Thus, \MMath{\mathcal{O} ^{\mathfrak{bos} }(\mathbf{v} ) \,  \mathcal{O} ^{\mathfrak{bos} }(\mathbf{w} ) = e^{{\nicefrac{i}{2} \,  \Omega (\mathbf{v} ,\mathbf{w} )}} \,  \mathcal{O} ^{\mathfrak{bos} }(\mathbf{v}  + \mathbf{w} )}.%
\Mpar \Mproof Let \MMath{\mathbf{v}  \in  V }, \MMath{\lambda _{1} = \left(\mathbf{e} _{1},\dots ,\mathbf{e} _{2n}\right) \in  \mathcal{L} ^{\mathfrak{bos} }_{}} and \MMath{\lambda _{2} = \left(\mathbf{f} _{1},\dots ,\mathbf{f} _{2m}\right) \in  \mathcal{L} ^{\mathfrak{bos} }_{}} such that \MMath{\mathbf{v}  \in  V _{{\lambda _{1}}}} and \MMath{\lambda _{1} \leqslant  \lambda _{2}}. Let \MMath{\mathbf{x}  \in  \mathds{R}^{2n}} and \MMath{\mathbf{y}  \in  \mathds{R}^{2m}} such that:%
\Mpar \MStartEqua \MMath{\mathbf{v}  = \sum_{i=1}^{2n} \mathbf{x} _{i} \mathbf{e} _{i} = \sum_{j=1}^{2m} \mathbf{y} _{j} \mathbf{f} _{j}}.%
\MStopEqua \Mpar Let \MMath{\mu  \in  \mathcal{L} ^{\mathbf{b} }_{{\lambda _{2} \rightarrow  \lambda _{1}}}} and let \MMath{\left(\mathbf{e} _{2n+1},\dots ,\mathbf{e} _{2m}\right) \in  V ^{{2(m-n)}}} such that \MMath{\mu  \rhd  \left(\mathbf{f} _{1},\dots ,\mathbf{f} _{2m}\right) = \left(\mathbf{e} _{1},\dots ,\mathbf{e} _{2m}\right)}.
We define \MMath{\mathbf{x} ' \in  \mathds{R}^{2m}} by:%
\Mpar \MStartEqua \MMath{\mathbf{x} '_{i} \mathrel{\mathop:}=  \alternative{\mathbf{x} _{i}}{\text{{ if }} i \leqslant  2n}{0}{\text{{ otherwise}}}},%
\MStopEqua \Mpar so that \MMath{\mathbf{v}  = \sum_{i=1}^{2m} \mathbf{x} _{i} \mathbf{e} _{i}}.
Using {\seqBBBana}, we get:%
\Mpar \MStartEqua \MMath{\Phi ^{{\mathfrak{bos} ,-1}}_{\mu } \circ  \big( \mathcal{O} ^{\mathfrak{bos} }_{{\lambda _{1}}}(\mathbf{v} ) \otimes  \mathds{1}  \big) \circ  \Phi ^{\mathfrak{bos} }_{\mu } = \exp\big(i \,  \widehat{\sigma ^{-1} \,  \mathbf{x} '}\big)},%
\MStopEqua \Mpar where \MMath{\sigma  = p ^{(m)}(\mu )} (with the covering map \MMath{p ^{(m)} : \text{Mp}^{(m)} \rightarrow  \text{Sp}^{(m)}} from {\seqBBBand}).
From the definition of the left action {$\rhd$} {\seqDDDane}, we get \MMath{\mathbf{y}  = \sigma ^{-1} \,  \mathbf{x} '}, hence \MMath{\iota _{{\lambda _{2} \leftarrow  \lambda _{1}}}\big( \mathcal{O} ^{\mathfrak{bos} }_{{\lambda _{1}}}(\mathbf{v} ) \big) = \iota _{\mu } \big( \mathcal{O} ^{\mathfrak{bos} }_{{\lambda _{1}}}(\mathbf{v} ) \big) = \mathcal{O} ^{\mathfrak{bos} }_{{\lambda _{2}}}(\mathbf{v} )}.%
\Mpar \vspace{\Sssaut}\hspace{\Alinea}Let \MMath{\mathbf{v} ,\mathbf{w}  \in  V } and \MMath{\lambda  = \left(\mathbf{e} _{1},\dots ,\mathbf{e} _{2n}\right) \in  \mathcal{L} ^{\mathfrak{bos} }_{}} such that \MMath{\mathbf{v} ,\mathbf{w}  \in  V _{\lambda }}. Let \MMath{\mathbf{x} ,\mathbf{y}  \in  \mathds{R}^{2n}} such that \MMath{\mathbf{v}  = \sum_{i=1}^{2n} \mathbf{x} _{i} \mathbf{e} _{i}} and \MMath{\mathbf{w}  = \sum_{i=1}^{2n} \mathbf{y} _{i} \mathbf{e} _{i}}. Since \MMath{\left(\mathbf{e} _{1},\dots ,\mathbf{e} _{2n}\right)} is a symplectic family, \MMath{\Omega (\mathbf{v} ,\mathbf{w} ) = {}^{\text{\sc t}} \mathbf{x}  \,  \Omega ^{(n)} \,  \mathbf{y} }, so {\seqBBBang} follows from {\seqBBBanh}.%
\MendOfProof \Mpar \MstartPhyMode\hspace{\Alinea}In the fermionic case, the classical linear observables are written as \MMath{\left(\mathbf{v}  \middlewithspace| \, \cdot \,  \right)}, and since they will be represented by \emph{bounded} quantum operators (all \MMath{\mathcal{H} ^{\mathfrak{ferm} }_{\lambda }} are finite-dimensional, as fermionic Fock spaces over finitely many modes), we do not need to exponentiate them.%
\Mpar \vspace{\Saut}\hspace{\Alinea}In order for the corresponding quantum observables to be well-defined as elements of \MMath{\mathcal{B} ^{\mathfrak{ferm} }} {\seqDDDaex}, it is necessary to adopt a particular choice of sign conventions in the definition of the fermionic creation and annihilation operators {\seqDDDanj} and of the tensor-product factorization of fermionic Fock spaces {\seqDDDami}.
On the other hand, one could argue that fermionic observables are not directly \emph{measurable} anyway, only \emph{bosonic} observables are (including composite ones, eg.~even products of elementary fermionic observables). Hence, only the latter should be expected to belong to \MMath{\mathcal{B} ^{\mathfrak{ferm} }}, which they do regardless of the choice of conventions.%
\MleavePhyMode \hypertarget{PARank}{}\Mpar \Mdefinition{3.9}Let \MMath{\mathbf{v}  \in  V } and let \MMath{\lambda  = \left(\mathbf{e} _{1},\dots ,\mathbf{e} _{2n}\right) \in  \mathcal{L} ^{\mathfrak{ferm} }_{}} such that \MMath{\mathbf{v}  \in  V _{\lambda }}. Let \MMath{\mathbf{x}  \in  \mathds{R}^{2n}} such that \MMath{\mathbf{v}  = \sum_{i=1}^{2n} \mathbf{x} _{i} \mathbf{e} _{i}}. We define a bounded operator \MMath{\mathcal{O} ^{\mathfrak{ferm} }_{\lambda }(\mathbf{v} ) \in  \mathcal{B} ^{\mathfrak{ferm} }_{\lambda }} by \MMath{\mathcal{O} ^{\mathfrak{ferm} }_{\lambda }(\mathbf{v} ) \mathrel{\mathop:}=  \hat{\mathbf{x} }} (where \MMath{\hat{\mathbf{x} }} is the essentially self-adjoint operator on \MMath{\mathcal{H} ^{\mathfrak{ferm} }_{\lambda }} defined in {\seqBBBanl}).%
\hypertarget{PARanm}{}\Mpar \Mproposition{3.10}Let \MMath{\mathbf{v}  \in  V }, \MMath{\lambda _{1} \in  \mathcal{L} ^{\mathfrak{ferm} }_{}} such that \MMath{\mathbf{v}  \in  V _{{\lambda _{1}}}}, and \MMath{\lambda _{2} \geqslant  \lambda _{1}}. Then, \MMath{\mathcal{O} ^{\mathfrak{ferm} }_{{\lambda _{2}}}(\mathbf{v} ) = \iota _{{\lambda _{2} \leftarrow  \lambda _{1}}} \big( \mathcal{O} ^{\mathfrak{ferm} }_{{\lambda _{1}}}(\mathbf{v} ) \big)}. Thus, we can associate to any \MMath{\mathbf{v}  \in  V } an observable \MMath{\mathcal{O} ^{\mathfrak{ferm} }(\mathbf{v} ) \in  \mathcal{B} ^{\mathfrak{ferm} }}.%
\Mpar \vspace{\Sssaut}\hspace{\Alinea}Let \MMath{\mathbf{v} ,\mathbf{w}  \in  V } and \MMath{\lambda  \in  \mathcal{L} ^{\mathfrak{ferm} }_{}} such that \MMath{\mathbf{v} ,\mathbf{w}  \in  V _{\lambda }}. We have:%
\hypertarget{PARann}{}\Mpar \MStartEqua \MMath{\left[ \mathcal{O} ^{\mathfrak{ferm} }_{\lambda }(\mathbf{v} ), \mathcal{O} ^{\mathfrak{ferm} }_{\lambda }(\mathbf{w} ) \right]_{+} = \left( \mathbf{v}  \middlewithspace| \mathbf{w}  \right) \mathds{1} ^{\mathfrak{bos} }_{\lambda }},%
\NumeroteEqua{3.10}{1}\MStopEqua \Mpar where \MMath{\left[ \, \cdot \, , \, \cdot \,  \right]_{+}} denotes the anti-commutator. Thus, \MMath{\left[ \mathcal{O} ^{\mathfrak{ferm} }(\mathbf{v} ), \mathcal{O} ^{\mathfrak{ferm} }(\mathbf{w} ) \right]_{+} = \left( \mathbf{v}  \middlewithspace| \mathbf{w}  \right) \mathds{1} ^{\mathfrak{bos} }}.%
\Mpar \Mproof The proof is the same as the one of {\seqBBBanq} using {\seqBBBanr}, {\seqBBBans} and {\seqBBBant} in place of the corresponding bosonic results.%
\MendOfProof \hypertarget{SECanu}{}\MsectionB{anu}{3.2}{Universal Label Subsets on a Separable Normable Vector Space}%
\vspace{\PsectionB}\Mpar \MstartPhyMode\hspace{\Alinea}We have a natural action of the group of automorphisms of {$V$} (symplectomorphisms, resp.~isometries) on the quantum state space constructed in {\seqBBBacm}, which in particular confirms its polarization-independence.%
\Mpar \vspace{\Saut}\hspace{\Alinea}Note that, in the definition below, the maps \MMath{U ^{\grpAct{\text{m}}}_{\lambda }} between the Hilbert spaces are all trivial, because the hard work has \emph{already} been done when constructing the system of factorized Hilbert spaces {\seqDDDanw}: the actual unitary transformations relating the Fock spaces on different symplectic frames have been \emph{hard-wired} in the various arrows, underpinning the \emph{physical interpretation} of the labels, on which we can now rely in a transparent way.%
\MleavePhyMode \hypertarget{PARanx}{}\Mpar \Mdefinition{3.11}Let \MMath{\mathcal{A} ^{\mathfrak{bos} } \mathrel{\mathop:}=  \mathcal{A} (V , \Omega )} {\seqDDDany}, resp.~\MMath{\mathcal{A} ^{\mathfrak{ferm} } \mathrel{\mathop:}=  \mathcal{A} _{o}(V , \left(\, \cdot \, \middlewithspace|\, \cdot \, \right))} {\seqDDDanz}.
Let \MMath{\mathfrak{lin}  \in  \left\{ \mathfrak{bos} , \mathfrak{ferm}  \right\}}.
For any \MMath{\text{m} \in  \mathcal{A} ^{\mathfrak{lin} }}, we define \MMath{\grpAct{\text{m}} \in  \mathcal{A} ^{\mathfrak{lin} }} {\seqDDDahm} by:%
\Mpar \MList{1}for any \MMath{\lambda  = \left(\mathbf{e} _{1},\dots ,\mathbf{e} _{2n}\right) \in  \mathcal{L} ^{\mathfrak{lin} }_{}}, \MMath{\grpAct{\text{m}} \lambda  \mathrel{\mathop:}=  \left(\text{m} \mathbf{e} _{1},\dots ,\text{m} \mathbf{e} _{2n}\right)};%
\MStopList \Mpar \MList{2}for any \MMath{\lambda  \in  \mathcal{L} ^{\mathfrak{lin} }_{}}, \MMath{U ^{\grpAct{\text{m}}}_{\lambda } \mathrel{\mathop:}=  \text{id}_{{\mathcal{F} ^{(d _{\lambda })}_{\mathfrak{lin} }}} : \mathcal{H} ^{\mathfrak{lin} }_{\lambda } \rightarrow  \mathcal{H} ^{\mathfrak{lin} }_{{\grpAct{\text{m}} \lambda }}} (note that \MMath{d _{\lambda } = d _{{\grpAct{\text{m}} \lambda }}}).%
\MStopList \hypertarget{PARaoc}{}\Mpar \Mproposition{3.12}\MMath{\text{m} \mapsto  \grpAct{\text{m}}} is a group action of \MMath{\mathcal{A} ^{\mathfrak{lin} }} on the system of factorized Hilbert spaces {$\mathfrak{lin}$} {\seqDDDalg}.%
\Mpar \Mproof For any \MMath{\lambda  \in  \mathcal{L} ^{\mathfrak{lin} }_{}} and any \MMath{\text{m} \in  \mathcal{A} ^{\mathfrak{lin} }}, \MMath{\grpAct{\text{m}} \lambda  \in  \mathcal{L} ^{\mathfrak{lin} }_{}} by definition of \MMath{\mathcal{A} ^{\mathfrak{lin} }}. For any \MMath{\lambda  \in  \mathcal{L} ^{\mathfrak{lin} }_{}} and any \MMath{\text{m},\text{m}' \in  \mathcal{A} ^{\mathfrak{lin} }}, we have \MMath{\grpAct{(\text{m} \,  \text{m}')} \lambda  = \grpAct{\text{m}} (\grpAct{\text{m}'} \lambda )}. In particular, \MMath{\lambda  \mapsto  \grpAct{\text{m}} \lambda } is a bijective map \MMath{\mathcal{L} ^{\mathfrak{lin} }_{} \rightarrow  \mathcal{L} ^{\mathfrak{lin} }_{}}.%
\Mpar \vspace{\Sssaut}\hspace{\Alinea}Let \MMath{\text{m} \in  \mathcal{A} ^{\mathfrak{lin} }}. Since m{} is a linear map on {$V$}, we have, for any \MMath{\lambda  \in  \mathcal{L} ^{\mathfrak{lin} }_{}}, \MMath{V _{{\grpAct{\text{m}} \lambda }} = \text{m} \left\langle  V _{\lambda } \right\rangle } and, for any \MMath{\mu  \in  \text{Mp}^{(d _{\lambda })}}, resp.~\MMath{\text{Spin}^{(d _{\lambda })}}, \MMath{\mu  \rhd  (\grpAct{\text{m}} \lambda ) = \grpAct{\text{m}} (\mu  \rhd  \lambda )}.
Thus, for any \MMath{\lambda _{1}, \lambda _{2} \in  \mathcal{L} ^{\mathfrak{lin} }_{}}, \MMath{\mathcal{L} ^{\mathfrak{lin} }_{{\lambda _{2} \rightarrow  \lambda _{1}}} \subseteq \mathcal{L} ^{\mathfrak{lin} }_{{\grpAct{\text{m}} \lambda _{2} \rightarrow  \grpAct{\text{m}} \lambda _{1}}}}, and applying the same argument to \MMath{\text{m}^{-1}},\MMath{\mathcal{L} ^{\mathfrak{lin} }_{{\lambda _{2} \rightarrow  \lambda _{1}}} = \mathcal{L} ^{\mathfrak{lin} }_{{\grpAct{\text{m}} \lambda _{2} \rightarrow  \grpAct{\text{m}} \lambda _{1}}}}.
This ensures that {\seqBBBaix} holds.
Moreover, for any \MMath{\lambda _{1} \leqslant  \lambda _{2} \in  \mathcal{L} ^{\mathfrak{lin} }_{}}, there exists \MMath{\mu  \in  \mathcal{L} ^{\mathfrak{lin} }_{{\lambda _{2} \rightarrow  \lambda _{1}}} = \mathcal{L} ^{\mathfrak{lin} }_{{\grpAct{\text{m}} \lambda _{2} \rightarrow  \grpAct{\text{m}} \lambda _{1}}}}, and, defining \MMath{U ^{\grpAct{\text{m}}}_{\mu | \mu } = \text{id}_{{\mathcal{F} ^{(d _{{\lambda _{2}}} - d _{{\lambda _{1}}})}_{\mathfrak{lin} }}} : \mathcal{H} ^{\mathfrak{lin} }_{\mu } \rightarrow  \mathcal{H} ^{\mathfrak{lin} }_{\mu }}, we have \MMath{\Phi ^{\mathfrak{lin} }_{\mu } = \left( U ^{\grpAct{\text{m}}}_{{\lambda _{1}}} \otimes  U ^{\grpAct{\text{m}}}_{\mu | \mu } \right) \circ  \Phi ^{\mathfrak{lin} }_{\mu } \circ  U ^{{\grpAct{\text{m}},-1}}_{{\lambda _{2}}}}, so {\seqBBBagm} holds.
In this way, we have checked that \MMath{\grpAct{\text{m}}} is indeed an automorphism of the system of factorized Hilbert spaces {$\mathfrak{lin}$}, for any \MMath{\text{m} \in  \mathcal{A} ^{\mathfrak{lin} }}.%
\Mpar \vspace{\Sssaut}\hspace{\Alinea}Finally, for any \MMath{\text{m}, \text{m}' \in  \mathcal{A} ^{\mathfrak{lin} }} and any \MMath{\lambda  \in  \mathcal{L} ^{\mathfrak{lin} }_{}}, \MMath{U ^{\grpAct{(\text{m} \,  \text{m}')}}_{\lambda } = U ^{\grpAct{\text{m}}}_{{\grpAct{\text{m}'} \lambda }} \circ  U ^{\grpAct{\text{m}'}}_{\lambda }}, so \MMath{\text{m} \mapsto  \grpAct{\text{m}}} is a group action.%
\MendOfProof \Mpar \MstartPhyMode\hspace{\Alinea}To check the validity of the previously defined action, we control that the quantum observables transform as they should (in accordance with the discussion in {\seqBBBaof}, {\seqBBBaog} leads to automorphisms acting directly on observables, and inversely on states).%
\MleavePhyMode \hypertarget{PARaoh}{}\Mpar \Mproposition{3.13}For any \MMath{\mathbf{v}  \in  V } and any \MMath{\text{m} \in  \mathcal{A} ^{\mathfrak{lin} }}, \MMath{\Lambda ^{\grpAct{\text{m}}} \left( \mathcal{O} ^{\mathfrak{lin} }(\mathbf{v} ) \right) = \mathcal{O} ^{\mathfrak{lin} }(\text{m} \mathbf{v} )} (where \MMath{\Lambda ^{\grpAct{\text{m}}}} is the action induced on \MMath{\mathcal{B} ^{\mathfrak{lin} }} by the automorphism \MMath{\grpAct{\text{m}}}, see {\seqBBBaoi}).%
\Mpar \Mproof This follows from the definition of \MMath{\grpAct{\text{m}}} {\seqDDDaog} and \MMath{\mathcal{O} ^{\mathfrak{lin} }(\mathbf{v} )} ({\seqBBBaok}, resp.~{\seqBBBaol}).%
\MendOfProof \Mpar \MstartPhyMode\hspace{\Alinea}We now want to apply the prescription from {\seqBBBafq}, to extract a \emph{dense} subset of \MMath{\mathcal{L} ^{\mathfrak{lin} }_{}}, over which projective states can be constructed \emph{explicitly} and \emph{systematically}.
The natural choice for the group of symmetries {$\mathcal{T}$} is to simply take the \emph{full} group of automorphisms of {$V$}: by {\seqBBBaon}, this will in particular ensure that any automorphism of {$V$} can be (approximately) implemented on the resulting state space. Moreover, since in the present {\seqBBBabe} the classical phase space {$V$} is kept completely \emph{general}, we do not have, beyond its symplectic or inner-product structure, any further insight from the physics of the problem to pick out a more specific {$\mathcal{T}$}.%
\Mpar \vspace{\Saut}\hspace{\Alinea}Next, we need a \emph{topology} on {$\mathcal{T}$}. In the fermionic case, we get it for free, since the classical inner-product defines a \emph{norm} on {$V$}, and thus a topology on its group of isometries {\seqDDDaoo}.
In the bosonic case, we demand a topology that is \emph{normable} and \emph{compatible} with the symplectic structure {$\Omega$}, as developed in {\seqBBBaop}. This is an \emph{additional} structure on {$V$}, which will have to come from our understanding of the particular theory at hand (see the detailed example worked out in {\seqBBBabn}).
It is important to stress that selecting a \emph{normable} topology on {$V$} is \emph{not} the same as selecting a \emph{specific} norm: a choice of the latter would, in some cases, boil down to the choice of a \emph{polarization} {\seqDDDaoq}. By contrast, by only relying on the \emph{induced topology}, we ensure that the construction will be invariant under a much larger class of transformations (at first glance, homeomorphisms rather than isometries; see {\seqBBBabn}, in particular {\seqBBBabp}, for a more refined analysis of this question).%
\Mpar \vspace{\Saut}\hspace{\Alinea}With this setup in place, we obtain a \emph{one-to-one correspondence} between universal label subsets for the {$\mathfrak{lin}$} projective structure of the previous subsection and \emph{dense}, \emph{countably-generated} subspaces in {$V$}.
This result is very satisfactory, in the sense that it fits exactly with the physical picture that the notion of universal label subsets was meant to formalize (as motivated in {\seqBBBafq}): as far as the derivation of actual predictions for real-world experiments is concerned, it is perfectly sufficient to only characterize the quantum states over a \emph{dense sub-algebra of observables} (keeping in mind that the space of "good" linear observables can be identified with {$V$} itself, via the symplectic structure, resp.~inner-product).%
\MleavePhyMode \Mpar \vspace{\Saut}\hspace{\Alinea}For the rest of this subsection, we assume that {$V$} is equipped with a \emph{normable} topology, which is compatible with {$\Omega$} in the sense of {\seqBBBaos}, resp.~that {$V$} is equipped with the topology induced by the real scalar product \MMath{\left(\, \cdot \, \middlewithspace|\, \cdot \, \right)}.%
\hypertarget{PARaot}{}\Mpar \Mtheorem{3.14}Let \MMath{\mathcal{A} ^{\mathfrak{lin} }_{o} \subseteq \mathcal{A} ^{\mathfrak{lin} }} be the topological group \MMath{\mathcal{A} _{o}(V , \Omega )} {\seqDDDaou}, resp.~\MMath{\mathcal{A} _{o}(V , \left(\, \cdot \, \middlewithspace|\, \cdot \, \right))} {\seqDDDanz}.
\MMath{\mathcal{A} ^{\mathfrak{lin} }_{o}} acts on {$\mathfrak{lin}$} through the restriction of the action of \MMath{\mathcal{A} ^{\mathfrak{lin} }}.%
\Mpar \vspace{\Sssaut}\hspace{\Alinea}Let {$D$} be a \emph{dense}, \emph{countably generated}, subspace of {$V$}.
Let \MMath{\mathcal{K} _{D } \mathrel{\mathop:}=  \left\{ \kappa  \in  \mathcal{L} ^{\mathfrak{lin} }_{} \middlewithspace| V _{\kappa } \subseteq D  \right\}}. Then, \MMath{\mathcal{K} _{D }} is a \MMath{\mathcal{A} ^{\mathfrak{lin} }_{o}}-universal label subset for {$\mathfrak{lin}$} {\seqDDDaht}.%
\Mpar \vspace{\Sssaut}\hspace{\Alinea}Reciprocally, if {$\mathcal{K}$} is a \MMath{\mathcal{A} ^{\mathfrak{lin} }_{o}}-universal label subset for {$\mathfrak{lin}$}, \MMath{D  \mathrel{\mathop:}=  \bigcup_{\kappa \in \mathcal{K} } V _{\kappa }} is a dense, countably generated subspace of {$V$} and \MMath{\mathcal{K}  = \mathcal{K} _{D }}.%
\Mpar \vspace{\Sssaut}\hspace{\Alinea}In particular, {$\mathfrak{lin}$} admits \MMath{\mathcal{A} ^{\mathfrak{lin} }_{o}}-universal label subsets if, and only if, {$V$} is separable.%
\Mpar \Mproof \italique{\MMath{\mathcal{K} _{D }} is a \MMath{\mathcal{A} ^{\mathfrak{lin} }_{o}}-universal label subset.}
{$D$} is a symplectic vector space (as stressed at the beginning of {\seqBBBaow}), resp.~a real, infinite-dimensional pre-Hilbert space (since it is dense in {$V$}, which is infinite-dimensional). Hence, using the characterization of the order in \MMath{\mathcal{L} ^{\mathfrak{lin} }_{}} (from {\seqBBBalc}, resp.~{\seqBBBalu}) together with {\seqBBBakh}, resp.~{\seqBBBaly}, \MMath{\mathcal{K} _{D }} is directed and of countable cofinality {\seqDDDaio}.
Moreover, the characterization of the order also implies that \MMath{\mathcal{K} _{D }} is a lower set of \MMath{\mathcal{L} ^{\mathfrak{lin} }_{}} {\seqDDDajb}.%
\Mpar \vspace{\Sssaut}\hspace{\Alinea}Let \MMath{\left|\, \cdot \, \right|} be a norm on {$V$} inducing its topology (hence making it a normed symplectic vector space, see {\seqBBBaox}), resp.~the norm induced by the scalar product \MMath{\left(\, \cdot \, \middlewithspace|\, \cdot \, \right)}. Let {$\mathcal{V}$} be a neighborhood of \MMath{\text{id}_{V }} in \MMath{\mathcal{A} ^{\mathfrak{lin} }_{o}} and let \MMath{\epsilon  > 0} be such that \MMath{\forall  \text{m} \in  \mathcal{A} ^{\mathfrak{lin} }_{o}, \left\| \text{m} - \text{id}_{V } \right\| < \epsilon  \Rightarrow  \text{m} \in  \mathcal{V} }.
Let \MMath{\lambda  \in  \mathcal{L} ^{\mathfrak{lin} }_{}}. Using {\seqBBBaoy}, resp.~{\seqBBBaoz}, there exists \MMath{\text{m} \in  \mathcal{A} ^{\mathfrak{lin} }_{o}} such that \MMath{V _{{\grpAct{\text{m}} \lambda }} = \text{m} \left\langle  V _{\lambda } \right\rangle  \subseteq D }, \MMath{\left. \text{m} \right|_{{V _{\lambda } \cap  D }} = \left. \text{id}_{V } \right|_{{V _{\lambda } \cap  D }}} and \MMath{\left\| \text{m} - \text{id}_{V } \right\| < \epsilon }. 
Thus, \MMath{\grpAct{\text{m}} \lambda  \in  \mathcal{K} _{D }}. Moreover, for any \MMath{\kappa  \in  \mathcal{K} _{D }} such that \MMath{\kappa  \leqslant  \lambda }, \MMath{V _{\kappa } \subseteq V _{\lambda } \cap  D }, so \MMath{\grpAct{\text{m}} \kappa  = \kappa }, and, by definition of \MMath{\grpAct{\text{m}}}, \MMath{U ^{\grpAct{\text{m}}}_{\kappa } = \mathds{1} ^{\mathfrak{lin} }_{\kappa }}.
In other words, \MMath{\mathcal{K} _{D }} has the property of universal label subsets {\seqDDDahy}.%
\Mpar \vspace{\Ssaut}\italique{\MMath{\mathcal{K}  = \mathcal{K} _{D }} with \MMath{D  = {\textstyle \bigcup_{\kappa \in \mathcal{K} } V _{\kappa }}}.} Let {$\mathcal{K}$} be a \MMath{\mathcal{A} ^{\mathfrak{lin} }_{o}}-universal label subset for {$\mathfrak{lin}$} and let \MMath{D  = {\textstyle \bigcup_{\kappa \in \mathcal{K} } V _{\kappa }}}. Since {$\mathcal{K}$} is directed {\seqDDDaio}, {$D$} is a vector subspace of {$V$}.
Let \MMath{\left(\kappa _{n}\right)_{n\in \mathds{N}}} be a sequence in {$\mathcal{K}$} such that \MMath{\forall  \kappa  \in  \mathcal{K} , \exists  n \in  \mathds{N} \mathrel{\big/}  \kappa  \leqslant  \kappa _{n}} {\seqDDDaio}. Then, \MMath{D  = {\textstyle \bigcup_{n\in \mathds{N}} V _{{\kappa _{n}}}}}. Since each \MMath{V _{{\kappa _{n}}}} is finite-dimensional, {$D$} is countably generated.
Let \MMath{\mathbf{v}  \in  V } and \MMath{\epsilon  > 0}. Using {\seqBBBakh}, resp.~{\seqBBBaly}, there exists \MMath{\lambda  \in  \mathcal{L} ^{\mathfrak{lin} }_{}} such that \MMath{\mathbf{v}  \in  V _{\lambda }}. Let \MMath{\text{m} \in  \mathcal{A} ^{\mathfrak{lin} }_{o}}, with \MMath{\left\| \text{m} - \text{id}_{V } \right\| < \nicefrac{\epsilon }{\left|\mathbf{v} \right|+1}}, such that \MMath{\grpAct{\text{m}} \lambda  \in  \mathcal{K} } {\seqDDDahy}, and let \MMath{\mathbf{w}  \mathrel{\mathop:}=  \text{m} \mathbf{v} }. We have \MMath{\left| \mathbf{w}  - \mathbf{v}  \right| < \epsilon } and \MMath{\mathbf{w}  \in  D }, hence {$D$} is dense in {$V$}.%
\Mpar \vspace{\Sssaut}\hspace{\Alinea}By definition of {$D$}, \MMath{\mathcal{K}  \subseteq \mathcal{K} _{D }}. Let \MMath{\lambda  \in  \mathcal{K} _{D }}. Let \MMath{F  = V _{\lambda } \subseteq D } (if \MMath{\mathfrak{lin}  = \mathfrak{bos} }), resp.~let {$F$} be a finite dimensional subset of {$D$} such that \MMath{V _{\lambda } \subsetneq F } (if \MMath{\mathfrak{lin}  = \mathfrak{ferm} }, using that {$D$} is infinite dimensional, see above). Using the directedness of {$\mathcal{K}$}, there exists \MMath{\kappa  \in  \mathcal{K} } such that \MMath{F  \subseteq V _{\kappa }}, which implies \MMath{\lambda  \leqslant  \kappa }, hence \MMath{\lambda  \in  \mathcal{K} } {\seqDDDajb}. Thus, \MMath{\mathcal{K}  = \mathcal{K} _{D }}.%
\MendOfProof \Mpar \MstartPhyMode\hspace{\Alinea}In the particular class of theories we are considering here, we can slightly \emph{improve} over the universality and stability results from {\seqBBBafq} ({\seqBBBapb} respectively). As underlined in the discussion before {\seqBBBabf}, the small deformations involved in these results are a priori not themselves \emph{elements} of {$\mathcal{T}$}. Rather, they are transformations which \emph{coincide}, on each label {$\kappa$}, with a certain element \MMath{\text{{\sc t}}_{\kappa }} of {$\mathcal{T}$}, but the latter may be {$\kappa$}-\emph{dependent}.
What the proposition below shows is that, in the case of universal label subsets of \MMath{\mathcal{L} ^{\mathfrak{lin} }_{}}, \MMath{\text{{\sc t}}_{\kappa }} can, in fact, be chosen to be \emph{independent} of {$\kappa$}.%
\Mpar \vspace{\Saut}\hspace{\Alinea}In other words, any two dense, countably-generated subspaces of {$V$} can be \emph{mapped into each other} by an automorphism of {$V$}, and we can arrange for this automorphism to be \emph{arbitrarily close} to the identity map on {$V$}: this gives a transparent explanation why physical predictions will never depend on the specific dense subset on which we happen to be working.
Similarly, not only can any element of \MMath{\mathcal{A} ^{\mathfrak{lin} }_{o}} (aka.~automorphism of {$V$}) be approximated over any given universal label subset, but this approximation can be chosen to belong to a certain \emph{subgroup} of \MMath{\mathcal{A} ^{\mathfrak{lin} }_{o}} (namely the subgroup consisting of those automorphisms which stabilize the given dense subspace).%
\MleavePhyMode \hypertarget{PARapc}{}\Mpar \Mproposition{3.15}For any \MMath{D , D '} dense, countably generated subspaces of {$V$}, and any neighborhood {$\mathcal{V}$} of \MMath{\text{id}_{V }} in \MMath{\mathcal{A} ^{\mathfrak{lin} }_{o}}, there exists \MMath{\text{m} \in  \mathcal{V} } such that \MMath{\text{m} \left\langle  D  \right\rangle  = D '} (hence \MMath{\grpAct{\text{m}} \left\langle \mathcal{K} _{D }\right\rangle  = \mathcal{K} _{D '}}).%
\Mpar \vspace{\Sssaut}\hspace{\Alinea}Moreover, for any dense, countably generated subspace {$D$} of {$V$}, the subgroup \MMath{\mathcal{A} ^{\mathfrak{lin} }_{o}(D )} of \MMath{\mathcal{A} ^{\mathfrak{lin} }_{o}} defined by:%
\Mpar \MStartEqua \MMath{\forall  \text{m} \in  \mathcal{A} ^{\mathfrak{lin} }_{o}, {\text{m} \in  \mathcal{A} ^{\mathfrak{lin} }_{o}(D ) \Leftrightarrow  \text{m} \left\langle D \right\rangle  = D  \Leftrightarrow  \grpAct{\text{m}} \left\langle \mathcal{K} _{D }\right\rangle  = \mathcal{K} _{D }}},%
\MStopEqua \Mpar is dense in \MMath{\mathcal{A} ^{\mathfrak{lin} }_{o}}.%
\Mpar \Mproof \italique{Mapping a dense subspace into another.} Let \MMath{D ,D '} be two dense, countably generated subspaces of {$V$}, and let {$\mathcal{V}$} be a neighborhood of \MMath{\text{id}_{V }} in \MMath{\mathcal{A} ^{\mathfrak{lin} }_{o}}. Let \MMath{\epsilon  \in  \left]0,\nicefrac{1}{2}\right[} such that, \MMath{\forall  \text{m} \in  \mathcal{A} ^{\mathfrak{lin} }_{o}, {\left\|\text{m} - \text{id}_{V }\right\| \leqslant  \epsilon  \Rightarrow  \text{m} \in  \mathcal{V} }}. Applying {\seqBBBabf} to the \MMath{\mathcal{A} ^{\mathfrak{lin} }_{o}}-universal label subsets \MMath{\mathcal{K} _{D }} and \MMath{\mathcal{K} _{D '}}, let {$\tau$} be an isomorphism from the restriction of {$\mathfrak{lin}$} on \MMath{\mathcal{K} _{D }} to its restriction on \MMath{\mathcal{K} _{D '}}, and, for any \MMath{\kappa  \in  \mathcal{K} _{D }}, let \MMath{\text{m}_{\kappa } \in  \mathcal{A} ^{\mathfrak{lin} }_{o}} such that \MMath{\left\|\text{m}_{\kappa } -\text{id}_{V }\right\| \leqslant  \epsilon }, \MMath{\tau \kappa  = \grpAct{{\text{m}_{\kappa }}} \kappa } and \MMath{U ^{\tau }_{\kappa } = U ^{\grpAct{{\text{m}_{\kappa }}}}_{\kappa } = \mathds{1} ^{\mathfrak{lin} }_{\kappa }}.%
\Mpar \vspace{\Sssaut}\hspace{\Alinea}Let \MMath{\mathbf{v}  \in  D }. Using {\seqBBBakh}, resp.~{\seqBBBaly}, in the symplectic vector space, resp.~real, infinite-dimensional pre-Hilbert space, {$D$}, there exists \MMath{\kappa  \in  \mathcal{K} _{D }} such that \MMath{\mathbf{v}  \in  V _{\kappa }}. For any \MMath{\kappa  \in  \mathcal{K} _{D }} such that \MMath{\mathbf{v}  \in  V _{\kappa }}, we have:%
\Mpar \MStartEqua \MMath{\forall  t \in  \mathds{R}, {\Lambda ^{\tau } \big( \mathcal{O} ^{\mathfrak{lin} }(t\mathbf{v} ) \big) = \left[ \Lambda ^{\tau }_{\kappa } \big( \mathcal{O} ^{\mathfrak{lin} }_{\kappa }(t\mathbf{v} ) \big) \right]^{\mathfrak{lin} } = \left[ \mathcal{O} ^{\mathfrak{lin} }_{{\grpAct{{\text{m}_{\kappa }}}\kappa }}(t \,  \text{m}_{\kappa } \mathbf{v} ) \right]^{\mathfrak{lin} }}}.%
\MStopEqua \Mpar Moreover, if \MMath{\kappa ,\kappa ' \in  \mathcal{K} _{D }} are such that \MMath{\mathbf{v}  \in  V _{\kappa }, V _{\kappa '}}, there exists \MMath{\kappa '' \in  \mathcal{K} _{D }} such that \MMath{\kappa ,\kappa ' \leqslant  \kappa ''} (for \MMath{\mathcal{K} _{D }} is directed), hence \MMath{\grpAct{{\text{m}_{\kappa }}}\kappa  = \tau \kappa , \grpAct{{\text{m}_{\kappa '}}}\kappa ' = \tau \kappa ' \leqslant  \tau \kappa ''} {\seqDDDaix}. Then, from:%
\Mpar \MStartEqua \MMath{\forall  t \in  \mathds{R}, {\left[ \mathcal{O} ^{\mathfrak{lin} }_{{\grpAct{{\text{m}_{\kappa }}}\kappa }}(t \,  \text{m}_{\kappa } \mathbf{v} ) \right]^{\mathfrak{lin} } = \left[ \mathcal{O} ^{\mathfrak{lin} }_{{\grpAct{{\text{m}_{\kappa '}}}\kappa '}}(t \,  \text{m}_{\kappa '} \mathbf{v} ) \right]^{\mathfrak{lin} }}},%
\MStopEqua \Mpar we get:%
\Mpar \MStartEqua \MMath{\forall  t \in  \mathds{R}, {\mathcal{O} ^{\mathfrak{lin} }_{{\tau \kappa ''}}(t \,  \text{m}_{\kappa } \mathbf{v} ) = \mathcal{O} ^{\mathfrak{lin} }_{{\tau \kappa ''}}(t \,  \text{m}_{\kappa '} \mathbf{v} )}},%
\MStopEqua \Mpar which implies \MMath{\text{m}_{\kappa } \mathbf{v}  = \text{m}_{\kappa '} \mathbf{v} } (as follows from the definition of \MMath{\mathcal{O} ^{\mathfrak{lin} }_{\lambda }(\mathbf{v} )} in {\seqBBBapm}, resp.~{\seqBBBapn}). Therefore, for any \MMath{\mathbf{v}  \in  D }, we can define \MMath{\text{m} \mathbf{v}  \mathrel{\mathop:}=  \text{m}_{\kappa } \mathbf{v} }, for some \MMath{\kappa  \in  \mathcal{K} _{D }} such that \MMath{\mathbf{v}  \in  V _{\kappa }}, and we have \MMath{\text{m} \mathbf{v}  \in  V _{\tau \kappa } \subseteq D '}. Since, for any \MMath{\kappa  \in  \mathcal{K} _{D }}, \MMath{\left\| \text{m}_{\kappa } - \text{id}_{V } \right\| \leqslant  \epsilon }, we have, for any \MMath{\mathbf{v}  \in  D }, \MMath{\left|\text{m} \mathbf{v}  - \mathbf{v} \right| \leqslant  \epsilon  \,  \left|\mathbf{v} \right|}.
Let \MMath{\mathbf{v} ,\mathbf{w}  \in  D }. Using again {\seqBBBakh}, resp.~{\seqBBBaly}, there exists \MMath{\kappa  \in  \mathcal{K} _{D }}, such that \MMath{\mathbf{v} ,\mathbf{w}  \in  V _{\kappa }}. Hence, \MMath{\text{m} \mathbf{v}  = \text{m}_{\kappa } \mathbf{v} } and \MMath{\text{m} \mathbf{w}  = \text{m}_{\kappa } \mathbf{w} }, with \MMath{\text{m}_{\kappa } \in  \mathcal{A} ^{\mathfrak{lin} }_{o}}. Therefore, m{} is a linear map \MMath{D  \rightarrow  D '}, and a symplectomorphism, resp.~an orthogonal transformation.
Let \MMath{\mathbf{v} ' \in  D '}. There exists \MMath{\kappa ' \in  \mathcal{K} _{D '}} such that \MMath{\mathbf{v} ' \in  V _{\kappa '}}, and there exists \MMath{\kappa  \in  \mathcal{K} _{D }} such that \MMath{\tau \kappa  = \kappa '} {\seqDDDapo}. Thus, \MMath{\mathbf{v}  \mathrel{\mathop:}=  \text{m}_{\kappa }^{-1} \mathbf{v} ' \in  V _{\kappa }}, so \MMath{\text{m} \mathbf{v}  = \mathbf{v} '}. This proves that \MMath{\text{m} \left\langle  D  \right\rangle  = D '}.%
\Mpar \vspace{\Sssaut}\hspace{\Alinea}Since m{} is bounded on the dense subspace {$D$}, it can be extended to a linear map \MMath{\tilde{\text{m}} : V  \rightarrow  V }, with \MMath{\tilde{\text{m}} \left\langle D \right\rangle  = D '}. Moreover, {$\Omega$}, resp.~\MMath{\left(\, \cdot \, \middlewithspace|\, \cdot \, \right)}, is continuous with respect to \MMath{\left|\, \cdot \, \right|} {\seqDDDapq}, so \MMath{\tilde{\text{m}}} is a symplectomorphism, resp.~an orthogonal transformation. Finally, \MMath{\left\| \tilde{\text{m}} - \text{id}_{V } \right\| \leqslant  \epsilon } with \MMath{\epsilon  < \nicefrac{1}{2}}, so \MMath{\tilde{\text{m}}} is a bijective, bi-continuous, map \MMath{V  \rightarrow  V } and \MMath{\tilde{\text{m}} \in  \mathcal{V} }.%
\Mpar \vspace{\Ssaut}\italique{Group of automorphisms stabilizing a dense subspace.} Let {$D$} be a dense, countably generated subspace of {$V$}. Let \MMath{\text{m} \in  \mathcal{A} ^{\mathfrak{lin} }_{o}} and let {$\mathcal{V}$} be a neighborhood of m{} in \MMath{\mathcal{A} ^{\mathfrak{lin} }_{o}}. Let \MMath{D ' \mathrel{\mathop:}=  \text{m} \left\langle  D  \right\rangle }. \MMath{D '} is dense in {$V$} (for {$D$} is and m{} is bijective and bi-continuous) and countably generated (for {$D$} is and m{} is linear). Let \MMath{\tilde{\mathcal{V} } \mathrel{\mathop:}=  \left\{ \tilde{\text{m}} \in  \mathcal{A} ^{\mathfrak{lin} }_{o} \middlewithspace| \tilde{\text{m}}^{-1} \,  \text{m} \in  \mathcal{V}  \right\}}. \MMath{\tilde{\mathcal{V} }} is a neighborhood of \MMath{\text{id}_{V }} in \MMath{\mathcal{A} ^{\mathfrak{lin} }_{o}}. Hence, there exists \MMath{\tilde{\text{m}} \in  \tilde{\mathcal{V} }} such that \MMath{\tilde{\text{m}} \left\langle  D  \right\rangle  = D '}. Thus, \MMath{\tilde{\text{m}}^{-1} \,  \text{m} \in  \mathcal{V}  \cap  \mathcal{A} ^{\mathfrak{lin} }_{o}(D )}. This proves that \MMath{\mathcal{A} ^{\mathfrak{lin} }_{o}(D )} is dense in \MMath{\mathcal{A} ^{\mathfrak{lin} }_{o}}.%
\MendOfProof \hypertarget{SECaps}{}\MsectionB{aps}{3.3}{Embedding of Fock Representations}%
\vspace{\PsectionB}\Mpar \MstartPhyMode\hspace{\Alinea}Through the choice of a \emph{complex polarization} {$I$} (aka.~complex structure, see {\seqBBBapu}), {$V$} can be turned into a \emph{complex} Hilbert space.
The latter can be taken as the 1-particle Hilbert space on which to build a Fock space \MMath{\mathcal{F} ^{(V ,I )}_{\mathfrak{lin} }}. This  provides a representation of the \emph{infinite-dimensional algebra} of linear observables of {$V$} {\seqDDDapv}.
Thus, density matrices on \MMath{\mathcal{F} ^{(V ,I )}_{\mathfrak{lin} }} yield \emph{algebraic states} \bseqHHHaaa{part III, def.~2.2.8} on this algebra, and so do the projective states from {\seqBBBacm} (as we mentioned before {\seqBBBafh}).
Forgetting for a moment about the refinement of the construction advocated in {\seqBBBapw}, we ask how these two spaces of quantum states -- the one arising from a Fock representation and the one built over the "unrestricted" projective structure from {\seqBBBacm} -- are related.
Specifically, we will exhibit an \emph{injective} (aka.\ one-to-one) \emph{embedding} of the former into the latter.%
\Mpar \vspace{\Saut}\hspace{\Alinea}The proof comports two steps:%
\Mpar \MPhyList First, we consider the subset \MMath{\mathcal{L} _{I }} of labels which are \emph{compatible} with the complex structure {$I$}, namely the symplectic, resp.~real-orthonormal, families which arise from \emph{complex-orthonormal families}.\footnote{\label{RealPolas}A similar argument could be made to demonstrate that the labels which respect a given \emph{real polarization} also form a cofinal part of \MMath{\mathcal{L} ^{\mathfrak{c} }_{}}. It follows that the projective state space from {\seqBBBacm} can actually be identified with the one originally proposed by Kijowski {\bseqJJJaab}, where the position representation was employed to quantize the partial theories.} Any \emph{real} vector subspace can be (strictly) embedded into a \emph{complex} vector subspace, and any complex vector subspace admits a complex orthonormal basis. Thus, recalling the characterization of the label ordering from {\seqBBBapy}, any label from \MMath{\mathcal{L} ^{\mathfrak{c} }_{}} is \emph{dominated} by a label from \MMath{\mathcal{L} _{I }}.
In other words, \MMath{\mathcal{L} _{I }} forms a \emph{cofinal} part of \MMath{\mathcal{L} ^{\mathfrak{c} }_{}}.
As explained in {\seqBBBafq}, this means that restricting ourselves to \MMath{\mathcal{L} _{I }} does \emph{not} alter the projective state space (rightmost part of {\seqBBBapz}).%
\MStopList \Mpar \MPhyList In a second step, we exploit the relation between projective state spaces and a different construction, namely \emph{inductive limits} of Hilbert spaces {\seqDDDaqb}. Given a label {$\lambda$} in \MMath{\mathcal{L} _{I }}, state vectors in the partial Hilbert space \MMath{\mathcal{H} ^{\mathfrak{lin} }_{\lambda }} can be promoted into \MMath{\mathcal{F} ^{(V ,I )}_{\mathfrak{lin} }} by putting all remaining modes in the {$I$}-vacuum. This allows to view \MMath{\mathcal{F} ^{(V ,I )}_{\mathfrak{lin} }} as an inductive limit over the collection of Hilbert spaces \MMath{\left(\mathcal{H} ^{\mathfrak{lin} }_{\lambda }\right)_{{\lambda  \in  \mathcal{L} _{I }}}}. Density matrices with support only on \MMath{\mathcal{H} ^{\mathfrak{lin} }_{\lambda }} (seen as a vector subspace of \MMath{\mathcal{F} ^{(V ,I )}_{\mathfrak{lin} }}), can easily be transported into the projective state space by padding them with the {$I$}-vacuum {\seqDDDapz}. This transport can then be extended to arbitrary density matrices on \MMath{\mathcal{F} ^{(V ,I )}_{\mathfrak{lin} }} through a limiting process, providing the desired embedding.%
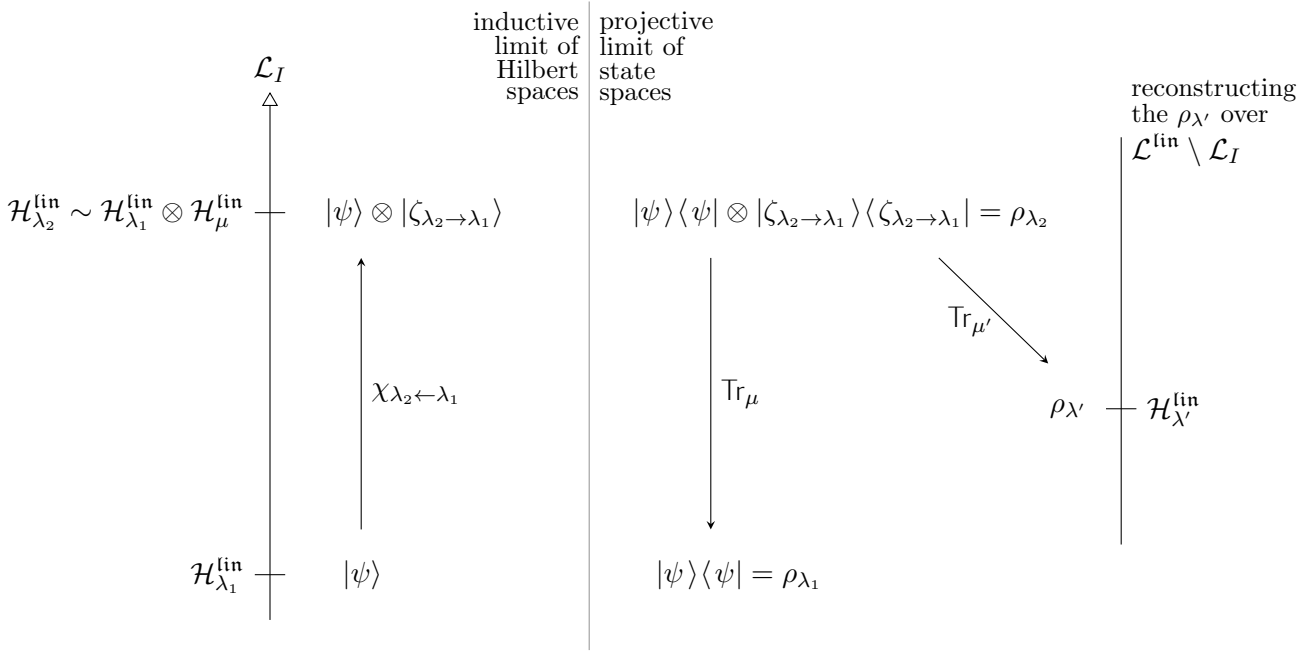
\Myfigure{3.3}{\hypertarget{PARaqc}{}}{Importing states from the Fock space into the projective state space over \MMath{\mathcal{L} _{I }}, before extending them to all \MMath{\mathcal{L} ^{\mathfrak{c} }_{}}}{{\InputFigPaqd}}%
\MStopList \Mpar \vspace{\Saut}\hspace{\Alinea}The understanding of Fock spaces as inductive limits in fact offers an interesting perspective on the \emph{polarization-dependence} of infinite-dimensional Fock representations.
The space of states on a "partial" Hilbert space \MMath{\mathcal{H} ^{\mathfrak{lin} }_{\lambda }} is, in this picture, understood as a \emph{subset} of the space of all states (since "partial" density matrices on \MMath{\mathcal{H} ^{\mathfrak{lin} }_{\lambda }} get identified with those "full" density matrices whose support is restricted to \MMath{\mathcal{H} ^{\mathfrak{lin} }_{\lambda } \subset \mathcal{F} ^{(V ,I )}_{\mathfrak{lin} }}). Crucially, the \emph{physical interpretation} of a given "partial" state on \MMath{\mathcal{H} ^{\mathfrak{lin} }_{\lambda }} \emph{depends} on the choice of {$I$}-vacuum, because the latter is used to fix the behavior of the corresponding "full" state along the remaining modes.
By contrast, in the projective approach, states on \MMath{\mathcal{H} ^{\mathfrak{lin} }_{\lambda }} are simply understood as reflecting \emph{partial information} about the full state: no attempt is made at "completing" them, thus their physical interpretations only depend on the interpretation of the label {$\lambda$}.
This is the reason why the Stone–von Neumann theorem can be lifted in the latter case, but not in the former.%
\MleavePhyMode \MleavePhyMode \hypertarget{PARaqf}{}\Mpar \Mtheorem{3.16}Let {$I$} be a compatible complex structure on \MMath{V ,\Omega }, resp.~\MMath{V ,\left(\, \cdot \, \middlewithspace|\, \cdot \, \right)} ({\seqBBBaqg}, resp.~{\seqBBBaqh}). Let \MMath{\mathcal{S} ^{\mathfrak{lin} }_{I }} denote the space of self-adjoint trace-class operators on \MMath{\mathcal{F} ^{(V ,I )}_{\mathfrak{lin} }} and \MMath{\mathcal{B} ^{\mathfrak{lin} }_{I }} the algebra of bounded operators on \MMath{\mathcal{F} ^{(V ,I )}_{\mathfrak{lin} }}. There exist two maps \MMath{\Sigma _{I } : \mathcal{S} ^{\mathfrak{lin} }_{I } \rightarrow  \mathcal{S} ^{\mathfrak{lin} }} (\MMath{\mathcal{S} ^{\mathfrak{lin} }} was defined in {\seqBBBaew}) and \MMath{\Lambda _{I } : \overline{\mathcal{B} ^{\mathfrak{lin} }} \rightarrow  \mathcal{B} ^{\mathfrak{lin} }_{I }} (\MMath{\overline{\mathcal{B} ^{\mathfrak{lin} }}} was defined in {\seqBBBaex}) such that:%
\hypertarget{PARaqi}{}\Mpar \MList{1}\MMath{\Lambda _{I }} is a \MMath{C^{*}}-algebra morphism and \MMath{\Sigma _{I }} is a linear, order-preserving map;%
\MStopList \hypertarget{PARaqj}{}\Mpar \MList{2}for any \MMath{\mathbf{v}  \in  V }, \MMath{\Lambda _{I } \big( \mathcal{O} ^{\mathfrak{lin} }(\mathbf{v} ) \big) = \mathcal{O} ^{\mathfrak{lin} }_{(I )}(\mathbf{v} )};%
\MStopList \hypertarget{PARaqk}{}\Mpar \MList{3}for any \MMath{\rho  \in  \mathcal{S} ^{\mathfrak{lin} }_{I }} and any \MMath{A \in  \overline{\mathcal{B} ^{\mathfrak{lin} }}}, \MMath{\mkop{Tr}  \big( \rho  \,  \Lambda _{I }(A) \big) = \mkop{Tr}  \big( \Sigma _{I }(\rho ) \,  A \big)};%
\MStopList \hypertarget{PARaql}{}\Mpar \MList{4}\MMath{\left. \Sigma _{I } \right|_{{\mathcal{S} ^{\mathfrak{lin} }_{{I ,+,1 }}}}} is injective and:%
\Mpar \MStartEqua \MMath{\Sigma _{I } \left\langle  \mathcal{S} ^{\mathfrak{lin} }_{{I ,+,1 }} \right\rangle  = \left\{ \rho  = \left(\rho _{\lambda }\right)_{{\lambda  \in  \mathcal{L} ^{\mathfrak{lin} }_{}}} \in  \mathcal{S} ^{\mathfrak{lin} }_{{+,1 }} \middlewithspace| \sup_{{\lambda _{1} \in  \mathcal{L} _{I }}} \inf_{{\deuxlignes{-1pt}{\lambda _{2} \in  \mathcal{L} _{I }}{\lambda _{1} \leqslant  \lambda _{2}}}} \mkop{Tr}  \left( \rho _{{\lambda _{2}}} \,  \Theta _{{\lambda _{2} | \lambda _{1}}}\right) = \mkop{Tr}  \rho  = 1 \right\}},%
\MStopEqua \Mpar where \MMath{\mathcal{S} ^{\mathfrak{lin} }_{{I ,+,1 }}} denotes the space of density matrices on \MMath{\mathcal{F} ^{(V ,I )}_{\mathfrak{lin} }} (ie.~trace-class, semi-definite positive operators of unit trace), \MMath{\mathcal{S} ^{\mathfrak{lin} }_{{+,1 }}} was defined in {\seqBBBaex}, \MMath{\mathcal{L} _{I }} is the label subset:%
\Mpar \MStartEqua \MMath{\mathcal{L} _{I } \mathrel{\mathop:}=  \left\{ \lambda  \in  \mathcal{L} ^{\mathfrak{lin} }_{} \middlewithspace| \exists  \left(\mathbf{b} _{1},\dots ,\mathbf{b} _{n}\right) I \text{{-orthonormal family}} \big/ \lambda  = \left(\mathbf{b} _{1},I \mathbf{b} _{1},\dots ,\mathbf{b} _{n},I \mathbf{b} _{n}\right) \right\}},%
\MStopEqua \Mpar and, for any \MMath{\lambda _{1}, \lambda _{2} \in  \mathcal{L} _{I }} with \MMath{\lambda _{1} \leqslant  \lambda _{2}}, \MMath{\Theta _{{\lambda _{2} | \lambda _{1}}} \in  \mathcal{B} ^{\mathfrak{lin} }_{{\lambda _{2}}}}.%
\MStopList \Mpar \Mproof \italique{Restriction to \MMath{\mathcal{L} _{I }}.} Let \MMath{\lambda  \in  \mathcal{L} ^{\mathfrak{lin} }_{}} and let \MMath{F  \mathrel{\mathop:}=  V _{\lambda } \oplus  I \left\langle  V _{\lambda } \right\rangle }. {$F$} is a \emph{complex} vector subspace of {$V$} (seen as a complex pre-Hilbert space, see {\seqBBBaqg}, resp.~{\seqBBBaqh}), hence {$F$} admits a {$I$}-orthonormal basis \MMath{\left(\mathbf{b} _{1},\dots ,\mathbf{b} _{n}\right)}. Let \MMath{\tilde{\lambda } \mathrel{\mathop:}=  \left(\mathbf{b} _{1},I \mathbf{b} _{1},\dots ,\mathbf{b} _{n},I \mathbf{b} _{n} \right)}. \MMath{\tilde{\lambda }} is a symplectic family, resp.~a \emph{real}-orthonormal family. If \MMath{\mathfrak{lin} =\mathfrak{ferm} } and \MMath{V _{\lambda } = F }, we choose \MMath{\mathbf{b} _{n+1} \in  F ^{\perp }} and we redefine \MMath{\tilde{\lambda } \mathrel{\mathop:}=  \left(\mathbf{b} _{1},I \mathbf{b} _{1},\dots ,\mathbf{b} _{n},I \mathbf{b} _{n},\mathbf{b} _{n+1},I \mathbf{b} _{n+1}\right)}. Then, we have, in any case, \MMath{\tilde{\lambda } \in  \mathcal{L} _{I } \subseteq \mathcal{L} ^{\mathfrak{lin} }_{}} and \MMath{\lambda  \leqslant  \tilde{\lambda }} (using the characterization from {\seqBBBapy}). In other words, \MMath{\mathcal{L} _{I }} is a cofinal part of \MMath{\mathcal{L} ^{\mathfrak{lin} }_{}}.%
\Mpar \vspace{\Sssaut}\hspace{\Alinea}Now, we define \MMath{(\mathfrak{lin} ,I )} to be the restriction of the  system of factorized Hilbert spaces {$\mathfrak{lin}$} to \MMath{\mathcal{L} _{I } \subset \mathcal{L} ^{\mathfrak{lin} }_{}}, and we have the maps:%
\Mpar \MStartEqua \MMath{\definitionFonction{\Sigma _{1}}{\mathcal{S} ^{\mathfrak{lin} }}{\mathcal{S} ^{(\mathfrak{lin} ,I )}}{\left(\rho _{\lambda }\right)_{{\lambda \in \mathcal{L} ^{\mathfrak{lin} }_{}}}}{\left(\rho _{\lambda }\right)_{{\lambda \in \mathcal{L} _{I }}}} \, \&\,  \definitionFonction{\Lambda _{1}}{\mathcal{B} ^{(\mathfrak{lin} ,I )}}{\mathcal{B} ^{\mathfrak{lin} }}{\left[ A_{\lambda } \right]^{(\mathfrak{lin} ,I )}}{\left[ A_{\lambda } \right]^{\mathfrak{lin} }}}.%
\MStopEqua \Mpar \MMath{\mathcal{L} _{I }} being \emph{cofinal} ensures that \MMath{\Sigma _{1}} is an isomorphism of ordered vector spaces and \MMath{\Lambda _{1}} can be extended to a \MMath{C^{*}}-algebra isomorphism \MMath{\overline{\mathcal{B} ^{(\mathfrak{lin} ,I )}} \rightarrow  \overline{\mathcal{B} ^{\mathfrak{lin} }}} (as can be shown by a straightforward adaptation of {\seqBBBaqt}). Moreover, for any \MMath{\rho  \in  \mathcal{S} ^{\mathfrak{lin} }} and any \MMath{A \in  \overline{\mathcal{B} ^{(\mathfrak{lin} ,I )}}}, we have \MMath{\mkop{Tr}  \big( \rho  \,  \Lambda _{1}(A) \big) = \mkop{Tr}  \big( \Sigma _{1}(\rho ) \,  A \big)} by construction.%
\Mpar \vspace{\Sssaut}\hspace{\Alinea}For any \MMath{\lambda _{1},\lambda _{2} \in  \mathcal{L} _{I }} with \MMath{\lambda _{1} = \left(\mathbf{b} _{1},I \mathbf{b} _{1},\dots ,\mathbf{b} _{n},I \mathbf{b} _{n}\right)}, \MMath{\lambda _{2} = \left(\mathbf{b} '_{1},I \mathbf{b} '_{1},\dots ,\mathbf{b} '_{m},I \mathbf{b} '_{m}\right)} and \MMath{\lambda _{1} \leqslant  \lambda _{2}}, we choose an {$I$}-orthonormal basis \MMath{\left(\mathbf{b} _{n+1},I \mathbf{b} _{n+1},\dots ,\mathbf{b} _{m},I \mathbf{b} _{m}\right)} of \MMath{V _{{\lambda _{1}}}^{\perp } \cap  V _{{\lambda _{2}}}}.
Next, we choose \MMath{\mu _{{\lambda _{2} \rightarrow  \lambda _{1}}} \in  \text{Mp}^{(m)}}, resp.~\MMath{\text{Spin}^{(m)}}, such that \MMath{\mu _{{\lambda _{2} \rightarrow  \lambda _{1}}} \rhd  \left(\mathbf{b} '_{1},I \mathbf{b} '_{1},\dots ,\mathbf{b} '_{m},I \mathbf{b} '_{m}\right) = \left(\mathbf{b} _{1},I \mathbf{b} _{1},\dots ,\mathbf{b} _{m},I \mathbf{b} _{m}\right)} and:%
\hypertarget{PARaqv}{}\Mpar \MStartEqua \MMath{\Gamma ^{(\mathbf{b} _{1},\dots ,\mathbf{b} _{m} ;V ,I )}_{\mathfrak{lin} } = \left( e^{{i\phi _{{\lambda _{2} \rightarrow  \lambda _{1}}}}} \,  \mathtt{T} ^{(m)}_{\mathfrak{lin} }(\mu _{{\lambda _{2} \rightarrow  \lambda _{1}}}) \otimes  \text{id}_{{\mathcal{F} ^{(W ,J )}_{\mathfrak{lin} }}}\right) \circ  \Gamma ^{(\mathbf{b} '_{1},\dots ,\mathbf{b} '_{m} ;V ,I )}_{\mathfrak{lin} }},%
\NumeroteEqua{3.16}{1}\MStopEqua \Mpar with \MMath{W  \mathrel{\mathop:}=  V _{{\lambda _{2}}}^{\perp }}, \MMath{J  \mathrel{\mathop:}=  \left.I \right|_{{W  \rightarrow  W }}}, and \MMath{\phi _{{\lambda _{2} \rightarrow  \lambda _{1}}} \in  \mathds{R}} {\seqDDDaqx}.
In particular, \MMath{\mu _{{\lambda _{2} \rightarrow  \lambda _{1}}} \in  \mathcal{L} ^{(\mathfrak{lin} ,I )}_{{\lambda _{2} \rightarrow  \lambda _{1}}} = \mathcal{L} ^{\mathfrak{lin} }_{{\lambda _{2} \rightarrow  \lambda _{1}}}} (and we can choose \MMath{\mu _{{\lambda _{2} \rightarrow  \lambda _{1}}} = \mathds{1} } if \MMath{\lambda _{1} = \lambda _{2}}).
Then, we can construct \MMath{\left( \mathcal{L} _{I },\,  \left(\mathcal{H} _{\lambda }\right)_{{\lambda  \in  \mathcal{L} _{I }}},\,  \left(\mathcal{H} _{{\lambda _{2} \rightarrow  \lambda _{1}}}\right)_{{\lambda _{1} \leqslant  \lambda _{2}}},\,  \left(\Phi _{{\lambda _{2} \rightarrow  \lambda _{1}}}\right)_{{\lambda _{1} \leqslant  \lambda _{2}}},\,  \left(\Phi _{{\lambda _{3} \rightarrow  \lambda _{2} \rightarrow  \lambda _{1}}}\right)_{{\lambda _{1} \leqslant  \lambda _{2} \leqslant  \lambda _{3}}} \right)} fulfilling {\seqBBBacq} as described in {\seqBBBabd}
(note that \MMath{\mathcal{L} _{I }} is directed, as a cofinal part of a directed set).%
\Mpar \vspace{\Ssaut}\italique{Vacuum state and inductive limit.} Let \MMath{\lambda _{1},\lambda _{2} \in  \mathcal{L} _{I }} with \MMath{\lambda _{1} = \left(\mathbf{b} _{1},I \mathbf{b} _{1},\dots ,\mathbf{b} _{n},I \mathbf{b} _{n}\right)}, \MMath{\lambda _{2} = \left(\mathbf{b} '_{1},I \mathbf{b} '_{1},\dots ,\mathbf{b} '_{m},I \mathbf{b} '_{m}\right)} and \MMath{\lambda _{1} \leqslant  \lambda _{2}}. Combining {\seqBBBaqz} with the definition of \MMath{\Phi ^{\mathfrak{lin} }_{\mu }} {\seqDDDara} and with {\seqBBBarb}, we obtain:%
\hypertarget{PARarc}{}\Mpar \MStartEqua \MMath{\left( \text{id}_{{\mathcal{F} ^{(n)}_{\mathfrak{lin} }}} \otimes  \Gamma ^{(\mathbf{b} _{n+1},\dots ,\mathbf{b} _{m} ;W _{1},J _{1})}_{\mathfrak{lin} } \right) \circ  \Gamma ^{(\mathbf{b} _{1},\dots ,\mathbf{b} _{n} ;V ,I )}_{\mathfrak{lin} } = \left( e^{{i\phi _{{\lambda _{2} \rightarrow  \lambda _{1}}}}} \,  \Phi ^{\mathfrak{lin} }_{{\mu _{{\lambda _{2} \rightarrow  \lambda _{1}}}}} \otimes  \text{id}_{{\mathcal{F} ^{(W _{2},J _{2})}_{\mathfrak{lin} }}}\right) \circ  \Gamma ^{(\mathbf{b} '_{1},\dots ,\mathbf{b} '_{m} ;V ,I )}_{\mathfrak{lin} }},%
\NumeroteEqua{3.16}{2}\MStopEqua \Mpar where \MMath{W _{{1,2}},\, J _{{1,2}}} denote the orthogonal complement of \MMath{V _{{\lambda _{{1,2}}}}} respectively.
Let \MMath{\left(\mathbf{b} _{j}\right)_{j\in J}} be an orthonormal basis of \MMath{W _{2}}, \MMath{\left(\mathbf{b} '_{j}\right)_{j\in J} \mathrel{\mathop:}=  \left(\mathbf{b} _{j}\right)_{j\in J}} and \MMath{I \mathrel{\mathop:}=  \left\{1,\dots ,m\right\} \sqcup J}.
Applying {\seqBBBare} to \MMath{\left| \left(0\right)_{i\in I;\, }\left(\mathbf{b} _{i}\right)_{i\in I} \right\rangle _{\mathfrak{lin} } = \left| \left(0\right)_{i\in I;\, }\left(\mathbf{b} '_{i}\right)_{i\in I} \right\rangle _{\mathfrak{lin} }} and using the definition of \MMath{\Phi _{{\lambda _{2} \rightarrow  \lambda _{1}}}} from {\seqBBBabd} yields:%
\hypertarget{PARarf}{}\Mpar \MStartEqua \MMath{\left| 0^{(n+1)} , \dots  , 0^{(m)} \right\rangle _{\mathfrak{lin} } \otimes  \left| 0^{(1)} , \dots  , 0^{(n)} \right\rangle _{\mathfrak{lin} } = e^{{i\phi _{{\lambda _{2} \rightarrow  \lambda _{1}}}}} \,  \Phi _{{\lambda _{2} \rightarrow  \lambda _{1}}} \left| 0^{(1)} , \dots  , 0^{(m)} \right\rangle _{\mathfrak{lin} }}. %
\NumeroteEqua{3.16}{3}\MStopEqua \Mpar We define, for any \MMath{\lambda _{1} \leqslant  \lambda _{2} \in  \mathcal{L} _{I }}, \MMath{\zeta _{{\lambda _{2} \rightarrow  \lambda _{1}}} \mathrel{\mathop:}=  e^{{-i\phi _{{\lambda _{2} \rightarrow  \lambda _{1}}}}} \,  \left| 0^{(d _{{\lambda _{1}}} +1)} , \dots  , 0^{(d _{{\lambda _{2}}})} \right\rangle _{\mathfrak{lin} } \in  \mathcal{H} _{{\lambda _{2} \rightarrow  \lambda _{1}}} \approx  \mathcal{F} ^{(d _{{\lambda _{2}}} - d _{{\lambda _{1}}})}_{\mathfrak{lin} }}.
For any \MMath{\lambda _{1} \leqslant  \lambda _{2} \leqslant  \lambda _{3} \in  \mathcal{L} _{I }}, we then have:%
\hypertarget{PARarh}{}\Mpar \MStartEqua \MMath{\Phi _{{\lambda _{3} \rightarrow  \lambda _{2} \rightarrow  \lambda _{1}}} \left(\zeta _{{\lambda _{3} \rightarrow  \lambda _{1}}}\right) = \zeta _{{\lambda _{3} \rightarrow  \lambda _{2}}} \otimes  \zeta _{{\lambda _{2} \rightarrow  \lambda _{1}}}},%
\NumeroteEqua{3.16}{4}\MStopEqua \Mpar where we have used {\seqBBBarj} together with the consistency relation fulfilled by \MMath{\Phi _{{\lambda _{3} \rightarrow  \lambda _{2} \rightarrow  \lambda _{1}}}} {\seqDDDaff}.%
\Mpar \vspace{\Sssaut}\hspace{\Alinea}We define a Hilbert space \MMath{\tilde{\mathcal{F} }^{(V ,I )}_{\mathfrak{lin} }} as (the completion of) the inductive limit of the system \MMath{\left( \left(\mathcal{H} _{\lambda }\right)_{{\lambda  \in  \mathcal{L} _{I }}}, \left(\chi _{{\lambda _{2} \leftarrow  \lambda _{1}}}\right)_{{\lambda _{1} \leqslant  \lambda _{2}}} \right)}, where, for any \MMath{\lambda _{1} \leqslant  \lambda _{2} \in  \mathcal{L} _{I }}, the isometric injection \MMath{\chi _{{\lambda _{2} \leftarrow  \lambda _{1}}}} is given by:%
\Mpar \MStartEqua \MMath{\definitionFonction{\chi _{{\lambda _{2} \leftarrow  \lambda _{1}}}}{\mathcal{H} _{{\lambda _{1}}}}{\mathcal{H} _{{\lambda _{2}}}}{\psi }{\Phi ^{-1}_{{\lambda _{2} \rightarrow  \lambda _{1}}} \left( \zeta _{{\lambda _{2} \rightarrow  \lambda _{1}}} \otimes  \psi  \right)}}.%
\MStopEqua \Mpar It follows from {\seqBBBarn} that, for any \MMath{\lambda _{1} \leqslant  \lambda _{2} \leqslant  \lambda _{3} \in  \mathcal{L} _{I }}, \MMath{\chi _{{\lambda _{3} \leftarrow  \lambda _{2}}} \circ  \chi _{{\lambda _{2} \leftarrow  \lambda _{1}}} = \chi _{{\lambda _{3} \leftarrow  \lambda _{1}}}}.
We denote by \MMath{\tilde{\mathcal{S} }^{\mathfrak{lin} }_{I }}, \MMath{\tilde{\mathcal{S} }^{\mathfrak{lin} }_{{I ,+,1 }}}, resp.~\MMath{\tilde{\mathcal{B} }^{\mathfrak{lin} }_{I }}, the space of self-adjoint trace-class operators, the space of density matrices, resp.~the algebra of bounded operators, on \MMath{\tilde{\mathcal{F} }^{(V ,I )}_{\mathfrak{lin} }}.
Using {\seqBBBaqb}\footnote{In the statement of {\seqBBBaqb}, the map \MMath{\Sigma _{2}} is only defined on the space of \emph{semi-definite positive}, trace-class operators. However, one can check from its construction in the proof that it can be extended to a linear, order-preserving map on the space of self-adjoint, trace-class operators. Similarly, the morphism properties of \MMath{\Lambda _{2}} can be checked from its definition.}, there exist a linear, order-preserving map \MMath{\Sigma _{2} : \tilde{\mathcal{S} }^{\mathfrak{lin} }_{I } \rightarrow  \mathcal{S} ^{(\mathfrak{lin} ,I )}} and a \MMath{C^{*}}-algebra morphism \MMath{\Lambda _{2} : \overline{\mathcal{B} ^{(\mathfrak{lin} ,I )}} \rightarrow  \tilde{\mathcal{B} }^{\mathfrak{lin} }_{I }} such that:%
\Mpar \MList{5}for any \MMath{\rho  \in  \tilde{\mathcal{S} }^{\mathfrak{lin} }_{I }} and any \MMath{A \in  \overline{\mathcal{B} ^{(\mathfrak{lin} ,I )}}}, \MMath{\mkop{Tr}  \big( \rho  \,  \Lambda _{2}(A) \big) = \mkop{Tr}  \big( \Sigma _{2}(\rho ) \,  A \big)};%
\MStopList \Mpar \MList{6}\MMath{\left. \Sigma _{2} \right|_{{\tilde{\mathcal{S} }^{\mathfrak{lin} }_{{I ,+,1 }}}}} is injective and:%
\Mpar \MStartEqua \MMath{\Sigma _{2} \left\langle  \tilde{\mathcal{S} }^{\mathfrak{lin} }_{{I ,+,1 }} \right\rangle  = \left\{ \rho  = \left(\rho _{\lambda }\right)_{{\lambda  \in  \mathcal{L} _{I }}} \in  \mathcal{S} ^{(\mathfrak{lin} ,I )}_{{+,1 }} \middlewithspace| \sup_{{\lambda _{1} \in  \mathcal{L} _{I }}} \inf_{{\deuxlignes{-1pt}{\lambda _{2} \in  \mathcal{L} _{I }}{\lambda _{1} \leqslant  \lambda _{2}}}} \mkop{Tr}  \left( \rho _{{\lambda _{2}}} \,  \Theta _{{\lambda _{2} | \lambda _{1}}}\right) = \mkop{Tr}  \rho  = 1 \right\}},%
\MStopEqua \Mpar where, for any \MMath{\lambda _{1} \leqslant  \lambda _{2} \in  \mathcal{L} _{I }}:%
\Mpar \MStartEqua \MMath{\Theta _{{\lambda _{2} | \lambda _{1}}} \mathrel{\mathop:}=  \Phi ^{-1}_{{\lambda _{2} \rightarrow  \lambda _{1}}} \, \circ \,  \left( \left| \zeta _{{\lambda _{2} \rightarrow  \lambda _{1}}} \right\rangle \!\!\left\langle  \zeta _{{\lambda _{2} \rightarrow  \lambda _{1}}} \right| \otimes  \mathds{1} ^{\mathfrak{c} }_{{\lambda _{1}}} \right) \, \circ \,  \Phi _{{\lambda _{2} \rightarrow  \lambda _{1}}} \in  \mathcal{B} ^{\mathfrak{lin} }_{{\lambda _{2}}}}.%
\MStopEqua \MStopList \Mpar \vspace{\Ssaut}\italique{Isomorphism between the inductive limit \MMath{\tilde{\mathcal{F} }^{(V ,I )}_{\mathfrak{lin} }} and the Fock space \MMath{\mathcal{F} ^{(V ,I )}_{\mathfrak{lin} }}.} For any \MMath{\lambda  = \left(\mathbf{b} _{1},I \mathbf{b} _{1},\dots ,\mathbf{b} _{n},I \mathbf{b} _{n}\right) \in  \mathcal{L} _{I }}, we define an isometric injection:%
\hypertarget{PARaru}{}\Mpar \MStartEqua \MMath{\definitionFonction{\chi _{{I \leftarrow \lambda }}}{\mathcal{H} _{\lambda }}{\mathcal{F} ^{(V ,I )}_{\mathfrak{lin} }}{\psi }{\left(\Gamma ^{(\mathbf{b} _{1},\dots ,\mathbf{b} _{n} ;V ,I )}_{\mathfrak{lin} }\right)^{-1} \left( \psi  \otimes  \left| \left(0\right)_{j\in J;\, }\left(\mathbf{b} _{j}\right)_{j\in J} \right\rangle _{\mathfrak{lin} } \right)}},%
\NumeroteEqua{3.16}{5}\MStopEqua \Mpar where \MMath{\left(\mathbf{b} _{j}\right)_{j\in J}} is some orthonormal basis of \MMath{V _{\lambda }^{\perp }}. Note that if \MMath{\left(\mathbf{b} '_{j}\right)_{j\in J}} is another orthonormal basis of \MMath{V _{\lambda }^{\perp }}, we have \MMath{\left| \left(0\right)_{j\in J;\, }\left(\mathbf{b} _{j}\right)_{j\in J} \right\rangle _{\mathfrak{lin} } = \left| \left(0\right)_{j\in J;\, }\left(\mathbf{b} '_{j}\right)_{j\in J} \right\rangle _{\mathfrak{lin} }}, hence \MMath{\chi _{I \leftarrow \lambda }} does not depend on the choice of the basis \MMath{\left(\mathbf{b} _{j}\right)_{j\in J}}.
Using {\seqBBBare}, together with the definitions of \MMath{\chi _{{\lambda _{2} \leftarrow  \lambda _{1}}}} and \MMath{\zeta _{{\lambda _{2} \rightarrow  \lambda _{1}}}}, we have, for any \MMath{\lambda _{1} \leqslant  \lambda _{2} \in  \mathcal{L} _{I }}:%
\Mpar \MStartEqua \MMath{\chi _{{I  \leftarrow  \lambda _{2}}} \circ  \chi _{{\lambda _{2} \leftarrow  \lambda _{1}}} = \chi _{{I  \leftarrow  \lambda _{1}}}}.%
\MStopEqua \Mpar Hence, by the universal property of the inductive limit, there exists a linear map \MMath{\tilde{\Phi }: \bigcup_{{\lambda \in \mathcal{L} _{I }}} \tilde{\chi }_{{I \leftarrow \lambda }} \left\langle  \mathcal{H} _{\lambda } \right\rangle  \rightarrow  \mathcal{F} ^{(V ,I )}_{\mathfrak{lin} }} such that, for any \MMath{\lambda  \in  \mathcal{L} _{I }}, \MMath{\chi _{{I  \leftarrow  \lambda }} = \tilde{\Phi } \circ  \tilde{\chi }_{{I  \leftarrow  \lambda }}}, where \MMath{\tilde{\chi }_{{I  \leftarrow  \lambda }}} denotes the natural injection \MMath{\mathcal{H} _{\lambda } \rightarrow  \tilde{\mathcal{F} }^{(V ,I )}_{\mathfrak{lin} }}.
Since both \MMath{\tilde{\chi }_{{I \leftarrow \lambda }}} and \MMath{\chi _{{I \leftarrow \lambda }}} are \emph{isometric} injections for all \MMath{\lambda  \in  \mathcal{L} _{I }}, so is \MMath{\tilde{\Phi }}, and since \MMath{\mathcal{F} ^{(V ,I )}_{\mathfrak{lin} }} is \emph{complete}, \MMath{\tilde{\Phi }} can be extended to an isometric injection \MMath{\tilde{\mathcal{F} }^{(V ,I )}_{\mathfrak{lin} } = \overline{\bigcup_{{\lambda \in \mathcal{L} _{I }}} \tilde{\chi }_{{I \leftarrow \lambda }} \left\langle  \mathcal{H} _{\lambda } \right\rangle } \rightarrow  \mathcal{F} ^{(V ,I )}_{\mathfrak{lin} }}.%
\Mpar \vspace{\Sssaut}\hspace{\Alinea}Let \MMath{N \geqslant  0} and let \MMath{\mathbf{v} _{1},\dots ,\mathbf{v} _{N} \in  V }. Let \MMath{\mathbf{v} _{1}^{*},\dots ,\mathbf{v} _{N}^{*}} be their dual vectors in \MMath{V ^{*}_{(I )}} (the dual of the Hilbert space \MMath{\overline{V }_{{\!(I )}}}, the latter being the completion of {$V$} with respect to the complex scalar product \MMath{\left\langle \, \cdot \, \middlewithspace|\, \cdot \, \right\rangle _{I }}, see {\seqBBBarz}) and let:%
\Mpar \MStartEqua \MMath{\psi  \mathrel{\mathop:}=  \left(\mathbf{v} ^{*}_{1} \otimes  \dots  \otimes  \mathbf{v} ^{*}_{N}\right)^{\text{sym}} \mathrel{\mathop:}=  \sum_{{\varepsilon  \in  S_{N}}} \hat{\varepsilon } \big( \mathbf{v} ^{*}_{1} \otimes  \dots  \otimes  \mathbf{v} ^{*}_{N} \big) \in  \left( V ^{*} \right)^{{\otimes N,\text{{sym}}}}},%
\MStopEqua \Mpar resp:%
\Mpar \MStartEqua \MMath{\psi  \mathrel{\mathop:}=  \left(\mathbf{v} ^{*}_{1} \otimes  \dots  \otimes  \mathbf{v} ^{*}_{N}\right)^{\text{alt}} \mathrel{\mathop:}=  \sum_{{\varepsilon  \in  S_{N}}} \text{sig}(\varepsilon ) \,  \hat{\varepsilon } \big( \mathbf{v} ^{*}_{1} \otimes  \dots  \otimes  \mathbf{v} ^{*}_{N} \big) \in  \left( V ^{*} \right)^{{\otimes N,\text{{alt}}}}},%
\MStopEqua \Mpar where \MMath{V ^{*}} denotes the dense subspace \MMath{\left\{ \mathbf{v} ^{*} \middlewithspace| \mathbf{v}  \in  V  \right\} \subseteq V ^{*}_{{\!(I )}}} and \MMath{\left( V ^{*} \right)^{{\otimes N,\text{{sym}}}}}, resp.~\MMath{\left( V ^{*} \right)^{{\otimes N,\text{{alt}}}}}, denotes the \emph{symmetric}, resp.~\emph{alternating}, subspace (as defined in {\seqBBBase}) of the tensor product \MMath{\left( V ^{*} \right)^{\otimes N}} (defined \emph{without} any completion, so that \MMath{\left( V ^{*} \right)^{\otimes N}} and \MMath{\left( V ^{*} \right)^{{\otimes N,\text{{sym}}}}}, resp.~\MMath{\left( V ^{*} \right)^{{\otimes N,\text{{alt}}}}}, are complex \emph{pre}-Hilbert spaces).
Using {\seqBBBasf}, together with the fact that \MMath{\mathcal{L} _{I }} is a cofinal part of \MMath{\mathcal{L} ^{\mathfrak{lin} }_{}}, there exists an {$I$}-orthonormal family \MMath{\left(\mathbf{b} _{1},\dots ,\mathbf{b} _{n}\right)} such that \MMath{\mkop{Span} _{\mathds{R}}\left\{\mathbf{v} _{1},\dots ,\mathbf{v} _{N}\right\} \subseteq \mkop{Span} _{\mathds{C}} \left\{ \mathbf{b} _{1},\dots ,\mathbf{b} _{n} \right\}}. Let \MMath{\lambda  \mathrel{\mathop:}=  \left(\mathbf{b} _{1},I \mathbf{b} _{1},\dots ,\mathbf{b} _{n},I \mathbf{b} _{n}\right) \in  \mathcal{L} _{I }} and let \MMath{\left(\mathbf{b} _{j}\right)_{j\in J}} be an orthonormal basis of the orthogonal complement of \MMath{V _{\lambda } = \mkop{Span} _{\mathds{C}} \left\{ \mathbf{b} _{1},\dots ,\mathbf{b} _{n} \right\}} in \MMath{\overline{V }_{{\!(I )}}}, so that \MMath{\left(\mathbf{b} _{i}\right)_{i\in I}} with \MMath{I \mathrel{\mathop:}=  \left\{1,\dots ,n\right\} \sqcup J} is an orthonormal basis of \MMath{\overline{V }_{{\!(I )}}}. For any \MMath{\left(N_{i}\right)_{i\in I} \in  \mathds{N}^{I}} such that \MMath{\forall  j \in  J, {N_{j} = 0}}, we have \MMath{\left| \left(N_{i}\right)_{i\in I;\, }\left(\mathbf{b} _{i}\right)_{i\in I} \right\rangle _{\mathfrak{lin} } \in  \chi _{{I \leftarrow \lambda }} \left\langle  \mathcal{H} _{\lambda } \right\rangle  \subseteq \tilde{\Phi } \left\langle  \tilde{\mathcal{F} }^{(V ,I )}_{\mathfrak{lin} } \right\rangle }. Moreover, we have, by construction:%
\Mpar \MStartEqua \MMath{\psi  \in  \mkop{Span} _{\mathds{C}} \left\{ \left| \left(N_{i}\right)_{i\in I;\, }\left(\mathbf{b} _{i}\right)_{i\in I} \right\rangle _{\mathfrak{lin} } \middlewithspace| \left(N_{i}\right)_{i\in I} \in  \mathds{N}^{I} \big/ \forall  j \in  J, {N_{j} = 0} \right\}}.%
\MStopEqua \Mpar Since:%
\Mpar \MStartEqua \MMath{\left( V ^{*} \right)^{{\otimes N,\text{{sym/alt}}}} = \mkop{Span} _{\mathds{C}} \left\{ \left(\mathbf{v} ^{*}_{1} \otimes  \dots  \otimes  \mathbf{v} ^{*}_{N}\right)^{\text{{sym/alt}}} \middlewithspace| \mathbf{v} _{1},\dots ,\mathbf{v} _{N} \in  V  \right\}},%
\MStopEqua \Mpar it follows that \MMath{\left( V ^{*} \right)^{{\otimes N,\text{{sym/alt}}}} \subseteq \tilde{\Phi } \left\langle  \tilde{\mathcal{F} }^{(V ,I )}_{\mathfrak{lin} } \right\rangle }.
Moreover, \MMath{\left( V ^{*} \right)^{{\otimes N,\text{{sym/alt}}}}} is dense in \MMath{\big( V ^{*}_{{\!(I )}} \big)^{{\otimes N,\text{{sym/alt}}}}}, hence \MMath{\bigoplus_{N\geqslant 0} \left( V ^{*} \right)^{{\otimes N,\text{{sym/alt}}}}} is dense in \MMath{\mathcal{F} ^{(V ,I )}_{\mathfrak{lin} }}. \MMath{\tilde{\mathcal{F} }^{(V ,I )}_{\mathfrak{lin} }} being \emph{complete}, this ensures that the isometry \MMath{\tilde{\Phi }} is surjective, and therefore a Hilbert space isomorphism.%
\Mpar \vspace{\Sssaut}\hspace{\Alinea}We define:%
\Mpar \MStartEqua \MMath{\definitionFonction{\Sigma _{3}}{\mathcal{S} ^{\mathfrak{lin} }_{I }}{\tilde{\mathcal{S} }^{\mathfrak{lin} }_{I }}{\rho }{\tilde{\Phi }^{-1} \, \rho \,  \tilde{\Phi }} \, \&\,  \definitionFonction{\Lambda _{3}}{\tilde{\mathcal{B} }^{\mathfrak{lin} }_{I }}{\mathcal{B} ^{\mathfrak{lin} }_{I }}{\tilde{A}}{\tilde{\Phi } \, \tilde{A}\,  \tilde{\Phi }^{-1}}},%
\MStopEqua \Mpar and \MMath{\Sigma _{I } \mathrel{\mathop:}=  \Sigma _{1}^{-1} \circ  \Sigma _{2} \circ  \Sigma _{3}}, \MMath{\Lambda _{I } \mathrel{\mathop:}=  \Lambda _{3} \circ  \Lambda _{2} \circ  \Lambda _{1}^{-1}}. \MMath{\Sigma _{I },\Lambda _{I }} then fulfill {\seqBBBasn}.%
\Mpar \vspace{\Ssaut}\italique{Observables.} Fetching from {\seqBBBasp} the expression for \MMath{\Lambda _{2}}, we have for any \MMath{\lambda  \in  \mathcal{L} _{I }} and any \MMath{A_{\lambda } \in  \mathcal{B} ^{\mathfrak{lin} }_{\lambda }}:%
\Mpar \MStartEqua \MMath{\Lambda _{I } \left( \left[ A_{\lambda } \right]^{\mathfrak{lin} } \right) = \tilde{\Phi } \, \tilde{\Phi }^{-1}_{{I \rightarrow \lambda }}\, \left(\tilde{\mathds{1} }_{{I \rightarrow \lambda }} \otimes  A_{\lambda }\right)\, \tilde{\Phi }_{{I \rightarrow \lambda }}\,  \tilde{\Phi }^{-1}}%
\MStopEqua \Mpar where:%
\Mpar \MList{7}\MMath{\tilde{\mathds{1} }_{{I \rightarrow \lambda }}} denotes the identity operator on the Hilbert space \MMath{\tilde{\mathcal{H} }_{{I \rightarrow \lambda }}}, the latter being defined as the (completion of) the inductive limit of the system \MMath{\left( \left(\mathcal{H} _{{\kappa  \rightarrow  \lambda }}\right)_{{\kappa  \geqslant  \lambda }}, \left(\chi _{{\kappa _{2} \leftarrow  \kappa _{1} \rightarrow  \lambda }}\right)_{{\kappa _{2} \geqslant  \kappa _{1} \geqslant  \lambda }} \right)} with, for any \MMath{\kappa _{1},\kappa _{2} \in  \mathcal{L} _{I }} such that \MMath{\kappa _{2} \geqslant  \kappa _{1} \geqslant  \lambda }:%
\Mpar \MStartEqua \MMath{\definitionFonction{\chi _{{\kappa _{2} \leftarrow  \kappa _{1} \rightarrow  \lambda }}}{\mathcal{H} _{{\kappa _{1} \rightarrow  \lambda }}}{\mathcal{H} _{{\kappa _{2} \rightarrow  \lambda }}}{\psi }{\Phi _{{\kappa _{2} \rightarrow  \kappa _{1} \rightarrow  \lambda }}^{-1} \left( \zeta _{{\kappa _{2} \rightarrow  \kappa _{1}}} \otimes  \psi  \right)}};%
\MStopEqua \MStopList \Mpar \MList{8}and \MMath{\tilde{\Phi }_{{I \rightarrow \lambda }}} is a Hilbert space isomorphism \MMath{\tilde{\mathcal{F} }^{(V ,I )}_{\mathfrak{lin} } \rightarrow  \tilde{\mathcal{H} }_{{I \rightarrow \lambda }} \otimes  \mathcal{H} ^{\mathfrak{lin} }_{\lambda }} satisfying:%
\Mpar \MStartEqua \MMath{\forall  \kappa  \in  \mathcal{L} _{I } \big/ \kappa  \geqslant  \lambda , {\tilde{\Phi }_{{I \rightarrow \lambda }} \circ  \tilde{\chi }_{{I \leftarrow \kappa }} = \left( \tilde{\chi }_{{I \leftarrow \kappa \rightarrow \lambda }} \otimes  \mathds{1} ^{\mathfrak{lin} }_{\lambda } \right) \circ  \Phi _{{\kappa  \rightarrow  \lambda }}}},%
\MStopEqua \Mpar with \MMath{\tilde{\chi }_{{I \leftarrow \kappa \rightarrow \lambda }}} the natural injection \MMath{\mathcal{H} _{{\kappa \rightarrow \lambda }} \rightarrow  \tilde{\mathcal{H} }_{{I \rightarrow \lambda }}}.%
\MStopList \Mpar Defining \MMath{\Phi _{{I \rightarrow \lambda }} \mathrel{\mathop:}=  \tilde{\Phi }_{{I \rightarrow \lambda }}\,  \tilde{\Phi }^{-1} : \mathcal{F} ^{(V ,I )}_{\mathfrak{lin} } \rightarrow  \tilde{\mathcal{H} }_{{I \rightarrow \lambda }} \otimes  \mathcal{H} ^{\mathfrak{lin} }_{\lambda }}, we have, for any \MMath{\lambda  \leqslant  \kappa  \in  \mathcal{L} _{I }}:%
\hypertarget{PARasy}{}\Mpar \MStartEqua \MMath{\Phi _{{I \rightarrow \lambda }}\,  \chi _{{I \leftarrow \kappa }} = \left( \tilde{\chi }_{{I \leftarrow \kappa \rightarrow \lambda }} \otimes  \mathds{1} ^{\mathfrak{lin} }_{\lambda } \right) \circ  \Phi _{{\kappa  \rightarrow  \lambda }}}.%
\NumeroteEqua{3.16}{6}\MStopEqua \Mpar \vspace{\Sssaut}\hspace{\Alinea}Let \MMath{\mathbf{v}  \in  V } and let \MMath{\lambda  \in  \mathcal{L} _{I }} such that \MMath{\mathbf{v}  \in  V _{\lambda }}. Let \MMath{\kappa  \in  \mathcal{L} _{I }} such that \MMath{\kappa  \geqslant  \lambda }. In particular, \MMath{\mathbf{v}  \in  V _{\kappa } \supseteq V _{\lambda }}. Let \MMath{\left(\mathbf{b} '_{1},\dots ,\mathbf{b} '_{m}\right)} be the {$I$}-orthonormal family such that \MMath{\kappa  = \left(\mathbf{b} '_{1},I \mathbf{b} '_{1},\dots ,\mathbf{b} '_{m},I \mathbf{b} '_{m}\right)}. Comparing the definition of \MMath{\mathcal{O} ^{\mathfrak{lin} }_{\kappa }(v)} {\seqDDData} with the one of \MMath{\mathcal{O} ^{\mathfrak{lin} }_{(I )}(\mathbf{v} )} {\seqDDDatb}, we have:%
\Mpar \MStartEqua \MMath{\Gamma ^{(\mathbf{b} '_{1},\dots ,\mathbf{b} '_{m} ;V ,I )}_{\mathfrak{lin} } \circ  \mathcal{O} ^{\mathfrak{lin} }_{(I )}(\mathbf{v} ) = \left( \mathcal{O} ^{\mathfrak{lin} }_{\kappa }(\mathbf{v} ) \otimes  \text{id}_{{\mathcal{F} ^{(W ,J )}_{\mathfrak{lin} }}}\right) \circ  \Gamma ^{(\mathbf{b} '_{1},\dots ,\mathbf{b} '_{m} ;V ,I )}_{\mathfrak{lin} }},%
\MStopEqua \Mpar with \MMath{W ,J } the {$I$}-orthogonal complement of \MMath{V _{\kappa }}.
Thus, we get, for any \MMath{\psi  \in  \mathcal{H} ^{\mathfrak{lin} }_{\kappa }}:%
\Mpar \MStartEqua \MMath{\mathcal{O} ^{\mathfrak{lin} }_{(I )}(\mathbf{v} ) \circ  \chi _{I \leftarrow \kappa }(\psi ) = \left(\Gamma ^{(\mathbf{b} '_{1},\dots ,\mathbf{b} '_{m} ;V ,I )}_{\mathfrak{lin} }\right)^{-1} \circ  \left( \mathcal{O} ^{\mathfrak{lin} }_{\kappa }(\mathbf{v} )(\psi ) \otimes  \left| \left(0\right)_{j\in J;\, }\left(\mathbf{b} _{j}\right)_{j\in J} \right\rangle _{\mathfrak{lin} } \right)\\
\hphantom{\mathcal{O} ^{\mathfrak{lin} }_{(I )}(\mathbf{v} ) \circ  \chi _{I \leftarrow \kappa }(\psi )} = \chi _{{I \leftarrow \kappa }} \circ  \mathcal{O} ^{\mathfrak{lin} }_{\kappa }(\mathbf{v} )(\psi )},%
\MStopEqua \Mpar with \MMath{\left(\mathbf{b} _{j}\right)_{j\in J}} some {$J$}-orthonormal basis of {$W$}.
Now, we also have \MMath{\mathcal{O} ^{\mathfrak{lin} }_{\kappa }(\mathbf{v} ) = \iota _{{\kappa  \leftarrow  \lambda }} \big( \mathcal{O} ^{\mathfrak{lin} }_{\lambda }(\mathbf{v} ) \big) = \Phi ^{-1}_{\kappa \rightarrow \lambda } \,  \left( \mathds{1} ^{\mathfrak{lin} }_{{\mu _{\kappa \rightarrow \lambda }}} \otimes  \mathcal{O} ^{\mathfrak{lin} }_{\lambda }(\mathbf{v} ) \right) \,  \Phi _{\kappa \rightarrow \lambda }} {\seqDDDatg}, so, using {\seqBBBath} twice, we get:%
\Mpar \MStartEqua \MMath{\mathcal{O} ^{\mathfrak{lin} }_{(I )}(\mathbf{v} ) \circ  \chi _{I \leftarrow \kappa } = \Phi _{{I \rightarrow \lambda }}^{-1\, } \left( \tilde{\chi }_{{I \leftarrow \kappa \rightarrow \lambda }} \otimes  \mathcal{O} ^{\mathfrak{lin} }_{\lambda }(\mathbf{v} ) \right) \,  \Phi _{\kappa \rightarrow \lambda } \\
\hphantom{\mathcal{O} ^{\mathfrak{lin} }_{(I )}(\mathbf{v} ) \circ  \chi _{I \leftarrow \kappa }} = \Phi _{{I \rightarrow \lambda }}^{-1\, } \left( \tilde{\mathds{1} }_{{I \rightarrow \lambda }} \otimes  \mathcal{O} ^{\mathfrak{lin} }_{\lambda }(\mathbf{v} ) \right)\, \Phi _{{I \rightarrow \lambda }}\,  \chi _{{I \leftarrow \kappa }}\\
\hphantom{\mathcal{O} ^{\mathfrak{lin} }_{(I )}(\mathbf{v} ) \circ  \chi _{I \leftarrow \kappa }} = \Lambda _{I } \left( \mathcal{O} ^{\mathfrak{lin} }(\mathbf{v} ) \right) \circ  \chi _{{I \leftarrow \kappa }}}.%
\MStopEqua \Mpar Since \MMath{\mathcal{F} ^{(V ,I )}_{\mathfrak{lin} } = \overline{\bigcup_{{\deuxlignes{-3pt}{\kappa \in \mathcal{L} _{I }}{\kappa \geqslant \lambda }}} \chi _{I \leftarrow \kappa } \left\langle  \mathcal{H} _{\kappa } \right\rangle }} (thanks to the directedness of \MMath{\mathcal{L} _{I }}), this ensures that {\seqBBBatk} holds.%
\MendOfProof \Mpar \MstartPhyMode\hspace{\Alinea}We now want to clarify the relation between Fock state spaces and the projective state space built, along the lines of {\seqBBBapw}, on a \emph{universal label subset} \MMath{\mathcal{K} _{D }} (for some dense, countably-generated subspace {$D$} of {$V$}).
When \emph{restricting} the label set from \MMath{\mathcal{L} ^{\mathfrak{lin} }_{}} to \MMath{\mathcal{K} _{D }}, the projective state space is affected in two different ways: the new state space may contain states that have no equivalent in the original one (ie.~states that cannot be \emph{extended} over the full label set \MMath{\mathcal{L} ^{\mathfrak{lin} }_{}}), while some projective states that were distinct over \MMath{\mathcal{L} ^{\mathfrak{lin} }_{}} may become \emph{identified} once restricted to \MMath{\mathcal{K} _{D }} (intuitively, we have now less observables to distinguish between states).
The first effect is what makes states much easier to construct over \MMath{\mathcal{K} _{D }}, but the second may break the embedding of Fock state spaces: while the embedding into the state space over \MMath{\mathcal{L} ^{\mathfrak{lin} }_{}}, established in {\seqBBBaao}, can certainly be post-composed with the restriction of projective states to \MMath{\mathcal{K} _{D }}, yielding a mapping into the state space over \MMath{\mathcal{K} _{D }}, the question is whether this mapping will still be \emph{injective}.%
\Mpar \vspace{\Saut}\hspace{\Alinea}{\seqCCCatm} below answers this question in the affirmative, \emph{provided} (in the bosonic case) that the complex structure {$I$} underlying the considered Fock space induces the \emph{same} topology on {$V$} as the one with respect to which {$D$} is \emph{dense}.
Keeping in mind that this topology is supposed to come from \emph{physical considerations} (see the discussion before {\seqBBBabi}, as well as the example studied in {\seqBBBatn}, in particular {\seqBBBato}), this appears to be a fairly reasonable assumption.
Note that we \emph{do not} need any further compatibility between {$I$} and {$D$}: in particular, the complex structure {$I$} is \emph{not} assumed to stabilize the vector subspace {$D$}.
This result demonstrates that projective state spaces on universal label subsets, while admitting a convenient constructive description, offer considerable flexibility for the construction of quantum states:
all Fock representations respecting the topology can be embedded \emph{side by side} into such a state space (as stressed in {\seqBBBatp}, there can be many distinct complex structures consistent with the same topology: in {\seqBBBatn}, we will show an example of a \emph{continuum} of \emph{inequivalent} Fock representations which are all compatible with the same normable topology).%
\Mpar \vspace{\Saut}\hspace{\Alinea}The proof relies on two facts:%
\Mpar \MPhyList A quantum state is completely \emph{determined} by its \emph{characteristic function} (in the bosonic case, where linear observables have to be exponentiated), resp.\ by its \emph{moments} (in the fermionic case): this holds for density matrices on a \emph{finite-dimensional} Fock representation (first part of {\seqBBBatr}), hence can be lifted to \emph{projective states} in the state space from {\seqBBBacm}, and therefore also holds for any \emph{infinite-dimensional} Fock representation, by virtue of the previous embedding result.%
\MStopList \Mpar \MPhyList Infinite-dimensional Fock representations are \emph{weakly-continuous} (with respect to any norm on {$V$} topologically compatible with their complex structure): this ensures that the characteristic function, resp.\ the \MMath{n}-points functions, of a Fock state will be \emph{continuous} (in this topology) and can be fully reconstructed once known on a \emph{dense} subset.
The weak-continuity of \emph{infinite-dimensional} Fock representations is, again, deduced from the corresponding property of \emph{finite-dimensional} ones (second part of {\seqBBBatr}), together with the characterization of Fock states from {\seqBBBatt} (roughly expressing that those states lie arbitrarily close to the vacuum over \emph{all but finitely many modes}: thus, the task of proving weak-continuity for such states can be reduced to the task of proving it for the vacuum, whose characteristic function is simply a Gaussian distribution).%
\MStopList \MleavePhyMode \hypertarget{PARatu}{}\Mpar \Mproposition{3.17}Let {$I$} be a compatible complex structure on \MMath{V ,\Omega }, resp.\ \MMath{V ,\left(\, \cdot \, \middlewithspace|\, \cdot \, \right)}, and let \MMath{\left|\, \cdot \, \right|} be the norm induced by {$I$} {\seqDDDaqg}, resp.\ by \MMath{\left(\, \cdot \, \middlewithspace|\, \cdot \, \right)}.
Let {$D$} be a \MMath{\left|\, \cdot \, \right|}-dense subspace of {$V$} and let:%
\Mpar \MStartEqua \MMath{\mathcal{K} _{D } \mathrel{\mathop:}=  \left\{ \kappa  \in  \mathcal{L} ^{\mathfrak{lin} }_{} \middlewithspace| V _{\kappa } \subseteq D  \right\}}.%
\MStopEqua \Mpar Let \MMath{(\mathfrak{lin} ,D )} denotes the restriction to \MMath{\mathcal{K} _{D }} of the system of factorized Hilbert spaces {$\mathfrak{lin}$} and let \MMath{\Sigma _{D } : \mathcal{S} ^{\mathfrak{lin} } \rightarrow  \mathcal{S} ^{(\mathfrak{lin} ,D )}} denotes the restriction of states.
Then, \MMath{\left. \Sigma _{D } \circ  \Sigma _{I } \right|_{{\mathcal{S} ^{\mathfrak{lin} }_{{I ,+,1 }}}}} is injective.%
\Mpar \Mproof Using the argument outlined at the beginning of {\seqBBBaty} (note that this argument did \emph{not} make use of {$D$} being countably generated), \MMath{\mathcal{K} _{D }} is directed, hence \MMath{\mathcal{S} ^{(\mathfrak{lin} ,D )}} and \MMath{\Sigma _{D }} are well defined.%
\Mpar \vspace{\Sssaut}\hspace{\Alinea}Let \MMath{\rho _{I },\rho '_{I } \in  \mathcal{S} ^{\mathfrak{lin} }_{{I ,+,1 }}} such that \MMath{\Sigma _{D } \circ  \Sigma _{I }(\rho ) = \Sigma _{D } \circ  \Sigma _{I }(\rho ')}. Let \MMath{\rho  = \left(\rho _{\lambda }\right)_{{\lambda \in \mathcal{L} ^{}_{\mathfrak{lin} }}} \mathrel{\mathop:}=  \Sigma _{I }(\rho _{I })}, resp.~\MMath{\rho ' = \left(\rho '_{\lambda }\right)_{{\lambda \in \mathcal{L} ^{}_{\mathfrak{lin} }}} \mathrel{\mathop:}=  \Sigma _{I }(\rho '_{I })}. By hypothesis, we have:%
\hypertarget{PARatz}{}\Mpar \MStartEqua \MMath{\forall  \kappa  \in  \mathcal{K} _{D }, {\rho _{\kappa } = \rho '_{\kappa }}}.%
\NumeroteEqua{3.17}{1}\MStopEqua \Mpar \vspace{\Ssaut}\italique{Bosonic case.} We first consider the case \MMath{\mathfrak{lin} =\mathfrak{bos} }. Using {\seqBBBaub} together with {\seqBBBauc}, we have:%
\hypertarget{PARaud}{}\Mpar \MStartEqua \MMath{\forall  \mathbf{v}  \in  D , {\mkop{Tr}  \left( \rho _{I } \, \mathcal{O} ^{\mathfrak{bos} }_{(I )}(\mathbf{v} ) \right) = \mkop{Tr}  \left( \rho '_{I } \, \mathcal{O} ^{\mathfrak{bos} }_{(I )}(\mathbf{v} ) \right)}}.%
\NumeroteEqua{3.17}{2}\MStopEqua \Mpar \vspace{\Sssaut}\hspace{\Alinea}Let \MMath{\epsilon >0}. Recalling from {\seqBBBafp} that \MMath{\mathcal{F} ^{(V ,I )}_{\mathfrak{bos} } = \overline{\bigcup_{{\lambda  \in  \mathcal{L} _{I }}} \chi _{{I \leftarrow \lambda }} \left\langle  \mathcal{H} ^{\mathfrak{bos} }_{\lambda } \right\rangle }}, and using {\seqBBBauf}, there exists \MMath{\lambda  \in  \mathcal{L} _{I }} such that:%
\hypertarget{PARaug}{}\Mpar \MStartEqua \MMath{\left\| \rho _{I } - \Theta _{\lambda } \,  \rho _{I } \,  \Theta _{\lambda } \right\|_{1} < \frac{\epsilon }{6}},%
\NumeroteEqua{3.17}{3}\MStopEqua \Mpar where \MMath{\left\|\, \cdot \, \right\|} denotes the trace-norm \bseqHHHaaq{theorem VI.20} and \MMath{\Theta _{\lambda }} the orthogonal projection on \MMath{\chi _{{I \leftarrow \lambda }} \left\langle  \mathcal{H} ^{\mathfrak{bos} }_{\lambda } \right\rangle }.
Let \MMath{\left(\mathbf{b} _{1},\dots ,\mathbf{b} _{n}\right)} be the {$I$}-orthonormal family such that \MMath{\lambda  = \left(\mathbf{b} _{1},I \mathbf{b} _{1},\dots ,\mathbf{b} _{n},I \mathbf{b} _{n}\right)}.%
\Mpar \vspace{\Sssaut}\hspace{\Alinea}Let \MMath{\mathbf{v}  \in  V } and let \MMath{\mathbf{v} _{\lambda }} and \MMath{\mathbf{v} _{\lambda }^{\perp }} be the {$I$}-orthogonal projections of {$\mathbf{v}$} on \MMath{V _{\lambda }} and \MMath{W  \mathrel{\mathop:}=  V _{\lambda }^{\perp }}, respectively.
If \MMath{\mathbf{v} _{\lambda }^{\perp } = 0}, we have, using {\seqBBBauj}:%
\Mpar \MStartEqua \MMath{\mathcal{O} ^{\mathfrak{bos} }_{(I )}(\mathbf{v} ) = \left(\Gamma ^{(\mathbf{b} _{1},\dots ,\mathbf{b} _{n} ;V ,I )}_{\mathfrak{bos} }\right)^{-1} \circ  \left( \mathcal{O} ^{\mathfrak{bos} }_{\lambda }(\mathbf{v} _{\lambda }) \otimes  \text{id}_{{\mathcal{F} ^{(W ,J )}_{\mathfrak{bos} }}}\right) \circ  \Gamma ^{(\mathbf{b} _{1},\dots ,\mathbf{b} _{n} ;V ,I )}_{\mathfrak{bos} }}.%
\MStopEqua \Mpar If \MMath{\mathbf{v} _{\lambda }^{\perp } \neq  0}, let \MMath{\mathbf{b} _{n+1} \mathrel{\mathop:}=  \nicefrac{\mathbf{v} _{\lambda }^{\perp }}{\left|\mathbf{v} _{\lambda }^{\perp }\right|}}. {\seqCCCauj} then yields:%
\hypertarget{PARaum}{}\Mpar \MStartEqua \MMath{\mathcal{O} ^{\mathfrak{bos} }_{(I )}(\mathbf{v} ) = \left(\Gamma ^{(\mathbf{b} _{1},\dots ,\mathbf{b} _{n},\mathbf{b} _{n+1} ;V ,I )}_{\mathfrak{bos} }\right)^{-1} \circ  \left( \exp \left( i\, \hat{\mathbf{x} } \right) \otimes  \text{id}_{{\mathcal{F} ^{(W ',J ')}_{\mathfrak{bos} }}}\right) \circ  \Gamma ^{(\mathbf{b} _{1},\dots ,\mathbf{b} _{n},\mathbf{b} _{n+1} ;V ,I )}_{\mathfrak{bos} }},%
\NumeroteEqua{3.17}{4}\MStopEqua \Mpar with \MMath{W ',J '} the {$I$}-orthogonal complement of \MMath{\mkop{Span} _{\mathds{C}} \left\{\mathbf{b} _{1},\dots ,\mathbf{b} _{n+1}\right\}} and \MMath{\mathbf{x}  \in  \mathds{R}^{2n+2}} such that \MMath{\mathbf{v}  = \sum_{p=1}^{n+1} \left(\mathbf{x} _{2p-1} + i\, \mathbf{x} _{2p}\right) \mathbf{b} _{p}}.
Defining \MMath{\mathbf{x} _{\lambda } \mathrel{\mathop:}=  \left(\mathbf{x} _{1},\dots ,\mathbf{x} _{2n},0,0\right)} and \MMath{\mathbf{x} _{\lambda }^{\perp } \mathrel{\mathop:}=  \left(0,\dots ,0,\big| \mathbf{v} _{\lambda }^{\perp } \big|,0\right)}, we have from {\seqBBBanh}:%
\Mpar \MStartEqua \MMath{\exp \left( i\, \hat{\mathbf{x} } \right) = \exp \left( i\, \hat{\mathbf{x} }_{\lambda } \right) \,  \exp \left( i\, \hat{\mathbf{x} }_{\lambda }^{\perp } \right)}.%
\MStopEqua \Mpar Next, {\seqBBBauq} gives:%
\Mpar \MStartEqua \MMath{\exp \left( i\, \hat{\mathbf{x} }_{\lambda } \right) = \left(\Gamma ^{(n+1,n)}_{\mathfrak{bos} }\right)^{-1} \circ  \left( \mathcal{O} ^{\mathfrak{bos} }_{\lambda }(\mathbf{v} _{\lambda }) \otimes  \mathds{1}  \right) \circ  \Gamma ^{(n+1,n)}_{\mathfrak{bos} }},%
\MStopEqua \Mpar while combining {\seqBBBaut} with {\seqBBBauq} gives:%
\Mpar \MStartEqua \MMath{\exp \left( i\, \hat{\mathbf{x} }_{\lambda }^{\perp } \right) = \left(\Gamma ^{(n+1,n)}_{\mathfrak{bos} }\right)^{-1} \circ  \Big( \mathds{1}  \otimes  \exp \big( i\, v\, \left( \mathtt{a}  + \mathtt{a} ^{+} \right) \big) \Big) \circ  \Gamma ^{(n+1,n)}_{\mathfrak{bos} }},%
\MStopEqua \Mpar where \MMath{v \mathrel{\mathop:}=  \nicefrac{\left|\mathbf{v} _{\lambda }^{\perp }\right|}{\sqrt{2}}}. Inserting into {\seqBBBauw} and using {\seqBBBaux}, we obtain:%
\hypertarget{PARauy}{}\Mpar \MStartEqua \MMath{\mathcal{O} ^{\mathfrak{bos} }_{(I )}(\mathbf{v} ) = \left(\Gamma ^{(\mathbf{b} _{1},\dots ,\mathbf{b} _{n} ;V ,I )}_{\mathfrak{bos} }\right)^{-1} \circ  \big( \mathcal{O} ^{\mathfrak{bos} }_{\lambda }(\mathbf{v} _{\lambda }) \otimes  \mathcal{O} ^{\mathfrak{bos} }_{{\lambda ,\perp }}(\mathbf{v} _{\lambda }^{\perp }) \big) \circ  \Gamma ^{(\mathbf{b} _{1},\dots ,\mathbf{b} _{n} ;V ,I )}_{\mathfrak{bos} }},%
\NumeroteEqua{3.17}{5}\MStopEqua \Mpar where \MMath{\mathcal{O} ^{\mathfrak{bos} }_{{\lambda ,\perp }}(\mathbf{v} _{\lambda }^{\perp })} is the unitary operator on \MMath{\mathcal{F} ^{(W ,J )}_{\mathfrak{bos} }} defined by:%
\Mpar \MStartEqua \MMath{\mathcal{O} ^{\mathfrak{bos} }_{{\lambda ,\perp }}(\mathbf{v} _{\lambda }^{\perp }) \mathrel{\mathop:}=  \left(\Gamma ^{(\mathbf{b} _{n+1;W },J )}_{\mathfrak{bos} }\right)^{-1} \circ  \left( \exp \big( i\, v\, \left( \mathtt{a}  + \mathtt{a} ^{+} \right) \big) \otimes  \text{id}_{{\mathcal{F} ^{(W ',J ')}_{\mathfrak{bos} }}} \right) \circ  \Gamma ^{(\mathbf{b} _{n+1;W },J )}_{\mathfrak{bos} }}.%
\MStopEqua \Mpar If \MMath{\mathbf{v} _{\lambda }^{\perp } = 0}, we define \MMath{\mathcal{O} ^{\mathfrak{bos} }_{{\lambda ,\perp }}(\mathbf{v} _{\lambda }^{\perp }) \mathrel{\mathop:}=  \text{id}_{{\mathcal{F} ^{(W ,J )}_{\mathfrak{bos} }}}}, so that {\seqBBBavc} holds in any case.%
\Mpar \vspace{\Sssaut}\hspace{\Alinea}Using the definition of \MMath{\chi _{I \leftarrow \lambda }} {\seqDDDave}, we then have, for any \MMath{\psi ,\psi ' \in  \mathcal{H} ^{\mathfrak{bos} }_{\lambda }}:%
\Mpar \MStartEqua \MMath{\left\langle  \chi _{I \leftarrow \lambda }(\psi ') \middlewithspace| \mathcal{O} ^{\mathfrak{bos} }_{(I )}(\mathbf{v} ) \middlewithspace| \chi _{I \leftarrow \lambda }(\psi ) \right\rangle _{\mathfrak{bos} } = \left\langle  \psi ' \middlewithspace| \mathcal{O} ^{\mathfrak{bos} }_{\lambda }(\mathbf{v} _{\lambda }) \middlewithspace| \psi  \right\rangle _{\mathfrak{bos} } \,  \left\langle  \left(0\right)_{j\in J;\, }\left(\mathbf{b} _{j}\right)_{j\in J} \middlewithspace| \mathcal{O} ^{\mathfrak{bos} }_{{\lambda ,\perp }}(\mathbf{v} _{\lambda }^{\perp }) \middlewithspace| \left(0\right)_{j\in J;\, }\left(\mathbf{b} _{j}\right)_{j\in J} \right\rangle _{\mathfrak{bos} }},%
\MStopEqua \Mpar where \MMath{\left(\mathbf{b} _{j}\right)_{{j\in J'}}} is some orthonormal basis of \MMath{W '}, resp.~{$W$}, and \MMath{J = J' \sqcup \left\{n+1\right\}} (if \MMath{\mathbf{v} _{\lambda }^{\perp } \neq  0}), resp.~\MMath{J = J'} (if \MMath{\mathbf{v} _{\lambda }^{\perp } = 0}).
Inserting the definition of \MMath{\mathcal{O} ^{\mathfrak{bos} }_{{\lambda ,\perp }}(\mathbf{v} _{\lambda }^{\perp })}, this becomes:%
\Mpar \MStartEqua \MMath{\left\langle  \chi _{I \leftarrow \lambda }(\psi ') \middlewithspace| \mathcal{O} ^{\mathfrak{bos} }_{(I )}(\mathbf{v} ) \middlewithspace| \chi _{I \leftarrow \lambda }(\psi ) \right\rangle _{\mathfrak{bos} } = \left\langle  \psi ' \middlewithspace| \mathcal{O} ^{\mathfrak{bos} }_{\lambda }(\mathbf{v} _{\lambda }) \middlewithspace| \psi  \right\rangle _{\mathfrak{bos} } \,  \left\langle  0 \middlewithspace| \exp \Big( i\, \big|\mathbf{v} _{\lambda }^{\perp } \big|\, \frac{\mathtt{a}  + \mathtt{a} ^{+}}{\sqrt{2}} \Big) \middlewithspace| 0 \right\rangle _{\mathfrak{bos} }}%
\MStopEqua \Mpar (note that this holds even if \MMath{\mathbf{v} _{\lambda }^{\perp } = 0}, for \MMath{\exp(i\, 0) = \mathds{1} }).
The first term can then be substituted using the equation for \MMath{\mathbf{v} _{\lambda }} (taking advantage of \MMath{\left(\mathbf{v} _{\lambda }\right)_{\lambda }^{\perp } = 0}), while the second term can be calculated eg.~in the Schrödinger representation {\seqDDDavj}, yielding:%
\Mpar \MStartEqua \MMath{\left\langle  \chi _{I \leftarrow \lambda }(\psi ') \middlewithspace| \mathcal{O} ^{\mathfrak{bos} }_{(I )}(\mathbf{v} ) \middlewithspace| \chi _{I \leftarrow \lambda }(\psi ) \right\rangle _{\mathfrak{bos} } = e^{{-\nicefrac{1}{4}\, \left|\mathbf{v} _{\lambda }^{\perp }\right|^{2}}} \,  \left\langle  \chi _{I \leftarrow \lambda }(\psi ') \middlewithspace| \mathcal{O} ^{\mathfrak{bos} }_{(I )}(\mathbf{v} _{\lambda }) \middlewithspace| \chi _{I \leftarrow \lambda }(\psi ) \right\rangle _{\mathfrak{bos} }}.%
\MStopEqua \Mpar In other words, we have:%
\Mpar \MStartEqua \MMath{\Theta _{\lambda } \,  \mathcal{O} ^{\mathfrak{bos} }_{(I )}(\mathbf{v} ) \,  \Theta _{\lambda } = e^{{-\nicefrac{1}{4}\, \left|\mathbf{v} _{\lambda }^{\perp }\right|^{2}}} \,  \Theta _{\lambda } \,  \mathcal{O} ^{\mathfrak{bos} }_{(I )}(\mathbf{v} _{\lambda }) \,  \Theta _{\lambda }}.%
\MStopEqua \Mpar Using {\seqBBBavo} and recalling that both \MMath{\mathcal{O} ^{\mathfrak{bos} }_{(I )}(\mathbf{v} )} and \MMath{\mathcal{O} ^{\mathfrak{bos} }_{(I )}(\mathbf{v} _{\lambda })} are unitary operators (hence of unit norm), we get:%
\Mpar \MStartEqua \MMath{\left| \mkop{Tr}  \left( \rho _{I } \, \mathcal{O} ^{\mathfrak{bos} }_{(I )}(\mathbf{v} ) \right) - e^{{-\nicefrac{1}{4}\, \left|\mathbf{v} _{\lambda }^{\perp }\right|^{2}}} \,  \mkop{Tr}  \left( \rho _{I } \, \mathcal{O} ^{\mathfrak{bos} }_{(I )}(\mathbf{v} _{\lambda }) \right) \right| < \frac{\epsilon }{3}}.%
\MStopEqua \Mpar Using {\seqBBBauc} in the reverse direction, and noting that \MMath{\mathbf{v} _{\lambda } \in  V _{\lambda }} by definition, we arrive at:%
\hypertarget{PARavr}{}\Mpar \MStartEqua \MMath{\left| \mkop{Tr}  \left( \rho _{I } \, \mathcal{O} ^{\mathfrak{bos} }_{(I )}(\mathbf{v} ) \right) - e^{{-\nicefrac{1}{4}\, \left|\mathbf{v} _{\lambda }^{\perp }\right|^{2}}} \,  \mkop{Tr}  \left( \rho _{\lambda } \, \mathcal{O} ^{\mathfrak{bos} }_{\lambda }(\mathbf{v} _{\lambda }) \right) \right| < \frac{\epsilon }{3}}.%
\NumeroteEqua{3.17}{6}\MStopEqua \Mpar \vspace{\Sssaut}\hspace{\Alinea}The functions \MMath{V  \rightarrow  \mathds{R},\,  \mathbf{v}  \mapsto  e^{{-\nicefrac{1}{4}\, \left|\mathbf{v} _{\lambda }^{\perp }\right|^{2}}}} and \MMath{V  \rightarrow  V _{\lambda },\,  \mathbf{v}  \mapsto  \mathbf{v} _{\lambda }} are \MMath{\left|\, \cdot \, \right|}-continuous. Moreover, from {\seqBBBavt}, the function \MMath{V _{\lambda } \rightarrow  \mathds{R},\,  \mathbf{v} _{\lambda } \mapsto  \mkop{Tr}  \left( \rho _{\lambda } \, \mathcal{O} ^{\mathfrak{bos} }_{\lambda }(\mathbf{v} _{\lambda }) \right)} is continuous as well. Hence, for any \MMath{\mathbf{v}  \in  V }, there exists a \MMath{\left|\, \cdot \, \right|}-open neighborhood \MMath{U_{\epsilon }} of {$\mathbf{v}$} such that:%
\Mpar \MStartEqua \MMath{\forall  \mathbf{w}  \in  U_{\epsilon }, {\left| e^{{-\nicefrac{1}{4}\, \left|\mathbf{w} _{\lambda }^{\perp }\right|^{2}}} \,  \mkop{Tr}  \left( \rho _{\lambda } \, \mathcal{O} ^{\mathfrak{bos} }_{\lambda }(\mathbf{w} _{\lambda }) \right) - e^{{-\nicefrac{1}{4}\, \left|\mathbf{v} _{\lambda }^{\perp }\right|^{2}}} \,  \mkop{Tr}  \left( \rho _{\lambda } \, \mathcal{O} ^{\mathfrak{bos} }_{\lambda }(\mathbf{v} _{\lambda }) \right) \right|} < \frac{\epsilon }{3}}.%
\MStopEqua \Mpar Combining with {\seqBBBavw}, this gives:%
\Mpar \MStartEqua \MMath{\forall  \mathbf{w}  \in  U_{\epsilon }, {\left| \mkop{Tr}  \left( \rho _{I } \, \mathcal{O} ^{\mathfrak{bos} }_{(I )}(\mathbf{w} ) \right) - \mkop{Tr}  \left( \rho _{I } \, \mathcal{O} ^{\mathfrak{bos} }_{(I )}(\mathbf{v} ) \right) \right|} < \epsilon }.%
\MStopEqua \Mpar In other words, the function \MMath{V  \rightarrow  \mathds{R},\,  \mathbf{v}  \mapsto  \mkop{Tr}  \left( \rho _{I } \, \mathcal{O} ^{\mathfrak{bos} }_{(I )}(\mathbf{v} ) \right)} is \MMath{\left|\, \cdot \, \right|}-continuous. Similarly, the same is true for \MMath{\rho '_{I }}. Since {$D$} is \MMath{\left|\, \cdot \, \right|}-dense, it follows that {\seqBBBavz} holds for every \MMath{\mathbf{v}  \in  V }.
Using {\seqBBBauc} once more, we then have, for any \MMath{\lambda  \in  \mathcal{L} ^{\mathfrak{bos} }_{}}:%
\Mpar \MStartEqua \MMath{\forall  \mathbf{v}  \in  V _{\lambda }, {\mkop{Tr}  \left( \rho _{\lambda } \, \mathcal{O} ^{\mathfrak{bos} }_{\lambda }(\mathbf{v} ) \right) = \mkop{Tr}  \left( \rho '_{\lambda } \, \mathcal{O} ^{\mathfrak{bos} }_{\lambda }(\mathbf{v} ) \right)}}.%
\MStopEqua \Mpar Using {\seqBBBavt}, this implies \MMath{\forall  \lambda  \in  \mathcal{L} ^{\mathfrak{bos} }_{}, {\rho _{\lambda } = \rho '_{\lambda }}}, so \MMath{\rho  = \rho '}. Since \MMath{\left. \Sigma _{I } \right|_{{\mathcal{S} ^{\mathfrak{bos} }_{{I ,+,1 }}}}} is injective {\seqDDDatt}, we conclude that \MMath{\rho _{I } = \rho '_{I }}.%
\Mpar \vspace{\Ssaut}\italique{Fermionic case.} We now consider the case \MMath{\mathfrak{lin} =\mathfrak{ferm} }. For any \MMath{k \geqslant  0} and any \MMath{\mathbf{v} _{1},\dots ,\mathbf{v} _{k} \in  D }, there exists \MMath{\kappa  \in  \mathcal{K} _{D }} such that \MMath{\mathbf{v} _{1},\dots ,\mathbf{v} _{k} \in  V _{\kappa }}. Hence, {\seqBBBaub} yields:%
\Mpar \MStartEqua \MMath{\forall  k\geqslant 0,\,  \forall  \mathbf{v} _{1},\dots ,\mathbf{v} _{k} \in  D , {\mkop{Tr}  \left( \rho \, \mathcal{O} ^{\mathfrak{ferm} }(\mathbf{v} _{1}) \dots  \mathcal{O} ^{\mathfrak{ferm} }(\mathbf{v} _{k}) \right) = \mkop{Tr}  \left( \rho '\, \mathcal{O} ^{\mathfrak{ferm} }(\mathbf{v} _{1}) \dots  \mathcal{O} ^{\mathfrak{ferm} }(\mathbf{v} _{k}) \right)}}.%
\MStopEqua \Mpar Using {\seqBBBawf}, we then get:%
\hypertarget{PARawg}{}\Mpar \MStartEqua \MMath{\forall  k\geqslant 0,\,  \forall  \mathbf{v} _{1},\dots ,\mathbf{v} _{k} \in  D , {\mkop{Tr}  \left( \rho _{I } \, \mathcal{O} ^{\mathfrak{ferm} }_{(I )}(\mathbf{v} _{1}) \dots  \mathcal{O} ^{\mathfrak{ferm} }_{(I )}(\mathbf{v} _{k}) \right) = \mkop{Tr}  \left( \rho '_{I } \, \mathcal{O} ^{\mathfrak{ferm} }_{(I )}(\mathbf{v} _{1}) \dots  \mathcal{O} ^{\mathfrak{ferm} }_{(I )}(\mathbf{v} _{k}) \right)}}.%
\NumeroteEqua{3.17}{7}\MStopEqua \Mpar \vspace{\Sssaut}\hspace{\Alinea}Let \MMath{\mathbf{v} ,\mathbf{w}  \in  V }. Applying {\seqBBBawi} for some {$I$}-orthonormal basis \MMath{\left(\mathbf{b} _{i}\right)_{i\leqslant n}} of \MMath{\mkop{Span} _{\mathds{C}} \left\{\mathbf{v} ,\mathbf{w} \right\}}, we have:%
\Mpar \MStartEqua \MMath{\mathcal{O} ^{\mathfrak{ferm} }_{(I )}(\mathbf{v} ) = \left(\Gamma ^{(\left(\mathbf{b} _{i}\right);V ,I )}_{\mathfrak{ferm} }\right)^{-1} \circ  \left(\hat{\mathbf{x} } \otimes  \text{id}_{{\mathcal{F} ^{(W ,J )}_{\mathfrak{ferm} }}}\right) \circ  \Gamma ^{(\left(\mathbf{b} _{i}\right);V ,I )}_{\mathfrak{ferm} } \\
\mathcal{O} ^{\mathfrak{ferm} }_{(I )}(\mathbf{w} ) = \left(\Gamma ^{(\left(\mathbf{b} _{i}\right);V ,I )}_{\mathfrak{ferm} }\right)^{-1} \circ  \left(\hat{\mathbf{y} } \otimes  \text{id}_{{\mathcal{F} ^{(W ,J )}_{\mathfrak{ferm} }}}\right) \circ  \Gamma ^{(\left(\mathbf{b} _{i}\right);V ,I )}_{\mathfrak{ferm} }},%
\MStopEqua \Mpar where \MMath{W  \mathrel{\mathop:}=  \left(\mkop{Span} _{\mathds{C}} \left\{\mathbf{v} ,\mathbf{w} \right\}\right)^{\perp }}, \MMath{J  \mathrel{\mathop:}=  \left.I \right|_{W }} and \MMath{\mathbf{x} ,\mathbf{y}  \in  \mathds{R}^{n}} are such that \MMath{\mathbf{v}  = \sum_{p=1}^{n} \left(\mathbf{x} _{2p-1} + i\, \mathbf{x} _{2p}\right) \mathbf{b} _{p}}, resp.~\MMath{\mathbf{w}  = \sum_{p=1}^{n} \left(\mathbf{y} _{2p-1} + i\, \mathbf{y} _{2p}\right) \mathbf{b} _{p}}. Using the definition of \MMath{\hat{\mathbf{x} }} {\seqDDDanl}, we thus have:%
\Mpar \MStartEqua \MMath{\left\|\mathcal{O} ^{\mathfrak{ferm} }_{(I )}(\mathbf{v} )\right\| = \sqrt{\left\|\hat{\mathbf{x} }^{\dag} \,  \hat{\mathbf{x} }\right\|} = \sqrt{{\frac{{}^{\text{\sc t}} \mathbf{x}  \,  \mathbf{x} }{2}}} = \frac{\left|\mathbf{v} \right|}{\sqrt{2}} \, \&\,  \left\|\mathcal{O} ^{\mathfrak{ferm} }_{(I )}(\mathbf{w} ) - \mathcal{O} ^{\mathfrak{ferm} }_{(I )}(\mathbf{v} )\right\| = \sqrt{{\frac{{}^{\text{\sc t}} (\mathbf{y} -\mathbf{x} ) \,  (\mathbf{y} -\mathbf{x} )}{2}}} = \frac{\left|\mathbf{w}  - \mathbf{v} \right|}{\sqrt{2}}}.%
\MStopEqua \Mpar Thus, for any \MMath{k\geqslant 0}, the map \MMath{V ^{k} \rightarrow  \mathcal{B} ^{\mathfrak{ferm} }_{I },\,  \mathbf{v} _{1},\dots ,\mathbf{v} _{k} \mapsto  \mathcal{O} ^{\mathfrak{ferm} }_{(I )}(\mathbf{v} _{1}) \dots  \mathcal{O} ^{\mathfrak{ferm} }_{(I )}(\mathbf{v} _{k})} is \MMath{\left|\, \cdot \, \right|}-continuous. Since {$D$} is \MMath{\left|\, \cdot \, \right|}-dense, and \MMath{\rho _{I },\rho '_{I }} are trace-class, this ensures that {\seqBBBawn} holds for any \MMath{\mathbf{v} _{1},\dots ,\mathbf{v} _{k} \in  V }.
Using {\seqBBBawf} in the reverse direction, we obtain, for any \MMath{\lambda  \in  \mathcal{L} ^{\mathfrak{ferm} }_{}}:%
\Mpar \MStartEqua \MMath{\forall  k\geqslant 0,\,  \forall  \mathbf{v} _{1},\dots ,\mathbf{v} _{k} \in  V _{\lambda }, {\mkop{Tr}  \left( \rho _{\lambda } \,  \mathcal{O} ^{\mathfrak{ferm} }_{\lambda }(\mathbf{v} _{1}) \dots  \mathcal{O} ^{\mathfrak{ferm} }_{\lambda }(\mathbf{v} _{k}) \right) = \mkop{Tr}  \left( \rho '_{\lambda } \,  \mathcal{O} ^{\mathfrak{ferm} }_{\lambda }(\mathbf{v} _{1}) \dots  \mathcal{O} ^{\mathfrak{ferm} }_{\lambda }(\mathbf{v} _{k}) \right)}}.%
\MStopEqua \Mpar Therefore, {\seqBBBawq} implies that \MMath{\rho _{\lambda } = \rho '_{\lambda }} for any \MMath{\lambda  \in  \mathcal{L} ^{\mathfrak{ferm} }_{}}, in other words, that \MMath{\rho  = \rho '}. Since \MMath{\left. \Sigma _{I } \right|_{{\mathcal{S} ^{\mathfrak{ferm} }_{{I ,+,1 }}}}} is injective {\seqDDDatt}, we conclude that \MMath{\rho _{I } = \rho '_{I }}.
\MendOfProof %
\Mnomdefichier{lin21}%
\hypertarget{SECaws}{}\MsectionA{aws}{4}{Application to QFT on Curved Spacetime}%
\vspace{\PsectionA}\Mpar \MstartPhyMode\hspace{\Alinea}In the previous section, we took as \emph{input} some linear \emph{classical phase space} (equipped with a suitable topology) and we established how to construct a corresponding \emph{quantum state space}. Then we proved various universality results. We now want to illustrate this procedure, by showing how a suitable phase space can be obtained for a \emph{concrete} classical field theory, and by discussing what the universality results grant us in this case.%
\MleavePhyMode \hypertarget{SECawu}{}\MsectionB{awu}{4.1}{Description of a Klein–Gordon Field}%
\vspace{\PsectionB}\Mpar \MstartPhyMode\hspace{\Alinea}We take a \emph{globally hyperbolic spacetime} {$\mathcal{M}$} (ie.~a spacetime that can be \emph{foliated} by \emph{Cauchy surfaces}, see {\bseqJJJaav}), and consider a \emph{free} Klein–Gordon scalar field thereon.
To any Cauchy surface \MMath{\Sigma _{\mathtt{c} }} in {$\mathcal{M}$} we can associate a classical phase space \MMath{V _{\mathtt{c} }}, consisting of the \emph{germs of solutions} of the field equations in the vicinity of this surface.
The \emph{time evolution} between two such Cauchy surfaces provides a \emph{linear} symplectomorphism between the corresponding phase spaces (left part of {\seqBBBaww}).%
\Mpar \vspace{\Saut}\hspace{\Alinea}By mapping each spatial slice into a \emph{model slice} {$\Sigma$}, we can map the phase spaces on them into a \emph{common} symplectic vector space {$V$}, carrying a linear, symplectomorphic representation of the time evolution (or, more generally, of the full groupoid of slice displacements and deformations), as represented on {\seqBBBaww}.
Then, using the framework from {\seqBBBabe}, we want to turn {$V$} into a quantum state space on which slice changes will act as \emph{automorphisms}. Since there is, in general, \emph{no} complex structure on {$V$} that would be \emph{preserved} under slice change (we show in {\seqBBBabp} how this fails even in a very simple example), relying on a Fock representation is unsuitable: most slice changes would not act as unitary transformations.%
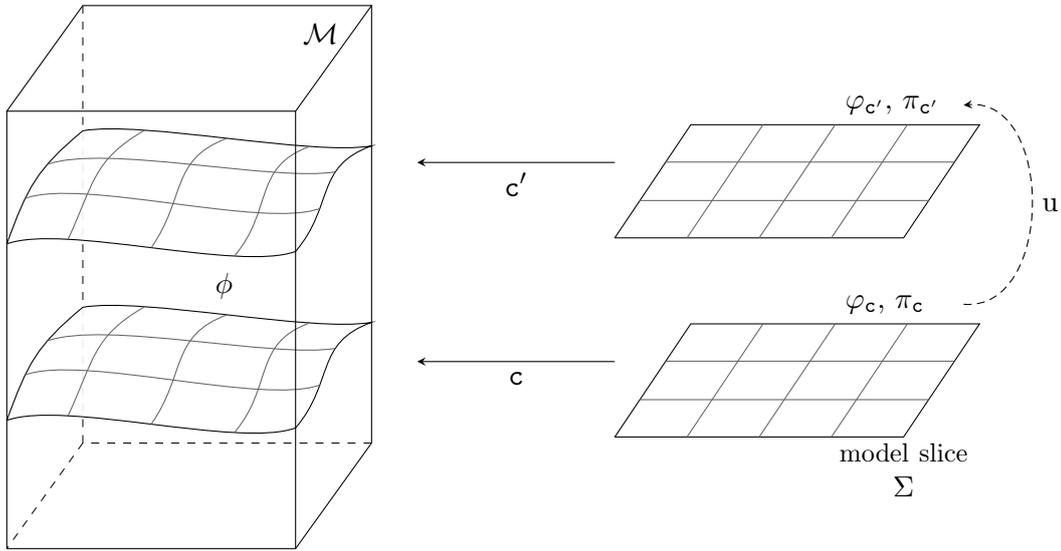
\Myfigure{4.1}{\hypertarget{PARawx}{}}{Time evolution between Cauchy surfaces, yielding an automorphism of the phase space built over a model slice}{{\InputFigPawy}}%
\Mpar \vspace{\Saut}\hspace{\Alinea}Alternatively, we could dispense from introducing a model slice, and construct an \emph{a priori distinct} quantum state space on each slice: the symplectomorphism relating the phase spaces \MMath{V _{\mathtt{c} }} on two different slices should then lead to an identification (isomorphism in the sense of {\seqBBBaha}) between the corresponding quantum state spaces and associated observables algebras.
While this would be equivalent to the previous approach (for the identifications would effectively solder together the state spaces over all possible slices into a single, common state space),
the advantage would be to pave the way from a more flexible formulation, and in particular for the use of partial, \emph{localized} spatial slices, rather than full, \emph{extended} ones (in the spirit of the General Boundary Formulation {\bseqJJJaaw}, as will be briefly discussed in {\seqBBBaax}).%
\MleavePhyMode \MleavePhyMode \hypertarget{PARaxa}{}\Mpar \Mdefinition{4.1}Let \MMath{\mathcal{M} } be an oriented \MMath{(d+1)}-dimensional (smooth) manifold and let {$g$} be a pseudo-Riemannian metric on {$\mathcal{M}$}, of signature \MMath{-,+,\dots ,+}. Let \MMath{m \geqslant  0}. We define:%
\Mpar \MStartEqua \MMath{\mathcal{S} _{\mathcal{M} } \mathrel{\mathop:}=  \left\{ \phi  \in  C^{\infty }(\mathcal{M}  \rightarrow  \mathds{R}) \middlewithspace| \square_{g } \phi  - m^{2} \,  \phi  = 0 \right\}},%
\MStopEqua \Mpar where \MMath{C^{\infty }(\mathcal{M}  \rightarrow  \mathds{R})} denotes the space of smooth, real-valued functions on {$\mathcal{M}$}, and \MMath{\square_{g }} the d'Alembert operator (aka.~Laplace–Beltrami operator, see \bseqHHHaax{section V.B.4}), whose expression in a coordinate chart is (using implicit summation convention):%
\Mpar \MStartEqua \MMath{\square_{g } \mathrel{\mathop:}=  \frac{1}{\sqrt{\left|g \right|}} \partial _{\mu } \left( \sqrt{\left|g \right|} \,  g ^{\mu \nu } \,  \partial _{\nu } \, \cdot \,  \right)},%
\MStopEqua \Mpar with \MMath{\left|g \right| \mathrel{\mathop:}=  - \det g _{\mu \nu }}.%
\Mpar \MstartPhyMode\hspace{\Alinea}The Klein–Gordon equation involves \emph{2nd-order} time-derivatives of the field, thus we expect the phase space attached to a Cauchy surface \MMath{\Sigma _{\mathtt{c} }} to consist of the value of the \emph{field} and its \emph{first time-derivative} on \MMath{\Sigma _{\mathtt{c} }}.
In order for the time evolution (as well as more general slice changes) to act as a \emph{symplectomorphism}, we need however to clarify the \emph{differential-geometric nature} of the objects involved. A careful analysis reveals that the time derivative of the field, aka.~its \emph{momentum}, should be thought as a \emph{volume-form}\footnote{To be even more precise, it is a \emph{density} \bseqHHHaay{section 11.4}: its sign is tied to the orientation of the \emph{normal vector} to the spatial slice, rather than to an intrinsic orientation \emph{along} this slice. In {\seqBBBaxg}, the orientation of this normal vector is inferred by comparing an orientation on {$\Sigma$} with a pre-determined \emph{global} orientation of the spacetime {$\mathcal{M}$}, which is how a dependence in the former is fortuitously introduced.} over \MMath{\Sigma _{\mathtt{c} }}.
Then, by contracting the (scalar-valued) field with the (top-form-valued) momentum, we obtain an object which can be integrated \emph{intrinsically}, so that the symplectic structure defined in this way will \emph{not} depend on any data \emph{local to the slice}, and will be preserved under slice change.%
\Mpar \vspace{\Saut}\hspace{\Alinea}For convenience, we will \emph{convert} the momentum \emph{back} into a simple scalar-valued function, through the use of a \emph{fiducial metric}: this is the reason for the particular scaling entering the definition of the normal vector {$\mathbf{n}$} below. By fixing the fiducial metric once and for all on the "model slice" {$\Sigma$}, we guarantee the invariance of the symplectic structure: it will be expressed as an integral against the fiducial volume element (see {\seqBBBaxh}).%
\MleavePhyMode \hypertarget{PARaxi}{}\Mpar \Mdefinition{4.2}Let {$\Sigma$} be an oriented \MMath{d}-dimensional manifold and let \MMath{h _{o}} be a positive-definite metric on {$\Sigma$}. A spatial {$\Sigma$}-slice in {$\mathcal{M}$} is a map \MMath{\mathtt{c}  : \Sigma  \rightarrow  \mathcal{M} } such that:%
\Mpar \MList{1}the image \MMath{\Sigma _{\mathtt{c} } \mathrel{\mathop:}=  \mathtt{c}  \left\langle  \Sigma  \right\rangle } of {$\mathtt{c}$} is a regular (aka.~embedded, \bseqHHHaaz{section 5}) submanifold of {$\mathcal{M}$};%
\MStopList \Mpar \MList{2}{$\mathtt{c}$} induces a \emph{diffeomorphism} between {$\Sigma$} and \MMath{\Sigma _{\mathtt{c} }} (the latter with its manifold structure inherited from {$\mathcal{M}$});%
\MStopList \Mpar \MList{3}\MMath{\Sigma _{\mathtt{c} }} is space-like, ie.~the pull-back \MMath{h _{\mathtt{c} } \mathrel{\mathop:}=  \mathtt{c} _{*} g } of {$g$} is a positive-definite metric on {$\Sigma$}.%
\MStopList \Mpar \vspace{\Sssaut}\hspace{\Alinea}Let \MMath{x \in  \Sigma } and \MMath{p \mathrel{\mathop:}=  \mathtt{c} (x) \in  \Sigma _{\mathtt{c} }}. A coordinate chart \MMath{(y^{\mu })_{{\mu =0,1,\dots ,d}}} in a neighborhood \MMath{U} of \MMath{p} in {$\mathcal{M}$} is \emph{adapted} to {$\mathtt{c}$} if:%
\Mpar \MList{4}\MMath{\Sigma _{c} \cap  U = \left\{ (y^{\mu })_{{\mu =0,1,\dots ,d}} \middlewithspace| y^{0} = 0 \right\}};%
\MStopList \Mpar \MList{5}the basis \MMath{(\partial _{0},\partial _{1},\dots ,\partial _{d})} of \MMath{T_{p}(\mathcal{M} )} is positively oriented with respect to the orientation of {$\mathcal{M}$};%
\MStopList \Mpar \MList{6}the basis \MMath{(\partial _{1},\dots ,\partial _{d})} of \MMath{T_{p}(\Sigma _{\mathtt{c} })} is the image by \MMath{[d\mathtt{c} ]_{x}} of a basis of \MMath{T_{x}(\Sigma )} positively oriented with respect to the orientation of {$\Sigma$}.%
\MStopList \Mpar Such adapted coordinate charts always exist (by definition of a regular submanifold). Given \MMath{(y^{\mu })_{{\mu =0,1,\dots ,d}}}, we define the \emph{associated} coordinate chart \MMath{(x^{i})_{{i=1,\dots ,d}}}, in the neighborhood \MMath{\mathtt{c} ^{-1} \left\langle U\right\rangle } of \MMath{x} in {$\Sigma$}, such that the coordinate representation of {$\mathtt{c}$} becomes \MMath{(x^{1},\dots ,x^{d}) \mapsto  (0,x^{1},\dots ,x^{d})}.%
\Mpar \vspace{\Sssaut}\hspace{\Alinea}For any smooth, real-valued function {$\phi$} on {$\mathcal{M}$} we define \MMath{\phi _{\mathtt{c} } \in  C^{\infty }(\Sigma  \rightarrow  \mathds{R}) \times  C^{\infty }(\Sigma  \rightarrow  \mathds{R})} by:%
\Mpar \MStartEqua \MMath{\forall  x \in  \Sigma , {\phi _{\mathtt{c} }(x) \mathrel{\mathop:}=  \left( \phi  \circ  \mathtt{c} (x);\,  \left[ d\phi  \right]_{{\mathtt{c} (x)}}(\mathbf{n} _{x}) \right)}},%
\MStopEqua \Mpar where, for any \MMath{x \in  \Sigma }, \MMath{\mathbf{n} _{x} \in  T_{{\mathtt{c} (x)}}(\mathcal{M} )} is the vector defined by:%
\Mpar \MStartEqua \MMath{\mathbf{n} ^{\nu }_{x} \mathrel{\mathop:}=  - \sqrt{{\frac{\left|g \right|}{\left|h _{o}\right|}}} \,  g ^{0\nu }},%
\MStopEqua \Mpar in a coordinate chart adapted to {$\mathtt{c}$} around \MMath{\mathtt{c} (x) \in  \Sigma _{\mathtt{c} }}, with the determinant \MMath{\left|h _{o}\right|} computed in the associated coordinate chart around \MMath{x \in  \Sigma }.
Note that \MMath{\mathbf{n} _{x}} is independent of the choice of adapted coordinate chart.%
\Mpar \MstartPhyMode\hspace{\Alinea}In a first approximation, the phase space can be parametrized by pairs of \emph{smooth, real-valued functions} on {$\Sigma$} (the field and its momentum, as defined by the \MMath{\phi  \mapsto  \phi _{\mathtt{c} }} mapping from {\seqBBBaxg}), and the symplectic structure thereon can be defined by \emph{integrating} those with respect to the fiducial metric \MMath{h _{o}}.
The form of the Klein–Gordon equation, together with the precise expression for \MMath{\phi _{\mathtt{c} }} (including the proper scaling of the normal vector, as explained in the comment preceding {\seqBBBaxg}), ensures that slice changes are indeed realized as \emph{linear symplectomorphisms}.
Thus, quantizing this phase space as described in {\seqBBBacm}, the implementation of the dynamics will follow immediately from the general action of linear symplectomorphisms presented in {\seqBBBaog}.%
\Mpar \vspace{\Saut}\hspace{\Alinea}However, to take advantage of universal label subsets, along the lines of {\seqBBBapw}, we would like to equip the classical phase space with a suitable \emph{topology}.
This topology need to be \emph{compatible} with the symplectic structure (as formalized in {\seqBBBaos}) and such that the symplectomorphisms representing slice changes are all \emph{bounded}.
Indeed, it is \emph{only} for bounded transformations (elements of the topological group \MMath{\mathcal{A} ^{\mathfrak{lin} }_{o}} used in {\seqBBBatp}) that {\seqBBBaxx} guarantee the existence of \emph{arbitrary good approximations}: we cannot expect \emph{unbounded} transformations to be represented in a satisfactory way over the projective quantum state space once we restrict ourselves to a universal label subset.
As we will see in the next subsection, getting the topology right is the most subtle part of the construction.%
\Mpar \vspace{\Saut}\hspace{\Alinea}Having chosen our topology, we may want to complete the phase space accordingly (eg.~turning it into a \MMath{L_{2}} space or, more generally, into some kind of Sobolev space): although this step is optional (since we have allowed for the symplectic form to be only weakly non-degenerate, see {\seqBBBamq}), it can make the resulting phase space {$W$} more convenient to work with.
Note that we have implicitly assumed that the symplectic form is well-defined for \emph{arbitrary} pairs of smooth functions (so that the phase space {$W$} \emph{extends} the space \MMath{C^{\infty }(\Sigma  \rightarrow  \mathds{R}) \times  C^{\infty }(\Sigma  \rightarrow  \mathds{R})}, or at least the image under \MMath{\phi _{\mathtt{c} }} of the space of \emph{global} smooth solutions): this is not an issue as long as {$\Sigma$} is \emph{compact} since smooth functions will be bounded in this case, and the integration domain will be finite.
As will become clear in {\seqBBBatn}, the framework set up in the present subsection is anyway not really appropriate for non-compact spatial slices (see also {\seqBBBaax}).%
\MleavePhyMode \hypertarget{PARaxy}{}\Mpar \Mdefinition{4.3}Let {$\mathcal{M}$} be as in {\seqBBBaxz} and let {$\Sigma$} be as in {\seqBBBaxg}. Let {$\mathcal{C}$} be a set of spatial {$\Sigma$}-slices in {$\mathcal{M}$}. Let \MMath{m \geqslant  0}. A phase space for \MMath{\mathcal{M} , \mathcal{C} } is a \emph{normable} symplectic vector space {$W$}, together with a \emph{linear} map \MMath{\Psi : C^{\infty }(\Sigma  \rightarrow  \mathds{R}) \times  C^{\infty }(\Sigma  \rightarrow  \mathds{R}) \rightarrow  W } satisfying:%
\hypertarget{PARaya}{}\Mpar \MList{1}for any \MMath{\mathtt{c}  \in  \mathcal{C} }, the map:%
\Mpar \MStartEqua \MMath{\definitionFonction{\Psi _{\mathtt{c} }}{\mathcal{S} _{\mathcal{M} }}{W }{\phi }{\Psi  \big( \phi _{\mathtt{c} } \big)}}%
\MStopEqua \Mpar is \emph{injective} (aka.~one-to-one) and its image is a \emph{dense} vector subspace of {$W$};%
\MStopList \hypertarget{PARayd}{}\Mpar \MList{2}for any \MMath{\mathtt{c} ,\mathtt{c} ' \in  \mathcal{C} }, there exists a \emph{bounded} linear symplectomorphism u{} on {$W$}, such that \MMath{\Psi _{\mathtt{c} '} = \text{u} \circ  \Psi _{\mathtt{c} }} (note that, due to {\seqBBBaye}, this uniquely characterizes u{}, and applying for \MMath{\mathtt{c} ',\mathtt{c} } implies that u{} is bijective and bi-continuous, ie.~it is an element of \MMath{\mathcal{A} _{o}(W )}, see {\seqBBBaou}).%
\MSStopList \hypertarget{SECayf}{}\MsectionB{ayf}{4.2}{Example: Spatially-compact Cosmological Spacetimes}%
\vspace{\PsectionB}\Mpar \MstartPhyMode\hspace{\Alinea}As a minimal example, we now specialize to the type of geometry often considered in cosmological models: the spacetime {$\mathcal{M}$} is assumed to admit a \emph{preferred foliation}, with the sole time-dependence being in the form of a \emph{scale factor} multiplying a constant spatial metric (and, for simplicity, we will restrict ourselves to spatial slices cut \emph{along} this preferred foliation, using the constant spatial metric as our fiducial metric).%
\Mpar \vspace{\Saut}\hspace{\Alinea}It turns out that the right topology for this system is \emph{not} the naive \MMath{L_{2}} one on \MMath{C^{\infty }(\Sigma  \rightarrow  \mathds{R}) \times  C^{\infty }(\Sigma  \rightarrow  \mathds{R})}. Instead, in order to allow for the time-evolution to be realized as a \emph{bounded} transformation, we need to introduce corrective weights at high-energy. In \emph{Fourier transform}, this amounts roughly to integrating the modulus squared of the field with respect to a \MMath{\left(p^{2} + m^{2}\right)^{{\nicefrac{1}{2}}} \,  d^{(d)}p} measure\footnote{As for the modulus squared of the field conjugate momentum, it is integrated with a \MMath{\left(p^{2} + m^{2}\right)^{{\heavyminus \nicefrac{1}{2}}} \,  d^{(d)}p} measure, to compensate for the implicit \MMath{\sqrt{{p^{2} + m^{2}}}} contribution coming from the time derivative: these two opposite rescalings cancel out when considering a product of the field with its conjugate momentum, to give the correct compatibility with the symplectic structure (which is defined without any special weighting).}, which corresponds to the \emph{Lorentz-invariant Dirac distribution} on the mass-shell of a particle of mass \MMath{m} \bseqHHHaaf{section 2.5}.
However, this does \emph{not} mean that we are dependent on the spacetime to be \emph{globally} Lorentz-invariant (it is not), as we would if we were using this modified \MMath{L_{2}}-structure to pick-out a preferred \emph{polarization} (keeping in mind that the choice of a \emph{norm} often prescribes an associated polarization, as shown in {\seqBBBaoq}).
Because we only care about the \emph{topology} (thanks to {\seqBBBayh}), only the \emph{asymptotic behavior} of the weight is important: any measure behaving asymptotically as \MMath{\Theta (\left|p\right|)} (ie.~bounded above and below by \MMath{\left|p\right|} up to a constant) induces the same topology.
This suggests that the topology we will be using is not only adequate for the particular class of non-Minkowski spacetimes considered here (as we will explicitly prove), but should more generally be so for any scalar field theory with the same \emph{high-energy} (aka.~UV) structure as the Klein–Gordon equation (namely \emph{local} Lorentz-invariance), in agreement with one the motivations put forward in the introduction {\seqDDDako}: it is better if we can avoid \emph{fine-tuning} the quantum state space to the dynamics.%
\Mpar \vspace{\Saut}\hspace{\Alinea}Indeed, the \emph{adiabatic} methods employed in the proof below can presumably be adapted, eg.~to more general spacetimes and generic spatial slices in them. The basic idea is that the boundedness of the time-evolution is governed by the growth of the high-energy modes (because there are only finitely many modes \emph{below} any given energy threshold), and, since those oscillate very fast, they are not \emph{sensible} to the variations of the metric: hence, we can simply adopt the topology that would be appropriate on Minkowski background (which, as a static spacetime, admits an invariant complex structure, and therefore a norm with respect to which the time-evolution is isometric; see {\seqBBBato} below).
On the other hand, the compactness hypothesis of {\seqBBBabo} cannot be so easily relaxed. Continuous functions on a non-compact space are not necessarily bounded, so that different norms are less likely to be \emph{equivalent} (aka.~induce the same topology) and operators are less likely to be bounded. In particular, there is not even a canonical \MMath{L_{2}} topology on a \emph{non-compact manifold}: by contrast, all \MMath{L_{2}} spaces defined from \emph{smooth measures} \bseqHHHaay{section 11.4} on a compact manifold are isomorphic \emph{as topological vector spaces} (see also \bseqHHHaba{chap.~3}).
A possible way out of this issue will be sketched in {\seqBBBaax}.%
\Mpar \vspace{\Saut}\hspace{\Alinea}Finally, some comments are warranted on the subject of \emph{modes}. The very simple example we chose to discuss here admits a preferred \emph{mode-decomposition}: its classical phase space splits into a large direct sum of two-dimensional vector subspaces, each of which holds one \dof (ie.~can be parametrized by a pair of canonically-conjugate variables), and this splitting is \emph{preserved} (in fact: selected) by the time-evolution. This should not be confused with a preferred \emph{polarization}: although typical field theories on Minkowski background possess both, we may have the former without the latter (as we do in the present case, due to the lack of preferred complex structures on the modes \emph{themselves}), or vice-versa (a polarization only entails a preferred structure of complex Hilbert space on the phase space, it does not a priori equip it with a preferred complex orthogonal \emph{basis}). Furthermore, while the availability of such a mode-decomposition certainly make for a huge simplification in all calculations below, this should not give the impression that it would be an essential ingredient in our construction of the quantum state space.
Nowhere in {\seqBBBabe} did we assume any further structure on the classical phase space beyond its linear and symplectic structures and its topology. In particular, the definition of the label set {\seqDDDayi} has no knowledge of preferred modes: labels do \emph{not} have to respect the boundaries of a mode-decomposition, should the phase space admit one, and their coarse-graining relations can be established \emph{intrinsically} (they follow directly from the symplectic structure of the phase space once the particular labels involved are given).%
\MleavePhyMode \hypertarget{PARayj}{}\Mpar \Mtheorem{4.4}Let {$\Sigma$} be a \emph{compact}, oriented, \MMath{d}-dimensional manifold (\emph{without} boundary), equipped with a positive-definite metric \MMath{h } and let \MMath{\alpha ,\beta  \in  C^{\infty } \big( \mathds{R} \rightarrow  \left]0, +\infty \right[ \big)}. Let \MMath{\mathcal{M}  \mathrel{\mathop:}=  \mathds{R} \times  \Sigma } and let {$g$} be the pseudo-Riemannian metric on {$\mathcal{M}$}:%
\Mpar \MStartEqua \MMath{g  \mathrel{\mathop:}=  \matrixTwo{-\beta (t)}{}{}{\alpha (t)\, h }}.%
\MStopEqua \Mpar Let \MMath{h _{o} = h } and let:%
\Mpar \MStartEqua \MMath{\mathcal{C}  \mathrel{\mathop:}=  \left\{ \mathtt{c} _{t:} \Sigma  \rightarrow  \mathcal{M} , x \mapsto  \left(t,\, x\right) \middlewithspace| t \in  \mathds{R} \right\}}.%
\MStopEqua \Mpar The system \MMath{\mathcal{M} , \mathcal{C} } admits a \emph{separable} phase space \MMath{W , \Psi }.%
\Mpar \Mproof \italique{Bounded evolution for one mode.} Let \MMath{T > 0}, \MMath{A,B \in  C^{\infty } \big(\left[-T,T\right] \rightarrow  \left]0,\, +\infty \right[\big)} and \MMath{F \in  C^{\infty } \big(\left[-T,T\right] \rightarrow  \mathds{R} \big)}. Let \MMath{\mu  > 0} and define the time-dependent real \MMath{2\times 2} matrix:%
\Mpar \MStartEqua \MMath{H_{\mu }(t) \mathrel{\mathop:}=  \matrixTwo{}{\mu \, B(t)}{-\mu \, A(t)+\frac{1}{\mu }F(t)}{}}.%
\MStopEqua \Mpar By compactness of \MMath{{[-T,\, T]}}, there exists \MMath{\mu _{T} > 0} such that:%
\Mpar \MStartEqua \MMath{\forall  \mu  > \mu _{T}, \forall  t \in  \left[-T,\, T\right], {\mu \, A(t)-\frac{1}{\mu }F(t) > 0}}.%
\MStopEqua \Mpar We now assume \MMath{\mu  > \mu _{T}}. Then, for any \MMath{t \in  \left[-T,\, T\right]}, \MMath{H(t)} is diagonalizable in \MMath{\text{M}_{2}(\mathds{C})} (the space of \MMath{2\times 2} complex matrices), with eigenvalues \MMath{\pm  i \,  \omega _{\mu }(t)}, where:%
\Mpar \MStartEqua \MMath{\omega _{\mu }(t) = \sqrt{{\mu ^{2} \, A(t)\, B(t) - F(t)\, B(t)}} > 0}.%
\MStopEqua \Mpar For any \MMath{t \in  \left[-T,\, T\right]}, we define the corresponding spectral projectors:%
\Mpar \MStartEqua \MMath{\Pi ^{\pm }_{\mu }(t) \mathrel{\mathop:}=  \frac{1}{2} \mp  i \frac{H_{\mu }(t)}{2\, \omega _{\mu }(t)}}.%
\MStopEqua \Mpar (Note that \MMath{\Pi ^{\pm }_{\mu }} are \emph{not} orthogonal projectors, since \MMath{H_{\mu }} is \emph{not} a normal operator.)%
\Mpar \vspace{\Sssaut}\hspace{\Alinea}We define the evolution operator \MMath{U_{\mu } \in  C^{\infty } \big(\left[-T,\, T\right] \rightarrow  \text{GL}_{2}(\mathds{R})\big)} as the solution of the differential equation:%
\Mpar \MStartEqua \MMath{\dot{U}_{\mu } = H_{\mu } \,  U_{\mu }}%
\MStopEqua \Mpar (where \MMath{\dot{(\, \, )}} denotes derivative with respect to \MMath{t}), with initial condition \MMath{U_{\mu }(0) = \mathds{1} } (note that \MMath{U_{\mu }(t)} is invertible for any \MMath{t}, since its inverse can be obtained as the solution of the time-reversed evolution).
We define \MMath{U^{\pm }_{\mu } = \Pi ^{\pm }_{\mu } \,  U_{\mu } \in  \text{M}_{2}(\mathds{C}) \setminus \left\{0\right\}}, so that \MMath{U_{\mu } = U^{+}_{\mu } + U^{-}_{\mu }}. Let \MMath{\left\|\, \cdot \, \right\|} denote the operator norm in \MMath{\text{M}_{2}(\mathds{C})}. We have:%
\Mpar \MStartEqua \MMath{\left\| U_{\mu } \right\| \leqslant  \left\| U^{+}_{\mu } \right\| + \left\| U^{-}_{\mu } \right\|},%
\MStopEqua \Mpar as well as:%
\Mpar \MStartEqua \MMath{\left\| U^{\pm }_{\mu } \right\|^{2} = \left\| U^{{\pm \dag}}_{\mu } \,  U^{\pm }_{\mu } \right\|}.%
\MStopEqua \Mpar Moreover, the differential equation satisfied by \MMath{U_{\mu }} together with the definition of \MMath{\Pi ^{\pm }_{\mu }} yields:%
\Mpar \MStartEqua \MMath{\frac{d}{dt} U^{{\pm \dag}}_{\mu } \,  U^{\pm }_{\mu } = U^{{\pm \dag}}_{\mu } \,  \dot{\Pi }^{\pm }_{\mu } \,  U_{\mu } \pm  i \, \omega _{\mu } \, U^{{\pm \dag}}_{\mu } \,  U^{\pm }_{\mu } + U_{\mu }^{\dag} \,  \dot{\Pi }^{{\pm \dag}}_{\mu } \,  U^{\pm }_{\mu } \mp  i \omega _{\mu }  U^{{\pm \dag}}_{\mu } \,  U^{\pm }_{\mu }}\\
\MMath{\hphantom{\frac{d}{dt} U^{{\pm \dag}}_{\mu } \,  U^{\pm }_{\mu }} = U^{{\pm \dag}}_{\mu } \,  \dot{\Pi }^{\pm }_{\mu } \,  U_{\mu } + U_{\mu }^{\dag} \,  \dot{\Pi }^{{\pm \dag}}_{\mu } \,  U^{\pm }_{\mu }}.%
\MStopEqua \Mpar Thus, we get:%
\Mpar \MStartEqua \MMath{\frac{d}{dt} \ln \left\| U_{\mu } \right\| \leqslant  \frac{\frac{d}{dt} \left\| U^{+}_{\mu } \right\|^{2}}{2\, \left\|U_{\mu }\right\|\, \left\|U^{+}_{\mu }\right\|} + \frac{\frac{d}{dt} \left\| U^{-}_{\mu } \right\|^{2}}{2\, \left\|U_{\mu }\right\|\, \left\|U^{-}_{\mu }\right\|}
\leqslant  \frac{\left\| \frac{d}{dt} U^{{+\dag}}_{\mu } \,  U^{+}_{\mu } \right\|}{2\, \left\|U_{\mu }\right\|\, \left\|U^{+}_{\mu }\right\|} + \frac{\left\| \frac{d}{dt} U^{{-\dag}}_{\mu } \,  U^{-}_{\mu } \right\|}{2\, \left\|U_{\mu }\right\|\, \left\|U^{-}_{\mu }\right\|}
\leqslant  \left\| \dot{\Pi }^{+}_{\mu } \right\| + \left\| \dot{\Pi }^{-}_{\mu } \right\|},%
\MStopEqua \Mpar where the second inequality follows from the triangular inequality.
Next, we have:%
\Mpar \MStartEqua \MMath{\dot{\Pi }^{\pm }_{\mu } = \mp  i \frac{\dot{H}_{\mu }}{2\, \omega _{\mu }} \pm  i \frac{H_{\mu }}{2\, \omega _{\mu }} \frac{\dot{\omega }_{\mu }}{\omega _{\mu }}},%
\MStopEqua \Mpar so we get:%
\Mpar \MStartEqua \MMath{\frac{d}{dt} \ln \left\| U_{\mu } \right\| \leqslant  \frac{\left\|\vphantom{H_{\mu }}\smash{\dot{H}_{\mu }}\right\|}{\omega _{\mu }} + \frac{\left\|H_{\mu }\right\|}{\omega _{\mu }} \frac{\left|\dot{\omega }_{\mu }\right|}{\omega _{\mu }}}.%
\MStopEqua \Mpar Therefore, there exists a constant \MMath{C_{T}} \emph{independent of {$\mu$}} such that:%
\Mpar \MStartEqua \MMath{\forall  t \in  \left[-T,\, T\right], {\left\| U_{\mu }(t) \right\| \leqslant  C_{T}}}.%
\MStopEqua \Mpar Since \MMath{U_{\mu }(t)} is real for any \MMath{t \in  \left[-T,\, T\right]}, its operator norm in \MMath{\text{M}_{2}(\mathds{R})} coincides with its operator norm in \MMath{\text{M}_{2}(\mathds{C})}.%
\Mpar \vspace{\Ssaut}\italique{Definition of {$W$} and {$\Psi$}.} Let \MMath{\widetilde{W } \mathrel{\mathop:}=  L_{2} \left(\Sigma  \rightarrow  \mathds{R}, \sqrt{\left|h \right|}d^{(d)}x\right)}. \MMath{\widetilde{W }} is a \emph{real} Hilbert space, whose scalar product will be denoted \MMath{\left(\, \cdot \, ,\, \cdot \, \right)}.
By the Sturm–Liouville decomposition \bseqHHHabb{theorem 44}, there exists a sequence \MMath{\left( \lambda _{k} \right)_{k\in \mathds{N}}} and an \emph{orthonormal basis} \MMath{\left( E_{k} \right)_{k\in \mathds{N}}} of \MMath{\widetilde{W }} such that:%
\hypertarget{PARazp}{}\Mpar \MList{1}\MMath{\left( \lambda _{k} \right)_{k\in \mathds{N}}} is \emph{increasing} and \MMath{\lambda _{k} \underset{k \rightarrow  \infty }{\longrightarrow} \infty };%
\MStopList \Mpar \MList{2}for any \MMath{k \in  \mathds{N}}, \MMath{E_{k} \in  C^{\infty }(\Sigma  \rightarrow  \mathds{R})} and \MMath{\Delta _{h } E_{k} = - \lambda _{k} \,  E_{k}}, where \MMath{\Delta _{h } \mathrel{\mathop:}=  \frac{1}{\sqrt{\left|h \right|}} \partial _{i} \left( \sqrt{\left|h \right|} \,  h ^{ij} \,  \partial _{j} \, \cdot \,  \right)}.%
\MStopList \Mpar For any \MMath{\psi ,\, \psi ' \in  C^{\infty }(\Sigma  \rightarrow  \mathds{R})}, integration by parts yields:%
\Mpar \MStartEqua \MMath{\left(\psi ,\,  \Delta _{h } \psi '\right) = - \int_{\Sigma } \sqrt{\left|h \right|} d^{(d)}x \hspace{0.25cm}  h ^{ij} \,  \partial _{i} \psi  \,  \partial _{j} \psi ' = \left(\Delta _{h } \psi ,\,  \psi '\right)}.%
\MStopEqua \Mpar In particular, \MMath{\forall  k,\,  \lambda _{k} \geqslant  0}.
For any \MMath{\psi  \in  C^{\infty }(\Sigma  \rightarrow  \mathds{R})}, we have \MMath{\psi  \in  \widetilde{W }} (for \MMath{\psi ^{2} \,  \sqrt{\left|h \right|} \,  d^{(d)}x} is a smooth measure on the compact manifold {$\Sigma$}, thus integrable), hence:%
\Mpar \MStartEqua \MMath{\sum_{k} \left(E_{k},\,  \psi \right)^{2} = \left(\psi ,\,  \psi \right) < \infty }%
\MStopEqua \Mpar Since \MMath{\Delta _{h } \psi  \in  C^{\infty }(\Sigma  \rightarrow  \mathds{R})}, we also get:%
\Mpar \MStartEqua \MMath{\sum_{k} \lambda _{k}^{2} \,  \left(E_{k},\,  \psi \right)^{2} < \infty },%
\MStopEqua \Mpar therefore, using that \MMath{\left(\lambda _{k}\right)_{k\in \mathds{N}}} is a positive, increasing sequence:%
\Mpar \MStartEqua \MMath{\sum_{k} \sqrt{{\lambda _{k} + 1}} \,  \left(E_{k},\,  \psi \right)^{2} < \infty  \& \sum_{k} \frac{1}{\sqrt{{\lambda _{k} + 1}}} \,  \left(E_{k},\,  \psi \right)^{2} < \infty }.%
\MStopEqua \Mpar We define \MMath{W  \mathrel{\mathop:}=  \ell _{2}(\mathds{N} \rightarrow  \mathds{R}) \times  \ell _{2}(\mathds{N} \rightarrow  \mathds{R})} and the linear map:%
\Mpar \MStartEqua \MMath{\definitionFonction{\Psi }{C^{\infty }(\Sigma  \rightarrow  \mathds{R}) \times  C^{\infty }(\Sigma  \rightarrow  \mathds{R})}{W }{\varphi ,\pi }{\Big( (\lambda _{k} + 1)^{{\nicefrac{1}{4}}} \,  \left(E_{k},\,  \varphi \right) \Big)_{k\in \mathds{N}},\,  \Big( (\lambda _{k} + 1)^{{-\nicefrac{1}{4}}} \,  \left(E_{k},\,  \pi \right) \Big)_{k\in \mathds{N}}}}.%
\MStopEqua \Mpar \MMath{\Psi } is injective since \MMath{\Psi (\varphi ,\pi ) = \Psi (\varphi ',\pi ')} implies that {$\varphi$} and \MMath{\varphi '}, resp.~{$\pi$} and \MMath{\pi '}, are almost everywhere equal, hence, being smooth, that they are equal.%
\Mpar \vspace{\Ssaut}\italique{Evolution between spatial slices.} For any \MMath{t \in  \mathds{R}}, \MMath{\mathtt{c} _{t}} is a spatial {$\Sigma$}-slice, and, for any \MMath{\phi  \in  C^{\infty }(\mathcal{M}  \rightarrow  \mathds{R})}, we define \MMath{\phi _{t} \in  C^{\infty }(\Sigma  \rightarrow  \mathds{R}) \times  C^{\infty }(\Sigma  \rightarrow  \mathds{R})} by:%
\Mpar \MStartEqua \MMath{\forall  x \in  \Sigma , {\phi _{t}(x) \mathrel{\mathop:}=  \phi _{{\mathtt{c} _{t}}}(x) = \left( \phi (t,x); \sqrt{{\frac{\alpha ^{d}(t)}{\beta (t)}}} \,  \partial _{t} \phi (t,x) \right)}}%
\MStopEqua \Mpar (where we have expressed \MMath{\mathbf{n} _{x}} using the expression for {$g$}), so that:%
\Mpar \MStartEqua \MMath{\Psi _{t}(\phi ) \mathrel{\mathop:}=  \Psi  \left( \phi _{t} \right) = \Big( \left(\lambda _{k} + 1\right)^{{\nicefrac{1}{4}}} \big(E_{k},\,  \phi (t,\, \cdot \, )\big) \Big)_{k\in \mathds{N}}, \bigg( \left(\lambda _{k} + 1\right)^{{-\nicefrac{1}{4}}} \,  \sqrt{{\frac{\alpha ^{d}(t)}{\beta (t)}}} \big(E_{k},\,  \partial _{t} \phi (t,\, \cdot \, ) \big) \bigg)_{k\in \mathds{N}}}.%
\MStopEqua \Mpar Let \MMath{\phi  \in  C^{\infty }(\mathcal{M}  \rightarrow  \mathds{R})} and \MMath{\forall t, {\big( u_{k}(t) \big)_{k\in \mathds{N}},\,  \big( v_{k}(t) \big)_{k\in \mathds{N}} \mathrel{\mathop:}=  \Psi _{t}(\phi )}}. Using the expression for \MMath{g }, we get:%
\hypertarget{PARbah}{}\Mpar \MStartEqua \MMath{\phi  \in  \mathcal{S} _{\mathcal{M} } \Leftrightarrow  \partial _{t} \left( \sqrt{{\frac{\alpha ^{d}}{\beta }}} \,  \partial _{t} \phi  \right) = \sqrt{{\beta \, \alpha ^{d-2}}} \,  \Delta _{h } \phi  - m^{2} \,  \sqrt{{\beta \, \alpha ^{d}}} \,  \phi }.%
\NumeroteEqua{4.4}{1}\MStopEqua \Mpar For any \MMath{k \in  \mathds{N}} and any \MMath{t \in  \mathds{R}}, the compactness of {$\Sigma$}, together with the smoothness of {$\phi$} and \MMath{E_{k}}, allows to differentiate under the integration, yielding:%
\hypertarget{PARbaj}{}\Mpar \MStartEqua \MMath{\frac{d}{dt} u_{k}(t) = \left(\lambda _{k} + 1\right)^{{\nicefrac{1}{2}}} \,  \sqrt{{\frac{\beta (t)}{\alpha ^{d}(t)}}} \,  v_{k}(t)},%
\NumeroteEqua{4.4}{2}\MStopEqua \Mpar as well as:%
\Mpar \MStartEqua \MMath{\left( E_{k},\,   \partial _{t} \left( \sqrt{{\frac{\alpha ^{d}}{\beta }}} \,  \partial _{t} \phi  \right)(t,\, \cdot \, ) \right) = \left(\lambda _{k} + 1\right)^{{\nicefrac{1}{4}}} \frac{d}{dt} v_{k}(t)},%
\MStopEqua \Mpar and integration by parts yields:%
\Mpar \MStartEqua \MMath{\left( E_{k},\,  \sqrt{{\beta \, \alpha ^{d-2}}} \,  \Delta _{h } \phi (t,\, \cdot \, ) - m^{2} \,  \sqrt{{\beta \, \alpha ^{d}}} \,  \phi (t,\, \cdot \, ) \right) = \left( - \sqrt{{\beta (t)\, \alpha ^{d-2}(t)}} \,  \lambda _{k} - m^{2} \,  \sqrt{{\beta (t)\, \alpha ^{d}(t)}} \right) \,  \left(\lambda _{k} + 1\right)^{{-\nicefrac{1}{4}}} u_{k}(t)}.%
\MStopEqua \Mpar Thus, {\seqBBBbap} becomes:%
\hypertarget{PARbaq}{}\Mpar \MStartEqua \MMath{\phi  \in  \mathcal{S} _{\mathcal{M} } \Leftrightarrow  \forall  k \in  \mathds{N}, \forall  t \in  \mathds{R}, {\frac{d}{dt} v_{k}(t) = \left(\lambda _{k} + 1\right)^{{-\nicefrac{1}{2}}} \left( - \sqrt{{\beta (t)\, \alpha ^{d-2}(t)}} \,  \lambda _{k} - m^{2} \,  \sqrt{{\beta (t)\, \alpha ^{d}(t)}} \right) u_{k}(t)}}%
\NumeroteEqua{4.4}{3}\MStopEqua \Mpar (the `$\Leftarrow$' directions follows, like the injectivity of \MMath{\Psi }, from the equality of almost everywhere equal smooth functions).%
\Mpar \vspace{\Sssaut}\hspace{\Alinea}Let \MMath{t \in  \mathds{R}}. We define, for any \MMath{s \in  \mathds{R}}:%
\hypertarget{PARbat}{}\Mpar \MStartEqua \MMath{A(s) \mathrel{\mathop:}=  \sqrt{{\beta (t+s)\, \alpha ^{d-2}(t+s)}} > 0,\hspace{0.25cm}  B(s) \mathrel{\mathop:}=  \sqrt{{\frac{\beta (t+s)}{\alpha ^{d}(t+s)}}} > 0\\\hspace*{0.6cm}\& F(s) \mathrel{\mathop:}=  \sqrt{{\beta (t+s)\, \alpha ^{d}(t+s)}} \left( \frac{1}{\alpha (t+s)} - m^{2} \right)},%
\NumeroteEqua{4.4}{4}\MStopEqua \Mpar and, using the notations of the first part of the present proof:%
\Mpar \MStartEqua \MMath{\forall  k \in  \mathds{N}, {\text{u}_{k}(t,t+s) \mathrel{\mathop:}=  U_{{\sqrt{{\lambda _{k} + 1}}}}(s) \in  \text{GL}_{2}(\mathds{R})}}.%
\MStopEqua \Mpar For any \MMath{k \in  \mathds{N}} such that \MMath{\sqrt{{\lambda _{k} + 1}} > \mu _{\left|s\right|}}, \MMath{\left\| \text{u}_{k}(t,t+s) \right\| \leqslant  C_{\left|s\right|}}. Moreover, as \MMath{\lambda _{k} \underset{k \rightarrow  \infty }{\longrightarrow} \infty }, there are only finitely many \MMath{k \in  \mathds{N}} with \MMath{\sqrt{{\lambda _{k} + 1}} \leqslant  \mu _{\left|s\right|}}. Thus, for any \MMath{s \in  \mathds{R}}, there exists a well-defined, \emph{bounded} (with respect to the canonical norm on \MMath{W  = \ell _{2}(\mathds{N} \rightarrow  \mathds{R}) \times  \ell _{2}(\mathds{N} \rightarrow  \mathds{R})}) linear operator \MMath{\text{u}(t,t+s): W  \rightarrow  W } such that:%
\hypertarget{PARbax}{}\Mpar \MStartEqua \MMath{\forall  u,v = \left(u_{k}\right)_{k\in \mathds{N}},\,   \left(v_{k}\right)_{k\in \mathds{N}} \in  W ,\\\hspace*{1.5cm} {\text{u}(t,t+s) \big(u,\, v\big) = \left( (1,0)\, .\, \text{u}_{k}(t,t+s)\, \left(\begin{array}{c}u_{k} \\v_{k} \end{array}\right)  \right)_{k\in \mathds{N}}, \left( (0,1)\, .\, \text{u}_{k}(t,t+s)\, \left(\begin{array}{c}u_{k} \\v_{k} \end{array}\right)  \right)_{k\in \mathds{N}}}}.%
\NumeroteEqua{4.4}{5}\MStopEqua \Mpar Then, {\seqBBBbaz} ensure that, for any \MMath{\phi  \in  C^{\infty }(\mathcal{M}  \rightarrow  \mathds{R})}:%
\hypertarget{PARbba}{}\Mpar \MStartEqua \MMath{\phi  \in  \mathcal{S} _{\mathcal{M} } \Leftrightarrow  \forall  s \in  \mathds{R}, {\Psi _{t+s}(\phi ) = \text{u}(t,t+s) \big( \Psi _{t}(\phi ) \big)}}.%
\NumeroteEqua{4.4}{6}\MStopEqua \Mpar Noting that, for any \MMath{t \in  \mathds{R}}, \MMath{\Psi _{{\mathtt{c} _{t}}} = \left. \Psi _{t} \right|_{{\mathcal{S} _{\mathcal{M} }}}}, we have proved that, for any \MMath{t,t' \in  \mathds{R}}, there exits a bounded linear operator \MMath{\text{u}(t,t'): W  \rightarrow  W } such that \MMath{\Psi _{{\mathtt{c} _{t'}}} = \text{u}(t,t') \circ  \Psi _{{\mathtt{c} _{t}}}}.%
\Mpar \vspace{\Ssaut}\italique{Injectivity and dense image of \MMath{\Psi _{{\mathtt{c} _{t}}}}.} Let \MMath{t \in  \mathds{R}}. Let \MMath{\phi  \neq  \phi ' \in  \mathcal{S} _{\mathcal{M} }}. Then, there exists \MMath{t' \in  \mathds{R}} such that \MMath{\phi (t',\, \cdot \, ) \neq  \phi '(t',\, \cdot \, )}, so \MMath{\phi _{t'} \neq  \phi '_{t'}}. As {$\Psi$} is injective, we thus have \MMath{\Psi _{t'}(\phi ) \neq  \Psi _{t'}(\phi ')}. Since \MMath{\Psi _{t'}(\phi ) = \text{u}(t,t') \big( \Psi _{t}(\phi ) \big)} and \MMath{\Psi _{t'}(\phi ') = \text{u}(t,t') \big( \Psi _{t}(\phi ') \big)}, this implies that \MMath{\Psi _{t}(\phi ) \neq  \Psi _{t}(\phi ')}. Hence, \MMath{\Psi _{{\mathtt{c} _{t}}} = \left. \Psi _{t} \right|_{{\mathcal{S} _{X}}}} is injective.%
\Mpar \vspace{\Sssaut}\hspace{\Alinea}Let \MMath{u = \left( u_{k} \right)_{k\in \mathds{N}},\,  v = \left( v_{k} \right)_{k\in \mathds{N}}} be sequences in \MMath{\mathds{R}} such that there exists \MMath{K \in  \mathds{N}} such that \MMath{\forall  k > K, {u_{k} = v_{k} = 0}}.
In particular, \MMath{\left(u,\, v\right) \in  W }.
For any \MMath{t' \in  \mathds{R}}, we define:%
\Mpar \MStartEqua \MMath{u(t'),\,  v(t') \mathrel{\mathop:}=  \text{u}(t,t') \big( u,\,  v \big)}.%
\MStopEqua \Mpar By definition of \MMath{\text{u}(t,t')}, we have, for any \MMath{t' \in  \mathds{R}} and any \MMath{k > K}, \MMath{u_{k}(t') = v_{k}(t') = 0}. Moreover, for any \MMath{k \leqslant  K}, \MMath{t' \mapsto  u_{k}(t') \in  C^{\infty }(\mathds{R} \rightarrow  \mathds{R})}. Thus, defining:%
\Mpar \MStartEqua \MMath{\forall  t' \in  \mathds{R},\,  \forall  x \in  \Sigma , {\phi (t',x) \mathrel{\mathop:}=  \sum_{k=0}^{\infty } \left(\lambda _{k} + 1\right)^{{-\nicefrac{1}{4}}} u_{k}(t') \,  E_{k}(x)}},%
\MStopEqua \Mpar we have \MMath{\phi  \in  C^{\infty }(\mathcal{M}  \rightarrow  \mathds{R})} (recalling that \MMath{E_{k} \in  C^{\infty }(\Sigma  \rightarrow  \mathds{R})} for any \MMath{k \in  \mathds{N}}). {\seqCCCbbh}, together with the definition of \MMath{\text{u}(t,t')}, ensures that, for any \MMath{t' \in  \mathds{R}}, \MMath{\Psi _{t'}(\phi ) = u(t'),\, v(t')}. In particular, \MMath{\Psi _{t}(\phi ) = u,\, v}. Moreover, {\seqBBBbbi} then implies that \MMath{\phi  \in  \mathcal{S} _{\mathcal{M} }}.
Hence, the space of eventually zero sequences is included in the image of \MMath{\Psi _{{\mathtt{c} _{t}}}}, and, since this space is dense in {$W$}, so is the image of \MMath{\Psi _{{\mathtt{c} _{t}}}}.
To summarize, we have proved that {\seqBBBaye} holds.%
\Mpar \vspace{\Ssaut}\italique{Symplectic structure.} We equip {$W$} with a weak symplectic structure {$\Omega$} {\seqDDDakg}, defined by:%
\Mpar \MStartEqua \MMath{\forall  (u,\, v),\,  (u',\, v') \in  W , {\Omega  \big(u,v;\, u',v'\big) \mathrel{\mathop:}=  \sum_{k=0}^{\infty } u_{k} \,  v'_{k} - u'_{k} \,  v_{k}}}.%
\MStopEqua \Mpar The Cauchy–Schwarz inequality ensures that {$\Omega$} is a well-defined, bilinear, antisymmetric form on \MMath{W  = \ell _{2}(\mathds{N} \rightarrow  \mathds{R}) \times  \ell _{2}(\mathds{N} \rightarrow  \mathds{R})}, and, for any \MMath{(u,\, v) \in  W  \setminus \left\{0\right\} }, we have:%
\Mpar \MStartEqua \MMath{\Omega  \left( u,v; \frac{-v}{\left|u,v\right|^{2}}, \frac{u}{\left|u,v\right|^{2}} \right) = 1},%
\MStopEqua \Mpar where \MMath{\left|\, \cdot \, \right|} denotes the canonical norm on {$W$}, namely \MMath{\left|u,v\right|^{2} \mathrel{\mathop:}=  \left|u\right|^{2}_{{\ell _{2}(\mathds{N} \rightarrow  \mathds{R})}} + \left|v\right|^{2}_{{\ell _{2}(\mathds{N} \rightarrow  \mathds{R})}}}.
More precisely, for any \MMath{(u,\, v),\,  (u',\, v') \in  W }, we have:%
\Mpar \MStartEqua \MMath{\left| \Omega  \big(u,v;\, u',v'\big) \right| \leqslant  \left|u\right|_{{\ell _{2}(\mathds{N} \rightarrow  \mathds{R})}} \,  \left|v'\right|_{{\ell _{2}(\mathds{N} \rightarrow  \mathds{R})}} + \left|u'\right|_{{\ell _{2}(\mathds{N} \rightarrow  \mathds{R})}} \,  \left|v\right|_{{\ell _{2}(\mathds{N} \rightarrow  \mathds{R})}} \leqslant  \left|u,v\right| \,  \left|u',v'\right|},%
\MStopEqua \Mpar and for any \MMath{(u,\, v) \in  W  \setminus \left\{0\right\}}:%
\Mpar \MStartEqua \MMath{\left|u,v\right| \,  \left|\frac{-v}{\left|u,v\right|^{2}}, \frac{u}{\left|u,v\right|^{2}}\right| \leqslant  1}.%
\MStopEqua \Mpar Thus, \MMath{W ,\, \Omega ,\, \left|\, \cdot \, \right|} is a normed symplectic vector space with \MMath{\left\|\Omega \right\| = \left\|\Omega ^{-1}\right\| = 1} {\seqDDDaos}, so in particular, \MMath{W ,\,  \Omega } equipped with the topology induced by \MMath{\left|\, \cdot \, \right|} is a \emph{normable} symplectic vector space. In addition, since the eventually zero sequences form a dense, countably generated subspace of {$W$}, {$W$} is \emph{separable}.%
\Mpar \vspace{\Sssaut}\hspace{\Alinea}Finally, using the notations of the first part of the present proof, we have, for any \MMath{s \in  \mathds{R}}:%
\hypertarget{PARbbt}{}\Mpar \MStartEqua \MMath{{}^{\text{\sc t}} H_{\mu }(s) \matrixTwo{}{1}{-1}{} + \matrixTwo{}{1}{-1}{} H_{\mu }(s) = 0},%
\NumeroteEqua{4.4}{7}\MStopEqua \Mpar ensuring that:%
\Mpar \MStartEqua \MMath{{}^{\text{\sc t}} U_{\mu }(s) \matrixTwo{}{1}{-1}{} U_{\mu }(s) = \matrixTwo{}{1}{-1}{}}.%
\MStopEqua \Mpar Using the definitions of \MMath{\text{u}(t,t')} and {$\Omega$}, it follows that, for any \MMath{t,t' \in  \mathds{R}}, \MMath{\text{u}(t,t')} is a symplectomorphism of \MMath{W ,\,  \Omega }, which completes the proof of {\seqBBBbbx}.%
\MendOfProof \Mpar \MstartPhyMode\hspace{\Alinea}As we will see, even on such a simple cosmological spacetime, there does \emph{not} exist any polarization that would be \emph{preserved} under time evolution (unless the spacetime happens to be \emph{static}, ie.~the scale factor \MMath{\alpha (t)} is constant).
The best that can be done if we insist on choosing a complex structure is to elect one with respect to which the \emph{instantaneous} Hamiltonian at \emph{some} time \MMath{t} is the generator of an isometry (aka.~an essentially (anti-)self-adjoint operator).
Unsurprisingly, a complex structure chosen in this way induces on {$W$} precisely the topology that we derived above (we derived this topology from the requirement for the time evolution to be \emph{bounded}: clearly, for the instantaneous Hamiltonian to be the generator of an isometry, it should at least be the derivative of a bounded evolution).
This means (by virtue of  {\seqBBBatm}) that the Fock spaces constructed on those "instantaneous" complex structures will be satisfactorily \emph{embedded} in our projective quantum state space, even after restricting ourselves to a universal label subset. It also offers further confirmation that the topology we are using is physically reasonable and appropriate given the time evolution.%
\MleavePhyMode \hypertarget{PARbbz}{}\Mpar \Mproposition{4.5}Let \MMath{W , \Psi } be the phase space constructed in {\seqBBBaxh} and assume \MMath{m > 0}. Let \MMath{t \in  \mathds{R}}, and, for any \MMath{t' \in  \mathds{R}}, let \MMath{\text{u}(t,t')} be the symplectomorphism of {$W$} satisfying \MMath{\Psi _{{\mathtt{c} _{t'}}} = \text{u}(t,t') \circ  \Psi _{{\mathtt{c} _{t}}}}.
There exists a compatible complex structure \MMath{I _{t}} on {$W$} {\seqDDDaqg} such that \MMath{\left. {\textstyle\frac{d}{dt'}} \text{u}(t,t') \right|_{{t'=t}}} is a densely-defined, essentially \MMath{I _{t}}-anti-self-adjoint operator on {$W$}.%
\Mpar \vspace{\Sssaut}\hspace{\Alinea}Moreover, the norm \MMath{\left|\, \cdot \, \right|_{t}} associated to \MMath{I _{t}} induces the topology of {$W$}.%
\Mpar \Mproof \italique{Complex structure for one mode.} Let \MMath{H_{\mu }(s)} be as in the first part of {\seqBBBaxh}, with \MMath{A(s),\, B(s) \esperluette\! F(s)} from {\seqBBBbcb} and \MMath{\mu  = \sqrt{{\lambda _{k} + 1}} \geqslant  1}.
We have \MMath{H_{\mu }^{2}(0) = - \omega _{\mu }^{2} \,  \mathds{1} }, with \MMath{\omega _{\mu }(0) > 0} (note that, since we assumed \MMath{m > 0}, \MMath{\mu ^{2} \,  A(0) - F(0) > 0} for any \MMath{\mu  \geqslant  1}), hence:%
\Mpar \MStartEqua \MMath{I _{\mu } \mathrel{\mathop:}=  \frac{-1}{\omega _{\mu }(0)} H_{\mu }(0)}%
\MStopEqua \Mpar is such that \MMath{I _{\mu }^{2} = - \mathds{1} }. Moreover:%
\Mpar \MStartEqua \MMath{\Omega ^{(1)}\,  I _{\mu } = \frac{1}{\omega _{\mu }(0)} \matrixTwo{\mu \, A(0)-\frac{1}{\mu }F(0)}{}{}{\mu \, B(0)} > 0}, where \MMath{\Omega ^{(1)} \mathrel{\mathop:}=  \matrixTwo{}{1}{-1}{}},%
\MStopEqua \Mpar and, using {\seqBBBbcg}, \MMath{{}^{\text{\sc t}} I _{\mu } \,  \Omega ^{(1)}\,  I _{\mu } = \Omega ^{(1)}}, so \MMath{I _{\mu }} defines a compatible complex structure on \MMath{\mathds{R}^{2},\,  \Omega ^{(1)}}.
The map:%
\Mpar \MStartEqua \MMath{\definitionFonction{z_{\mu }}{\mathds{R}^{2}}{\mathds{C}}{u,v}{\left\langle 1,1\middlewithspace|u,v\right\rangle _{{I _{\mu }}} = \left( \xi _{\mu } - i \right) u + \left( \nicefrac{1}{\xi _{\mu }} + i \right) v}}, where \MMath{\xi _{\mu } \mathrel{\mathop:}=  \sqrt{{\frac{A(0) - \nicefrac{1}{\mu ^{2}}\, F(0)}{B(0)}}}},%
\MStopEqua \Mpar then defines a Hilbert space isomorphism from \MMath{\mathds{R}^{2}} equipped with the complex scalar product \MMath{\left\langle \, \cdot \, \middlewithspace|\, \cdot \, \right\rangle _{{I _{\mu }}}} {\seqDDDaqg} into \MMath{\mathds{C}} with its canonical Hilbert space structure.
In particular, the norm \MMath{\left|\, \cdot \, \right|_{\mu } \mathrel{\mathop:}=  \sqrt{{\left\langle \, \cdot \, \middlewithspace|\, \cdot \, \right\rangle _{{I _{\mu }}}}} = \left|z_{\mu }(\, \cdot \, )\right|_{\mathds{C}}} associated to \MMath{I _{\mu }} satisfies:%
\Mpar \MStartEqua \MMath{\sqrt{{1 + \min\left(\xi _{\mu }^{2},\, \nicefrac{1}{\xi _{\mu }^{2}}\right)}} \,  \left|\, \cdot \, \right|_{{\mathds{R}^{2}}} \leqslant  \left|\, \cdot \, \right|_{\mu } \leqslant  \sqrt{{1 + \max\left(\xi _{\mu }^{2},\, \nicefrac{1}{\xi _{\mu }^{2}}\right)}} \,  \left|\, \cdot \, \right|_{{\mathds{R}^{2}}}},%
\MStopEqua \Mpar where \MMath{\left|\, \cdot \, \right|_{{\mathds{R}^{2}}}} denotes the Euclidean norm of \MMath{\mathds{R}^{2}}.%
\Mpar \vspace{\Ssaut}\italique{Complex structure on {$W$}.} We define \MMath{I _{t} : W  \rightarrow  W } by:%
\Mpar \MStartEqua \MMath{\forall  u,v = \left(u_{k}\right)_{k\in \mathds{N}},\,   \left(v_{k}\right)_{k\in \mathds{N}} \in  W ,\\\hspace*{1.5cm} {I _{t} \big(u,\, v\big) = \left( (1,0)\, .\, I _{\sqrt{{\lambda _{k} + 1}}} \, \left(\begin{array}{c}u_{k} \\v_{k} \end{array}\right)  \right)_{k\in \mathds{N}}, \left( (0,1)\, .\, I _{\sqrt{{\lambda _{k} + 1}}} \, \left(\begin{array}{c}u_{k} \\v_{k} \end{array}\right)  \right)_{k\in \mathds{N}}}}.%
\MStopEqua \Mpar By construction, \MMath{I _{t}} is a compatible complex structure on \MMath{W ,\,  \Omega }.
Moreover, the norm \MMath{\left|\, \cdot \, \right|_{t}} associated to \MMath{I _{t}} satisfies:%
\Mpar \MStartEqua \MMath{\sqrt{{1 + \min\left(\xi _{\text{inf}}^{2},\, \nicefrac{1}{\xi _{\text{sup}}^{2}}\right)}} \,  \left|\, \cdot \, \right| \leqslant  \left|\, \cdot \, \right|_{t} \leqslant  \sqrt{{1 + \max\left(\xi _{\text{sup}}^{2},\, \nicefrac{1}{\xi _{\text{inf}}^{2}}\right)}} \,  \left|\, \cdot \, \right|},%
\MStopEqua \Mpar where \MMath{\xi _{\text{inf}} \mathrel{\mathop:}=  \inf_{k\in \mathds{N}} \xi _{\sqrt{{\lambda _{k} + 1}}}}, resp.~\MMath{\xi _{\text{sup}} \mathrel{\mathop:}=  \sup_{k\in \mathds{N}} \xi _{\sqrt{{\lambda _{k} + 1}}}}, and \MMath{\left|\, \cdot \, \right|} denotes the canonical norm of \MMath{W  = \ell _{2}(\mathds{N} \rightarrow  \mathds{R}) \times  \ell _{2}(\mathds{N} \rightarrow  \mathds{R})}.
We have \MMath{\xi _{\sqrt{{\lambda _{k} + 1}}} > 0} for any \MMath{k \in  \mathds{N}} and, since \MMath{\lambda _{k} \underset{k \rightarrow  \infty }{\longrightarrow} \infty } {\seqDDDbcq}, \MMath{\xi _{\sqrt{{\lambda _{k} + 1}}} \underset{k \rightarrow  \infty }{\longrightarrow} \sqrt{{\nicefrac{A(0)}{B(0)}}} > 0}, hence \MMath{0 < \xi _{\text{inf}},\,  \xi _{\text{sup}} < \infty }, so \MMath{\left|\, \cdot \, \right|_{t}} and \MMath{\left|\, \cdot \, \right|} are equivalent.
In particular, \MMath{W ,\,  \left\langle \, \cdot \, \middlewithspace|\, \cdot \, \right\rangle _{{I _{t}}}} is a Hilbert space (it is complete with respect to \MMath{\left|\, \cdot \, \right|_{t}}) and the map \MMath{Z_{t:} W  \rightarrow  \ell _{2}(\mathds{N} \rightarrow  \mathds{C})} such that:%
\Mpar \MStartEqua \MMath{\forall  u,v = \left(u_{k}\right)_{k\in \mathds{N}},\,   \left(v_{k}\right)_{k\in \mathds{N}} \in  W , {Z_{t} \big(u,\, v\big) = \Big( z_{\sqrt{{\lambda _{k} + 1}}} \, \left(u_{k}, v_{k}\right)  \Big)_{k\in \mathds{N}}}},%
\MStopEqua \Mpar is well-defined and a isomorphism of complex Hilbert spaces.%
\Mpar \vspace{\Ssaut}\italique{Self-adjoint Hamiltonian.} We denote by {$D$} the subspace of {$W$} consisting of the \MMath{(u,v) \in  W } such that \MMath{\frac{\text{u}(t,t') (u,v) - (u,v)}{t' - t}} converges in norm when \MMath{t' \rightarrow  t}, and, for any \MMath{(u,v) \in  D }, we define:%
\Mpar \MStartEqua \MMath{\left. \frac{d}{dt'} \text{u}(t,t') \right|_{t'=t} \,  (u,v) \mathrel{\mathop:}=  \lim_{{t' \rightarrow  t}} \frac{\text{u}(t,t') (u,v) - (u,v)}{t' - t}}.%
\MStopEqua \Mpar {$D$} is a dense subspace of {$W$} for it contains all pairs of eventually-zero sequences, as follows from the definition of \MMath{\text{u}(t',t)} {\seqDDDbcw}.%
\Mpar \vspace{\Sssaut}\hspace{\Alinea}We denote by \MMath{D '} the subspace of \MMath{\ell _{2}(\mathds{N} \rightarrow  \mathds{C})} consisting of the \MMath{z = \left(z_{k}\right)_{k\in \mathds{N}}} such that:%
\Mpar \MStartEqua \MMath{\sum_{k=0}^{\infty } \omega _{\sqrt{{\lambda _{k} + 1}}}^{2} \,  \left|z_{k}\right|_{\mathds{C}}^{2} < \infty }.%
\MStopEqua \Mpar For any \MMath{z = \left(z_{k}\right)_{k\in \mathds{N}} \in  D '}, we define:%
\Mpar \MStartEqua \MMath{\mathtt{H} \,  z \mathrel{\mathop:}=  \left( \omega _{\sqrt{{\lambda _{k} + 1}}} \,  z_{k} \right) \in  \ell _{2}(\mathds{N} \rightarrow  \mathds{C})}.%
\MStopEqua \Mpar We can check that the domain of the adjoint of {$\mathtt{H}$} is \MMath{D '}, and that it is equal to {$\mathtt{H}$}, in other words {$\mathtt{H}$} is self-adjoint.%
\Mpar \vspace{\Sssaut}\hspace{\Alinea}By definition of \MMath{\text{u}(t,t')} and \MMath{Z_{t}}, we have, for any \MMath{\left(u,v\right) \in  D }, \MMath{Z_{t}(u,v) \in  D '} and:%
\Mpar \MStartEqua \MMath{Z_{t} \left[ I _{t} \,  \left(\left. \frac{d}{dt'} \text{u}(t,t') \right|_{t'=t}\right) \,  (u,v) \right] = \mathtt{H} \,  Z_{t}(u,v)}.%
\MStopEqua \Mpar Hence, \MMath{I _{t} \,  \left(\left. {\textstyle\frac{d}{dt'}} \text{u}(t,t') \right|_{t'=t}\right)} admits a self-adjoint extension (namely \MMath{Z_{t}^{-1} \,  \mathtt{H}  \,  Z_{t}} defined on \MMath{Z_{t}^{-1} \left\langle  D ' \right\rangle  \supseteq D }), ie.~\MMath{\left. {\textstyle\frac{d}{dt'}} \text{u}(t,t') \right|_{t'=t}} is essentially anti-self-adjoint.%
\MendOfProof \Mpar \MstartPhyMode\hspace{\Alinea}Finally, we want to examine to what extend the use of projective techniques \emph{improves} the \emph{universality} of the quantum state space. By universality we mean \emph{limited reliance} on data \emph{beyond} the linear and symplectic structures of the phase space (both of which being assumed, in the scope of the present study, to be physically unobjectionable; this is further discussed at the end of {\seqBBBaaw} and in {\seqBBBaax}). Thus, the question can be equivalently rephrased as relating to the implementation of \emph{linear symplectomorphisms} in the quantum theory, and a family of particular interest is the time evolution itself (because, even if there were good reasons to rely on some extra data, ideally this data should be \emph{preserved} under the evolution).%
\Mpar \vspace{\Saut}\hspace{\Alinea}So we first need to understand which linear symplectomorphisms can be \emph{unitarily} represented on a Fock space. Obviously, transformations that preserve the complex polarization on which the Fock space is built are straightforward to implement (for they already act as unitary transformations on the 1-particle Hilbert space, see eg.~{\seqBBBbdg} and its proof).
But, as shown by Shale {\bseqJJJaag}, the class of unitarily implementable symplectomorphisms is actually \emph{larger}: in addition to polarization-preserving ones, it also includes those which differ from the identity only by a \emph{Hilbert–Schmidt} operator (with respect to the \emph{real} inner-product induced by the complex structure), as well as compositions of both groups. Further explanations, together with a self-contained proof of this result (based on the characterization of Fock states from {\seqBBBatt}), can be found in {\seqBBBabr}.
Nevertheless, in the model we are considering here (further specialized to a \emph{flat torus} spatial geometry and a \emph{non-constant} scale factor), the result below implies that there is \emph{no choice} of polarization that would make the time evolution implementable on a Fock representation (at least, complex structures that would induce the topology of {$W$} are excluded, which, in light of {\seqBBBato}, should be sufficient to settle the question).%
\Mpar \vspace{\Saut}\hspace{\Alinea}Now, when working with projective state spaces, and restricting ourselves to a universal label subset, as presented in {\seqBBBapw}, the time evolution may also fail to \emph{strictly} be an automorphism in our quantum theory.
Instead, in the spirit of {\seqBBBaon}, and thanks to the topology having been chosen so that this evolution is bounded, we plan to approximate it \emph{at an arbitrary good precision} by a transformation that \emph{does} induces an automorphism both of our quantum state space and of our observable algebra.
Then, to be fair in our comparison, we should allow such an approximating process to take place on the Fock side as well, although the \emph{reason} why some approximation would be needed in the first place would be quite different: in the former case, it is the restricted algebra of observables on which the state space is constructed which is not stabilized by the evolution (unless the dense subspace {$D$} happens to be chosen to support it), while in the latter case, the algebra is the "complete" one, and is certainly stabilized, but the particular subset of states we are confining ourselves to is too small, and, as states evolves, they immediately leave it.%
\Mpar \vspace{\Saut}\hspace{\Alinea}Still, as the following proposition shows, relaxing what we mean by an "implementation of the time evolution" is not enough to make the use of Fock spaces satisfactory. Indeed, it is not possible to arrange for arbitrarily good approximations: there is an incompressible \emph{error}, bounded below by a term that may well get prohibitively large over cosmological times (eg.~for spacetimes that undergo periods of de Sitter-like expansion or contraction).%
\MleavePhyMode \hypertarget{PARbdh}{}\Mpar \Mproposition{4.6}We consider the phase space \MMath{W , \Psi } constructed in {\seqBBBaxh} and we specialize to the case of {$\Sigma$} being a \emph{flat} \MMath{d}-torus (\MMath{d > 1}).
Let {$I$} be a compatible complex structure on {$W$} {\seqDDDaqg} and suppose that the norm \MMath{\left|\, \cdot \, \right|_{I }} associated to {$I$} induces the topology of {$W$}.%
\Mpar \vspace{\Sssaut}\hspace{\Alinea}Let \MMath{t,t' \in  \mathds{R}}. Let m{} be a bijective linear symplectomorphism of {$W$} and suppose that there exists a Hilbert space automorphism \MMath{\hat{\text{m}}} of \MMath{\mathcal{F} ^{(W ,I )}_{\mathbf{b} }} such that:%
\Mpar \MStartEqua \MMath{\forall  \mathbf{v}  \in  W , {\hat{\text{m}} \,  \mathcal{O} ^{\mathfrak{bos} }_{(I )}(\mathbf{v} ) \,  \hat{\text{m}}^{-1} = \mathcal{O} ^{\mathfrak{bos} }_{(I )}(\text{m} \mathbf{v} )}}.%
\MStopEqua \Mpar Then:%
\hypertarget{PARbdk}{}\Mpar \MStartEqua \MMath{\left\| \text{u}(t,t') - \text{m} \right\|_{I } \geqslant  \frac{\big|\alpha ^{c}(t') - \alpha ^{c}(t)\big|}{\min\big(\alpha ^{c}(t),\alpha ^{c}(t')\big)}},%
\NumeroteEqua{4.6}{1}\MStopEqua \Mpar where \MMath{\left\|\, \cdot \, \right\|_{I }} denotes the operator norm corresponding to the norm \MMath{\left|\, \cdot \, \right|_{I }}, and \MMath{c \mathrel{\mathop:}=  \frac{d-1}{4}}.%
\Mpar \Mproof \italique{Time-dependent field redefinition.} Let \MMath{t \in  \mathds{R}}. Let \MMath{H_{\mu }(s)} and \MMath{U_{\mu }(s)} be as in the first part of {\seqBBBaxh}, with \MMath{A(s),\, B(s) \esperluette\! F(s)} from {\seqBBBbcb} and \MMath{\mu  = \sqrt{{\lambda _{k} + 1}} \geqslant  1}. We define, for any \MMath{s \in  \mathds{R}}:%
\Mpar \MStartEqua \MMath{N_{\mu }(s) \mathrel{\mathop:}=  \renewcommand{\arraystretch}{2}\matrixTwo{\raisebox{0.2cm}{\hbox{$n(s)$}}}{}{\displaystyle\frac{1}{\mu }\frac{\dot{n}(s)}{n^{2}(s)\, B(s)}}{\displaystyle\hspace{0.2cm}\frac{1}{n(s)}}\renewcommand{\arraystretch}{1}},%
\MStopEqua \Mpar where \MMath{n(s) \mathrel{\mathop:}=  \alpha ^{{\frac{d-1}{4}}}(t+s) > 0}, as well as:%
\Mpar \MStartEqua \MMath{U^{(N)}_{\mu }(s) \mathrel{\mathop:}=  N_{\mu }(s) \,  U_{\mu }(s) \,  N_{\mu }^{-1}(0)}.%
\MStopEqua \Mpar We have \MMath{U^{(N)}_{\mu }(0) = \mathds{1} }, and \MMath{\dot{U}^{(N)}_{\mu } = H^{(N)}_{\mu } \,  U^{(N)}_{\mu }}, with, for any \MMath{s \in  \mathds{R}}:%
\Mpar \MStartEqua \MMath{H^{(N)}_{\mu }(s) \mathrel{\mathop:}=  \dot{N}_{\mu }(s) \,  N_{\mu }^{-1}(s) + N_{\mu }(s) \,  H_{\mu }(s) \,  N_{\mu }^{-1}(s) = \mu  \,  \eta (s) \matrixTwo{}{1}{-1}{} + \frac{1}{\mu } \varepsilon ^{(N)}(s)},%
\MStopEqua \Mpar where \MMath{\eta (s) \mathrel{\mathop:}=  \sqrt{{\nicefrac{\beta (t+s)}{\alpha (t+s)}}}} and \MMath{\varepsilon ^{(N)}(s)} is a matrix \emph{independent} of {$\mu$}. For any \MMath{s \in  \mathds{R}}, we define \MMath{\theta (s) \mathrel{\mathop:}=  \int_{0}^{{s}} \!\eta (\tau )\, d\tau }.%
\Mpar \vspace{\Sssaut}\hspace{\Alinea}For any \MMath{s \in  \mathds{R}}, we have:%
\Mpar \MStartEqua \MMath{\left\| N_{\mu }(s) \right\|^{2} = \left\| N_{\mu }^{-1}(s) \right\|^{2} = p_{\mu }(s) + \sqrt{{p_{\mu }^{2}(s) - 1}}}, where \MMath{p_{\mu }(s) \mathrel{\mathop:}=  \frac{1}{2} \left( n^{2}(s) + \frac{1}{n^{2}(s)} + \frac{1}{\mu ^{2}}\frac{\dot{n}^{2}(s)}{n^{4}(s)\, B^{2}(s)} \right) \geqslant  1}.%
\MStopEqua \Mpar Thus, we get, for any \MMath{s \in  \mathds{R}} and any \MMath{\mu \geqslant 1}:%
\Mpar \MStartEqua \MMath{\left\|N_{\mu }(s)\right\| = \left\|N_{\mu }^{-1}(s)\right\| \leqslant  P(s) \mathrel{\mathop:}=  \sqrt{{p_{1}(s) + \sqrt{{p_{1}^{2}(s) -1}}}}}.%
\MStopEqua \Mpar Let \MMath{\delta  \geqslant  0} and define:%
\Mpar \MStartEqua \MMath{U^{(\delta )}_{\mu }(s) \mathrel{\mathop:}=  R^{-1} \big( \mu \, \theta (s) - \delta \, s \big) \,  U^{(N)}_{\mu }(s)},%
\MStopEqua \Mpar where, for any \MMath{\theta  \in  \mathds{R}}:%
\Mpar \MStartEqua \MMath{R(\theta ) \mathrel{\mathop:}=  \matrixTwo{\cos(\theta )}{\sin(\theta )}{-\sin(\theta )}{\cos(\theta )}}.%
\MStopEqua \Mpar We have \MMath{U^{(\delta )}_{\mu }(0) = \mathds{1} }, and \MMath{\dot{U}^{(\delta )}_{\mu } = H^{(\delta )}_{\mu } \,  U^{(\delta )}_{\mu }}, with, for any \MMath{s \in  \mathds{R}}:%
\Mpar \MStartEqua \MMath{\left\| H^{(\delta )}_{\mu }(s) \right\| \leqslant  \delta  + \frac{1}{\mu }\left\| \varepsilon ^{(N)}(s) \right\|}.%
\MStopEqua \Mpar Hence, we get, for any \MMath{s \in  \mathds{R}}:%
\hypertarget{PARbee}{}\Mpar \MStartEqua \MMath{\left\| U_{\mu }(s) - \matrixTwo{n(s)}{}{}{\frac{1}{n(s)}} \,  R \big( \mu \, \theta (s) - \delta \, s \big) \,  \matrixTwo{\frac{1}{n(0)}}{}{}{n(0)}\right\| \leqslant  P(0)\,  P(s)\,  \left( e^{{\delta \, \left|s\right| + \frac{1}{\mu }\,  E(s)}} - 1 \right) + \frac{1}{\mu }\,  Q(s)},%
\NumeroteEqua{4.6}{2}\MStopEqua \Mpar where:%
\Mpar \MStartEqua \MMath{E(s) \mathrel{\mathop:}=  \left| \int_{0}^{s} \!\left\|\varepsilon ^{(N)}(\tau )\right\|\, d\tau  \right| \, \&\,  Q(s) \mathrel{\mathop:}=  \frac{P(0)\,  \dot{n}(s)}{n^{2}(s)\, B(s)} + \frac{P(s)\,  \dot{n}(0)}{n^{2}(0)\, B(0)}}.%
\MStopEqua \Mpar \vspace{\Ssaut}\italique{Modes with a nearly diagonal evolution.} Let {$\Gamma$} be the \MMath{d}-dimensional lattice in \MMath{\mathds{R}^{d}} such that \MMath{\Sigma  \mathrel{\mathop:}=  \nicefrac{\textstyle\mathds{R}^{d}}{\textstyle\Gamma }}. Then, \MMath{\left\{ \lambda _{k} \middlewithspace| k\geqslant 0 \right\} = \left\{ 4\pi ^{2} \,  {}^{\text{\sc t}} \mathbf{x} .\mathbf{x}  \middlewithspace| \mathbf{x}  \in  \Gamma ^{*} \right\}}, where \MMath{\Gamma ^{*} \mathrel{\mathop:}=  \left\{ \mathbf{x}  \in  \mathds{R}^{d} \middlewithspace| \forall  \gamma  \in  \Gamma , {{}^{\text{\sc t}} \mathbf{x} .\gamma  \in  \mathds{Z}} \right\}} \bseqHHHabb{section 5.2}. Since \MMath{\Gamma ^{*}} is again a \MMath{d}-dimensional lattice, there exists \MMath{\mathbf{x}  \in  \Gamma ^{*}} with \MMath{\lambda  \mathrel{\mathop:}=  4\pi ^{2} \,  {}^{\text{\sc t}} \mathbf{x} .\mathbf{x}  \neq  0}, and, for any \MMath{n \in  \mathds{N}}, there exists \MMath{k_{n} \in  \mathds{N}} such that \MMath{\lambda _{{k_{n}}} = \lambda \,  n^{2}}.%
\Mpar \vspace{\Sssaut}\hspace{\Alinea}Let \MMath{t' \in  \mathds{R} \setminus \left\{t\right\}} and \MMath{\epsilon  > 0}. We define:%
\Mpar \MStartEqua \MMath{\epsilon _{1} \mathrel{\mathop:}=  \ln \left( 1 + \frac{\epsilon }{2\, P(0)\, P(t'-t)} \right) > 0 \, \&\,  \mu _{o} \mathrel{\mathop:}=  \max \left( \frac{2\, Q(t'-t)}{\epsilon },\,  \frac{3\, E(t'-t)}{\epsilon _{1}} \right) < \infty },%
\MStopEqua \Mpar as well as:%
\Mpar \MStartEqua \MMath{n_{o} \mathrel{\mathop:}=  \max \left( \frac{\mu _{o}}{\sqrt{\lambda }},\,  \frac{3\, \left|\theta (t'-t)\right|}{2\,  \sqrt{\lambda } \,  \epsilon _{1}} \right)}.%
\MStopEqua \Mpar The set:%
\Mpar \MStartEqua \MMath{K \mathrel{\mathop:}=  \left\{ n \in  \mathds{N} \middlewithspace| n \geqslant  n_{o} \& \exists  N \in  \mathds{N} \big/ \left| \sqrt{\lambda } \,  \theta (t'-t)\,  n - 2\pi \, N \right| \leqslant  \frac{\epsilon _{1}}{3} \right\}}%
\MStopEqua \Mpar is infinite. Indeed, if \MMath{\frac{\sqrt{\lambda } \,  \theta (t'-t)}{2\pi }} is rational, there exists \MMath{q \geqslant  1} such that \MMath{q\,  \mathds{N} \cap  \left[n_{o},\,  +\infty \right[ \subseteq K}. If it is irrational, \MMath{\left\{ \sqrt{\lambda } \,  \theta (t'-t)\,  n \mkop{mod} 2\pi  \middlewithspace| n \in  \mathds{N} \right\}} is dense in the circle, and, in particular, \MMath{0} is an accumulation point; since \MMath{\left\{0,\dots ,n_{o}\right\}} is finite, \MMath{K} is infinite.%
\Mpar \vspace{\Sssaut}\hspace{\Alinea}Let \MMath{n \in  K} and let \MMath{\mu  \mathrel{\mathop:}=  \sqrt{{\lambda _{{k_{n}}} + 1}} = \sqrt{{\lambda \,  n^{2} + 1}} \geqslant  \mu _{o}}. Let \MMath{N \in  \mathds{N}} such that \MMath{\left| \sqrt{\lambda } \,  \theta (t'-t)\,  n - 2\pi \, N \right| \leqslant  \frac{\epsilon _{1}}{3}}. Then, defining \MMath{\delta  \mathrel{\mathop:}=  {\textstyle\frac{1}{t'-t}} \left( \mu \, \theta (t'-t) - 2\pi \, N \right)}, we have:%
\Mpar \MStartEqua \MMath{\left| \delta \, (t'-t) \right| \leqslant  \frac{\epsilon _{1}}{3} + \frac{\left|\theta (t'-t)\right|}{2\,  n\,  \sqrt{\lambda }} \leqslant  \frac{2\,  \epsilon _{1}}{3}}.%
\MStopEqua \Mpar Hence, by definition of \MMath{\epsilon _{1}} and \MMath{\mu _{o}}, {\seqBBBber} becomes:%
\Mpar \MStartEqua \MMath{\left\| U_{\mu }(t'-t) - \matrixTwo{\frac{n(t'-t)}{n(0)}}{}{}{\frac{n(0)}{n(t'-t)}} \right\| \leqslant  \epsilon }.%
\MStopEqua \Mpar \vspace{\Sssaut}\hspace{\Alinea}We now suppose \MMath{\alpha (t') \geqslant  \alpha (t)} (the case \MMath{\alpha (t') \leqslant  \alpha (t)} is similar). For any \MMath{n \in  K}, we define \MMath{u^{(n)},v^{(n)} \in  W } by:%
\Mpar \MStartEqua \MMath{\forall  k \in  \mathds{N}, {u^{(n)}_{k} = \delta _{{kk_{n}}} \& v^{(n)}_{k} = 0}}.%
\MStopEqua \Mpar By definition of \MMath{\text{u}(t',t)} {\seqDDDbcw}, we obtain:%
\Mpar \MStartEqua \MMath{\left| \text{u}(t',t) \big( u^{(n)},\,  v^{(n)} \big) - \frac{n(t'-t)}{n(0)} \,  \big( u^{(n)},\,  v^{(n)} \big) \right| \leqslant  \epsilon },%
\MStopEqua \Mpar where \MMath{\left|\, \cdot \, \right|} denotes the canonical norm of \MMath{W  = \ell _{2}(\mathds{N} \rightarrow  \mathds{R}) \times  \ell _{2}(\mathds{N} \rightarrow  \mathds{R})}. Moreover, for any \MMath{n \neq  n' \in  K}, \MMath{\mkop{Span} \left\{\big( u^{(n)},\,  v^{(n)} \big),\,  \text{u}(t',t) \big( u^{(n)},\,  v^{(n)} \big)\right\} \perp  \mkop{Span} \left\{\big( u^{(n')},\,  v^{(n')} \big),\,  \text{u}(t',t) \big( u^{(n')},\,  v^{(n')} \big)\right\}} (with respect to the canonical real scalar product of {$W$}). Hence, defining:%
\Mpar \MStartEqua \MMath{W _{\epsilon } \mathrel{\mathop:}=  \mkop{Span}  \left\{ \big( u^{(n)},\,  v^{(n)} \big) \middlewithspace| n \in  K \right\}}, %
\MStopEqua \Mpar \MMath{W _{\epsilon }} is an \emph{infinite-dimensional} vector subspace of {$W$}, and we have, for any \MMath{\mathbf{v}  \in  W _{\epsilon }}:%
\Mpar \MStartEqua \MMath{\left| \text{u}(t',t)(\mathbf{v} ) - \frac{n(t'-t)}{n(0)} \,  \mathbf{v}  \right| \leqslant  \epsilon  \,  \left|\mathbf{v} \right|}.%
\MStopEqua \Mpar \vspace{\Ssaut}\italique{Lower-bound on \MMath{\left\| \text{u}(t,t') - \text{m} \right\|_{I }}.} Let \MMath{\left\langle \, \cdot \, \middlewithspace|\, \cdot \, \right\rangle _{I }} be the complex scalar product associated to {$I$} {\seqDDDaqg} and let \MMath{\left(\, \cdot \, \middlewithspace|\, \cdot \, \right)_{I }} be the corresponding real scalar product, ie.~\MMath{\left(\, \cdot \, \middlewithspace|\, \cdot \, \right)_{I } \mathrel{\mathop:}=  \mkop{Re} \left\langle \, \cdot \, \middlewithspace|\, \cdot \, \right\rangle _{I }}. Let \MMath{\overline{W }} be the \emph{real} Hilbert space obtained by completing {$W$} with respect to the norm \MMath{\left|\, \cdot \, \right|_{I }} induced by \MMath{\left(\, \cdot \, \middlewithspace|\, \cdot \, \right)_{I }}.
Since \MMath{\left|\, \cdot \, \right|} and \MMath{\left|\, \cdot \, \right|_{I }} define the same topology on {$W$}, there exist \MMath{C_{1},\, C_{2} > 0} such that:%
\Mpar \MStartEqua \MMath{C_{1} \,  \left|\, \cdot \, \right| \leqslant  \left|\, \cdot \, \right|_{I } \leqslant  C_{2} \,  \left|\, \cdot \, \right|},%
\MStopEqua \Mpar so, for any \MMath{\mathbf{v}  \in  W _{\epsilon }}, we have:%
\hypertarget{PARbfe}{}\Mpar \MStartEqua \MMath{\left| \text{u}(t',t)(\mathbf{v} ) - \frac{n(t'-t)}{n(0)} \,  \mathbf{v}  \right|_{I } \leqslant  \epsilon  \,  \frac{C_{2}}{C_{1}} \,  \left|\mathbf{v} \right|_{I }}.%
\NumeroteEqua{4.6}{3}\MStopEqua \Mpar \vspace{\Sssaut}\hspace{\Alinea}Now, from \bseqHHHaag{corollary 6.1.1} (of which we give an alternative proof in {\seqBBBabr}), there exists an orthogonal automorphism {$O$} and a symmetric Hilbert–Schmidt operator {$T$} on \MMath{\overline{W }} such that \MMath{\text{m} = \left. \left(\text{id}_{{\overline{W }}} + T \right)\, O  \right|_{W }}.
Let \MMath{\left(\mathbf{b} _{n}\right)_{n\in K}} be a \MMath{\left(\, \cdot \, \middlewithspace|\, \cdot \, \right)_{I }}-orthonormal basis of \MMath{W _{\epsilon }}. Then, \MMath{\left(O \,  \mathbf{b} _{n}\right)_{n\in K}} is an orthonormal family in \MMath{\overline{W }}, so, {$T$} being Hilbert–Schmidt, we have:%
\Mpar \MStartEqua \MMath{\sum_{n\in K} \left|\text{m}\,  \mathbf{b} _{n} - O \,  \mathbf{b} _{n}\right|^{2}_{I } < \infty }.%
\MStopEqua \Mpar Thus, \MMath{K} being infinite, there exists \MMath{n \in  K} such that \MMath{\left|\text{m}\,  \mathbf{b} _{n} - O \,  \mathbf{b} _{n}\right|_{I } \leqslant  \epsilon }, and, in particular, \MMath{\left|\text{m}\,  \mathbf{b} _{n}\right|_{I } \leqslant  1 + \epsilon }.
On the other hand, {\seqBBBbfi} requires \MMath{\left| \text{u}(t',t)(\mathbf{b} _{n}) \right|_{I } \geqslant  \frac{n(t'-t)}{n(0)} - \epsilon  \,  \frac{C_{2}}{C_{1}}}.
So, we arrive at:%
\Mpar \MStartEqua \MMath{\left\| \text{u}(t',t) - \text{m} \right\|_{I } \geqslant  \frac{n(t'-t) - n(0)}{n(0)} - \epsilon  \,  \left(1 + \frac{C_{2}}{C_{1}}\right)}.%
\MStopEqua \Mpar Since this holds for any \MMath{\epsilon  > 0}, this proves {\seqBBBbfl} (remember we focused on the case \MMath{\alpha (t') \geqslant  \alpha (t)}).%
\Mpar \vspace{\Ssaut}\italique{Note.} If we had taken {$\Sigma$} to be the \MMath{d}-sphere instead of a \MMath{d}-torus, we would have \MMath{\left\{ \lambda _{k} \middlewithspace| k\geqslant 0 \right\} = \left\{ n\, (n + d-1) \right\}} \bseqHHHabb{section 5.4}. Asymptotically for large \MMath{n}, \MMath{\sqrt{{n\, (n+d-1) + 1}} \sim n + {\textstyle\frac{d-1}{2}} + o(1)}. Thus, in the case of an \emph{even} spatial dimension \MMath{d}, the argument in the second part of the proof would fail whenever \MMath{\nicefrac{\theta (t'-t)}{2\pi }} is a rational with an odd numerator (because the linear diophantine equation determining the modes with a diagonal evolution would have no solution in this case). Still, it would still hold for generic \MMath{\nicefrac{\theta (t'-t)}{2\pi }}. It does not seem unreasonable to conjecture that for less symmetric manifolds the \MMath{\left\{\sqrt{{\lambda _{k} + 1}} \,  \theta (t'-t) \mkop{mod} 2\pi  \middlewithspace| \lambda _{k} \geqslant  0\right\}} should be even more likely to be densely distributed over the circle. This suggests that the lower bound {\seqBBBbfl}, which relies on having infinitely many \MMath{\sqrt{{\lambda _{k} + 1}}} arbitrarily close to \MMath{0 \mkop{mod} \nicefrac{2\pi }{\theta (t'-t)}}, may in fact hold \emph{generically}. Unfortunately, while some results are known about the asymptotic distribution of the \MMath{\lambda _{k}} {\bseqJJJabc}, these results do not seem currently sufficient to settle this conjecture.%
\MendOfProof \Mpar \MstartPhyMode\hspace{\Alinea}{\seqCCCbfo} actually suggests a way in which the time-evolution \emph{could} be made unitarily implementable on a suitable Fock representation: namely, at the price of a \emph{time-dependent} rescaling of the field by \MMath{\alpha ^{{\nicefrac{(d-1)}{4}}}(t)} (the field conjugate momentum is then not only rescaled correspondingly, but also pick an extra contribution from the \emph{derivative} of this rescaling, hence the form of the time-dependent transformation \MMath{N_{\mu }(t)} applied in the first part of the proof above; see also {\bseqJJJabd}).%
\Mpar \vspace{\Saut}\hspace{\Alinea}One can check that this field redefinition matches precisely the one we would get if the mapping {$\Psi$} in {\seqBBBaxh} were defined using the \emph{instantaneous} spatial metric \MMath{\alpha (t)\,  h } instead of the \emph{constant} one \MMath{h }.
Except that, since {$\Psi$} is defined at the level of the \emph{common} phase space (according to the picture presented in the introduction of {\seqBBBabw}), it cannot depend on the metric at a \emph{particular time}: we no longer know from \emph{which} spatial slice the field is coming at the point where {$\Psi$} is applied (cf.~{\seqBBBbfp}), which is the reason why we had to use the fiducial metric in its definition (besides, it was anyway good enough for our purpose, as the \emph{topology} is \emph{not} sensible to the particular metric we are using).
So, if we would like to take advantage of such a field rescaling, it would have to be incorporated already into the mapping \MMath{\phi  \mapsto  \phi _{\mathtt{c} }} from {\seqBBBaxg} (going much beyond the simple adjustment which was necessary to allow for a \emph{slice-independent} symplectic structure on the common phase space).%
\Mpar \vspace{\Saut}\hspace{\Alinea}Such a redefinition would effectively absorb the non-trivial part of the dynamics (through which the \emph{curvature} of the background spacetime manifested itself).
Of course, this is always possible: it can be understood as a variation on \emph{reduced phase space} quantization, where the dynamics is completely solved \emph{beforehand}, at the \emph{classical} level, rather than trying to \emph{implement} it in the \emph{quantum} theory.
However, doing so would go against our stated goal to avoid, as much as possible, fine-tuning the quantum state space to every detail of the dynamics.
From a more conceptual perspective, it would be somewhat in tension with the physical intuition of \emph{locality}:
if one believes that it is directly the field that is measured in practice (and not whatever combination of field and metric observables happens to trivialize the time evolution), it is difficult to justify why it should obey different, inequivalent representations at different spacetime locations.
\MleavePhyMode %
\Mnomdefichier{lin30}%
\hypertarget{SECbfr}{}\MsectionA{bfr}{5}{Outlook}%
\hypertarget{SECbfs}{}\MsectionB{bfs}{5.1}{Summary}%
\vspace{\PsectionB}\Mpar \MstartPhyMode\hspace{\Alinea}We set out to construct, starting from a classical \emph{linear} field theory, a quantum state space that would depend on nothing but the linear and symplectic (resp.~Euclidean) structures of the classical phase space.
The first part of the solution is to build such a quantum state space by gluing together small \emph{building blocks}: by having each building block only hold \emph{finitely many \dofs}, we can hope to benefit from the Stone–von Neumann theorem.
However, this alone does not \emph{guarantee} the polarization-independence of the resulting state space: an infinite-dimensional Fock representation can also been understood as linking together finite-dimensional ones (when viewing it as an inductive limit, as described in {\seqBBBbfu}), yet it fails to be polarization-independent because the \emph{linking itself} depends crucially on a \emph{global} choice of \emph{vacuum state}, and vacuum states are \emph{not} preserved under changes of polarization.%
\Mpar \vspace{\Saut}\hspace{\Alinea}It is the particular form of the coarse-graining relations needed in projective state spaces {\seqDDDahe} that allows us to fully exploit the finite-dimensional \emph{metaplectic representation}: the latter (whose existence is largely a consequence of the Stone–von Neumann theorem) turns out to be exactly the right tool to \emph{consistently} put \emph{side by side}, in an extended label set, \emph{all} possible truncations of the theory, quantized using \emph{all} possible choices of frames {\seqDDDacm}.
The thus established quantum theory is then \emph{strictly} and \emph{manifestly universal}: its polarization-independence is further confirmed both by its ability to support an \emph{action} of the \emph{full} automorphism group of the classical phase space, as well as by the existence of \emph{natural embeddings} of \emph{arbitrary} infinite-dimensional Fock state spaces.%
\Mpar \vspace{\Saut}\hspace{\Alinea}However, it is not yet \emph{constructive}: quantum states defined as \emph{projective families} may well be just as hard to construct as the \emph{abstract} states used in algebraic QFT (recall {\seqBBBahd}).
To ensure that our state space will be of practical use, we want it to be \emph{systematically} and \emph{explicitly} parametrized, and this requires to make some concessions in terms of universality.
First, we need an additional structure on the classical phase space: namely, a normable \emph{topology} (at least in the bosonic case; in the fermionic case, a suitable topology can be derived directly from the inner-product structure).
We can then restrict the algebra of observables to a \emph{dense}, \emph{countably-generated} subspace, which can legitimately be considered sufficient as far as experimental predictivity is concerned (see also the next subsection), and yields a \emph{countably-cofinal} label set, on which projective states can be constructed in a straightforward \emph{recursive} manner ({\seqBBBapw}; this is assuming that the classical phase space is separable: the physical argument developed before {\seqBBBaht} suggests that it indeed should be).
Importantly, the \emph{particular choice} of this dense subspace can be shown to be effectively irrelevant: universality is \emph{restored} in an \emph{approximate} way, by proving that any two such dense subspaces can be mapped into each other by a symplectomorphism \emph{arbitrarily close} to the identity.%
\Mpar \vspace{\Saut}\hspace{\Alinea}The cosmological model examined in {\seqBBBabn} validates this approach.
It confirms both, that \emph{concrete} physical theories do in fact possess a \emph{natural} topology, and that our slightly \emph{relaxed} notion of universality \emph{still} constitute a marked improvement compared to what could be achieved over a traditional Fock representation.%
\MleavePhyMode \hypertarget{SECbfv}{}\MsectionB{bfv}{5.2}{Notions of Density and Universality}%
\vspace{\PsectionB}\Mpar \MstartPhyMode\hspace{\Alinea}An important feature of the proposed quantization scheme is its focus on an \emph{operational} perspective, where one strives to only include in the mathematical formalism what is strictly needed to compute predictions for actual physical experiments.
In this sense, it can be seen as simply formalizing into a rigorous and consistent framework the pragmatic approaches which are routinely (and often implicitly) used in QFT calculations.
A typical example is the textbook computation of the CMB power spectrum from inflationary models: issues relating to the lack of a unitary implementation of the time evolution (which was stressed in {\seqBBBatn}) do not surface, because the calculation can simply be done separately \emph{mode by mode}.
This is consistent with the realization underpinning projective quantum states spaces, namely, that we never need to consider \emph{at once} \emph{all} the infinitely many \dofs of a (quantum) field theory.%
\Mpar \vspace{\Saut}\hspace{\Alinea}In this spirit, it is sometimes argued that the universality limitations of Fock representations are sufficiently addressed by Fell's theorem {\bseqJJJaau}.
Given finitely many observables \MMath{A_{1},\dots ,A_{n}} and associated error thresholds \MMath{\epsilon _{1},\dots ,\epsilon _{n}}, this result ensures that any state {$\rho$} on the observable algebra (in the sense of \bseqHHHaaa{part III, def.~2.2.8}) can be approximated by a density matrix \MMath{\rho '} on the Fock space, satisfying \MMath{\big| \mkop{Tr}  \rho \,  A_{i} - \mkop{Tr}  \rho '\,  A_{i} \big| \leqslant  \epsilon _{i}} for any \MMath{i \in  \left\{1,\dots ,n\right\}}.
The argument is that this is good enough, as far as universality is concerned, since, as we stressed many times, only finitely many observables are ever measured in practice.
However, when relying on Fell's theorem to restore the universality of the quantum state space, one is \emph{effectively} constructing quantum states as families of successive approximations, each such approximation capturing, within certain error margins, the properties of the state over a certain finite \emph{truncation} of the theory, yet being encoded in a density matrix \MMath{\rho '} on the \emph{full} Fock space.
As we have demonstrated in the present work, the same strategy can be carried out much more economically by taking each \MMath{\rho '} to be a density matrix on a small, \emph{partial} Hilbert space, holding only the truncation on which \MMath{\rho '} is significant.%
\Mpar \vspace{\Saut}\hspace{\Alinea}In addition, the construction of quantum states can then be further simplified by restricting the set of truncations to a \emph{dense subset} {$\mathcal{K}$}. The universality property fulfilled by such subsets ensures that, for any two choices {$\mathcal{K}$}, \MMath{\mathcal{K} '}, and any error threshold {$\epsilon$}, there exists a bijective mapping associating a state \MMath{\rho '} on \MMath{\mathcal{K} '} to each state {$\rho$} on {$\mathcal{K}$}, together with a dual mapping \MMath{A \mapsto  A'} between observables, with \MMath{A} and \MMath{A'} \emph{{$\epsilon$}-close} and satisfying \MMath{\mkop{Tr}  \rho \,  A = \mkop{Tr}  \rho '\,  A'}. While this notion of universality may seem somewhat reminiscent of the just discussed Fell's theorem, its meaning and applicability is quite different.
First, the density property it establishes is of a much stronger nature, as the bound {$\epsilon$} is \emph{uniform} across all states {$\rho$} and observables \MMath{A}.
Second, and more importantly, while Fell's theorem is about approximating \emph{states} (a given Fock representation spans a restricted subspace of states over the \emph{full} algebra of observables), this result should be thought as being about approximating \emph{observables} (building states over a universal label subset confines the observables to a dense sub-algebra but, in a certain sense, enlarges the space of states; see the discussion preceding {\seqBBBatm}).
Because of this, our measure of \emph{closeness} is at the \emph{classical} rather than \emph{quantum} level: we are putting a bound directly on the distance between the observables \MMath{A} and \MMath{A'} (with respect to a topology inherited from the \emph{classical theory}), rather than trying to put a bound on \MMath{\big| \mkop{Tr}  \rho \,  A - \mkop{Tr}  \rho '\,  A \big| = \big| \mkop{Tr}  \rho '\,  A' - \mkop{Tr}  \rho '\,  A \big|} (in fact, \MMath{\mkop{Tr}  \rho '\,  A} does not make sense since \MMath{A} a priori does \emph{not} belong to the sub-algebra on which the state \MMath{\rho '} is defined).%
\Mpar \vspace{\Saut}\hspace{\Alinea}This is justified because the classical theory we are starting from \emph{is} relevant for the resulting quantum theory. Indeed, being the language in which experimental protocols will be written, it serves to label quantum observables with an actual interpretation, thus providing the \emph{interface} between the mathematical formalism and the physical world.
More precisely, since realistic measurements have a limited resolution, what an experimental protocol specifies are certain small \emph{opens} in the classical observable algebra.
For all practical purposes it is therefore sufficient to pick in each of these opens an observable that happens to be implemented in our quantum theory: our notion of density is precisely meant to ensures that this will always be possible.%
\MleavePhyMode \hypertarget{SECbfx}{}\MsectionB{bfx}{5.3}{Toward Non-spatially-compact Spacetimes and Interacting Theories}%
\vspace{\PsectionB}\Mpar \MstartPhyMode\hspace{\Alinea}For the notions of density and universality discussed in the previous subsection to be physically appropriate, it is important to get the classical topology right, and, as we have seen in {\seqBBBabn}, this is in fact the only non-trivial ingredient we need to determine in order to apply the formalism of {\seqBBBabe} to a specific (free) field theory.
However, it was underlined at the beginning of {\seqBBBatn} that, if the classical phase space is parametrized by fields defined over a \emph{non-compact} spatial slice, it is unlikely to support the kind of topology we need\footnote{Actually, it may be difficult to even define the phase space \MMath{V , \Omega } itself in this case, since some \emph{integrability} condition is needed to ensure the convergence of {$\Omega$}, and it is not clear how to arrange for this condition to be preserved under the time evolution.} (namely, a distinguished structure of normable symplectic vector space, as described in {\seqBBBaos}).%
\Mpar \vspace{\Saut}\hspace{\Alinea}On the other hand, and in the spirit of the operational focus mentioned above, it would be sufficient, instead of considering the whole spacetime at once, to consider arbitrary \emph{compact} regions thereof, which are the regions in which actual experiments take place. A way to achieve this would be to combine the formalism discussed in the present article with the General Boundary Formulation of QFT {\bseqJJJabe}. The latter is built on a collection of \emph{partial} (spacial) slices, which may be chosen to be compact ones. We can then study the evolution between two (or more) such partial slices, glued along their boundaries to delimit a compact region of spacetime.
Going over to the quantum theory, the idea is to attach an \emph{a priori distinct} quantum state space to each partial slice: while the approach we followed in {\seqBBBabn} consisted in identifying the phase spaces over different spatial slices with the one on a "model" slice, which was then quantized, allowing us to represent the time evolution as an automorphism of the resulting quantum state space, we mentioned already at the beginning of {\seqBBBabw} that it would work just as well to represent the evolution by morphisms relating quantum state spaces built on different slices.%
\Mpar \vspace{\Saut}\hspace{\Alinea}Using a priori distinct phase/state spaces could also be a strategy to handle non-linear (aka.~interacting) theories. When exploiting the time evolution to solder together the phase spaces over different slices (aka.~kinematical spaces), the resulting joint phase space (aka.~reduced or dynamical space) does not carry a preferred linear structure, unless this time evolution is linear. But the individual kinematical phase spaces do carry such a linear structure\footnote{More generally, the techniques developed in {\seqBBBabe} can presumably be adapted to any classical structure that gives rise to a result of the Stone–von Neumann kind in the finite-dimensional case (eg.~left-invariant structures on Lie group cotangent bundles, see {\bseqJJJabf}).}, which could be used to quantize them along the lines of {\seqBBBabe}.
The role of this kinematical linear structure would just be to select the \emph{elementary} observables of the theory (see {\seqBBBaaw}; note that these elementary observables are, by nature, kinematical: see the discussion in {\seqBBBbfz}).
Even though we would then not get for free the implementation of the dynamics in the quantum theory (since the time evolution would no longer be linear, its quantization would not simply follow from {\seqBBBaog}), this does not preclude the existence of suitable morphisms relating the state spaces on different (partial) slices.%
\Mpar \vspace{\Saut}\hspace{\Alinea}In fact, this is nothing but the standard strategy when dealing with \emph{finitely} many \dofs (ie.~quantum mechanics of one or finitely many particles): one relies on a (kinematical) linear structure to perform the quantization (see the comment at the end of {\seqBBBaaw}), yielding eg.~the ordinary Schrödinger representation, which turns out to support the dynamics of \emph{non-linear} systems (such as the hydrogen atom).
Going over to the case of \emph{infinitely} many \dofs, it is thus not unreasonable to expect that the added \emph{flexibility} afforded by the use of projective state spaces, which suffices to rescue the Stone–von Neumann theorem, may also be enough to \emph{escape} Haag's no-go theorem {\bseqJJJaaj}.
\MleavePhyMode %
\Mnomdefichier{lin40}%
\Mpar \Acknowledge%
\hypertarget{SECbgd}{}\MsectionA{bgd}{A}{Classical Linear Geometry}%
\vspace{\PsectionA}\Mpar \MstartPhyMode\hspace{\Alinea}We review in details the quantization of finite- and infinite-dimensional linear systems.
The aim is to fix the notations and conventions, as well as to derive various auxiliary results that will be needed in {\seqBBBabe}.
The present appendix is devoted to the classical structure of linear systems (symplectic, resp.~Euclidean, vector spaces), while their quantization will be discussed in {\seqBBBbgf} (bosonic, resp.~fermionic, Fock spaces).%
\MleavePhyMode \hypertarget{SECbgg}{}\MsectionB{bgg}{A.1}{Symplectic Vector Spaces}%
\hypertarget{SECbgh}{}\MsectionC{bgh}{A.1.1}{Finite Symplectic Families}%
\vspace{\PsectionC}\Mpar \MstartPhyMode\hspace{\Alinea}A symplectic structure on a \emph{finite-dimensional} vector space {$V$} is a \emph{non-degenerate} \emph{anti-symmetric} \emph{bilinear} form {$\Omega$}, where non-degeneracy means that the mapping \MMath{\mathbf{v}  \mapsto  \Omega (\mathbf{v} ,\, \cdot \, )} should provide a bijective identification between {$V$} and its dual.
Given a function \MMath{f} on {$V$}, we can then define its Hamiltonian vector field \MMath{X_{f}} through \MMath{\Omega (X_{f},\, \cdot \, ) \mathrel{\mathop:}=  df}, and the Poisson-brackets between two functions given by \MMath{\left\{f,\, g\right\} \mathrel{\mathop:}=  \Omega (X_{g},\,  X_{f})}.%
\Mpar \vspace{\Saut}\hspace{\Alinea}If the vector space {$V$} is \emph{infinite-dimensional}, we have two options regarding the non-degeneracy property \bseqHHHaax{VII.A.2}: we can either demand that the mapping \MMath{\mathbf{v}  \mapsto  \Omega (\mathbf{v} ,\, \cdot \, )} be simply injective (aka.\ one-to-one), in which case {$\Omega$} is said to be \emph{weakly} non-degenerate, or we can demand it to be bijective into the \emph{topological dual} of {$V$} (obviously, this requires some kind of topology on {$V$}), making {$\Omega$} \emph{strongly} non-degenerate.
The only linear forms to which a Hamiltonian vector field can be associated (and for which Poisson-brackets can be computed) will be the ones that lies in the \emph{image} of the \MMath{\mathbf{v}  \mapsto  \Omega (\mathbf{v} ,\, \cdot \, )} mapping: this defines, in the algebraic dual of {$V$}, a subspace of admissible linear observables (which, in the case of a strong symplectic form, will correspond precisely to the topological dual).%
\Mpar \vspace{\Saut}\hspace{\Alinea}For the purpose of the construction in {\seqBBBabe}, weak non-degeneracy is sufficient (and in fact, until {\seqBBBbgj} included, we do not even assume {$V$} to be equipped with a topology).%
\MleavePhyMode \hypertarget{PARbgk}{}\Mpar \Mdefinition{A.1}A symplectic vector space \MMath{V , \Omega } is a (possibly infinite dimensional) \emph{real} vector space equipped with a (weak) symplectic form \bseqHHHaax{VII.A.2}, ie.\ a bilinear antisymmetric \MMath{\Omega : V  \times  V  \rightarrow  \mathds{R}} such that:%
\hypertarget{PARbgl}{}\Mpar \MStartEqua \MMath{\forall  \mathbf{v}  \in  V  \mathrel{\big/}  \mathbf{v}  \neq  0, \exists  \mathbf{w}  \in  V  \mathrel{\big/}  \Omega (\mathbf{v} ,\mathbf{w} ) = 1}.%
\NumeroteEqua{A.1}{1}\MStopEqua \Mpar \vspace{\Saut}\hspace{\Alinea}For the rest of this {\seqBBBabk}, {$V$} will denote a (possibly infinite dimensional) symplectic vector space.%
\hypertarget{PARbgn}{}\Mpar \Mdefinition{A.2}A finite symplectic family in \MMath{V } is a family \MMath{\left(\mathbf{e} _{1},\dots ,\mathbf{e} _{2n}\right)} of vectors in {$V$} (for some n{$\geqslant$}0) such that:%
\hypertarget{PARbgo}{}\Mpar \MStartEqua \MMath{\forall  i,j \leqslant  2n, {\Omega (\mathbf{e} _{i}, \mathbf{e} _{j}) = \sum_{p=1}^{n} \delta _{{i,2p-1}}\delta _{{j,2p}} - \delta _{{i,2p}}\delta _{{j,2p-1}} =\mathrel{\mathop:}  \Omega ^{(n)}_{ij}}}.%
\NumeroteEqua{A.2}{1}\MStopEqua \hypertarget{PARbgp}{}\Mpar \Mproposition{A.3}Let {$F$} be a finite dimensional vector subspace of {$V$}, and let \MMath{\left(\mathbf{e} _{1},\dots ,\mathbf{e} _{2n}\right)} be a finite symplectic family in {$F$}. Then, there exists \MMath{m \geqslant  n} and a family \MMath{\left(\mathbf{e} _{2n+1},\dots ,\mathbf{e} _{2m}\right)} of vectors in {$V$} such that \MMath{\left(\mathbf{e} _{1},\dots ,\mathbf{e} _{2m}\right)} is a symplectic family and \MMath{F  \subseteq \mkop{Span}  \left\{\mathbf{e} _{1},\dots ,\mathbf{e} _{2m}\right\}}.%
\Mpar \vspace{\Sssaut}\hspace{\Alinea}In particular, for any finite dimensional vector subspace {$F$} of {$V$}, there exists a finite symplectic family \MMath{\left(\mathbf{e} _{1},\dots ,\mathbf{e} _{2m}\right)} in \MMath{V } such that \MMath{F  \subseteq \mkop{Span}  \left\{\mathbf{e} _{1},\dots ,\mathbf{e} _{2m}\right\}}.%
\Mpar \Mproof We proceed by recursion on \MMath{\mkop{dim}  F  - 2n}. Since \MMath{\Omega ^{(n)}} is an \emph{invertible} \MMath{2n \times  2n} matrix, the vectors in a finite symplectic family are linearly independent. Hence, \MMath{2n \leqslant  \mkop{dim}  F } and, if \MMath{\mkop{dim}  F  = 2n}, \MMath{F  = \mkop{Span}  \left\{\mathbf{e} _{1},\dots ,\mathbf{e} _{2n}\right\}}.%
\Mpar \vspace{\Sssaut}\hspace{\Alinea}We now assume \MMath{\mkop{dim}  F  > 2n}. Let \MMath{\tilde{F } \mathrel{\mathop:}=  \left\{ \mathbf{v}  \in  F  \middlewithspace| \forall  i \leqslant  2n, {\Omega (\mathbf{v} , \mathbf{e} _{i}) = 0} \right\}}. Using again the invertibility of \MMath{\Omega ^{(n)}}, we have \MMath{F  = \mkop{Span}  \left\{\mathbf{e} _{1},\dots ,\mathbf{e} _{2n}\right\} \oplus  \tilde{F }}, so in particular \MMath{\mkop{dim}  \tilde{F } > 0}. Let \MMath{\mathbf{v}  \in  \tilde{F } \setminus \left\{0\right\}}.
Let \MMath{\mathbf{w}  \in  V} such that \MMath{\Omega (\mathbf{v} , \mathbf{w} ) = 1} {\seqDDDbgr}. We define \MMath{\mathbf{e} _{2n+1} \mathrel{\mathop:}=  \mathbf{v} , \mathbf{e} _{2n+2} \mathrel{\mathop:}=  \mathbf{w}  - {\textstyle \sum_{{i,j \leqslant  2n}}} \Omega (\mathbf{w} , \mathbf{e} _{i}) \Omega ^{{(n),-1}}_{ij} \mathbf{e} _{j}}. By definition of \MMath{\tilde{F }}, \MMath{\left(\mathbf{e} _{1},\dots ,\mathbf{e} _{2n+2}\right)} is a symplectic family in \MMath{F ' \mathrel{\mathop:}=  F  + \mkop{Span}  \left\{\mathbf{w} \right\}}.
Noting that \MMath{\mkop{dim}  F ' - (2n+2) < \mkop{dim}  F  - 2n}, we can then apply the recursion hypothesis.%
\MendOfProof \hypertarget{SECbgs}{}\MsectionC{bgs}{A.1.2}{Finite-dimensional Symplectic and Metaplectic Groups}%
\vspace{\PsectionC}\Mpar \MstartPhyMode\hspace{\Alinea}The symplectic group is the group of automorphisms of a symplectic vector space. It can in particular be defined on \MMath{\mathds{R}^{2n}} equipped with its canonical symplectic structure.%
\MleavePhyMode \hypertarget{PARbgu}{}\Mpar \Mdefinition{A.4}Let \MMath{n\geqslant 0}. We denote by \MMath{\text{Sp}^{(n)}} the linear symplectic group over \MMath{\mathds{R}^{2n}}:%
\Mpar \MStartEqua \MMath{\text{Sp}^{(n)} \mathrel{\mathop:}=  \left\{ \sigma  \in  \text{GL}_{2n}(\mathds{R}) \middlewithspace| {}^{\text{\sc t}} \sigma  \,  \Omega ^{(n)} \,  \sigma  = \Omega ^{(n)} \right\}}%
\MStopEqua \Mpar (where \MMath{\text{GL}_{2n}(\mathds{R})} denotes the group of invertible \MMath{2n \times  2n} real matrices), and define a (left) action {$\rhd$} of \MMath{\text{Sp}^{(n)}} on the space of symplectic \MMath{(2n)}-families in {$V$} by:%
\Mpar \MStartEqua \MMath{\forall  \sigma  \in  \text{Sp}^{(n)}, \forall  \left(\mathbf{e} _{1},\dots ,\mathbf{e} _{2n}\right) \text{{ symplectic family in }} V , {\sigma  \rhd  \left(\mathbf{e} _{1},\dots ,\mathbf{e} _{2n}\right) \mathrel{\mathop:}=  \left( \sigma ^{-1}_{ji} \,  \mathbf{e} _{j} \right)_{{i\leqslant 2n}}}}.%
\MStopEqua \Mpar \vspace{\Sssaut}\hspace{\Alinea}As a closed subgroup of the Lie group \MMath{\text{GL}_{2n}(\mathds{R})}, \MMath{\text{Sp}^{(n)}} is a Lie group \bseqHHHabg{theorem 3.42}, with Lie algebra:%
\hypertarget{PARbgz}{}\Mpar \MStartEqua \MMath{\mathfrak{sp} ^{(n)} = \left\{ \text{h} \in  \text{M}_{2n}(\mathds{R}) \middlewithspace| {}^{\text{\sc t}} \text{h} \,  \Omega ^{(n)} + \Omega ^{(n)} \,  \text{h} = 0 \right\}}%
\NumeroteEqua{A.4}{1}\MStopEqua \Mpar (where \MMath{\text{M}_{2n}(\mathds{R})} denotes the vector space of \MMath{2n \times  2n} real matrices).%
\Mpar \MstartPhyMode\hspace{\Alinea}Identifying \MMath{\mathds{R}^{2n} \approx  \left(\mathds{R}^{2}\right)^{n}} with \MMath{\mathds{C}^{n}}, all \emph{symplectic} transformations which happen to be \emph{\MMath{\mathds{C}}-linear} (and not just \MMath{\mathds{R}}-linear) are actually \emph{unitary} and, vice-versa, all unitary transformations are symplectic (see \bseqHHHaae{section 5.2} and/or {\seqBBBbhc} for further insight on why it is so). In other words, the unitary group is a subgroup of the symplectic group, and the latter can be decomposed (at least as a topological space) into unitary and non-unitary components (in the spirit of polar decomposition, see the first part of {\seqBBBbhd}).%
\Mpar \vspace{\Saut}\hspace{\Alinea}We have a corresponding decomposition of the symplectic Lie algebra: indeed, any \MMath{2n\times 2n} real matrix can be written as the sum of a \MMath{\mathds{C}}-linear part and an anti-\MMath{\mathds{C}}-linear one.%
\MleavePhyMode \hypertarget{PARbhe}{}\Mpar \Mproposition{A.5}For any \MMath{\text{h} \in  \mathfrak{sp} ^{(n)}}, we define \MMath{\beta (\text{h}), \gamma (\text{h}) \in  \text{M}_{n}(\mathds{C})} by:%
\Mpar \MStartEqua \MMath{\forall  p,q \leqslant n, {\beta _{pq}(\text{h}) \mathrel{\mathop:}=  \frac{1}{2} \left( \text{h}_{{2p-1,2q-1}} + \text{h}_{{2p,2q}} \right) + \frac{i}{2} \left( \text{h}_{{2p-1,2q}} - \text{h}_{{2p,2q-1}} \right)}\\
\hphantom{\forall  p,q \leqslant n, }\llap{$\&$}{\gamma _{pq}(\text{h}) \mathrel{\mathop:}=  \frac{1}{2} \left( \text{h}_{{2p-1,2q-1}} - \text{h}_{{2p,2q}} \right) + \frac{i}{2} \left( \text{h}_{{2p-1,2q}} + \text{h}_{{2p,2q-1}} \right)}}.%
\MStopEqua \Mpar We have:%
\hypertarget{PARbhh}{}\Mpar \MList{1}for any \MMath{\text{h} \in  \mathfrak{sp} ^{(n)}}, \MMath{\beta ^{\dag}(\text{h}) = -\beta (\text{h})} and \MMath{{}^{\text{\sc t}} \gamma (\text{h}) = \gamma (\text{h})} (where \MMath{{}^{\dag}} denotes the adjoint matrix);%
\MStopList \hypertarget{PARbhi}{}\Mpar \MList{2}for any \MMath{\text{h},\text{h}' \in  \mathfrak{sp} ^{(n)}}:%
\Mpar \MStartEqua \MMath{\beta \big([\text{h}, \text{h}']\big) = \big[ \beta (\text{h}), \beta (\text{h}') \big] + \gamma ^{*}(\text{h}) \,  \gamma (\text{h}') - \gamma ^{*}(\text{h}') \,  \gamma (\text{h}) \\[3pt]
\llap{$\&$} \gamma \big([\text{h}, \text{h}']\big) = \gamma (\text{h}) \,  \beta (\text{h}') - \gamma (\text{h}') \,  \beta (\text{h}) + \beta ^{*}(\text{h}) \,  \gamma (\text{h}') - \beta ^{*}(\text{h}') \,  \gamma (\text{h})}%
\MStopEqua \Mpar (where \MMath{{}^{*}} denotes the complex conjugate).%
\MStopList \Mpar \Mproof Using the characterization of \MMath{\mathfrak{sp} ^{(n)}} {\seqDDDbhm}, we get, for any \MMath{\text{h} \in  \mathfrak{sp} ^{(n)}} and any \MMath{p,q \leqslant n}:%
\Mpar \MStartEqua \MMath{\text{h}_{{2p-1,2q-1}} = - \text{h}_{{2q,2p}},\hspace{0.25cm}  \text{h}_{{2p-1,2q}} = \text{h}_{{2q-1,2p}} \& \text{h}_{{2p,2q-1}} = \text{h}_{{2q,2p-1}}},%
\MStopEqua \Mpar hence statement {\seqBBBbhp} holds.%
\Mpar \vspace{\Sssaut}\hspace{\Alinea}For any \MMath{\text{h} \in  \mathfrak{sp} ^{(n)}} and any \MMath{p,q \leqslant n}, we have:%
\Mpar \MStartEqua \MMath{\beta _{pq}(\text{h}) + \gamma _{pq}(\text{h}) = \text{h}_{{2p-1,2q-1}} + i \,  \text{h}_{{2p-1,2q}}\\
\llap{$\&$} \beta _{pq}(\text{h}) - \gamma _{pq}(\text{h}) = \text{h}_{{2p,2q}} - i \,  \text{h}_{{2p,2q-1}}}.%
\MStopEqua \Mpar Thus, for any \MMath{\text{h},\text{h}' \in  \mathfrak{sp} ^{(n)}}, expanding the commutator and reorganizing the terms yields:%
\Mpar \MStartEqua \MMath{\beta \big([\text{h}, \text{h}']\big) + \gamma \big([\text{h}, \text{h}']\big) = (\beta (\text{h}) + \gamma (\text{h})) \,  \beta (\text{h}') + (\beta ^{*}(\text{h}) + \gamma ^{*}(\text{h})) \,  \gamma (\text{h}') +\\
\hphantom{\beta \big([\text{h}, \text{h}']\big) + \gamma \big([\text{h}, \text{h}']\big) =} - (\beta (\text{h}') + \gamma (\text{h}')) \,  \beta (\text{h}) - (\beta ^{*}(\text{h}') + \gamma ^{*}(\text{h}')) \,  \gamma (\text{h})}.%
\MStopEqua \Mpar and:%
\Mpar \MStartEqua \MMath{\beta \big([\text{h}, \text{h}']\big) - \gamma \big([\text{h}, \text{h}']\big) = (\beta (\text{h}) - \gamma (\text{h})) \,  \beta (\text{h}') - (\beta ^{*}(\text{h}) - \gamma ^{*}(\text{h})) \,  \gamma (\text{h}') +\\
\hphantom{\beta \big([\text{h}, \text{h}']\big) - \gamma \big([\text{h}, \text{h}']\big) =} - (\beta (\text{h}') - \gamma (\text{h}')) \,  \beta (\text{h}) + (\beta ^{*}(\text{h}') - \gamma ^{*}(\text{h}')) \,  \gamma (\text{h})},%
\MStopEqua \Mpar Combining these two equations, we obtain statement {\seqBBBbhx}.%
\MendOfProof \Mpar \MstartPhyMode\hspace{\Alinea}The symplectic group is \emph{not simply-connected}, as its unitary subgroup (mentioned above) comports a \emph{topologically non-trivial loop} (the determinant of unitary matrices can describe a full circle).
In particular, it admits a \emph{double-covering} group, the so-called \emph{metaplectic group}, whose importance comes from the fact that it is this metaplectic group, and not directly the symplectic group, that will be realized in the quantum theory (see \bseqHHHaae{chap.~10} and/or {\seqBBBaaq}, as well as the introduction of {\seqBBBacm}): this is analogous to the way particles of half-integer spin exhibit a representation of the double-cover of the 3-dimensional rotation group (see also \bseqHHHaaf{section 2.7 and appendix 2.B}).%
\Mpar \vspace{\Saut}\hspace{\Alinea}For the benefit of {\seqBBBbhz},  we detail here the construction of this double covering, following \bseqHHHabh{section 1.3}.%
\MleavePhyMode \hypertarget{PARbia}{}\Mpar \Mproposition{A.6}For \MMath{n> 0}, \MMath{\text{Sp}^{(n)}} is path-connected and its fundamental group \MMath{\pi _{1} \big( \text{Sp}^{(n)} \big)} \bseqHHHabh{chap.~1} is isomorphic to \MMath{\mathds{Z}}. Hence, \MMath{\text{Sp}^{(n)}} admits a unique (up to isomorphism) connected double cover, called the metaplectic group over \MMath{\mathds{R}^{2n}} and denoted \MMath{\text{Mp}^{(n)}}, which can be constructed as \bseqHHHabh{theorem 1.38 and prop.~1.32}:%
\Mpar \MStartEqua \MMath{\text{Mp}^{(n)} \mathrel{\mathop:}=  \setquot{\left\{ \gamma  \text{{ path from {$\mathds{1}$} to }} \sigma  \in  \text{Sp}^{(n)} \right\}}{\equiv _{2}}},%
\MStopEqua \Mpar the equivalence relation \MMath{\equiv _{2}} being defined as:%
\Mpar \MStartEqua \MMath{\forall  \gamma ,\gamma ' \text{{ paths from {$\mathds{1}$} to }} \sigma ,\sigma ' \in  \text{Sp}^{(n)}, {\gamma  \equiv _{2} \gamma ' \Leftrightarrow  \left( \sigma  = \sigma ' \& \hmtpCls{{\gamma ^{-1} \mathrel{.}  \gamma '}} \in  2 \mathds{Z} \subset  \mathds{Z} \approx  \pi _{1} \big( \text{Sp}^{(n)} \big) \right)}}%
\MStopEqua \Mpar where, respectively, \MMath{\gamma ^{-1}} denotes the reversed path {$\gamma$}, \MMath{\gamma ^{-1} \mathrel{.}  \gamma '} the composed path \MMath{\gamma '} followed by \MMath{\gamma ^{-1}}, and \MMath{\hmtpCls{{\, \cdot \, }}} the homotopy class of a path (with fixed endpoints).
The covering map is then defined as:%
\hypertarget{PARbif}{}\Mpar \MStartEqua \MMath{\definitionFonction{p ^{(n)}}{\text{Mp}^{(n)}}{\text{Sp}^{(n)}}{\left[\gamma \right]_{{\equiv _{2}}}}{\gamma (1)}}%
\NumeroteEqua{A.6}{1}\MStopEqua \Mpar \vspace{\Sssaut}\hspace{\Alinea}For \MMath{n=0}, \MMath{\text{Sp}^{(0)} = \left\{\mathds{1} \right\}} and we \emph{define} \MMath{\text{Mp}^{(0)} \mathrel{\mathop:}=  \left\{\mathds{1} ,\, \mathds{1} ^{-}\right\} \approx  \mathds{Z}_{2}}.%
\Mpar \vspace{\Sssaut}\hspace{\Alinea}The left action {$\rhd$} can be lifted as an action of \MMath{\text{Mp}^{(n)}} on the space of symplectic \MMath{(2n)}-families in {$V$}.%
\Mpar \Mproof \italique{Fundamental group.} To study the topology of \MMath{\text{Sp}^{(n)}} for \MMath{n>0}, one can use the following isomorphism (see \bseqHHHabi{section 4.4} and \bseqHHHabj{prop.~34}):%
\Mpar \MStartEqua \MMath{\definitionFonctionCrtsPrd{\phi }{\text{U}^{(n)}}{\text{H}^{(n)}}{\text{Sp}^{(n)}}{\text{u}}{\text{h}}{\text{u} \,  \exp(\text{h})}},%
\MStopEqua \Mpar where:%
\Mpar \MStartEqua \MMath{\text{U}^{(n)} \mathrel{\mathop:}=  \left\{ \text{u} \in  \text{GL}_{2n}(\mathds{R}) \middlewithspace| {}^{\text{\sc t}} \text{u} \,  \Omega ^{(n)} \,  \text{u} = \Omega ^{(n)} \& I ^{(n)} \,  \text{u} = \text{u} \,  I ^{(n)} \right\}}%
\MStopEqua \Mpar (with \MMath{\forall  i,j \leqslant  2n, {I ^{(n)}_{ij} = \smallsum_{p=1}^{n} - \delta _{{i,2p-1}}\delta _{{j,2p}} + \delta _{{i,2p}}\delta _{{j,2p-1}}}}\footnote{While the \emph{matrices} \MMath{\Omega ^{(n)}} and \MMath{I ^{(n)}} only differ by a sign prefactor, we use different notations to underline that these are the matrix representations of \emph{different objects}, with different transformation properties under change of basis: \MMath{\Omega ^{(n)}} represents a bilinear form, while \MMath{I ^{(n)}} represents an endomorphism (namely the canonical complex structure of \MMath{\mathds{R}^{2n} \approx  \mathds{C}^{n}}, see {\seqBBBbhc}).}) and:%
\Mpar \MStartEqua \MMath{\text{H}^{(n)} \mathrel{\mathop:}=  \left\{ \text{h} \in  \text{M}_{2n}(\mathds{R}) \middlewithspace| {}^{\text{\sc t}} \text{h} = \text{h} \& \text{h} \,  \Omega ^{(n)} + \Omega ^{(n)} \,  \text{h} = 0 \right\}}%
\MStopEqua \Mpar The invertibility of {$\phi$} follows from polar decomposition (with \MMath{\text{h} = \smallfrac{1}{2} \log ({}^{\text{\sc t}} \sigma  \,  \sigma )} and \MMath{\text{u} = \sigma  \,  \exp(-\text{h})}, see \bseqHHHaaq{theorem VI.10} as well as the previously mentioned references). \MMath{\text{H}^{(n)}} is a vector subspace of \MMath{\text{M}_{2n}(\mathds{R})}, hence is simply-connected. One can check that \MMath{\text{U}^{(n)}} is isomorph to the group \MMath{\text{U}_{n}(\mathds{C})} of unitary \MMath{n \times  n} complex matrices through the natural identification of \MMath{\mathds{C}^{n}} with \MMath{\mathds{R}^{2n}} (in agreement with the logic from {\seqBBBbhc}), and \MMath{\text{U}_{n}(\mathds{C})} is isomorph to \MMath{\text{U}_{1}(\mathds{C}) \times  \text{SU}_{n}(\mathds{C})} via:%
\Mpar \MStartEqua \MMath{\definitionFonctionCrtsPrd{\psi }{\text{U}_{1}(\mathds{C})}{\text{SU}_{n}(\mathds{C})}{\text{U}_{n}(\mathds{C})}{\lambda }{\text{r}}{\text{u}_{n}(\lambda ) \,  \text{r}}} where \MMath{\text{u}_{n}(\lambda ) \mathrel{\mathop:}=  \matrixDiag{\lambda }{1}{1}}%
\MStopEqua \Mpar (whose inverse is given by \MMath{\lambda  = \det \text{u}} and \MMath{\text{r} = \text{u}_{n}(\nicefrac{1}{\lambda }) \,  \text{u}}). \MMath{\text{SU}_{n}(\mathds{C})} is simply-connected \bseqHHHabk{appendix E} and \MMath{\text{U}_{1}(\mathds{C})} is path-connected with \MMath{\pi _{1} \big( \text{U}_{1}(\mathds{C}) \big) \approx  \mathds{Z}} \bseqHHHabh{theorem 1.7}. Putting everything together, \MMath{\text{Sp}^{(n)}} is path-connected and \MMath{\pi _{1} \big( \text{Sp}^{(n)} \big)} is isomorphic to \MMath{\mathds{Z}}.%
\Mpar \vspace{\Ssaut}\italique{Smooth double covering.} \MMath{\equiv _{2}} being an equivalence relation (ie.\ a reflexive, symmetric and transitive binary relation) follows from \MMath{2 \mathds{Z}} being a subgroup of \MMath{\mathds{Z}}, and the map \MMath{p ^{(n)}} given by {\seqBBBand} is well-defined by definition of \MMath{\equiv _{2}}.
Let \MMath{\sigma  \in  \text{Sp}^{(n)}} and let {$\gamma$} be a path from {$\mathds{1}$} to {$\sigma$}. Let \MMath{\tilde{\gamma }} be a path from {$\mathds{1}$} to {$\mathds{1}$} such that \MMath{\hmtpCls{{\tilde{\gamma }}} = 1 \in  \mathds{Z} \approx  \pi _{1} \big( \text{Sp}^{(n)} \big)} and let \MMath{\gamma ' = \gamma  \mathrel{.}  \tilde{\gamma }}. By construction, \MMath{\left[\gamma \right]_{{\equiv _{2}}} \neq  \left[\gamma '\right]_{{\equiv _{2}}}} and \MMath{p ^{(n)} \big( \left[\gamma \right]_{{\equiv _{2}}} \big) = p ^{(n)} \big( \left[\gamma '\right]_{{\equiv _{2}}} \big) = \sigma }. Moreover, if \MMath{\gamma ''} is a path from {$\mathds{1}$} to {$\sigma$}, we either have \MMath{\hmtpCls{{\gamma ^{-1} \mathrel{.}  \gamma ''}} \in  2 \mathds{Z}}, implying \MMath{\gamma '' \equiv _{2} \gamma }, or \MMath{\hmtpCls{{\gamma ^{-1} \mathrel{.}  \gamma ''}} \in  \mathds{Z} \setminus 2 \mathds{Z}}, implying \MMath{\gamma '' \equiv _{2} \gamma '}. Thus, every point in \MMath{\text{Sp}^{(n)}} has exactly 2 pre-images under \MMath{p ^{(n)}}. Equipping \MMath{\text{Mp}^{(n)}} with a suitable topology \bseqHHHabh{section 1.3}, \MMath{p ^{(n)}} can be made a double covering of \MMath{\text{Sp}^{(n)}}.
Furthermore, there exists a unique smooth structure on \MMath{\text{Mp}^{(n)}} making \MMath{p ^{(n)}} smooth \bseqHHHaaz{prop.~2.8}.%
\Mpar \vspace{\Ssaut}\italique{Group structure.} The pointwise multiplication of paths define a binary operation\footnote{This multiplication operation should \emph{not} be confused with the \emph{composition} of paths. In particular, two paths starting from {$\mathds{1}$} are not composable unless the first one also end at {$\mathds{1}$}, ie.~is a loop based at one.} on the space of paths in \MMath{\text{Sp}^{(n)}} starting from {$\mathds{1}$}:%
\Mpar \MStartEqua \MMath{\forall  \gamma _{1},\gamma _{2} \text{{ paths starting from }} \mathds{1} , {\gamma _{1} \,  \gamma _{2} =\mathrel{\mathop:}  \left( \gamma  : \left[0,\,  1\right] \rightarrow  \text{Sp}^{(n)},\,  t \mapsto  \gamma _{1}(t) \,  \gamma _{2}(t) \right)}}.%
\MStopEqua \Mpar Let \MMath{\gamma _{1},\gamma '_{1}} be paths from {$\mathds{1}$} to \MMath{\sigma _{1} \in  \text{Sp}^{(n)}} such that \MMath{\gamma _{1} \equiv _{2} \gamma '_{1}}, and \MMath{\gamma _{2}} be a path from {$\mathds{1}$} to \MMath{\sigma _{2} \in  \text{Sp}^{(n)}}. Let \MMath{\gamma  \mathrel{\mathop:}=  \gamma _{1} \,  \gamma _{2}}, resp.~\MMath{\gamma ' \mathrel{\mathop:}=  \gamma '_{1} \,  \gamma _{2}}. We have \MMath{\gamma (1) = \sigma _{1} \,  \sigma _{2} = \gamma '(1)}, as well as \MMath{\hmtpCls{{\gamma ^{-1} \mathrel{.}  \gamma '}} = \hmtpCls{{\gamma _{1}^{-1} \mathrel{.}  \gamma '_{1}}}}, eg.~via the homotopy map \MMath{\eta _{s}} defined by:%
\Mpar \MStartEqua \MMath{\forall  s \in  \left[0,\,  1\right],\,  \forall  t \in  \left[0,\,  1\right], {\eta _{s}(t) \mathrel{\mathop:}=  \left\{ \withAS{1.3}{ll}{\gamma '_{1} \big(2t\big) \,  \gamma _{2} \big(2st\big) \, &\,  \text{{ if }} t \leqslant  \nicefrac{1}{2}\\ \gamma _{1} \big(2(1-t)\big) \,  \gamma _{2} \big(2s(1-t)\big) \, &\,  \text{{ if }} t \geqslant  \nicefrac{1}{2}} \right.}}.%
\MStopEqua \Mpar Thus, \MMath{\hmtpCls{{\gamma ^{-1} \mathrel{.}  \gamma '}} \in  2 \mathds{Z}}, ie.~\MMath{\gamma  \equiv _{2} \gamma '}. In other words, the pointwise multiplication of paths is compatible with the equivalence relation \MMath{\equiv _{2}} with respect to its first argument. Similarly, it is compatible with \MMath{\equiv _{2}} with respect to its second argument, and, by transitivity of \MMath{\equiv _{2}}, it is also compatible with \MMath{\equiv _{2}} with respect to both arguments simultaneously. This allows the pointwise multiplication of paths to be lifted to a binary operation on \MMath{\text{Mp}^{(n)}},
which turns \MMath{\text{Mp}^{(n)}} into a topological group (associativity follows from the associativity of the matrix multiplication in \MMath{\text{Sp}^{(n)}}; the identity element of \MMath{\text{Mp}^{(n)}} is the equivalence class of the constant path \MMath{t \mapsto  \mathds{1} }; the inverse of any element can be obtained by pointwise inversion of a representative path; and the compatibility with the topology of \MMath{\text{Mp}^{(n)}} can be checked from the definition of the latter together with the continuity of the matrix multiplication and inverse in \MMath{\text{Sp}^{(n)}}). Moreover, \MMath{p ^{(n)}} is then a group homomorphism \MMath{\text{Mp}^{(n)} \rightarrow  \text{Sp}^{(n)}}.%
\Mpar \vspace{\Sssaut}\hspace{\Alinea}Since \MMath{p ^{(n)}} is also a smooth covering map, \MMath{\text{Mp}^{(n)}} is in particular a Lie group, and its Lie algebra can be identified with \MMath{\mathfrak{sp} ^{(n)}} via \MMath{\left[dp ^{(n)}\right]_{\mathds{1} } \approx  \text{id}_{{\mathfrak{sp} ^{(n)}}}}.
Specifically, we have, for any \MMath{\text{h} \in  \mathfrak{sp} ^{(n)}}:%
\hypertarget{PARbiw}{}\Mpar \MStartEqua \MMath{\exp_{{\text{Mp}^{(n)}}}(\text{h}) = \left[t \mapsto  \exp_{{\text{Sp}^{(n)}}} \big( t \,  \text{h} \big)\right]_{{\equiv _{2}}}},%
\NumeroteEqua{A.6}{2}\MStopEqua \Mpar where \MMath{\exp_{{\text{Sp}^{(n)}}}}, resp.~\MMath{\exp_{{\text{Mp}^{(n)}}}} denotes the exponential map in \MMath{\text{Sp}^{(n)}}, resp.~\MMath{\text{Mp}^{(n)}} (indeed \MMath{s \mapsto  \left[t \mapsto  \exp_{{\text{Sp}^{(n)}}} \big( s\,  t\,  \text{h} \big)\right]_{{\equiv _{2}}}} is a one-parameter subgroup \bseqHHHabg{def.~3.29} and its tangential vector at {$\mathds{1}$} is h{}).%
\Mpar \vspace{\Ssaut}\italique{Uniqueness.} Let \MMath{\widetilde{\text{Mp}}^{(n)}} be a \emph{connected} double cover of \MMath{\text{Sp}^{(n)}} with covering map \MMath{\tilde{p }^{(n)} : \widetilde{\text{Mp}}^{(n)} \rightarrow  \text{Sp}^{(n)}}. Let \MMath{\tilde{\mathds{1} } \in  \tilde{p }^{{(n),-1}} \left\langle  \mathds{1}  \right\rangle }.
For any path {$\gamma$} in \MMath{\text{Sp}^{(n)}} starting from {$\mathds{1}$}, there exists a unique path \MMath{\tilde{\gamma }} in \MMath{\widetilde{\text{Mp}}^{(n)}} starting from \MMath{\tilde{\mathds{1} }} such that \MMath{\tilde{p }^{(n)} \circ  \tilde{\gamma } = \gamma }; moreover if two paths \MMath{\gamma ,\gamma '} starting from {$\mathds{1}$} are homotopic (with fixed endpoints), so are their lifts \MMath{\tilde{\gamma }, \tilde{\gamma }'} in \MMath{\widetilde{\text{Mp}}^{(n)}} and, in particular, \MMath{\tilde{\gamma }(1) = \tilde{\gamma }'(1)} \bseqHHHabh{prop.~1.30 and subsequent comment}.
For any homotopy class \MMath{\hmtpCls{{\gamma }}} of paths starting from {$\mathds{1}$}, we define \MMath{\chi (\hmtpCls{{\gamma }}) \mathrel{\mathop:}=  \tilde{\gamma }(1)}.
Let \MMath{H \mathrel{\mathop:}=  \chi ^{-1} \left\langle  \tilde{\mathds{1} } \right\rangle  \subset  \pi _{1} \big( \text{Sp}^{(n)} \big)}
(the set of homotopy classes of loops based at {$\mathds{1}$} whose lifts are loops based at \MMath{\tilde{\mathds{1} }}).
\MMath{H} is a subgroup of \MMath{\pi _{1} \big( \text{Sp}^{(n)} \big)} and, for any pair of paths \MMath{\gamma ,\gamma '} from {$\mathds{1}$} to \MMath{\sigma  \in  \text{Sp}^{(n)}}, \MMath{\chi (\hmtpCls{{\gamma }}) = \chi (\hmtpCls{{\gamma '}})} iff.\ \MMath{\hmtpCls{{\gamma ^{-1} \mathrel{.}  \gamma '}} \in  H}.
In particular, {$\chi$} induces a bijective mapping from \MMath{\textstyle\setquot{\pi _{1} \big( \text{Sp}^{(n)} \big)}{H}} into \MMath{\tilde{p }^{{(n),-1}} \left\langle  \mathds{1}  \right\rangle } (the surjectivity follows from \MMath{\widetilde{\text{Mp}}^{(n)}} being path-connected, as connected cover of a locally path-connected space). Since \MMath{\tilde{p }^{{(n),-1}} \left\langle  \mathds{1}  \right\rangle } is a set of two elements, we get \MMath{H = 2 \mathds{Z}} and {$\chi$} thus induces a bijective mapping \MMath{\tilde{\chi } : \text{Mp}^{(n)} \rightarrow  \widetilde{\text{Mp}}^{(n)}} satisfying \MMath{\tilde{p }^{(n)} \circ  \tilde{\chi } = p ^{(n)}}.
Using the definition of the topology of \MMath{\text{Mp}^{(n)}}, together with the covering map property of \MMath{\tilde{p }^{(n)}}, one can check that \MMath{\tilde{\chi }} is an homeomorphism.%
\Mpar \vspace{\Sssaut}\hspace{\Alinea}Moreover, if \MMath{\widetilde{\text{Mp}}^{(n)}} is equipped with a topological group structure such that \MMath{\tilde{p }^{(n)}} is a group homomorphism, and \MMath{\tilde{\mathds{1} }} is chosen to be the group unit, the maps:%
\Mpar \MStartEqua \MMath{\definitionFonctionCrtsPrd{m}{\text{Mp}^{(n)}}{\text{Mp}^{(n)}}{\text{Mp}^{(n)}}{\mu _{1}}{\mu _{2}}{\mu _{1} \,  \mu _{2}} \, \&\,  \definitionFonctionCrtsPrd{\tilde{m}}{\text{Mp}^{(n)}}{\text{Mp}^{(n)}}{\text{Mp}^{(n)}}{\mu _{1}}{\mu _{2}}{\tilde{\chi }^{-1} \big( \tilde{\chi }(\mu _{1}) \,  \tilde{\chi }(\mu _{2}) \big)}}%
\MStopEqua \Mpar are two lifts of the map:%
\Mpar \MStartEqua \MMath{\definitionFonctionCrtsPrd{m'}{\text{Mp}^{(n)}}{\text{Mp}^{(n)}}{\text{Sp}^{(n)}}{\mu _{1}}{\mu _{2}}{p ^{(n)}(\mu _{1}) \,  p ^{(n)}(\mu _{2})}},%
\MStopEqua \Mpar and coincide at \MMath{(\mathds{1} ,\mathds{1} )}, hence they are equal \bseqHHHabh{prop.~1.34}. In other words, \MMath{\tilde{\chi }} is then a group isomorphism.%
\Mpar \vspace{\Ssaut}\italique{Action on finite symplectic families.} \MMath{p ^{(n)}} being a group homomorphism, the action {$\rhd$} can be lifted to \MMath{\text{Mp}^{(n)}} by precomposing it with \MMath{p ^{(n)}} (this also holds in the \MMath{n=0} case, where {$\rhd$} is simply the trivial action on the empty symplectic family).%
\MendOfProof \Mpar \MstartPhyMode\hspace{\Alinea}In {\seqBBBabe}, finite-dimensional truncations of the classical theory will be labeled by symplectic families, and a family \MMath{\lambda _{1}} will be considered \emph{coarser} that a family \MMath{\lambda _{2}} if the \emph{span} of \MMath{\lambda _{1}} is \emph{included} in the span of \MMath{\lambda _{2}}.
If the family \MMath{\lambda _{2}} happens to \emph{extend} the family \MMath{\lambda _{1}}, it is easy to write down the corresponding coarse-graining relation between the partial quantum theories (thanks to {\seqBBBama}). The following result shows that we can always go back to this case using the natural action of the symplectic group (and, hence, of the metaplectic group) on finite symplectic families (from {\seqBBBane}).%
\MleavePhyMode \hypertarget{PARbjf}{}\Mpar \Mproposition{A.7}Let \MMath{n \leqslant  m} and let \MMath{\left(\mathbf{e} _{1},\dots ,\mathbf{e} _{2n}\right), \left(\mathbf{f} _{1},\dots ,\mathbf{f} _{2m}\right)} be two finite symplectic families in {$V$} such that \MMath{\mkop{Span}  \left\{\mathbf{e} _{1},\dots ,\mathbf{e} _{2n}\right\} \subseteq \mkop{Span}  \left\{\mathbf{f} _{1},\dots ,\mathbf{f} _{2m}\right\}}. Then, there exists a family \MMath{\left(\mathbf{e} _{2n+1},\dots ,\mathbf{e} _{2m}\right)} of vectors in {$V$} and an element \MMath{\mu  \in  \text{Mp}^{(m)}} such that \MMath{\mu  \rhd  \left(\mathbf{f} _{1},\dots ,\mathbf{f} _{2m}\right) = \left(\mathbf{e} _{1},\dots ,\mathbf{e} _{2m}\right)}.%
\Mpar \Mproof Let \MMath{F  \mathrel{\mathop:}=  \mkop{Span}  \left\{\mathbf{f} _{1},\dots ,\mathbf{f} _{2m}\right\}}. Using the invertibility of \MMath{\Omega ^{(m)}}, {$F$} equipped with the restriction of {$\Omega$} is a symplectic vector space, so we can apply {\seqBBBakh} to the finite dimensional vector subspace {$F$} in {$F$}. Hence, there exists a family \MMath{\left(\mathbf{e} _{2n+1},\dots ,\mathbf{e} _{2m}\right)} of vectors in {$F$} such that \MMath{\left(\mathbf{e} _{1},\dots ,\mathbf{e} _{2m}\right)} is a symplectic family and \MMath{\mkop{Span}  \left\{\mathbf{e} _{1},\dots ,\mathbf{e} _{2m}\right\} = \mkop{Span}  \left\{\mathbf{f} _{1},\dots ,\mathbf{f} _{2m}\right\}}. Let {$\sigma$} be the \MMath{2m \times  2m} matrix such that \MMath{\forall  i \leqslant  2m, \mathbf{f} _{i} = \sigma _{ji} \,  \mathbf{e} _{j}}. Since both \MMath{\left(\mathbf{e} _{1},\dots ,\mathbf{e} _{2m}\right)} and \MMath{\left(\mathbf{f} _{1},\dots ,\mathbf{f} _{2m}\right)} are symplectic families, \MMath{\sigma  \in  \text{Sp}^{(m)}} and we can choose \MMath{\mu  \in  \text{Mp}^{(m)}} above {$\sigma$}, so that \MMath{\mu  \rhd  \left(\mathbf{f} _{1},\dots ,\mathbf{f} _{2m}\right) = \left(\mathbf{e} _{1},\dots ,\mathbf{e} _{2m}\right)}.%
\MendOfProof \Mpar \MstartPhyMode\hspace{\Alinea}To discuss the \emph{composition} of coarse-graining relations {\seqDDDbji}, we will need to identify lower-dimensional metaplectic groups as subgroups of higher-dimensional ones. This identification is derived from the action of the metaplectic group on finite symplectic families, by considering what happen when such families can be extended.%
\MleavePhyMode \hypertarget{PARbjj}{}\Mpar \Mdefinition{A.8}Let \MMath{m \geqslant  n \geqslant  0}. We define an injection \MMath{\iota _{m\leftarrow n} : \text{Sp}^{(n)} \rightarrow  \text{Sp}^{(m)}} via:%
\Mpar \MStartEqua \MMath{\forall  \sigma  \in  \text{Sp}^{(n)}, \forall  i,j \leqslant  2m, {{\iota _{m\leftarrow n}(\sigma )}_{ij} = \alternative{\sigma _{ij}}{\text{{ if }} i,j \leqslant  2n}{\delta _{ij}}{\text{{ otherwise}}}}}.%
\MStopEqua \hypertarget{PARbjl}{}\Mpar \Mproposition{A.9}For any \MMath{m \geqslant  n \geqslant  0}, \MMath{\iota _{m\leftarrow n}} can be lifted to an injective group homomorphism \MMath{\ell _{m\leftarrow n} : \text{Mp}^{(n)} \rightarrow  \text{Mp}^{(m)}}. For any \MMath{l \geqslant  m \geqslant  n \geqslant  0}, we have \MMath{\ell _{l\leftarrow m} \circ  \ell _{m\leftarrow n} = \ell _{l\leftarrow n}}.%
\Mpar \vspace{\Sssaut}\hspace{\Alinea}For any \MMath{m \geqslant  n \geqslant  0}, any \MMath{\mu  \in  \text{Mp}^{(n)}} and any symplectic \MMath{(2m)}-family \MMath{\left(\mathbf{e} _{1},\dots ,\mathbf{e} _{2m}\right)} in {$V$}, we have:%
\Mpar \MStartEqua \MMath{\ell _{m\leftarrow n}(\mu ) \rhd  \left(\mathbf{e} _{1},\dots ,\mathbf{e} _{2m}\right) = \left(\mathbf{f} _{1},\dots ,\mathbf{f} _{2m}\right)}%
\MStopEqua \Mpar with \MMath{\left(\mathbf{f} _{1},\dots ,\mathbf{f} _{2n}\right) \mathrel{\mathop:}=  \mu  \rhd  \left(\mathbf{e} _{1},\dots ,\mathbf{e} _{2n}\right)} and \MMath{\left(\mathbf{f} _{2n+1},\dots ,\mathbf{f} _{2m}\right) \mathrel{\mathop:}=  \left(\mathbf{e} _{2n+1},\dots ,\mathbf{e} _{2m}\right)}.%
\Mpar \Mproof For any \MMath{l \geqslant  m \geqslant  n \geqslant  0}, we have \MMath{\iota _{l\leftarrow m} \circ  \iota _{m\leftarrow n} = \iota _{l\leftarrow n}}.%
\Mpar \vspace{\Sssaut}\hspace{\Alinea}For any \MMath{m \geqslant  0}, we define \MMath{\ell _{m\leftarrow 0}(\mathds{1} ) \mathrel{\mathop:}=  \mathds{1} } and \MMath{\ell _{m\leftarrow 0}(\mathds{1} ^{-}) = \mathds{1} ^{-}}, where \MMath{\mathds{1} ^{-} \in  \text{Mp}^{(m)}} is the element such that \MMath{p ^{{(m),-1}} \left\langle  \mathds{1}  \right\rangle  = \left\{\mathds{1} ,\, \mathds{1} ^{-}\right\}}. Then, we have \MMath{p ^{(m)} \circ  \ell _{m\leftarrow 0} = \mathds{1}  = \iota _{m\leftarrow 0} \circ  p ^{(0)}}.%
\Mpar \vspace{\Sssaut}\hspace{\Alinea}Let \MMath{n > 0}. From {\seqBBBbhd}, a representative of the homotopy class \MMath{1 \in  \mathds{Z} \approx  \pi _{1} \big( \text{Sp}^{(n)} \big)} is provided by the loop \MMath{\gamma ^{(n)}_{1} : [0,1] \rightarrow  \text{Sp}^{(n)}, t \mapsto  \iota _{{n\leftarrow 1}}\big(\text{r}(2\pi t)\big)} where:%
\Mpar \MStartEqua \MMath{\forall  \theta  \in  \mathds{R}, {\text{r}(\theta ) \mathrel{\mathop:}=  \matrixTwo{\cos \theta }{-\sin \theta }{\sin \theta }{\cos \theta }}}.%
\MStopEqua \Mpar Now, for any \MMath{m \geqslant  n > 0}, the push-forward map \MMath{\iota _{{m\leftarrow n,*}}} induces a group homomorphism:%
\Mpar \MStartEqua \MMath{\definitionFonction{h }{\pi _{1} \big( \text{Sp}^{(n)} \big)}{\pi _{1} \big( \text{Sp}^{(m)} \big)}{\hmtpCls{{\gamma }}}{\hmtpCls{{\iota _{m\leftarrow n} \circ  \gamma }}}}%
\MStopEqua \Mpar (because homotopy maps can be push-forwarded as well). Since \MMath{h (1) = \hmtpCls{{\iota _{m\leftarrow n} \circ  \gamma ^{(n)}_{1}}} = \hmtpCls{{\gamma ^{(m)}_{1}}} = 1}, we get \MMath{h  = \text{id}_{\mathds{Z}}}. Thus, \MMath{\iota _{{m\leftarrow n,*}}} induces an injection \MMath{\ell _{m\leftarrow n} : \text{Mp}^{(n)} \rightarrow  \text{Mp}^{(m)}, \left[\gamma \right]_{{\equiv _{2}}} \mapsto  \left[\iota _{m\leftarrow n} \circ  \gamma \right]_{{\equiv _{2}}}}. Moreover, we have \MMath{p ^{(m)} \circ  \ell _{m\leftarrow n} = \iota _{m\leftarrow n} \circ  p ^{(n)}} (with the covering map \MMath{p ^{(n)}: \text{Mp}^{(n)} \rightarrow  \text{Sp}^{(n)}}, resp.\ \MMath{p ^{(m)}: \text{Mp}^{(m)} \rightarrow  \text{Sp}^{(m)}}, from {\seqBBBand}).%
\Mpar \vspace{\Sssaut}\hspace{\Alinea}The compatibility with the group structure, the composition property and the expression for the action follow from the corresponding properties of \MMath{\iota _{m\leftarrow n}}.%
\MendOfProof \hypertarget{SECbju}{}\MsectionC{bju}{A.1.3}{Infinite-dimensional Norms and Automorphisms}%
\vspace{\PsectionC}\Mpar \MstartPhyMode\hspace{\Alinea}For the characterization of \emph{universal label subsets} {\seqDDDapw}, we will need a topology on {$V$}, which should be suitably \emph{compatible} with the symplectic structure.
{\seqCCCapq} express the \emph{continuity} of {$\Omega$}, while {\seqBBBbjw} can be understood as the continuity of the \emph{inverse} map (which, as described in {\seqBBBamq}, is only defined on a \emph{subspace} of the topological dual).%
\Mpar \vspace{\Saut}\hspace{\Alinea}Note that we restrict ourselves to \emph{normable topologies}: the proof of {\seqBBBaoy} relies quite heavily on this hypothesis. However, the example studied in {\seqBBBatn} suggests that this should not be a problem in practice (it arguably fails in the non-compact case, yet it is not even totally clear that the symplectic vector space {$V$} \emph{itself} could be defined unambiguously in this case, which calls for an altogether different approach, as advocated in {\seqBBBaax}).%
\MleavePhyMode \hypertarget{PARbjx}{}\Mpar \Mdefinition{A.10}A \emph{normed} symplectic vector space is a symplectic vector space \MMath{V ,\Omega } equipped with a norm \MMath{\left|\, \cdot \, \right|} such that there exist two real numbers \MMath{\left\|\Omega \right\|,\left\|\Omega ^{-1}\right\| > 0} satisfying:%
\hypertarget{PARbjy}{}\Mpar \MList{1}\MMath{\forall  \mathbf{v} ,\mathbf{w}  \in  V , \left|\Omega (\mathbf{v} ,\mathbf{w} )\right| \leqslant  \left\|\Omega \right\| \,  \left|\mathbf{v} \right| \,  \left|\mathbf{w} \right|};%
\MStopList \hypertarget{PARbjz}{}\Mpar \MList{2}\MMath{\forall  \mathbf{v}  \in  V  \setminus \left\{0\right\}, \exists  \mathbf{w}  \in  V  \big/ \Omega (\mathbf{v} ,\mathbf{w} ) = 1 \& \left|\mathbf{v} \right| \,  \left|\mathbf{w} \right| \leqslant  \left\|\Omega ^{-1}\right\|}.%
\MStopList \Mpar A \emph{normable} symplectic vector space is a symplectic vector space \MMath{V ,\Omega } equipped with a topology which can be induced by a norm that makes {$V$} into a normed symplectic vector space.%
\Mpar \MstartPhyMode\hspace{\Alinea}A topology as prescribed in {\seqBBBaos} allows to define a \emph{topological} group of automorphisms on {$V$}.
Importantly, the definition of this group (and of its topological structure) can be shown to be independent of the particular \emph{norm} we are using on {$V$}: it only depends on the topology \emph{induced} by this norm.
This property will be crucial for the \emph{universality} of the construction considered in {\seqBBBatp} (see also the comments preceding {\seqBBBaoq}, as well as the extensive discussion revolving around this question in {\seqBBBatn}).%
\MleavePhyMode \hypertarget{PARbkc}{}\Mpar \Mdefinition{A.11}We define the group \MMath{\mathcal{A} (V , \Omega )} of automorphisms of \MMath{V , \Omega } as the group of \emph{bijective linear} mappings \MMath{\text{m} : V  \rightarrow  V } satisfying:%
\Mpar \MStartEqua \MMath{\forall  \mathbf{v} ,\mathbf{w}  \in  V , \Omega (\text{m} \mathbf{v} ,\text{m} \mathbf{w} ) = \Omega (\mathbf{v} , \mathbf{w} )}.%
\MStopEqua \Mpar \vspace{\Sssaut}\hspace{\Alinea}If \MMath{\left| \, \cdot \,  \right|} is a norm making {$V$} into a normed symplectic vector space, we equip the space \MMath{\mathcal{B} (V ,\left|\, \cdot \, \right|)} of continuous linear mappings \MMath{V  \rightarrow  V } with the norm \MMath{\left\|\, \cdot \, \right\|} defined by:%
\Mpar \MStartEqua \MMath{\forall  \text{m} \in  \mathcal{B} (V , \left|\, \cdot \, \right|), {\left\|\text{m}\right\| \mathrel{\mathop:}=  \sup_{{|\mathbf{v} | = 1}} |\text{m} \mathbf{v} |}}%
\MStopEqua \Mpar (aka.~the operator norm). 
We define the topological group \MMath{\mathcal{A} _{o}(V , \Omega , \left|\, \cdot \, \right|)}, as the group of \emph{bi-continuous} automorphisms of \MMath{V , \Omega , \left| \, \cdot \,  \right|}, equipped with the topology induced by \MMath{\left\|\, \cdot \, \right\|}.%
\hypertarget{PARbkh}{}\Mpar \Mproposition{A.12}If \MMath{\left|\, \cdot \, \right|_{1}} is a norm making {$V$} into a normed symplectic vector space {\seqDDDaos}, and \MMath{\left|\, \cdot \, \right|_{2}} is a norm that defines the \emph{same} topology on {$V$}, \MMath{V ,\Omega ,\left|\, \cdot \, \right|_{2}} is also a normed symplectic vector space.
Moreover, \MMath{\left\|\, \cdot \, \right\|_{1}}, \MMath{\left\|\, \cdot \, \right\|_{2}} define the same topology on \MMath{\mathcal{A} _{o}(V , \Omega , \left|\, \cdot \, \right|_{1}) = \mathcal{A} _{o}(V , \Omega , \left|\, \cdot \, \right|_{2})} {\seqDDDany}.%
\Mpar \vspace{\Sssaut}\hspace{\Alinea}Hence, if {$V$} is equipped with a topology making it a \emph{normable} symplectic vector space {\seqDDDaos}, we can define the topological group \MMath{\mathcal{A} _{o}(V , \Omega )} as \MMath{\mathcal{A} _{o}(V , \Omega , \left|\, \cdot \, \right|)} (for some norm \MMath{\left|\, \cdot \, \right|} inducing the topology of {$V$}).%
\Mpar \Mproof Since \MMath{\left|\, \cdot \, \right|_{1}}, \MMath{\left|\, \cdot \, \right|_{2}} define the same topology, there exists \MMath{c,c' > 0} such that:%
\Mpar \MStartEqua \MMath{c \,  \left|\, \cdot \, \right|_{1} \leqslant   \left|\, \cdot \, \right|_{2} \leqslant  c' \,  \left|\, \cdot \, \right|_{1}}.%
\MStopEqua \Mpar Thus, we get:%
\Mpar \MStartEqua \MMath{\forall  \mathbf{v} ,\mathbf{w}  \in  V , {c^{2} \,  \left|\mathbf{v} \right|_{1} \,  \left|\mathbf{w} \right|_{1} \leqslant   \left|\mathbf{v} \right|_{2} \,  \left|\mathbf{w} \right|_{2} \leqslant  c^{{\prime 2}} \,  \left|\mathbf{v} \right|_{1} \,  \left|\mathbf{w} \right|_{1}}},%
\MStopEqua \Mpar so {\seqBBBbkn} hold with \MMath{\left\|\Omega \right\|_{2} = \frac{\left\|\Omega \right\|_{1}}{c^{2}}} and \MMath{\left\|\Omega ^{-1}\right\|_{2} = c^{{\prime 2}} \,  \left\|\Omega ^{-1}\right\|_{1}}. We also get:%
\Mpar \MStartEqua \MMath{\frac{c}{c'} \left\|\, \cdot \, \right\|_{1} \leqslant  \left\|\, \cdot \, \right\|_{2} \leqslant  \frac{c'}{c} \,  \left\|\, \cdot \, \right\|_{1}},%
\MStopEqua \Mpar so \MMath{\left\|\, \cdot \, \right\|_{1}} and \MMath{\left\|\, \cdot \, \right\|_{2}} define the same topology on \MMath{\mathcal{A} _{o}(V , \Omega , \left|\, \cdot \, \right|_{1}) = \mathcal{A} _{o}(V , \Omega , \left|\, \cdot \, \right|_{2})}.%
\MendOfProof \Mpar \MstartPhyMode\hspace{\Alinea}The following auxiliary result is at the core of the characterization of universal label subsets in \MMath{\mathcal{L} ^{\mathfrak{lin} }_{}} {\seqDDDabi}.%
\MleavePhyMode \hypertarget{PARbkr}{}\Mpar \Mproposition{A.13}Let \MMath{\left|\, \cdot \, \right|} be a norm on {$V$} making it a normed symplectic vector space {\seqDDDaos} and let {$D$} be a dense vector subspace in {$V$}.
For any \MMath{\epsilon  > 0} and any \emph{finite-dimensional} vector subspace {$F$} of {$V$}, there exists \MMath{\text{m} \in  \mathcal{A} _{o}(V , \Omega , \left|\, \cdot \, \right|)} such that:%
\hypertarget{PARbks}{}\Mpar \MList{1}\MMath{\text{m} \left\langle  F  \right\rangle  \subseteq D };%
\MStopList \hypertarget{PARbkt}{}\Mpar \MList{2}\MMath{\left. \text{m} \right|_{{F  \cap  D }} = \left. \text{id}_{V } \right|_{{F  \cap  D }}};%
\MStopList \hypertarget{PARbku}{}\Mpar \MList{3}\MMath{\left\| \text{m} - \text{id}_{V } \right\| < \epsilon }.%
\MStopList \hypertarget{PARbkv}{}\Mpar \Mlemma{A.14}Let \MMath{V , \Omega , \left|\, \cdot \, \right|} be a normed symplectic vector space. Let \MMath{n > 0}, \MMath{E \geqslant  0} and \MMath{\epsilon  > 0}. There exists \MMath{\epsilon _{o} > 0} such that, for any finite symplectic family \MMath{\left(e_{1},\dots ,e_{2n}\right)} and any family \MMath{\left(\mathbf{v} _{1},\dots ,\mathbf{v} _{2n}\right)} of vectors in {$V$}, satisfying:%
\Mpar \MStartEqua \MMath{\forall  i \leqslant  2n, \left|\mathbf{e} _{i}\right| \leqslant  E \& \left|\mathbf{e} _{i} - \mathbf{v} _{i}\right| < \epsilon _{o}},%
\MStopEqua \Mpar there exists a \emph{symplectic} family \MMath{\left(\mathbf{f} _{1},\dots ,\mathbf{f} _{2n}\right)} in {$V$} satisfying:%
\Mpar \MStartEqua \MMath{\forall  i \leqslant  2n, \mathbf{f} _{i} \in  \mkop{Span}  \left\{\mathbf{v} _{j} \middlewithspace| j \leqslant  i\right\} \& \left|\mathbf{e} _{i} - \mathbf{f} _{i}\right| < \epsilon }.%
\MStopEqua \Mpar \Mproof We proceed by recursion over \MMath{n \geqslant  1}.%
\Mpar \vspace{\Sssaut}\hspace{\Alinea}We first consider the case \MMath{n=1}. Let \MMath{\epsilon >0} and \MMath{E\geqslant 0}. We define:%
\Mpar \MStartEqua \MMath{\epsilon _{o} \mathrel{\mathop:}=  \frac{\min(1,\epsilon )}{1 + 2 \,  \left\| \Omega  \right\| \left( 2E + 1 \right)^{2}} > 0}.%
\MStopEqua \Mpar Let \MMath{\left(e_{1},e_{2}\right)} be a symplectic family in {$V$} such that \MMath{\forall  i \leqslant  2, \left|\mathbf{e} _{i}\right| \leqslant  E}. Let \MMath{\left(\mathbf{v} _{1},\mathbf{v} _{2}\right)} be a family of vectors in {$V$} such that \MMath{\forall  i \leqslant  2, \left| \mathbf{e} _{i} - \mathbf{v} _{i} \right| < \epsilon _{o}}. Then,%
\Mpar \MStartEqua \MMath{\left| \Omega (\mathbf{v} _{1},\mathbf{v} _{2}) - 1 \right| < \epsilon _{o} \,  \left\|\Omega \right\| \left( \left| \mathbf{e} _{1} \right| + \left| \mathbf{e} _{2} \right| + \epsilon _{o} \right) \leqslant  \epsilon _{o} \,  \left\|\Omega \right\| \left( 2E + 1 \right)}.%
\MStopEqua \Mpar In particular, \MMath{\Omega (\mathbf{v} _{1},\mathbf{v} _{2}) > \nicefrac{1}{2}}, so defining \MMath{\alpha  \mathrel{\mathop:}=  \nicefrac{1}{\sqrt{{\Omega (\mathbf{v} _{1},\mathbf{v} _{2})}}} > 0}, we have \MMath{\left| \alpha  - 1 \right| < \left| \alpha ^{2} - 1 \right| < 2 \,  \epsilon _{o} \,  \left\|\Omega \right\| \left( 2E + 1 \right)}. Thus, defining \MMath{\mathbf{f} _{i} \mathrel{\mathop:}=  \alpha  \,  \mathbf{v} _{i}} for \MMath{i \leqslant  2}, \MMath{\left(\mathbf{f} _{1},\mathbf{f} _{2}\right)} is a symplectic family with, for any \MMath{i\leqslant 2}, \MMath{\mathbf{f} _{i} \in  \mkop{Span}  \left\{\mathbf{v} _{j} \middlewithspace| j \leqslant  i\right\}} and:%
\Mpar \MStartEqua \MMath{\left| \mathbf{e} _{i} - \mathbf{f} _{i} \right| < \epsilon _{o} \left[ 1 + 2 \,  \left\|\Omega \right\| \left( 2E + 1 \right) \left( \left| \mathbf{e} _{i} \right| + \epsilon _{o} \right) \right] \leqslant  \epsilon }.%
\MStopEqua \Mpar \vspace{\Sssaut}\hspace{\Alinea}We now consider the case \MMath{n > 1}, assuming that the result holds for \MMath{m \leqslant  n-1}.
Let \MMath{\epsilon  > 0} and \MMath{E \geqslant  0}. Let \MMath{\epsilon _{1} > 0} be obtained by applying the case \MMath{m=n-1} to \MMath{\epsilon } and \MMath{E}, and define:%
\Mpar \MStartEqua \MMath{\epsilon ' \mathrel{\mathop:}=  \min \left( 1, \epsilon , \frac{\epsilon _{1}}{4 \,  \left\| \Omega  \right\|\,  \left( E + 1 \right)^{2}} \right) > 0}.%
\MStopEqua \Mpar Let \MMath{\epsilon _{2} > 0} be obtained by applying the case \MMath{m=1} to \MMath{\epsilon '} and \MMath{E}, and define:%
\Mpar \MStartEqua \MMath{\epsilon _{o} \mathrel{\mathop:}=  \min \left( \epsilon _{2}, \frac{\epsilon _{1}}{2 + 4\, \left\| \Omega  \right\| \left( E + 1 \right)^{2}} \right) > 0}.%
\MStopEqua \Mpar Let \MMath{\left(\mathbf{e} _{1},\dots ,\mathbf{e} _{2n}\right)} be a symplectic family in {$V$} such that \MMath{\forall  i \leqslant  2n, \left|\mathbf{e} _{i}\right| \leqslant  E}. Let \MMath{\left(\mathbf{v} _{1},\dots ,\mathbf{v} _{2n}\right)} be a family of vectors in {$V$} such that \MMath{\forall  i \leqslant  2n, \left| \mathbf{e} _{i} - \mathbf{v} _{i} \right| < \epsilon _{o}}. By definition of \MMath{\epsilon _{2} \geqslant  \epsilon _{o}}, there exists a symplectic family \MMath{\left(\mathbf{f} _{1},\mathbf{f} _{2}\right)} with, for any \MMath{i\leqslant 2}, \MMath{\mathbf{f} _{i} \in  \mkop{Span}  \left\{\mathbf{v} _{j} \middlewithspace| j \leqslant  i\right\}} and \MMath{\left| \mathbf{e} _{i} - \mathbf{f} _{i} \right| < \epsilon '}.
In particular, \MMath{\left| \mathbf{e} _{i} - \mathbf{f} _{i} \right| < \epsilon } and \MMath{\left| \mathbf{f} _{i} \right| < \left| \mathbf{e} _{i} \right| + \epsilon ' \leqslant  E + 1}.
For any \MMath{j>2}, we define:%
\Mpar \MStartEqua \MMath{\mathbf{v} '_{j} \mathrel{\mathop:}=  \mathbf{v} _{j} - \Omega (\mathbf{f} _{1}, \mathbf{v} _{j}) \,  \mathbf{f} _{2} + \Omega (\mathbf{f} _{2}, \mathbf{v} _{j}) \,  \mathbf{f} _{1}},%
\MStopEqua \Mpar so that, for any \MMath{i\leqslant 2} and any \MMath{j>2}, \MMath{\Omega (\mathbf{f} _{i}, \mathbf{v} '_{j}) = 0}.
We have, for any \MMath{j>2}:%
\Mpar \MStartEqua \MMath{\left| \mathbf{e} _{j} - \mathbf{v} '_{j} \right| < \epsilon _{o} + \left| \Omega (\mathbf{f} _{1}, \mathbf{v} _{j}) \right| \,  \left| \mathbf{f} _{2} \right| + \left| \Omega (\mathbf{f} _{2}, \mathbf{v} _{j}) \right| \,  \left| \mathbf{f} _{1} \right|\\
\hphantom{\left| \mathbf{e} _{j} - \mathbf{v} '_{j} \right|} < \epsilon _{o} + 2\, \left\|\Omega \right\|\,  \epsilon _{o} \,  \left| \mathbf{f} _{1} \right| \,  \left| \mathbf{f} _{2} \right| + \left\|\Omega \right\|\,  \epsilon '\,  \left| \mathbf{e} _{j} \right| \left( \left|\mathbf{f} _{1}\right| + \left|\mathbf{f} _{2}\right| \right)\\
\hphantom{\left| \mathbf{e} _{j} - \mathbf{v} '_{j} \right|} < \epsilon _{1}}.%
\MStopEqua \Mpar Thus, by definition of \MMath{\epsilon _{1}}, there exists a symplectic family \MMath{\left(\mathbf{f} _{3},\dots ,\mathbf{f} _{2n}\right)} with, for any \MMath{i>2}, \MMath{\mathbf{f} _{i} \in  \mkop{Span}  \left\{\mathbf{v} '_{j} \middlewithspace| 2 < j \leqslant  i\right\}} and \MMath{\left| \mathbf{e} _{i} - \mathbf{f} _{i} \right| < \epsilon }.
Then, by construction, \MMath{\left(\mathbf{f} _{1},\dots ,\mathbf{f} _{2n}\right)} is a symplectic family with the desired properties.%
\MendOfProof \hypertarget{PARblo}{}\Mpar \Mlemma{A.15}Let \MMath{V , \Omega , \left|\, \cdot \, \right|} be a normed symplectic vector space. Let \MMath{k \geqslant  0} and \MMath{E_{o} \geqslant  0}. There exists \MMath{E\geqslant 0} such that, for any finite-dimensional vector subspace {$F$} of {$V$}, of dimension \MMath{\mkop{dim}  F  = d}, and any finite symplectic family \MMath{\left(\mathbf{e} _{1},\dots ,\mathbf{e} _{2n}\right)} in {$F$}, with:%
\Mpar \MStartEqua \MMath{d = 2n+k \& \forall  i \leqslant  2n, \left|\mathbf{e} _{i}\right| \leqslant  E_{o}},%
\MStopEqua \Mpar \MMath{\left(\mathbf{e} _{1},\dots ,\mathbf{e} _{2n}\right)} can be extended into a finite symplectic family \MMath{\left(\mathbf{e} _{1},\dots ,\mathbf{e} _{2m}\right)} with \MMath{m \leqslant  d-n} satisfying:%
\Mpar \MStartEqua \MMath{F  \subseteq \mkop{Span} \left\{ \mathbf{e} _{1},\dots ,\mathbf{e} _{2m} \right\} \& \forall i \leqslant  2m, \left|\mathbf{e} _{i}\right| \leqslant  E}.%
\MStopEqua \Mpar \Mproof Let \MMath{E_{o} \geqslant  0}. We proceed by recursion over \MMath{d-2n = k \geqslant  0}. The case \MMath{d=2n} is trivially satisfied with \MMath{E=E_{o}}.
We consider the case \MMath{d-2n = k > 1}, assuming that the result holds for \MMath{d-2n = k-1} with \MMath{E'\geqslant 0}.
We define:%
\Mpar \MStartEqua \MMath{E \mathrel{\mathop:}=  \max \left( E', \sqrt{{\big(1+2d\, \left\|\Omega \right\|\, E^{{\prime 2}}\big) \left\|\Omega ^{-1}\right\|}} \right) \geqslant  0}%
\MStopEqua \Mpar Let {$F$} with \MMath{\mkop{dim}  F  = d}, \MMath{\left(\mathbf{e} _{1},\dots ,\mathbf{e} _{2n}\right)} be a finite symplectic family in {$F$}, with \MMath{\forall  i \leqslant  2n, \left|\mathbf{e} _{i}\right| \leqslant  E_{o}}, and let \MMath{F '} be a \MMath{(d-1)}-dimensional vector subspace of {$F$} containing \MMath{\left(\mathbf{e} _{1},\dots ,\mathbf{e} _{2n}\right)}. By definition of \MMath{E'}, \MMath{\left(\mathbf{e} _{1},\dots ,\mathbf{e} _{2n}\right)} can be extended into a finite symplectic family \MMath{\left(\mathbf{e} _{1},\dots ,\mathbf{e} _{2m}\right)} with \MMath{m\leqslant d-n-1} satisfying:%
\Mpar \MStartEqua \MMath{F ' \subseteq \mkop{Span} \left\{ \mathbf{e} _{1},\dots ,\mathbf{e} _{2m} \right\} \& \forall i \leqslant  2m, \left|\mathbf{e} _{i}\right| \leqslant  E' \leqslant  E}.%
\MStopEqua \Mpar Let \MMath{\tilde{F } \mathrel{\mathop:}=  \mkop{Span} \left\{ \mathbf{e} _{1},\dots ,\mathbf{e} _{2m} \right\}} and \MMath{\tilde{F }^{\perp } \mathrel{\mathop:}=  \left\{ \mathbf{v}  \in  V  \middlewithspace| \forall  \mathbf{w}  \in  \tilde{F }, \Omega (\mathbf{v} ,\mathbf{w} ) = 0 \right\}}. If \MMath{F  \subseteq \tilde{F }}, we are done.
Otherwise, there exists \MMath{\tilde{\mathbf{v} } \neq  0} such that \MMath{F  \subseteq \tilde{F } \oplus  \mkop{Span} \left\{\tilde{\mathbf{v} }\right\}}.
We define:%
\Mpar \MStartEqua \MMath{\definitionFonction{\Pi }{V }{V }{\mathbf{w} }{\mathbf{w}  - \sum_{{i,j=1}}^{2m} \Omega ^{{(m),-1}}_{ij} \,  \Omega (\mathbf{w} ,\mathbf{e} _{i}) \,  \mathbf{e} _{j}}}.%
\MStopEqua \Mpar Then, \MMath{\mathbf{v}  \mathrel{\mathop:}=  \Pi (\tilde{\mathbf{v} }) \in  \tilde{F }^{\perp } \setminus \left\{0\right\}} and \MMath{F  \subseteq \tilde{F } \oplus  \mkop{Span} \left\{\mathbf{v} \right\}}.
Applying {\seqBBBbjw}, there exists \MMath{\tilde{\mathbf{w} } \in  V } such that \MMath{\Omega (\mathbf{v} ,\tilde{\mathbf{w} }) = 1} and \MMath{\left|\mathbf{v} \right| \,  \left|\tilde{\mathbf{w} }\right| \leqslant  \left\|\Omega ^{-1}\right\|}.
Let \MMath{\mathbf{w}  \mathrel{\mathop:}=  \Pi  (\tilde{\mathbf{w} }) \in  \tilde{F }^{\perp }}. Since \MMath{\mathbf{v}  \in  \tilde{F }^{\perp }}, \MMath{\Omega (\mathbf{v} ,\mathbf{w} ) = \Omega (\mathbf{v} ,\tilde{\mathbf{w} }) = 1}, and, using the expression for {$\Pi$}, we get:%
\Mpar \MStartEqua \MMath{\left|\mathbf{w} \right| \leqslant  \big| \tilde{\mathbf{w} } \big| \left( 1 + 2m\, \left\|\Omega \right\|\, E^{{\prime 2}} \right)
          \leqslant  \frac{E^{2}}{\left|\mathbf{v} \right|}}.%
\MStopEqua \Mpar Defining \MMath{\mathbf{e} _{2m+1} = \sqrt{{\nicefrac{\left|\mathbf{w} \right|}{\left|\mathbf{v} \right|}}} \,  \mathbf{v} } and \MMath{\mathbf{e} _{2m+2} = \sqrt{{\nicefrac{\left|\mathbf{v} \right|}{\left|\mathbf{w} \right|}}} \,  \mathbf{w} }, \MMath{\left(\mathbf{e} _{1},\dots ,\mathbf{e} _{2m+2}\right)} is a finite symplectic family with the desired properties.%
\MendOfProof \Mpar \MproofProposition{A.13}{\seqCCCapq} holds for any \MMath{\mathbf{v} ,\mathbf{w}  \in  V }, hence it in particular holds for any \MMath{\mathbf{v} ,\mathbf{w}  \in  D }. Let \MMath{\mathbf{v}  \in  D  \setminus \left\{0\right\}}. There exists \MMath{\tilde{\mathbf{w} } \in  V } such that \MMath{\Omega (\mathbf{v} ,\tilde{\mathbf{w} }) = 1} and \MMath{\left|\mathbf{v} \right| \,  \left|\tilde{\mathbf{w} }\right| \leqslant  \left\|\Omega ^{-1}\right\|}. Since {$D$} is dense, there exists \MMath{\mathbf{w}  \in  D } such that \MMath{\left| \mathbf{w}  - \tilde{\mathbf{w} } \right| \leqslant  \nicefrac{1}{2\, \left\|\Omega \right\|\, \left|\mathbf{v} \right|}}. Then, \MMath{\left| \Omega (\mathbf{v} ,\mathbf{w} ) - 1 \right| \leqslant  \nicefrac{1}{2}}, so \MMath{\Omega (\mathbf{v} ,\mathbf{w} ) \geqslant  \nicefrac{1}{2} > 0}, and therefore \MMath{\left|\mathbf{v} \right| \,  \left|\nicefrac{\mathbf{w} }{\Omega (\mathbf{v} ,\mathbf{w} )}\right| \leqslant  2\left\|\Omega ^{-1}\right\| + \left\|\Omega \right\|^{-1}}. Thus, {\seqBBBbjw} holds on {$D$} with \MMath{2\left\|\Omega ^{-1}\right\| + \left\|\Omega \right\|^{-1}}. Since {\seqBBBbjw} implies {\seqBBBbgr}, \MMath{D , \left.\Omega \right|_{D }} is in particular a symplectic vector space, hence \MMath{D , \left.\Omega \right|_{D }, \left|\, \cdot \, \right|} is a normed symplectic vector space.%
\Mpar \vspace{\Sssaut}\hspace{\Alinea}Applying {\seqBBBbmc} to the subspace \MMath{F  \cap  D } of {$D$}, there exist \MMath{E_{1} \geqslant  0} and a finite symplectic family \MMath{\left(\mathbf{e} _{1},\dots ,\mathbf{e} _{{2n_{1}}}\right)} in {$D$}, such that \MMath{F  \cap  D  \subseteq \mkop{Span} \left\{\mathbf{e} _{1},\dots ,\mathbf{e} _{{2n_{1}}}\right\} =\mathrel{\mathop:}  F _{1}} and \MMath{\forall  i \leqslant  2n_{1}, \left|e_{i}\right| \leqslant  E_{1}}.
Let \MMath{\tilde{F }_{1} \mathrel{\mathop:}=  F  + F _{1}}. Applying {\seqBBBbmc} to the subspace \MMath{\tilde{F }_{1}} of {$V$}, there exist \MMath{E_{2} \geqslant  0} and a family \MMath{\left(\mathbf{e} _{{2n_{1} + 1}},\dots ,\mathbf{e} _{{2n_{2}}}\right)} of vectors in {$V$}, such that \MMath{\left(\mathbf{e} _{1},\dots ,\mathbf{e} _{{2n_{2}}}\right)} is a symplectic family, \MMath{\tilde{F }_{1} \subseteq \mkop{Span} \left\{\mathbf{e} _{1},\dots ,\mathbf{e} _{{2n_{2}}}\right\} =\mathrel{\mathop:}  F _{2}} and \MMath{\forall  i \leqslant  2n_{2}, \left|e_{i}\right| \leqslant  E_{2}}.
If \MMath{F  \subseteq D }, \MMath{\text{m} = \text{id}_{V } \in  \mathcal{A} _{o}(V , \Omega , \left|\, \cdot \, \right|)} fulfills {\seqBBBbmd}, so in the following we assume \MMath{F  \not\subseteq D }, hence \MMath{n_{2} > n_{1}}.%
\Mpar \vspace{\Sssaut}\hspace{\Alinea}Let \MMath{K \mathrel{\mathop:}=  2(n_{2} - n_{1})} and \MMath{N \mathrel{\mathop:}=  3n_{2} - 2n_{1}}. For any \MMath{k \leqslant  K}, let \MMath{E_{3}(k)} be the bound provided by {\seqBBBbmc} given \MMath{k} and \MMath{E_{2}}, and define \MMath{E_{3} \mathrel{\mathop:}=  \max_{0\leqslant k\leqslant K} E_{3}(k) \geqslant  0}. Let:%
\Mpar \MStartEqua \MMath{\epsilon _{1} \mathrel{\mathop:}=  \frac{\min\left(\epsilon ,\nicefrac{1}{2}\right)}{2\left\|\Omega \right\|\, (N-n_{1})\, E_{3}} > 0}.%
\MStopEqua \Mpar For any \MMath{n \leqslant  N}, let \MMath{\epsilon _{o}(n)} be the bound provided by {\seqBBBbmg} given \MMath{n}, \MMath{E_{3}} and \MMath{\epsilon _{1}}, and define \MMath{\epsilon _{o} \mathrel{\mathop:}=  \min_{0<n\leqslant N} \epsilon _{o}(n) > 0}.%
\Mpar \vspace{\Sssaut}\hspace{\Alinea}Since {$D$} is dense, there exists, for any \MMath{2n_{1} < i \leqslant  2n_{2}}, \MMath{\mathbf{v} _{i} \in  D } such that \MMath{\left| \mathbf{e} _{i} - \mathbf{v} _{i} \right| < \epsilon _{o}}. Let \MMath{\tilde{F }_{2} \mathrel{\mathop:}=  F _{2} + \mkop{Span} \left\{\mathbf{v} _{{2n_{1} +1}},\dots ,\mathbf{v} _{{2n_{2}}}\right\}}. We have \MMath{\mkop{dim}  \tilde{F }_{2} \leqslant  2n_{2} + K}, hence, by definition of \MMath{E_{3}}, \MMath{\left(\mathbf{e} _{1},\dots ,\mathbf{e} _{{2n_{2}}}\right)} can be extended into a symplectic family \MMath{\left(\mathbf{e} _{1},\dots ,\mathbf{e} _{{2n_{3}}}\right)} with \MMath{n_{3} \leqslant  N}, such that \MMath{\tilde{F }_{2} \subseteq \mkop{Span} \left\{\mathbf{e} _{1},\dots ,\mathbf{e} _{{2n_{3}}}\right\} =\mathrel{\mathop:}  F _{3}} and \MMath{\forall  i \leqslant  2n_{3}, \left|\mathbf{e} _{i}\right| \leqslant  E_{3}}.
For any \MMath{i \in  \left\{1,\dots ,2n_{1}\right\} \cup  \left\{2n_{2} +1,\dots ,2n_{3}\right\}}, we define \MMath{\mathbf{v} _{i} \mathrel{\mathop:}=  \mathbf{e} _{i}}. Thus, by definition of \MMath{\epsilon _{o}}, there exists a finite symplectic family \MMath{\left(\mathbf{f} _{1},\dots ,\mathbf{f} _{{2n_{3}}}\right)} in {$V$} such that:%
\Mpar \MStartEqua \MMath{\forall  i \leqslant  2n_{3}, \mathbf{f} _{i} \in  \mkop{Span}  \left\{\mathbf{v} _{j} \middlewithspace| j \leqslant  i\right\} \& \left|\mathbf{e} _{i} - \mathbf{f} _{i}\right| < \epsilon _{1}}.%
\MStopEqua \Mpar The first condition implies \MMath{\forall  i \leqslant  2n_{3}, {\mkop{Span}  \left\{\mathbf{f} _{1},\dots ,\mathbf{f} _{i}\right\} \subseteq \mkop{Span}  \left\{\mathbf{v} _{1},\dots ,\mathbf{v} _{i}\right\}}}, hence:%
\Mpar \MStartEqua \MMath{\mkop{Span}  \left\{\mathbf{f} _{1},\dots ,\mathbf{f} _{{2n_{1}}}\right\} \subseteq F _{1}}, \MMath{\mkop{Span}  \left\{\mathbf{f} _{1},\dots ,\mathbf{f} _{{2n_{2}}}\right\} \subseteq D } and \MMath{\mkop{Span}  \left\{\mathbf{f} _{1},\dots ,\mathbf{f} _{{2n_{3}}}\right\} \subseteq \tilde{F }_{2} + F _{3} = F _{3}}.%
\MStopEqua \Mpar Since the vectors in a symplectic family are linearly independent, a dimension argument yields \MMath{\mkop{Span} \left\{\mathbf{f} _{1},\dots ,\mathbf{f} _{{2n_{1}}}\right\} = F _{1}} and \MMath{\mkop{Span} \left\{\mathbf{f} _{1},\dots ,\mathbf{f} _{{2n_{3}}}\right\} = F _{3}}. Moreover, using the invertibility of \MMath{\Omega ^{(n_{1})}}:%
\Mpar \MStartEqua \MMath{\mkop{Span} \left\{\mathbf{e} _{{2n_{1} +1}},\dots ,\mathbf{e} _{{2n_{3}}}\right\} = \left\{\mathbf{w}  \in  F _{3} \middlewithspace| \forall  \mathbf{v}  \in  F _{1}, \Omega (\mathbf{v} ,\mathbf{w} )=0\right\} = \mkop{Span} \left\{\mathbf{f} _{{2n_{1} +1}},\dots ,\mathbf{f} _{{2n_{3}}}\right\}}.%
\MStopEqua \Mpar Thus, there exists \MMath{\sigma  \in  \text{Sp}^{(n_{3} - n_{1})}} such that \MMath{\sigma  \rhd  \left(\mathbf{e} _{{2n_{1} +1}},\dots ,\mathbf{e} _{{2n_{3}}}\right) = \left(\mathbf{f} _{{2n_{1} +1}},\dots ,\mathbf{f} _{{2n_{3}}}\right)}.%
\Mpar \vspace{\Sssaut}\hspace{\Alinea}Now, we define \MMath{\text{m}: V  \rightarrow  V } by:%
\Mpar \MStartEqua \MMath{\forall  \mathbf{w}  \in  V , {\text{m} \mathbf{w}  \mathrel{\mathop:}=  \mathbf{w}  + \sum_{{i,j=2n_{1} +1}}^{{2n_{3}}} \Omega ^{{(n_{3} -n_{1}),-1}}_{ij} \,  \Omega (\mathbf{e} _{j},\mathbf{w} ) \,  (\mathbf{f} _{i} - \mathbf{e} _{i})}}.%
\MStopEqua \Mpar From the definition of \MMath{\epsilon _{1}}, we get:%
\Mpar \MStartEqua \MMath{\left\| \text{m} - \text{id}_{V } \right\| < \min\left(\epsilon ,\nicefrac{1}{2}\right)},%
\MStopEqua \Mpar so \MMath{\text{m}} is bijective and bi-continuous (using the power series expansion of \MMath{\nicefrac{1}{1-x}}), and it fulfills {\seqBBBbmt}.
Moreover, we have:%
\Mpar \MStartEqua \MMath{\forall  \mathbf{w} ,\mathbf{w} ' \in  V , \Omega (\text{m} \mathbf{w} ,\text{m} \mathbf{w} ') - \Omega (\mathbf{w} , \mathbf{w} ') = \sum_{{i,k=2n_{1} +1}}^{{2n_{3}}} w_{i} \,  w'_{k} \left[ {}^{\text{\sc t}} \sigma ^{-1} \,  \Omega ^{(n_{3} -n_{1})} \,  \sigma ^{-1} - \Omega ^{(n_{3} -n_{1})} \right]_{ik}},%
\MStopEqua \Mpar where \MMath{\forall  i \in  \left\{2n_{1} +1,\dots ,2n_{3}\right\}, w_{i} \mathrel{\mathop:}=  \sum_{{j=2n_{1} +1}}^{{2n_{3}}} \Omega ^{{(n_{3} -n_{1}),-1}}_{ij} \,  \Omega (\mathbf{e} _{j},\mathbf{w} )}.
Hence, \MMath{\text{m} \in  \mathcal{A} _{o}(V , \Omega , \left|\, \cdot \, \right|)} follows from \MMath{\sigma  \in  \text{Sp}^{(n_{3} - n_{1})}}.
Finally, for any \MMath{\mathbf{w}  \in  F  \subseteq F _{2}}, \MMath{\mathbf{w}  = \sum_{i=1}^{{2n_{2}}} w_{i} \,  \mathbf{e} _{i}}, so \MMath{\text{m} \mathbf{w}  = \sum_{i=1}^{{2n_{1}}} w_{i} \,  \mathbf{e} _{i} + \sum_{i=}{2n_{1} +1}^{{2n_{2}}} w_{i} \,  \mathbf{f} _{i} \in  F _{1} + \mkop{Span} \left\{\mathbf{f} _{1},\dots ,\mathbf{f} _{{2n_{2}}}\right\} \subseteq D }, so {\seqBBBbmw} holds, and, similarly, for any \MMath{\mathbf{w}  \in  F  \cap  D  \subseteq F _{1}}, \MMath{\text{m} \mathbf{w}  = \mathbf{w} }, so {\seqBBBbmx} holds. %
\MendOfProof \hypertarget{SECbmy}{}\MsectionC{bmy}{A.1.4}{Complex Structures}%
\vspace{\PsectionC}\Mpar \MstartPhyMode\hspace{\Alinea}A complex structure on a \emph{real} vector space is a prescription to turn it into a \emph{complex} vector space of \emph{half} the dimension: it can be specified as an \MMath{\mathds{R}}-linear bijective mapping, representing the scalar multiplication by the imaginary unit.%
\Mpar \vspace{\Saut}\hspace{\Alinea}On a symplectic vector space {$V$}, a complex structure \emph{suitably compatible} with the symplectic structure {$\Omega$} will not only turn {$V$} into a complex vector space, but also equip it with a structure of \emph{pre-Hilbert space} (aka.~inner-product space), the completion of which can be used as the 1-particle Hilbert space of an infinite-dimensional Fock representation, as we will describe in {\seqBBBbna} (see also {\seqBBBako}: \emph{polarizations} in QFT are precisely the complex structures obeying {\seqBBBaqg}).%
\MleavePhyMode \hypertarget{PARbnb}{}\Mpar \Mdefinition{A.16}A compatible complex structure on a symplectic vector space \MMath{V ,\Omega } is a linear map \MMath{I : V  \rightarrow  V } satisfying:%
\Mpar \MList{1}\MMath{I ^{2} = - \text{id}_{V }};%
\MStopList \Mpar \MList{2}\MMath{\forall  \mathbf{v} ,\mathbf{w}  \in  V , \Omega (I  \mathbf{v} ,I  \mathbf{w} ) = \Omega (\mathbf{v} ,\,  \mathbf{w} )};%
\MStopList \Mpar \MList{3}\MMath{\forall  \mathbf{v}  V  \setminus \left\{0\right\}, \Omega (\mathbf{v} ,\, I  \mathbf{v} ) > 0}.%
\MStopList \Mpar \vspace{\Sssaut}\hspace{\Alinea}\MMath{V } can then be equipped with a structure of \emph{complex} pre-Hilbert space, with the scalar multiplication defined by:%
\Mpar \MStartEqua \MMath{\forall  z \in  \mathds{C},\,  \forall  \mathbf{v}  \in  V , {z\, \mathbf{v}  \mathrel{\mathop:}=  \mkop{Re}(z)\, \mathbf{v}  + \mkop{Im}(z)\, I \mathbf{v} }},%
\MStopEqua \Mpar and the scalar product defined by:%
\Mpar \MStartEqua \MMath{\forall  \mathbf{v} ,\mathbf{w}  \in  V , {\left\langle  \mathbf{v}  \middlewithspace| \mathbf{w}  \right\rangle _{I } \mathrel{\mathop:}=  \Omega (\mathbf{v} ,\, I \mathbf{w} ) + i\, \Omega (\mathbf{v} ,\, \mathbf{w} )}}.%
\MStopEqua \Mpar \vspace{\Sssaut}\hspace{\Alinea}Moreover, the norm \MMath{\left|\, \cdot \, \right|} induced by the scalar product \MMath{\left\langle \, \cdot \, \middlewithspace|\, \cdot \, \right\rangle _{I }} makes {$V$} into a normed symplectic vector space (with \MMath{\left\|\Omega \right\| = \left\|\Omega ^{-1}\right\| = 1}).%
\Mpar \MstartPhyMode\hspace{\Alinea}It is a well-known result on \emph{finite-dimensional} \emph{unitary} vector spaces that, considering the three structures that can be recovered from the complex Hermitian product, viz.~the symplectic form {$\Omega$}, the complex structure {$I$}, and the real inner-product \MMath{\left(\, \cdot \, \middlewithspace|\, \cdot \, \right) \mathrel{\mathop:}=  \mkop{Re} \left\langle \, \cdot \, \middlewithspace|\, \cdot \, \right\rangle _{I }}, any two of them \emph{uniquely} determine the third one.
In addition, if a norm happens to arise from a real scalar product, this scalar product can be retrieved from the norm (although not all norms do: only those that satisfy the so-called "parallelogram identity").%
\Mpar \vspace{\Saut}\hspace{\Alinea}Analogous correspondences hold in the infinite-dimensional case: we already mentioned in {\seqBBBaqg} how the complex scalar product, hence in particular its real part, can be reconstructed from {$\Omega$} and {$I$}; the result below shows that, for a certain class of norms on symplectic vector spaces, a complex structure can be extracted; and the last part of the equivalence will be established in {\seqBBBaqh}.%
\Mpar \vspace{\Saut}\hspace{\Alinea}This is the reason why it was important to establish in {\seqBBBayh} that the structure we will be using in {\seqBBBatp} does \emph{not} depend on the choice of a \emph{norm}, but only of a \emph{normable topology} on {$V$}: while the former would, in many cases, amounts to a choice of \emph{polarization}, the latter affords a much greater flexibility, as demonstrated by the results from {\seqBBBatn}, in particular {\seqBBBbnl}.%
\MleavePhyMode \hypertarget{PARbnm}{}\Mpar \Mproposition{A.17}Let \MMath{V ,\Omega ,\left|\, \cdot \, \right|} be a normed symplectic vector space such that:%
\Mpar \MList{1}\MMath{\left\|\Omega \right\| = \left\|\Omega ^{-1}\right\| = 1};%
\MStopList \Mpar \MList{2}\MMath{\forall  \mathbf{v} ,\mathbf{w}  \in  V , {\left|\mathbf{v}  + \mathbf{w} \right|^{2} + \left|\mathbf{v}  - \mathbf{w} \right|^{2} = 2\, \left|\mathbf{v} \right|^{2} + 2\, \left|\mathbf{w} \right|^{2}}} \italique{(parallelogram identity)}.%
\MStopList \Mpar Then, there exists a \emph{unique} compatible complex structure {$I$} such that \MMath{\left|\, \cdot \, \right|} is the norm induced by {$I$}.%
\Mpar \Mproof Let \MMath{\mathbf{v}  \in  V  \setminus \left\{0\right\}}. From {\seqBBBbjw}, there exists \MMath{\mathbf{w}  \in  V } such that \MMath{\Omega (\mathbf{v} ,\, \mathbf{w} ) = 1} and \MMath{\left|\mathbf{v} \right|\, \left|\mathbf{w} \right| \leqslant  1}. Let \MMath{\mathbf{w} ,\, \mathbf{w} ' \in  V } such that:%
\Mpar \MStartEqua \MMath{\Omega (\mathbf{v} ,\, \mathbf{w} ) = \Omega (\mathbf{v} ,\, \mathbf{w} ') = 1 \& \left|\mathbf{w} \right|,\, \left|\mathbf{w} '\right| \leqslant  \frac{1}{\left|\mathbf{v} \right|}}.%
\MStopEqua \Mpar From {\seqBBBapq}, we have \MMath{2 \leqslant  \left|\mathbf{v} \right|\, \left|\mathbf{w} +\mathbf{w} '\right|}, hence:%
\Mpar \MStartEqua \MMath{\left|\mathbf{w}  - \mathbf{w} '\right|^{2} = 2\, \left|\mathbf{w} \right|^{2} + 2\, \left|\mathbf{w} '\right|^{2} - \left|\mathbf{w}  + \mathbf{w} '\right|^{2} \leqslant  \frac{4}{\left|\mathbf{v} \right|^{2}} - \frac{4}{\left|\mathbf{v} \right|^{2}} = 0}.%
\MStopEqua \Mpar Thus, for any \MMath{\mathbf{v}  \in  V  \setminus \left\{0\right\}}, we can define \MMath{I  \mathbf{v} } as the unique vector in {$V$} such that:%
\Mpar \MStartEqua \MMath{\Omega (\mathbf{v} ,\, I \mathbf{v} ) = \left|\mathbf{v} \right|^{2} \& \left|I \mathbf{v} \right| \leqslant  \left|\mathbf{v} \right|}.%
\MStopEqua \Mpar Using {\seqBBBapq}, this implies \MMath{\left|I \mathbf{v} \right| = \left|\mathbf{v} \right|}. Moreover, the definition can be extended to \MMath{\mathbf{v}  = 0}, since \MMath{I 0 \mathrel{\mathop:}=  0} is the unique vector in {$V$} such that \MMath{\left|I 0\right| \leqslant  0}.
The characterization of \MMath{I \mathbf{v} } ensures that, for any \MMath{\mathbf{v}  \in  V }, \MMath{I (I \mathbf{v} ) = -\mathbf{v} }, and, for any \MMath{\lambda  \in  \mathds{R}}, \MMath{I (\lambda \mathbf{v} ) = \lambda \, I \mathbf{v} }.%
\Mpar \vspace{\Sssaut}\hspace{\Alinea}Let \MMath{\mathbf{v} ,\, \mathbf{w}  \in  V } and let \MMath{F  \mathrel{\mathop:}=  \mkop{Span} \left\{\mathbf{v} ,I \mathbf{v} ,\mathbf{w} ,I \mathbf{w} \right\}}. The parallelogram identity ensures that there exists a real scalar product \MMath{\left(\, \cdot \, ,\, \cdot \, \right)_{F }} on {$F$} whose associated norm is the restriction of \MMath{\left|\, \cdot \, \right|} to {$F$}. Hence, there exists a vector \MMath{\mathbf{v} ' \in  F } such that \MMath{\left. \Omega (\mathbf{v} ,\, \cdot \, ) \right|_{F } = \left(\mathbf{v} ',\, \cdot \, \right)_{F }}. In particular, we have \MMath{\left|\mathbf{v} '\right|^{2} = \Omega (\mathbf{v} ,\, \mathbf{v} ') \leqslant  \left|\mathbf{v} \right|\, \left|\mathbf{v} '\right|}, as well as \MMath{\left|\mathbf{v} \right|^{2} = \left(\mathbf{v} ',\, I \mathbf{v} \right)_{F } \leqslant  \left|\mathbf{v} '\right|\, \left|I \mathbf{v} \right| = \left|\mathbf{v} '\right|\, \left|\mathbf{v} \right|}. The uniqueness of \MMath{I \mathbf{v} } then requires \MMath{\mathbf{v} ' = I \mathbf{v} }, so \MMath{\Omega (\mathbf{v} ,\, I \mathbf{w} ) = \left(I \mathbf{v} ,\, I \mathbf{w} \right)}. Applying the same reasoning to {$\mathbf{w}$} yields \MMath{\Omega (\mathbf{v} ,\, I \mathbf{w} ) = \Omega (\mathbf{w} ,\, I \mathbf{v} )}.
Since this holds for any \MMath{\mathbf{v} ,\, \mathbf{w}  \in  V }, it follows from \MMath{I (I \mathbf{v} ) = -\mathbf{v} } that \MMath{\Omega (I \mathbf{v} ,\, I \mathbf{w} ) = \Omega (\mathbf{v} ,\, \mathbf{w} )}.%
\Mpar \vspace{\Sssaut}\hspace{\Alinea}Let \MMath{\mathbf{w} _{1},\mathbf{w} _{2} \in  V }. For any \MMath{\mathbf{v}  \in  V }, we have \MMath{\Omega \big(\mathbf{v} ,\, I (\mathbf{w} _{1} + \mathbf{w} _{2})\big) = \Omega (\mathbf{w} _{1} + \mathbf{w} _{2},\, I \mathbf{v} ) = \Omega (\mathbf{v} ,\, I \mathbf{w} _{1} + I \mathbf{w} _{2})}, hence {\seqBBBbgr} implies that \MMath{I (\mathbf{w} _{1} + \mathbf{w} _{2}) = I \mathbf{w} _{1} + I \mathbf{w} _{2}}. This concludes the proof that {$I$} is a compatible complex structure inducing the norm \MMath{\left|\, \cdot \, \right|}.%
\Mpar \vspace{\Sssaut}\hspace{\Alinea}If \MMath{I '} is another compatible complex structure inducing the same norm, we have:%
\Mpar \MStartEqua \MMath{\left|\mathbf{v} \right|^{2} = \Omega (\mathbf{v} ,\, I '\mathbf{v} ) = \Omega (I '\mathbf{v} ,\, I 'I '\mathbf{v} ) = \left|I '\mathbf{v} \right|^{2}},%
\MStopEqua \Mpar hence, by uniqueness of \MMath{I \mathbf{v} }, \MMath{I ' = I }.%
\MendOfProof \hypertarget{PARboa}{}\Mpar \Mproposition{A.18}Let \MMath{V ^{(\mathds{C})},\, \left\langle \, \cdot \, \middlewithspace|\, \cdot \, \right\rangle } be a \emph{complex} pre-Hilbert space, and let {$V$} be the underlying real vector space. Define \MMath{\Omega : V  \times  V  \rightarrow  \mathds{R}} by:%
\Mpar \MStartEqua \MMath{\forall  \mathbf{v} ,\mathbf{w}  \in  V , {\Omega (\mathbf{v} ,\, \mathbf{w} ) \mathrel{\mathop:}=  \mkop{Im} \left\langle  \mathbf{v}  \middlewithspace| \mathbf{w}  \right\rangle }}.%
\MStopEqua \Mpar Then, \MMath{V ,\, \Omega } is a symplectic vector space, and \MMath{I : V  \rightarrow  V ,\,  \mathbf{v}  \mapsto  i\, \mathbf{v} } is a compatible complex structure.%
\Mpar \Mproof {$\Omega$} is \MMath{\mathds{R}}-bilinear and antisymmetric by construction. {\seqCCCbgr} is fulfilled with \MMath{\mathbf{w}  = \nicefrac{i\, \mathbf{v} }{\left|\mathbf{v} \right|^{2}}}.
{$I$} is \MMath{\mathds{R}}-linear, and fulfills {\seqBBBaqg} by construction.
\MendOfProof %
\Mnomdefichier{lin41}%
\hypertarget{SECbof}{}\MsectionB{bof}{A.2}{Real pre-Hilbert Spaces}%
\vspace{\PsectionB}\Mpar \MstartPhyMode\hspace{\Alinea}The elementary observables in \emph{fermionic} QFTs obey \emph{anti-commutator} relations, rather than commutator relations. Accordingly, the classical \emph{precursors} of such theories should be defined on a "phase space" carrying a \emph{symmetric} bilinear form, rather than an \emph{anti-symmetric} one.
Since symplectic forms were defined in {\seqBBBakg} as (weakly) \emph{non-degenerate}, bilinear, anti-symmetric forms, the natural analogue in the symmetric case is to consider a structure of \emph{real pre-Hilbert space} (aka.~inner-product space; the analogue of a strongly non-degenerate symplectic form would naturally be a real Hilbert space, in which, by virtue of the Riesz representation theorem, the scalar product provides a bijective identification between {$V$} and its dual).
The analogue of Hamiltonian vector fields will then be \emph{gradient vector fields}, and the analogue of the Poisson brackets of two functions \MMath{f,\, g} will be the scalar product of their gradients \MMath{\left( \nabla f \middlewithspace| \nabla g \right)} (which is, as requested, symmetric in \MMath{f,\, g}).%
\Mpar \vspace{\Saut}\hspace{\Alinea}The present appendix is mostly a straightforward repetition of {\seqBBBabk} in the case of a real scalar product, and we will only comment on the few differences.
Note that for the quantization to exist, the phase space should be \emph{even-dimensional} (this was automatically ensured in the symplectic case, since no non-degenerate anti-symmetric form can be constructed on an odd-dimensional space).%
\MleavePhyMode \Mpar \vspace{\Saut}\hspace{\Alinea}In this sub-section, \MMath{V , \left(\, \cdot \, \middlewithspace|\, \cdot \, \right)} is a \emph{real} pre-Hilbert space. {$V$} is assumed to be either \emph{even} or infinite dimensional.%
\hypertarget{SECboi}{}\MsectionC{boi}{A.2.1}{Finite Orthonormal Families}%
\vspace{\PsectionC}\Mpar \MstartPhyMode\hspace{\Alinea}Finite truncations of a field theory will still be labeled by partial frames, aka.~finite families. In agreement with the comment above, we only consider \emph{even-dimensional} families: like in the bosonic case, 1 \dof or "mode" corresponds to a 2-dimensional subspace of (the dual of) {$V$}.%
\MleavePhyMode \hypertarget{PARbok}{}\Mpar \Mproposition{A.19}Let {$F$} be a finite dimensional vector subspace of {$V$}, and let \MMath{\left(\mathbf{e} _{1},\dots ,\mathbf{e} _{2n}\right)} be a finite orthonormal family in {$F$}. Then, there exists \MMath{m \in  \mathds{N}}, with \MMath{n \leqslant  m \leqslant  \lceil \nicefrac{\mkop{dim}  F }{2} \rceil} (where \MMath{\lceil \, \cdot \,  \rceil} denotes the ceiling function), and a family \MMath{\left(\mathbf{e} _{2n+1},\dots ,\mathbf{e} _{2m}\right)} of vectors in {$V$} such that \MMath{\left(\mathbf{e} _{1},\dots ,\mathbf{e} _{2m}\right)} is an orthonormal family and \MMath{F  \subseteq \mkop{Span}  \left\{\mathbf{e} _{1},\dots ,\mathbf{e} _{2m}\right\}}.%
\Mpar \vspace{\Sssaut}\hspace{\Alinea}In particular, for any finite dimensional vector subspace {$F$} of {$V$}, there exists a finite orthonormal family \MMath{\left(\mathbf{e} _{1},\dots ,\mathbf{e} _{2m}\right)} in \MMath{V } such that \MMath{F  \subseteq \mkop{Span}  \left\{\mathbf{e} _{1},\dots ,\mathbf{e} _{2m}\right\}}.%
\Mpar \Mproof Let \MMath{\tilde{F } \mathrel{\mathop:}=  F  \cap  \left( \mkop{Span}  \left\{\mathbf{e} _{1},\dots ,\mathbf{e} _{2m}\right\} \right)^{\perp}}. Let \MMath{\left(\mathbf{e} _{2n+1},\dots ,\mathbf{e} _{p}\right)} be an orthonormal basis in \MMath{\tilde{F }}. Then, \MMath{\left(\mathbf{e} _{1},\dots ,\mathbf{e} _{p}\right)} is an orthonormal basis in \MMath{F } and, in particular, \MMath{p = \mkop{dim}  F }. If \MMath{p} is even, let \MMath{m = \nicefrac{p}{2}}. If \MMath{p} is odd, \MMath{F  \subsetneq V } (since {$V$} is even or infinite dimensional), hence there exists \MMath{\mathbf{e} _{p+1} \in  F ^{\perp}} with \MMath{\left( \mathbf{e} _{p+1} \middlewithspace| \mathbf{e} _{p+1} \right) = 1} and we let \MMath{m = \nicefrac{p+1}{2}}.%
\MendOfProof \hypertarget{SECbom}{}\MsectionC{bom}{A.2.2}{Finite-dimensional (Special) Orthogonal and Spin Group}%
\vspace{\PsectionC}\Mpar \MstartPhyMode\hspace{\Alinea}While symplectomorphisms are \emph{automatically} \emph{orientation-preserving} (because a volume form, prescribing an orientation, can be constructed directly from the symplectic form {$\Omega$}, see \bseqHHHaae{section 8.2}), this is \emph{not} the case of orthogonal transformations (viz.~reflections are orientation-reversing). However, as we will see in {\seqBBBamb}, arrows can only be associated to \emph{orientation-preserving} transformations, ie.~elements of the \emph{special} orthogonal group.
Fortunately, {\seqBBBalx} below ensures that, for {$V$} \emph{infinite-dimensional} (which is the case we are actually interested in), this restriction on the admissible arrows does not jeopardize the directedness of the label set {\seqDDDboo}: this is because a choice of orientation in a vector space \MMath{F _{2}} does \emph{not} fix the orientation of a \emph{strictly} lower-dimensional subspace \MMath{F _{1}} of \MMath{F _{2}} (we can always \emph{compensate} a change of orientation in \MMath{F _{1}} by a reflexion in its orthogonal complement).%
\Mpar \vspace{\Saut}\hspace{\Alinea}Also, beware that, since we are only interested in \emph{even-dimensional} phase spaces, and in agreement with the notations used in {\seqBBBabk}, we identify all groups and Lie algebras by the \emph{half-dimension} of the space on which they act (ie.~the number of \dofs).%
\MleavePhyMode \hypertarget{PARbop}{}\Mpar \Mdefinition{A.20}We denote by \MMath{\text{SO}^{(n)}} the special orthogonal group over \MMath{\mathds{R}^{2n}}:%
\Mpar \MStartEqua \MMath{\text{SO}^{(n)} \mathrel{\mathop:}=  \left\{ \sigma  \in  \text{GL}_{2n}(\mathds{R}) \middlewithspace| {}^{\text{\sc t}} \sigma  = \sigma ^{-1} \& \det \sigma  = 1 \right\}},%
\MStopEqua \Mpar and define a (left) action {$\rhd$} of \MMath{\text{SO}^{(n)}} on the space of orthonormal \MMath{(2n)}-families in {$V$} by:%
\Mpar \MStartEqua \MMath{\forall  \sigma  \in  \text{SO}^{(n)}, \forall  \left(\mathbf{e} _{1},\dots ,\mathbf{e} _{2n}\right) \text{{ orthonormal family in }} V , {\sigma  \rhd  \left(\mathbf{e} _{1},\dots ,\mathbf{e} _{2n}\right) \mathrel{\mathop:}=  \left( \sigma ^{-1}_{ji} \,  \mathbf{e} _{j} \right)_{{i\leqslant 2n}} = \left( \sigma _{ij} \,  \mathbf{e} _{j} \right)_{{i\leqslant 2n}}}}.%
\MStopEqua \Mpar \vspace{\Sssaut}\hspace{\Alinea}\MMath{\text{SO}^{(n)}} is a Lie group with Lie algebra:%
\Mpar \MStartEqua \MMath{\mathfrak{so} ^{(n)} \mathrel{\mathop:}=  \left\{ \text{h} \in  \text{M}_{2n}(\mathds{R}) \middlewithspace| {}^{\text{\sc t}} \text{h} + \text{h} = 0 \right\}}.%
\MStopEqua \Mpar \vspace{\Sssaut}\hspace{\Alinea}For any \MMath{m \geqslant  n \geqslant  0}, we define an injection \MMath{\iota _{m\leftarrow n} : \text{SO}^{(n)} \rightarrow  \text{SO}^{(m)}} as in {\seqBBBbow} (note that \MMath{\iota _{m\leftarrow n}} is indeed well-defined as it preserves the determinant).%
\Mpar \MstartPhyMode\hspace{\Alinea}Like in the symplectic case (see the comments before {\seqBBBboy}), the (even-dimensional) special orthogonal group contains a \emph{unitary} subgroup. In fact, since \MMath{\mathds{C}}-linear transformations which preserve the real inner-product preserve the associated symplectic structure, and vice-versa (in the spirit of the correspondence discussed before {\seqBBBaoq}), this unitary subgroup is precisely the \emph{intersection} of the symplectic and orthogonal groups.
This is the reason why the signs in {\seqBBBboz} match for the \MMath{\beta (\text{h})} part, but not for the \MMath{\gamma (\text{h})} one (the sign difference is emphasized in bold below).%
\MleavePhyMode \hypertarget{PARbpa}{}\Mpar \Mproposition{A.21}For any \MMath{\text{h} \in  \mathfrak{so} ^{(n)}}, we define \MMath{\beta (\text{h}), \gamma (\text{h}) \in  \text{M}_{n}(\mathds{C})} as in {\seqBBBboy}.%
\Mpar We have:%
\hypertarget{PARbpb}{}\Mpar \MList{1}for any \MMath{\text{h} \in  \mathfrak{so} ^{(n)}}, \MMath{\beta ^{\dag}(\text{h}) = -\beta (\text{h})} and \MMath{{}^{\text{\sc t}} \gamma (\text{h}) = \heavyminus \gamma (\text{h})};%
\MStopList \hypertarget{PARbpc}{}\Mpar \MList{2}for any \MMath{\text{h},\text{h}' \in  \mathfrak{so} ^{(n)}}:%
\Mpar \MStartEqua \MMath{\beta \big([\text{h}, \text{h}']\big) = \big[ \beta (\text{h}), \beta (\text{h}') \big] + \gamma ^{*}(\text{h}) \,  \gamma (\text{h}') - \gamma ^{*}(\text{h}') \,  \gamma (\text{h}) \\[3pt]
\llap{$\&$} \gamma \big([\text{h}, \text{h}']\big) = \gamma (\text{h}) \,  \beta (\text{h}') - \gamma (\text{h}') \,  \beta (\text{h}) + \beta ^{*}(\text{h}) \,  \gamma (\text{h}') - \beta ^{*}(\text{h}') \,  \gamma (\text{h})}.%
\MStopEqua \MStopList \Mpar \Mproof The proof is similar to the one of {\seqBBBboy}.%
\MendOfProof \hypertarget{PARbpf}{}\Mpar \Mproposition{A.22}For \MMath{n> 0}, \MMath{\text{SO}^{(n)}} is path-connected and its fundamental group is isomorphic to \MMath{\mathds{Z}} (if \MMath{n=1}) or \MMath{\mathds{Z}_{2}} (if \MMath{n>1}). Its connected double cover, called the spin group over \MMath{\mathds{R}^{2n}}, and denoted \MMath{\text{Spin}^{(n)}}, can be constructed like in {\seqBBBaka}, with covering map \MMath{p ^{(n)} : \text{Spin}^{(n)} \rightarrow  \text{SO}^{(n)}}.%
\Mpar \vspace{\Sssaut}\hspace{\Alinea}For \MMath{n=0}, \MMath{\text{SO}^{(0)} = \left\{\mathds{1} \right\}} and we \emph{define} \MMath{\text{Spin}^{(0)} \mathrel{\mathop:}=  \left\{\mathds{1} ,\, \mathds{1} ^{-}\right\} \approx  \mathds{Z}_{2}}.%
\Mpar \vspace{\Sssaut}\hspace{\Alinea}The left action {$\rhd$} of \MMath{\text{SO}^{(n)}} on the space of orthonormal \MMath{(2n)}-families in {$V$} can be lifted to an action of \MMath{\text{Spin}^{(n)}} by precomposing it with \MMath{p ^{(n)}}.%
\Mpar \Mproof Let \MMath{\text{SO}_{d}(\mathds{R}) \mathrel{\mathop:}=  \left\{ \sigma  \in  \text{GL}_{d}(\mathds{R}) \middlewithspace| {}^{\text{\sc t}} \sigma  = \sigma ^{-1} \!\&\! \det \sigma  = 1 \right\}}, so that, for any \MMath{n \geqslant  1}, \MMath{\text{SO}^{(n)} = \text{SO}_{2n}(\mathds{R})}. We first consider the case \MMath{n = 1}. \MMath{\text{SO}^{(1)} = \text{SO}_{2}(\mathds{R}) \approx  \text{U}_{1}(\mathds{C})}, hence it is path-connected with \MMath{\pi _{1} \big( \text{SO}^{(1)} \big) \approx  \mathds{Z}}.%
\Mpar \vspace{\Sssaut}\hspace{\Alinea}For any \MMath{d \geqslant  2}, \MMath{\text{SO}_{d}(\mathds{R})} can be written as a fiber bundle with base \MMath{S^{d-1}} (the unit sphere in \MMath{\mathds{R}^{d}}) via \bseqHHHabl{example I.5.1}:%
\Mpar \MStartEqua \MMath{\definitionFonction{\rho }{\text{SO}_{d}(\mathds{R})}{S^{d-1}}{\text{r}}{\text{r}\, \mathbf{u} _{d}}},%
\MStopEqua \Mpar where \MMath{\mathbf{u} _{d} \mathrel{\mathop:}=  \left(0,\dots ,0,1\right) \in  \mathds{R}^{d}}. The fibers can be identified with \MMath{\text{SO}_{d-1}(\mathds{R})} via:%
\Mpar \MStartEqua \MMath{\definitionFonction{\tilde{\iota }}{\text{SO}_{d-1}(\mathds{R})}{\rho ^{-1} \left\langle  \mathbf{u} _{d} \right\rangle  \subset  \text{SO}_{d}(\mathds{R})}{\text{r}}{\text{r}'}} where \MMath{\forall  i,j \leqslant  d, {\text{r}'_{ij} = \alternative{\text{r}_{ij}}{\text{{ if }} i,j \leqslant  d-1}{\delta _{ij}}{\text{{ otherwise}}}}}.%
\MStopEqua \Mpar Since \MMath{S^{d-1}} is path-connected for \MMath{d \geqslant  2}, the path-connectedness of \MMath{\text{SO}_{d}(\mathds{R})} for \MMath{d \geqslant  2} follows recursively from the path-connectedness of \MMath{\text{SO}_{2}(\mathds{R})}.
Moreover, the theory of homotopy groups on fiber bundles \bseqHHHabh{prop.~4.48, theorems 4.3 and 4.41} provides an exact sequence (a sequence of group homomorphisms, such that the image of a given morphism is the kernel of the next one):%
\Mpar \MStartEqua \MMath{\pi _{2} \big( S^{d-1} \big) \rightarrow  \pi _{1} \big( \text{SO}_{d-1}(\mathds{R}) \big) \stackrel{\tilde{\iota }_{*}}{\rightarrow } \pi _{1} \big( \text{SO}_{d}(\mathds{R}) \big) \rightarrow  \pi _{1} \big( S^{d-1} \big)}.%
\MStopEqua \Mpar For \MMath{d \geqslant  4}, \MMath{\pi _{2} \big( S^{d-1} \big) \approx  \pi _{1} \big( S^{d-1} \big) \approx  \left\{ 0 \right\}}, hence the push-forward map \MMath{\tilde{\iota }_{*}} is then an isomorphism \MMath{\pi _{1} \big( \text{SO}_{d-1}(\mathds{R}) \big) \rightarrow  \pi _{1} \big( \text{SO}_{d}(\mathds{R}) \big)}. Now, we have the double covering map \bseqHHHabk{section 1.6.1}:%
\Mpar \MStartEqua \MMath{\definitionFonction{\tilde{p }_{3}}{\text{SU}_{2}(\mathds{C})}{\text{SO}_{3}(\mathds{R})}{\cos\frac{\left\| \boldtheta  \right\|}{2}\, \mathds{1}  - i\, \sin\frac{\left\| \boldtheta  \right\|}{2}\, \sigma (\mathbf{u} _{\boldtheta })}{\mathds{1}  + (1-\cos \left\| \boldtheta  \right\|)\, J^{2}(\mathbf{u} _{\boldtheta }) + \sin \left\| \boldtheta  \right\| \, J(\mathbf{u} _{\boldtheta })}},%
\MStopEqua \Mpar where \MMath{\boldtheta  \in  \mathds{R}^{3}}, \MMath{\mathbf{u} _{\boldtheta } \mathrel{\mathop:}=  \nicefrac{\boldtheta }{\left\| \boldtheta  \right\|}} and, for any \MMath{\mathbf{u}  = (u_{1},u_{2},u_{3}) \in  \mathds{R}^{3}}:%
\Mpar \MStartEqua \MMath{\sigma (\mathbf{u} ) \mathrel{\mathop:}=  \matrixTwo{u_{3}}{u_{1} - i\, u_{2}}{u_{1} + i\, u_{2}}{-u_{3}} \& \forall  \mathbf{x}  \in  \mathds{R}^{3},\,  J(\mathbf{u} )\, \mathbf{x}  \mathrel{\mathop:}=  \mathbf{u}  \wedge \mathbf{x} }.%
\MStopEqua \Mpar Since \MMath{\text{SU}_{2}(\mathds{C})} is simply-connected \bseqHHHabk{appendix E}, we infer \MMath{\pi _{1} \big( \text{SO}_{3}(\mathds{R}) \big) \approx  \mathds{Z}_{2}} (as a consequence of the homotopy lifting property of covering spaces, \bseqHHHabh{prop.~1.32}), so \MMath{\forall  d \geqslant  3, {\pi _{1} \big( \text{SO}_{d}(\mathds{R}) \big) \approx  \mathds{Z}_{2}}}, and, in particular, \MMath{\forall  n \geqslant  2, {\pi _{1} \big( \text{SO}^{(n)} \big) \approx  \mathds{Z}_{2}}}. Specifically, for any \MMath{n \geqslant  2}, a representative of \MMath{1 \in  \mathds{Z}_{2} \approx  \pi _{1} \big( \text{SO}^{(n)} \big)} is given by the loop \MMath{\gamma ^{(n)}_{1} : [0,1] \rightarrow  \text{SO}^{(n)}, t \mapsto  \tilde{\text{r}}^{(n)}(2\pi t)}, where:%
\Mpar \MStartEqua \MMath{\forall  i,j \leqslant  2n, \forall  \theta  \in  \mathds{R}, {\tilde{\text{r}}^{(n)}_{ij}(\theta ) \mathrel{\mathop:}=  \alternative{\tilde{\text{r}}_{ij}(\theta )}{\text{{ if }} i,j \leqslant  3}{\delta _{ij}}{\text{{ otherwise}}}}\\
\llap{$\&$} \forall  \theta  \in  \mathds{R}, {\tilde{\text{r}}(\theta ) \mathrel{\mathop:}=  \mathds{1}  + (1-\cos \theta )\, J^{2}(\mathbf{u} _{3}) + \sin \theta  \, J(\mathbf{u} _{3}) = \matrixThree{\cos \theta }{-\sin \theta }{0}{\sin \theta }{\cos \theta }{0}{0}{0}{1}}}.%
\MStopEqua \Mpar Thus, for any \MMath{n \geqslant  1}, a representative of \MMath{1 \in  \pi _{1} \big( \text{SO}^{(n)} \big)} is given by the loop \MMath{\gamma ^{(n)}_{1} : [0,1] \rightarrow  \text{SO}^{(n)}, t \mapsto  \iota _{{n\leftarrow 1}}\big(\text{r}(2\pi t)\big)}, with \MMath{\iota _{{n\leftarrow 1}}} from {\seqBBBbpt} and \MMath{\text{r}(\theta )} as in {\seqBBBbpu}.%
\Mpar \vspace{\Sssaut}\hspace{\Alinea}For any \MMath{n\geqslant 1}, we can therefore obtain the unique connected double cover \MMath{\text{Spin}^{(n)}} of \MMath{\text{SO}^{(n)}}, with covering map \MMath{p ^{(n)} : \text{Spin}^{(n)} \rightarrow  \text{SO}^{(n)}}, by the same construction as in {\seqBBBbhd} (except that, for \MMath{n\geqslant 2}, the equivalence relation \MMath{\equiv _{2}} should be defined with respect to the subgroup \MMath{\left\{0\right\}} of \MMath{\mathds{Z}_{2}}, hence corresponds to homotopy equivalence of paths; the double cover coincides in this case with the universal cover, as follows from \bseqHHHabh{props.~1.31 and 1.32}).
Since \MMath{p ^{(n)}} is a group homomorphism (with the group structure on \MMath{\text{Spin}^{(n)}} defined as in {\seqBBBbhd}), the action {$\rhd$} can be lifted by precomposition.%
\MendOfProof \hypertarget{PARbpw}{}\Mpar \Mproposition{A.23}Let \MMath{n < m} and let \MMath{\left(\mathbf{e} _{1},\dots ,\mathbf{e} _{2n}\right), \left(\mathbf{f} _{1},\dots ,\mathbf{f} _{2m}\right)} be two finite orthonormal families in {$V$} such that \MMath{\mkop{Span}  \left\{\mathbf{e} _{1},\dots ,\mathbf{e} _{2n}\right\} \subsetneq \mkop{Span}  \left\{\mathbf{f} _{1},\dots ,\mathbf{f} _{2m}\right\}}. Then, there exists a family \MMath{\left(\mathbf{e} _{2n+1},\dots ,\mathbf{e} _{2m}\right)} of vectors in {$V$} and an element \MMath{\mu  \in  \text{Spin}^{(n)}} such that \MMath{\mu  \rhd  \left(\mathbf{f} _{1},\dots ,\mathbf{f} _{2m}\right) = \left(\mathbf{e} _{1},\dots ,\mathbf{e} _{2m}\right)}.%
\Mpar \Mproof Proceeding like in {\seqBBBald}, there exists a family \MMath{\left(\mathbf{e} _{2n+1},\dots ,\mathbf{e} _{2m}\right)} of vectors in {$V$} such that \MMath{\left(\mathbf{e} _{1},\dots ,\mathbf{e} _{2m}\right)} is an orthonormal family and \MMath{\mkop{Span}  \left\{\mathbf{e} _{1},\dots ,\mathbf{e} _{2m}\right\} = \mkop{Span}  \left\{\mathbf{f} _{1},\dots ,\mathbf{f} _{2m}\right\}}. 
If \MMath{\left(\mathbf{e} _{1},\dots ,\mathbf{e} _{2m}\right)} and \MMath{\left(\mathbf{f} _{1},\dots ,\mathbf{f} _{2m}\right)} have opposite orientations, we redefine \MMath{\mathbf{e} _{2m}} as \MMath{ -\mathbf{e} _{2m}} (note that we took the precaution to assume \MMath{n < m}). Then, there exists \MMath{\sigma  \in  \text{SO}^{(n)}} such that \MMath{\forall  i \leqslant  2m, \mathbf{f} _{i} = \sigma _{ji} \,  \mathbf{e} _{j}}, and we can choose \MMath{\mu  \in  \text{Spin}^{(m)}} above {$\sigma$}, so that \MMath{\mu  \rhd  \left(\mathbf{f} _{1},\dots ,\mathbf{f} _{2m}\right) = \left(\mathbf{e} _{1},\dots ,\mathbf{e} _{2m}\right)}.%
\MendOfProof \hypertarget{PARbpy}{}\Mpar \Mproposition{A.24}For any \MMath{m \geqslant  n \geqslant  0}, \MMath{\iota _{m\leftarrow n}} (from {\seqBBBbpt}) can be lifted to an injective group homomorphism \MMath{\ell _{m\leftarrow n} : \text{Spin}^{(n)} \rightarrow  \text{Spin}^{(m)}}. For any \MMath{l \geqslant  m \geqslant  n \geqslant  0}, we have \MMath{\ell _{l\leftarrow m} \circ  \ell _{m\leftarrow n} = \ell _{l\leftarrow n}}.%
\Mpar \vspace{\Sssaut}\hspace{\Alinea}For any \MMath{m \geqslant  n \geqslant  0}, any \MMath{\mu  \in  \text{Spin}^{(n)}} and any orthonormal \MMath{(2m)}-family \MMath{\left(\mathbf{e} _{1},\dots ,\mathbf{e} _{2m}\right)} in {$V$}, we have:%
\Mpar \MStartEqua \MMath{\ell _{m\leftarrow n}(\mu ) \rhd  \left(\mathbf{e} _{1},\dots ,\mathbf{e} _{2m}\right) = \left(\mathbf{f} _{1},\dots ,\mathbf{f} _{2m}\right)}%
\MStopEqua \Mpar with \MMath{\left(\mathbf{f} _{1},\dots ,\mathbf{f} _{2n}\right) \mathrel{\mathop:}=  \mu  \rhd  \left(\mathbf{e} _{1},\dots ,\mathbf{e} _{2n}\right)} and \MMath{\left(\mathbf{f} _{2n+1},\dots ,\mathbf{f} _{2m}\right) \mathrel{\mathop:}=  \left(\mathbf{e} _{2n+1},\dots ,\mathbf{e} _{2m}\right)}.%
\Mpar \Mproof Since we have the same expression for the representative loop \MMath{\gamma ^{(n)}_{1}} as in the symplectic case {\seqDDDbpu}, we obtain in the same way that, for any \MMath{m \geqslant  n > 0} the push-forward map \MMath{\pi _{1} \big( \text{SO}^{(n)} \big) \rightarrow  \pi _{1} \big( \text{SO}^{(m)} \big)} induced by \MMath{\iota _{m\leftarrow n}} is \MMath{\text{id}_{\mathds{Z}}} (if \MMath{m = n = 1}), or the quotient map \MMath{\mathds{Z} \rightarrow  \mathds{Z}_{2} \approx  \mathds{Z} / 2\mathds{Z}} (if \MMath{m > n = 1}), or \MMath{\text{id}_{{\mathds{Z}_{2}}}} (if \MMath{m \geqslant  n > 1}).
The rest of the proof is then similar.%
\MendOfProof \hypertarget{SECbqc}{}\MsectionC{bqc}{A.2.3}{Infinite-dimensional Automorphisms}%
\vspace{\PsectionC}\Mpar \MstartPhyMode\hspace{\Alinea}The question of norms and topology is much easier in the inner-product case than in the symplectic case {\seqDDDaop}, since the inner-product directly provides us with a \emph{preferred} norm.%
\MleavePhyMode \hypertarget{PARbqe}{}\Mpar \Mdefinition{A.25}We define the group \MMath{\mathcal{A} _{o}(V , \left(\, \cdot \, \middlewithspace|\, \cdot \, \right))} of automorphisms of \MMath{V , \left(\, \cdot \, \middlewithspace|\, \cdot \, \right)} as the group of \emph{bijective linear} mappings \MMath{\text{m} : V  \rightarrow  V } satisfying:%
\Mpar \MStartEqua \MMath{\forall  \mathbf{v} ,\mathbf{w}  \in  V , \left(\text{m} \mathbf{v} \middlewithspace|\text{m} \mathbf{w} \right) = \left(\mathbf{v} \middlewithspace|\mathbf{w} \right)}.%
\MStopEqua \Mpar \vspace{\Sssaut}\hspace{\Alinea}Equipping {$V$} with the norm \MMath{\left|\, \cdot \, \right|} induced by the scalar product \MMath{\left(\, \cdot \, \middlewithspace|\, \cdot \, \right)}, we note that any \MMath{\text{m} \in  \mathcal{A} _{o}(V , \left(\, \cdot \, \middlewithspace|\, \cdot \, \right))} is bi-continuous (as an isometry), so we can equip \MMath{\mathcal{A} _{o}(V , \left(\, \cdot \, \middlewithspace|\, \cdot \, \right))} with a structure of topological group like in {\seqBBBany}.%
\hypertarget{PARbqh}{}\Mpar \Mproposition{A.26}Equipping {$V$} with the norm \MMath{\left|\, \cdot \, \right|} induced by the scalar product \MMath{\left(\, \cdot \, \middlewithspace|\, \cdot \, \right)}, let {$D$} be a dense vector subspace in {$V$}.
For any \MMath{\epsilon  > 0} and any \emph{finite-dimensional} vector subspace {$F$} of {$V$}, there exists \MMath{\text{m} \in  \mathcal{A} _{o}(V , \left(\, \cdot \, \middlewithspace|\, \cdot \, \right))} such that:%
\hypertarget{PARbqi}{}\Mpar \MList{1}\MMath{\text{m} \left\langle  F  \right\rangle  \subseteq D };%
\MStopList \hypertarget{PARbqj}{}\Mpar \MList{2}\MMath{\left. \text{m} \right|_{{F  \cap  D }} = \left. \text{id}_{V } \right|_{{F  \cap  D }}};%
\MStopList \hypertarget{PARbqk}{}\Mpar \MList{3}\MMath{\left\| \text{m} - \text{id}_{V } \right\| < \epsilon }.%
\MStopList \hypertarget{PARbql}{}\Mpar \Mlemma{A.27}Let \MMath{n > 0} and \MMath{\epsilon  > 0}. There exists \MMath{\epsilon _{o} > 0} such that, for any finite orthonormal family \MMath{\left(\mathbf{e} _{1},\dots ,\mathbf{e} _{2n}\right)}, and any family \MMath{\left(\mathbf{v} _{1},\dots ,\mathbf{v} _{2n}\right)} of vectors in {$V$} satisfying:%
\Mpar \MStartEqua \MMath{\forall  i \leqslant  2n, \left|\mathbf{e} _{i} - \mathbf{v} _{i}\right| < \epsilon _{o}},%
\MStopEqua \Mpar there exists an \emph{orthonormal} family \MMath{\left(\mathbf{f} _{1},\dots ,\mathbf{f} _{2n}\right)} in {$V$} satisfying:%
\Mpar \MStartEqua \MMath{\forall  i \leqslant  2n, \mathbf{f} _{i} \in  \mkop{Span}  \left\{\mathbf{v} _{j} \middlewithspace| j \leqslant  i\right\} \& \left|\mathbf{e} _{i} - \mathbf{f} _{i}\right| < \epsilon }.%
\MStopEqua \Mpar \Mproof The proof is similar to the one of {\seqBBBbmg}, using standard Gram–Schmidt orthonormalization rather than its symplectic counterpart, and taking advantage of the fact that \MMath{\left|\mathbf{e} _{i}\right| = 1} by definition (which allows to dispense from the bound \MMath{E} that was needed for {\seqBBBbmg}).%
\MendOfProof \Mpar \MproofProposition{A.26}Since {$D$} is dense in {$V$}, which is even or infinite dimensional, {$D$} itself is even or infinite dimensional. The rest of the proof is similar to the one of {\seqBBBaoy}, but easier, since we do not need to keep track of a bound on the norm of the frame vectors \MMath{\mathbf{e} _{i}} (which is \MMath{1} by definition), so we can directly use {\seqBBBaly} in place of {\seqBBBbmc}.%
\MendOfProof \hypertarget{SECbqr}{}\MsectionC{bqr}{A.2.4}{Complex Structures}%
\vspace{\PsectionC}\Mpar \MstartPhyMode\hspace{\Alinea}Like in the bosonic/symplectic case, a choice of complex structure (aka.~polarization) is needed to construct an infinite-dimensional fermionic Fock space from a real pre-Hilbert space {\seqDDDbqt}.
The reconstruction of a \emph{complex} inner-product from a \emph{real} one together with a choice of complex structure constitutes the last part of the correspondence announced in {\seqBBBbhc}.%
\MleavePhyMode \hypertarget{PARbqu}{}\Mpar \Mdefinition{A.28}A compatible complex structure on a real pre-Hilbert space \MMath{V , \left(\, \cdot \, \middlewithspace|\, \cdot \, \right)} is a linear map \MMath{I : V  \rightarrow  V } satisfying:%
\Mpar \MList{1}\MMath{I ^{2} = - \text{id}_{V }};%
\MStopList \Mpar \MList{2}\MMath{\forall  \mathbf{v} ,\mathbf{w}  \in  V , \left( I  \mathbf{v}  \middlewithspace| I  \mathbf{w}  \right) = \left( \mathbf{v}  \middlewithspace| \mathbf{w}  \right)}.%
\MStopList \Mpar \vspace{\Sssaut}\hspace{\Alinea}\MMath{V } can then be equipped with a structure of \emph{complex} pre-Hilbert space, with the scalar multiplication defined by:%
\Mpar \MStartEqua \MMath{\forall  z \in  \mathds{C},\,  \forall  \mathbf{v}  \in  V , {z\, \mathbf{v}  \mathrel{\mathop:}=  \mkop{Re}(z)\, \mathbf{v}  + \mkop{Im}(z)\, I \mathbf{v} }},%
\MStopEqua \Mpar and the (complex) scalar product defined by:%
\Mpar \MStartEqua \MMath{\forall  \mathbf{v} ,\mathbf{w}  \in  V , {\left\langle  \mathbf{v}  \middlewithspace| \mathbf{w}  \right\rangle _{I } \mathrel{\mathop:}=  \left(\mathbf{v} \middlewithspace|\mathbf{w} \right) - i\, \left(\mathbf{v} \middlewithspace|I \mathbf{w} \right)}}.%
\MStopEqua \Mpar Note that, for any \MMath{\mathbf{v}  \in  V }:%
\Mpar \MStartEqua \MMath{\left(\mathbf{v} \middlewithspace|I \mathbf{v} \right) = \left(I \mathbf{v} \middlewithspace|I I \mathbf{v} \right) = - \left(I \mathbf{v} \middlewithspace|\mathbf{v} \right) = - \left(\mathbf{v} \middlewithspace|I \mathbf{v} \right) = 0},%
\MStopEqua \Mpar so \MMath{\left\langle \, \cdot \, \middlewithspace|\, \cdot \, \right\rangle _{I }} being positive-definite follows from the corresponding property of \MMath{\left(\, \cdot \, \middlewithspace|\, \cdot \, \right)}.
\Mnomdefichier{lin50}%
\hypertarget{SECbrf}{}\MsectionA{brf}{B}{Linear Holomorphic Quantization}%
\vspace{\PsectionA}\Mpar \MstartPhyMode\hspace{\Alinea}We introduce Fock spaces using their explicit representation as multi-particle spaces.
See \bseqHHHaae{section 9.2} for how to arrive at these spaces through an holomorphic polarization in the context of geometric quantization.
See also our treatment of Fock spaces on infinite-dimensional classical phase spaces (in {\seqBBBbna}) for additional insight regarding the role of complex structures {\seqDDDbhc}.%
\MleavePhyMode \hypertarget{SECbrh}{}\MsectionB{brh}{B.1}{Bosonic Fock Spaces}%
\hypertarget{SECbri}{}\MsectionC{bri}{B.1.1}{Finite-dimensional Case -- Metaplectic Representation}%
\vspace{\PsectionC}\Mpar \MstartPhyMode\hspace{\Alinea}We define here the Fock quantization of a finite-dimensional symplectic vector space in a chosen \emph{symplectic frame} (ie.~identifying the classical phase space with \MMath{\mathds{R}^{2n}}, equipped with its canonical symplectic structure). This gives a preferred orthonormal basis (in terms of occupation numbers) in the resulting Fock space: this very explicit description will be the one used for the partial Hilbert space that we will attach to a given label (as labels correspond to finite symplectic families in a possibly infinite-dimensional symplectic vector space, cf.~{\seqBBBacm} and {\seqBBBajy}).
We could alternatively start from just a complex structure (aka.~complex polarization) on the phase space (as we will do in {\seqBBBbna}), yielding a Fock space with an unambiguous notion of \emph{total} occupation number (hence, a preferred vacuum state), but no preferred mode basis. The two definitions coincide, noting that a frame in particular determines a complex structure (through the identification \MMath{\mathds{R}^{2n} \approx  \left(\mathds{R}^{2}\right)^{n} \approx  \mathds{C}^{n}}).%
\Mpar \vspace{\Saut}\hspace{\Alinea}Thanks to the Stone–von Neumann theorem, it is possible to relate Fock spaces constructed over different symplectic frames (even if said frames are associated to different complex structures). Equivalently (as explained in the introduction of {\seqBBBacm}), we have (projective) unitary representation of the symplectic group {\seqDDDane}, which we will construct in the present subsection.%
\MleavePhyMode \hypertarget{PARbrk}{}\Mpar \Mdefinition{B.1}Let \MMath{N \geqslant  0} and let {$\mathcal{H}$} be a Hilbert space. For any \MMath{\varepsilon  \in  S_{N}} (the permutation group over \MMath{\left\{1,\dots ,N\right\}}), we define a unitary operator \MMath{\hat{\varepsilon }} over \MMath{\mathcal{H} ^{{\otimes N}}} by:%
\Mpar \MStartEqua \MMath{\forall  \mathbf{e} _{1},\dots ,\mathbf{e} _{N} \in  \mathcal{H} , {\hat{\varepsilon } \big( \mathbf{e} _{1} \otimes  \dots  \otimes  \mathbf{e} _{N} \big) \mathrel{\mathop:}=  \mathbf{e} _{{\varepsilon (1)}} \otimes  \dots  \otimes  \mathbf{e} _{{\varepsilon (N)}}}}%
\MStopEqua \Mpar (note that \MMath{\varepsilon  \mapsto  \hat{\varepsilon }} is a \emph{right} group action of \MMath{S_{N}} on \MMath{\mathcal{H} ^{{\otimes N}}}).%
\Mpar \vspace{\Sssaut}\hspace{\Alinea}We define the Hilbert space \MMath{\mathcal{H} ^{{\otimes N,\text{{sym}}}}} by:%
\Mpar \MStartEqua \MMath{\mathcal{H} ^{{\otimes N,\text{{sym}}}} \mathrel{\mathop:}=  \left\{ \Psi  \in  \mathcal{H} ^{{\otimes N}} \middlewithspace| \forall  \varepsilon  \in  S_{N}, \hat{\varepsilon } \Psi  = \Psi  \right\}}.%
\MStopEqua \Mpar By convention, \MMath{\mathcal{H} ^{{\otimes 0,\text{{sym}}}} \approx  \mathds{C}}.%
\hypertarget{PARbrq}{}\Mpar \Mdefinition{B.2}Let \MMath{n \geqslant  0}. We define the bosonic Fock space over \MMath{n} states as \MMath{\mathcal{F} ^{(n)}_{\mathfrak{bos} } \mathrel{\mathop:}=  \overline{\mathcal{D} ^{(n)}_{\mathfrak{bos} }}} where:%
\Mpar \MStartEqua \MMath{\mathcal{D} ^{(n)}_{\mathfrak{bos} } \mathrel{\mathop:}=  \bigoplus_{N\geqslant 0} \left( \mathds{C}^{n} \right)^{{\otimes N,\text{{sym}}}}},%
\MStopEqua \Mpar and \MMath{\overline{\big(\, \cdot \, \big)}} denotes the completion with respect to the inner product norm. \MMath{\mathcal{F} ^{(n)}_{\mathfrak{bos} }} is a complex separable Hilbert space.%
\Mpar \vspace{\Sssaut}\hspace{\Alinea}Let \MMath{\left( \mathbf{b} _{1},\dots ,\mathbf{b} _{n} \right)} denotes the canonical basis of \MMath{\mathds{C}^{n}}. For any \MMath{\left( N_{1},\dots ,N_{n} \right) \in  \mathds{N}^{n}}, we define:%
\Mpar \MStartEqua \MMath{\left| N_{1},\dots ,N_{n} \right\rangle _{\mathfrak{bos} } \mathrel{\mathop:}=  \frac{1}{\sqrt{{N! N_{1} ! \dots  N_{n} !}}} \sum_{{\varepsilon  \in  S_{N}}} \hat{\varepsilon } \big( \underbrace{\mathbf{b} _{1} \otimes  \dots  \otimes  \mathbf{b} _{1}}_{{N_{1}}} \otimes  \dots  \otimes  \underbrace{\mathbf{b} _{n} \otimes  \dots  \otimes  \mathbf{b} _{n}}_{{N_{n}}} \big) \in  \left( \mathds{C}^{n} \right)^{{\otimes N,\text{{sym}}}}},%
\MStopEqua \Mpar with \MMath{N = N_{1} + \dots  + N_{n}}. \MMath{\big( \left| N_{1},\dots ,N_{n} \right\rangle _{\mathfrak{bos} } \big)_{{\left( N_{1},\dots ,N_{n} \right) \in  \mathds{N}^{n}}}} is an orthonormal basis of \MMath{\mathcal{F} ^{(n)}_{\mathfrak{bos} }}.%
\Mpar \vspace{\Sssaut}\hspace{\Alinea}For any \MMath{p \leqslant  n}, we define the linear operators \MMath{\mathtt{a} _{p}, \mathtt{a} ^{+}_{p} : \mathcal{D} ^{(n)}_{\mathfrak{bos} } \rightarrow  \mathcal{D} ^{(n)}_{\mathfrak{bos} }} by:%
\Mpar \MStartEqua \MMath{\forall  \left( N_{1},\dots ,N_{n} \right) \in  \mathds{N}^{n}, {\mathtt{a} _{p} \left| N_{1},\dots ,N_{n} \right\rangle _{\mathfrak{bos} } \mathrel{\mathop:}=  \sqrt{{N_{p}}} \left| N_{1},\dots ,N_{p} -1,\dots ,N_{n} \right\rangle _{\mathfrak{bos} }}\\
\hphantom{\forall  \left( N_{1},\dots ,N_{n} \right) \in  \mathds{N}^{n}, }\llap{$\&$}{\mathtt{a} ^{+}_{p} \left| N_{1},\dots ,N_{n} \right\rangle _{\mathfrak{bos} } \mathrel{\mathop:}=  \sqrt{{N_{p} +1}} \left| N_{1},\dots ,N_{p} +1,\dots ,N_{n} \right\rangle _{\mathfrak{bos} }}}.%
\MStopEqua \Mpar \MstartPhyMode\hspace{\Alinea}We give the quantization of both linear and quadratic observables.
The Hamiltonian vector fields of these observables generate affine symplectic transformations on the phase space, viz.~respectively, translations and linear symplectomorphisms.
Checking that the quantum commutators match the classical Poisson brackets, we ensure that we have a representation of the Lie algebra of the group of affine symplectic transformations (up to a central charge \bseqHHHaaf{section 2.7}, since the constant functions appearing as the Poisson brackets of linear observables act as trivial transformations on the phase space).%
\Mpar \vspace{\Saut}\hspace{\Alinea}The quantization of the generators of linear symplectomorphisms uses the decomposition of the symplectic Lie algebra into \MMath{\mathds{C}}-linear (unitary) and anti-\MMath{\mathds{C}}-linear parts {\seqDDDboy}.
In accordance with the comments at the beginning of the present subsection, the unitary part of a transformation, which preserves the complex structure determined by a symplectic frame, preserves the total occupation number (and in particular the vacuum), and, as we will see, this part is still well-defined over infinite-dimensional Fock representations ({\seqBBBbdg} and its proof, as well as {\seqBBBabr}). It is the polarization-changing, non-unitary component that turns out to be problematic when going over to infinite-dimensional classical phase spaces, and lead to the breakdown of the Stone–von Neumann theorem.%
\Mpar \vspace{\Saut}\hspace{\Alinea}Finally, the additional constant term at the end of the expression for \MMath{\hat{\text{h}}} generalizes the well-known 0-point energy correction of the quantum harmonic oscillator, and we will further discuss below why it is needed from the point of view of building a unitary representation of the symplectic group.%
\MleavePhyMode \hypertarget{PARbrz}{}\Mpar \Mdefinition{B.3}For any \MMath{\mathbf{x}  \in  \mathds{R}^{2n}}, we define a linear operator \MMath{\hat{\mathbf{x} }: \mathcal{D} ^{(n)}_{\mathfrak{bos} } \rightarrow  \mathcal{D} ^{(n)}_{\mathfrak{bos} }} by:%
\Mpar \MStartEqua \MMath{\hat{\mathbf{x} } \mathrel{\mathop:}=  \sum_{p=1}^{n} \frac{\mathbf{x} _{2p-1} + i \,  \mathbf{x} _{2p}}{\sqrt{2}} \mathtt{a} _{p} + \frac{\mathbf{x} _{2p-1} - i \,  \mathbf{x} _{2p}}{\sqrt{2}} \mathtt{a} ^{+}_{p}}.%
\MStopEqua \Mpar \vspace{\Sssaut}\hspace{\Alinea}For any \MMath{\text{h} \in  \mathfrak{sp} ^{(n)}}, we define a linear operator \MMath{\hat{\text{h}}: \mathcal{D} ^{(n)}_{\mathfrak{bos} } \rightarrow  \mathcal{D} ^{(n)}_{\mathfrak{bos} }} by:%
\Mpar \MStartEqua \MMath{\hat{\text{h}} \mathrel{\mathop:}=  \sum_{{p,q=1}}^{n} \left[ - i \,  \beta _{pq}(\text{h}) \,  \mathtt{a} ^{+}_{p} \mathtt{a} _{q} - \frac{i}{2} \left( \gamma _{pq}(\text{h}) \,  \mathtt{a} _{p} \mathtt{a} _{q} - \gamma ^{*}_{pq}(\text{h}) \,  \mathtt{a} ^{+}_{p} \mathtt{a} ^{+}_{q} \right) \right] - \frac{i}{2} \big( \mkop{Tr}  \beta (\text{h}) \big) \,  \mathds{1} }%
\MStopEqua \Mpar (with \MMath{\beta (\text{h}), \gamma (\text{h})} from {\seqBBBboy}).%
\hypertarget{PARbse}{}\Mpar \Mproposition{B.4}For any \MMath{\mathbf{x}  \in  \mathds{R}^{2n}}, \MMath{\hat{\mathbf{x} }} is a symmetric operator, and, for any \MMath{\text{h} \in  \mathfrak{sp} ^{(n)}}, \MMath{\hat{\text{h}}} is a symmetric operator.%
\Mpar \vspace{\Sssaut}\hspace{\Alinea}Moreover, for any \MMath{\mathbf{x} ,\mathbf{x} ' \in  \mathds{R}^{2n}} and any \MMath{\text{h},\text{h}' \in  \mathfrak{sp} ^{(n)}}, we have the following commutators:%
\Mpar \MStartEqua \MMath{\left[ \hat{\mathbf{x} }, \hat{\mathbf{x} '} \right] = - i \left({}^{\text{\sc t}} \mathbf{x}  \,  \Omega ^{(n)} \,  \mathbf{x}  '\right) \mathds{1} ,\hspace{0.25cm} 
\left[ \hat{\text{h}}, \hat{\mathbf{x} } \right] = - i \,  \widehat{\text{h} \,  \mathbf{x}  } \&
\left[ \hat{\text{h}}, \hat{\text{h}'} \right] = - i \,  \widehat{[\text{h}, \text{h}']}}.%
\MStopEqua \Mpar \Mproof For any \MMath{p \leqslant  n} and any \MMath{\psi ,\psi ' \in  \mathcal{D} ^{(n)}_{\mathfrak{bos} }}, \MMath{\left\langle  \psi ', \mathtt{a} _{p} \,  \psi  \right\rangle  = \left\langle  \mathtt{a} ^{+}_{p} \,  \psi ', \psi  \right\rangle }, hence, for any \MMath{\mathbf{x}  \in  \mathds{R}^{2n}}, \MMath{\hat{\mathbf{x} }} is symmetric, and, for any \MMath{\text{h} \in  \mathfrak{sp} ^{(n)}}, \MMath{\hat{\text{h}}} is symmetric (using \abbrevProposition{\seqBBBbhp}).%
\Mpar \vspace{\Sssaut}\hspace{\Alinea}For any \MMath{p,q \leqslant  n}, \MMath{\left[ \mathtt{a} _{p}, \mathtt{a} ^{+}_{q} \right] = \delta _{pq} \,  \mathds{1} } and \MMath{\left[ \mathtt{a} _{p}, \mathtt{a} _{q} \right] = \left[ \mathtt{a} ^{+}_{p}, \mathtt{a} ^{+}_{q} \right] = 0} (these commutators are well defined as linear operators \MMath{\mathcal{D} ^{(n)}_{\mathfrak{bos} } \rightarrow  \mathcal{D} ^{(n)}_{\mathfrak{bos} }} since \MMath{\mathcal{D} ^{(n)}_{\mathfrak{bos} }} is a common invariant domain for \MMath{\mathtt{a} _{p}, \mathtt{a} _{q}, \mathtt{a} ^{+}_{p}, \mathtt{a} ^{+}_{q}}). Defining, for any \MMath{u \in  \mathds{C}^{n}} and any \MMath{\beta ,\gamma  \in  \text{M}_{n}(\mathds{C})}:%
\Mpar \MStartEqua \MMath{X(u) \mathrel{\mathop:}=  \sum_{p=1}^{n} u_{p} \,  \mathtt{a} _{p},\hspace{0.25cm}  X^{+}(u) \mathrel{\mathop:}=  \sum_{p=1}^{n} u^{*}_{p} \,  \mathtt{a} ^{+}_{p}, \\[3pt]
A(\gamma ) \mathrel{\mathop:}=  \sum_{{p,q=1}}^{n} \gamma _{pq} \,  \mathtt{a} _{p} \mathtt{a} _{q},\hspace{0.25cm}  A^{+}(\gamma ) \mathrel{\mathop:}=  \sum_{{p,q=1}}^{n} \gamma ^{*}_{pq} \,  \mathtt{a} ^{+}_{p} \mathtt{a} ^{+}_{q}, \& B(\beta ) \mathrel{\mathop:}=  \sum_{{p,q=1}}^{n} \beta _{pq} \,  \mathtt{a} ^{+}_{p} \mathtt{a} _{q}},%
\MStopEqua \Mpar we get, for any \MMath{u, u' \in  \mathds{C}^{n}} and any \MMath{\beta ,\beta ',\gamma ,\gamma ' \in  \text{M}_{n}(\mathds{C})}:%
\Mpar \MStartEqua \MMath{\left[ X(u), X^{+}(u') \right] = \left\langle  u', u \right\rangle  \,  \mathds{1} \\[3pt]
\left[ B(\beta ), X(u) \right] = - X \left({}^{\text{\sc t}} \beta  u\right) \&
\left[ B(\beta ), X^{+}(u) \right] = X^{+} \left(\beta ^{*} u\right)\\[3pt]
\left[ A(\gamma ), X^{+}(u) \right] = X \big( \left(\gamma  + {}^{\text{\sc t}} \gamma \right) u^{*} \big) \&
\left[ A^{+}(\gamma ), X(u) \right] = - X^{+} \big( \left(\gamma  + {}^{\text{\sc t}} \gamma \right) u^{*} \big)\\[3pt]
\left[ B(\beta ), B(\beta ') \right] = B \big( \left[ \beta , \beta ' \right] \big)\\[3pt]
\left[ B(\beta ), A(\gamma ') \right] = - A \left( \gamma ' \beta  + {}^{\text{\sc t}} \beta  \gamma ' \right) = - A \big( \left(\gamma ' + {}^{\text{\sc t}} \gamma '\right) \beta  \big)\\[3pt]
\left[ B(\beta ), A^{+}(\gamma ') \right] = A^{+} \left( \beta ^{*} \gamma ' + \gamma ' \beta ^{\dag} \right) = A^{+} \big( \left( \gamma ' + {}^{\text{\sc t}} \gamma ' \right) \beta ^{\dag} \big)\\[3pt]
\left[ A(\gamma ), A^{+}(\gamma ') \right] = \mkop{Tr}  \big( \gamma  \left(\gamma ^{{\prime*}} + {}^{\text{\sc t}} \gamma ^{{\prime*}}\right) \big) \mathds{1}  + B \big( (\gamma ^{{\prime*}} + {}^{\text{\sc t}} \gamma ^{{\prime*}}) (\gamma  + {}^{\text{\sc t}} \gamma ) \big)\\
\hphantom{\left[ A(\gamma ), A^{+}(\gamma ') \right]} = \frac{1}{2} \mkop{Tr}  \big( \left(\gamma ^{{\prime*}} + {}^{\text{\sc t}} \gamma ^{{\prime*}}\right) \left(\gamma  + {}^{\text{\sc t}} \gamma \right) \big) + B \big( (\gamma ^{{\prime*}} + {}^{\text{\sc t}} \gamma ^{{\prime*}}) (\gamma  + {}^{\text{\sc t}} \gamma ) \big)}%
\MStopEqua \Mpar where \MMath{\left\langle  u', u \right\rangle  \mathrel{\mathop:}=  \sum_{p=1}^{n} u'^{*}_{p} u_{p}}. We obtain the desired commutators using {\seqBBBboy} together with the relation:%
\Mpar \MStartEqua \MMath{\forall \mathbf{x}  \in  \mathds{R}^{2n}, \forall  \text{h} \in  \mathfrak{sp} ^{(n)}, {\beta ^{*}(\text{h}) \,  u(\mathbf{x} ) + \gamma (\text{h}) \,  u^{*}(\mathbf{x} ) = u \left( \text{h} \,  \mathbf{x}  \right)}},%
\MStopEqua \Mpar where \MMath{\forall \mathbf{x}  \in  \mathds{R}^{2n}, \forall  p\leqslant n, {u_{p}(\mathbf{x} ) \mathrel{\mathop:}=  \frac{\mathbf{x} _{2p-1} + i \,  \mathbf{x} _{2p}}{\sqrt{2}}}}.%
\MendOfProof \Mpar \MstartPhyMode\hspace{\Alinea}We now need to justify that the previous Lie algebra representation can be \emph{exponentiated} yielding a (projective) unitary representation of the group of affine symplectic transformations: from this representation, we can then extract the advertised unitary representation of the metaplectic group {\seqDDDaka}, its action on the quantized linear observables, as well as the exponentiated commutation relations of the latter.%
\Mpar \vspace{\Saut}\hspace{\Alinea}Note that is over its \emph{unitary} subgroup that the representation of the symplectic group is \emph{projective}: as explained before {\seqBBBaka}, this unitary subgroup comports a topologically non-trivial loop, and, when going around this loop, we catch an extra phase factor, coming for the additional \MMath{ -\nicefrac{i}{2} \mkop{Tr}  \beta (\text{h})} term that was included in the expression of \MMath{\hat{\text{h}}}.
This is the reason why we do not actually have a representation of the symplectic group, but rather of its \emph{metaplectic} double cover.%
\Mpar \vspace{\Saut}\hspace{\Alinea}If we were to drop the constant term from the definition of \MMath{\hat{\text{h}}}, the representation would no longer be projective over the unitary part of the symplectic group (see {\seqBBBbso}), but the contributions from the non-unitary parts \MMath{\gamma (\text{h})} in the commutators would also no longer be correctly compensated (as can be seen from {\seqBBBbsp}): in other words, the representation would again be projective, this time because it would present a central charge (see also the introduction of {\seqBBBacm} for a different perspective on this issue).%
\MleavePhyMode \hypertarget{PARbsq}{}\Mpar \Mproposition{B.5}For any \MMath{\mathbf{x} ,\mathbf{y}  \in  \mathds{R}^{2n}}, \MMath{\hat{\mathbf{x} }, \hat{\mathbf{y} }} are essentially self-adjoint and we have:%
\hypertarget{PARbsr}{}\Mpar \MStartEqua \MMath{\exp(i \,  \hat{\mathbf{x} }) \,  \exp(i \,  \hat{\mathbf{y} }) = e^{{\nicefrac{i}{2} \,  {}^{\text{\sc t}} \mathbf{x}  \,  \Omega ^{(n)} \,  \mathbf{y} }} \,  \exp(i \,  \hat{\mathbf{x} } + i \,  \hat{\mathbf{y} })}%
\NumeroteEqua{B.5}{1}\MStopEqua \Mpar (where \MMath{\exp} denotes the exponential of an anti-self-adjoint operator, as defined through spectral resolution \bseqHHHaaq{theorem VIII.5}).%
\Mpar \vspace{\Sssaut}\hspace{\Alinea}Moreover, for \MMath{n > 0}, there exists a unique unitary representation \MMath{\mathtt{T} ^{(n)}_{\mathfrak{bos} }} of the metaplectic group \MMath{\text{Mp}^{(n)}} {\seqDDDaka} on \MMath{\mathcal{F} ^{(n)}_{\mathfrak{bos} }} such that:%
\hypertarget{PARbsu}{}\Mpar \MStartEqua \MMath{\forall  \text{h} \in  \mathfrak{sp} ^{(n)}, \forall  \psi  \in  \mathcal{D} ^{(n)}_{\mathfrak{bos} }, {\frac{\mathtt{T} ^{(n)}_{\mathfrak{bos} } \big( \exp_{{\text{Mp}^{(n)}}}(t\, \text{h}) \big) \psi  - \psi }{t} \underset{t\rightarrow 0}{\rightarrow } i \,  \hat{\text{h}} \,  \psi }}%
\NumeroteEqua{B.5}{2}\MStopEqua \Mpar (where \MMath{\exp_{{\text{Mp}^{(n)}}}} denotes the exponential mapping \MMath{\mathfrak{sp} ^{(n)} \rightarrow  \text{Mp}^{(n)}}), and it satisfies:%
\hypertarget{PARbsw}{}\Mpar \MStartEqua \MMath{\forall  \mu  \in  \text{Mp}^{(n)}, \forall  \mathbf{x}  \in  \mathds{R}^{2n}, {\mathtt{T} ^{(n)}_{\mathfrak{bos} } \big( \mu  \big) \,  \exp\big(i \,  \hat{\mathbf{x} }\big) \,  \mathtt{T} ^{(n)}_{\mathfrak{bos} } \big( \mu ^{-1} \big) = \exp\big(i \,  \widehat{p ^{(n)}(\mu ) \,  \mathbf{x} }\big)}}%
\NumeroteEqua{B.5}{3}\MStopEqua \Mpar (with the covering map \MMath{p ^{(n)} : \text{Mp}^{(n)} \rightarrow  \text{Sp}^{(n)}} from {\seqBBBand}).%
\Mpar \vspace{\Sssaut}\hspace{\Alinea}For \MMath{n=0}, we \emph{define} \MMath{\mathtt{T} ^{(0)}_{\mathfrak{bos} }} by:%
\Mpar \MStartEqua \MMath{\mathtt{T} _{\mathfrak{bos} }^{(0)}(\mathds{1} ) = \text{id}_{{\mathcal{F} ^{(0)}_{\mathfrak{bos} }}} \, \&\,  \mathtt{T} _{\mathfrak{bos} }^{(0)}(\mathds{1} ^{-}) = -\text{id}_{{\mathcal{F} ^{(0)}_{\mathfrak{bos} }}}}.%
\MStopEqua \Mpar \Mproof We define:%
\Mpar \MStartEqua \MMath{\mathfrak{g} \mathrel{\mathop:}=  \left\{ i \,  \theta  \,  \mathds{1}  + i \,  \hat{\mathbf{x} } + i \,  \hat{\text{h}} \middlewithspace| \theta  \in  \mathds{R},\,  \mathbf{x}  \in  \mathds{R}^{2n} \!\&\! \text{h} \in  \mathfrak{sp} ^{(n)} \right\}}.%
\MStopEqua \Mpar From {\seqBBBbtd}, \MMath{\mathfrak{g}, \left[\, \cdot \, ,\, \cdot \, \right]} is a Lie algebra. We denote by \MMath{G} the corresponding simply-connected Lie group \bseqHHHabg{theorem 3.28}.%
\Mpar \vspace{\Sssaut}\hspace{\Alinea}For any \MMath{M > 0}, we define:%
\Mpar \MStartEqua \MMath{\mathcal{D} ^{(n,M)}_{\mathfrak{bos} } \mathrel{\mathop:}=  \bigoplus_{N=0}^{M} \left( \mathds{C}^{n} \right)^{{\otimes N,\text{{sym}}}} \subset  \mathcal{D} ^{(n)}_{\mathfrak{bos} }}.%
\MStopEqua \Mpar From the definition of \MMath{\mathtt{a} _{p}, \mathtt{a} ^{+}_{p}} for \MMath{p\leqslant n} {\seqDDDame} and of the operators in \MMath{\mathfrak{g}} {\seqDDDaba}, for any \MMath{X \in  \mathfrak{g}}, there exists a constant \MMath{c(X) > 0} such that:%
\Mpar \MStartEqua \MMath{\forall  M > 0, \forall  \psi  \in  \mathcal{D} ^{(n,M)}_{\mathfrak{bos} }, {X \psi  \in  \mathcal{D} ^{(n,M+2)}_{\mathfrak{bos} } \& \left\| X \psi  \right\| \leqslant  c(X) \,  (M+2) \,  \left\| \psi  \right\|}}.%
\MStopEqua \Mpar Let \MMath{\left( X_{k} \right)_{k\leqslant K}} be a basis of the vector space \MMath{\mathfrak{g}} and let \MMath{c \mathrel{\mathop:}=  \max_{{k\leqslant K}} c(X_{k})}. Let \MMath{s \in  \left]0,\nicefrac{1}{2Kc}\right[}. For any \MMath{M > 0} and any \MMath{\psi  \in  \mathcal{D} ^{(n,M)}_{\mathfrak{bos} }}, we have:%
\Mpar \MStartEqua \MMath{\sum_{m=0}^{\infty } \frac{s^{m}}{m!} \sum_{{k_{1},\dots ,k_{m} \leqslant  K}} \left\| X_{{k_{1}}} \dots  X_{{k_{m}}} \psi  \right\| \leqslant  \sum_{m=0}^{\infty } \frac{(M+2) (M+4) \dots  (M+2m)}{m!} \,  (K c \,  s)^{m} \,  \left\| \psi  \right\|\\
\hphantom{\sum_{m=0}^{\infty } \frac{s^{m}}{m!} \sum_{{k_{1},\dots ,k_{m} \leqslant  K}} \left\| X_{{k_{1}}} \dots  X_{{k_{m}}} \psi  \right\| \leqslant \hspace{0.25cm} } = \left(1 - 2Kc\, s\right)^{{-\nicefrac{M}{2}-1}} \,  \left\| \psi  \right\| < \infty }.%
\MStopEqua \Mpar Using the notations of \bseqHHHabm{section 2} and defining \MMath{\xi  \mathrel{\mathop:}=  \left| X_{1} \right| + \dots  + \left| X_{K} \right|}, we thus have:%
\Mpar \MStartEqua \MMath{\forall  \psi  \in  \mathcal{D} ^{(n)}_{\mathfrak{bos} }, {\left\| e^{{s\xi }} \,  \psi  \right\| \mathrel{\mathop:}=  \sum_{m=0}^{\infty } \frac{s^{m}}{m!} \sum_{{k_{1},\dots ,k_{m} \leqslant  K}} \left\| X_{{k_{1}}} \dots  X_{{k_{m}}} \psi  \right\| < \infty }}.%
\MStopEqua \Mpar Hence, by Nelson's criterion for the integrability of a Lie algebra representation by unbounded skew-symmetric operators \bseqHHHabm{lemma 9.1}, every \MMath{-i\, X} for \MMath{X \in  \mathfrak{g}} is essentially self-adjoint, and there exists a unitary representation \MMath{\tilde{\mathtt{T} }^{(n)}_{\mathfrak{bos} }} of \MMath{G} on \MMath{\mathcal{F} ^{(n)}_{\mathfrak{bos} }} such that:%
\hypertarget{PARbtn}{}\Mpar \MStartEqua \MMath{\forall  X \in  \mathfrak{g}, \tilde{\mathtt{T} }^{(n)}_{\mathfrak{bos} } \big( \exp_{G}(X) \big) = \exp \big( X \big)}%
\NumeroteEqua{B.5}{4}\MStopEqua \Mpar (where \MMath{\exp_{G}} denotes the exponential mapping \MMath{\mathfrak{g} \rightarrow  G} and \MMath{\exp} denotes the exponential of operators defined through spectral resolution).%
\Mpar \vspace{\Sssaut}\hspace{\Alinea}In particular, for any \MMath{\mathbf{x}  \in  \mathds{R}^{2n}}, \MMath{\hat{\mathbf{x} }} is essentially self-adjoint and {\seqBBBanh} follows from the Baker–Campbell–Hausdorff formula in \MMath{G} (with \MMath{\left[ \hat{\mathbf{x} }, \hat{\mathbf{y} } \right]} from {\seqBBBbtd} for \MMath{\mathbf{x} ,\mathbf{y}  \in  \mathds{R}^{2n}}).%
\Mpar \vspace{\Sssaut}\hspace{\Alinea}From {\seqBBBbtd}, \MMath{\text{h} \mapsto  i\, \hat{\text{h}}} is a Lie algebra morphism \MMath{\mathfrak{sp} ^{(n)} \mapsto  \mathfrak{g}}. Hence, for any \MMath{n > 0}, there exists a Lie group homomorphism \MMath{\varphi } from the universal cover \MMath{\widetilde{\text{Sp}}^{(n)}} of \MMath{\text{Sp}^{(n)}} into \MMath{G} such that \MMath{\forall  \text{h} \in  \mathfrak{sp} ^{(n)},\,  \left[d\varphi \right]_{\mathds{1} }(\text{h}) = i\, \hat{\text{h}}} \bseqHHHabg{theorem 3.27}. In a way similar to {\seqBBBaka}, \MMath{\widetilde{\text{Sp}}^{(n)}} can be constructed as \bseqHHHabh{prop.~1.36}:%
\Mpar \MStartEqua \MMath{\widetilde{\text{Sp}}^{(n)} \mathrel{\mathop:}=  \left\{ \hmtpCls{{\gamma }} \middlewithspace| \gamma  \text{{ path on }} \text{Sp}^{(n)} \text{{ starting from }} \mathds{1}  \right\}},%
\MStopEqua \Mpar with the group structure being defined via pointwise multiplication of the paths, and we have the projection:%
\Mpar \MStartEqua \MMath{\definitionFonction{\tilde{p }^{(n)}}{\widetilde{\text{Sp}}^{(n)}}{\text{Mp}^{(n)}}{\hmtpCls{{\gamma }}}{\left[\gamma \right]_{{\equiv _{2}}}}}%
\MStopEqua \Mpar (\MMath{\tilde{p }^{(n)}} is well-defined since the relation \MMath{\equiv _{2}} is coarser than homotopy equivalence).
\MMath{\tilde{p }^{(n)}} is a covering map and a group homomorphism, with \MMath{\left[d\tilde{p }^{(n)}\right]_{\mathds{1} } = \text{id}_{{\mathfrak{sp} ^{(n)}}}}. Moreover, we have:%
\Mpar \MStartEqua \MMath{\tilde{p }^{{(n),-1}} \left\langle  \mathds{1}  \right\rangle  = \left\{ \hmtpCls{{\gamma }} \middlewithspace| \gamma  \text{{ loop based at }} \mathds{1} , \hmtpCls{{\gamma }} \in  2\mathds{Z} \subset  \mathds{Z} \approx  \pi _{1} \big( \text{Sp}^{(n)} \big) \right\}}.%
\MStopEqua \Mpar As underlined in {\seqBBBbpu}, a representative of the homotopy class \MMath{1 \in  \mathds{Z} \approx  \pi _{1} \big( \text{Sp}^{(n)} \big)} is provided by the loop \MMath{\gamma ^{(n)}_{1} : [0,1] \rightarrow  \text{Sp}^{(n)}, t \mapsto  \exp_{{\text{Sp}^{(n)}}} \left[ 2\pi t \,  \text{h}^{(n)}_{1} \right]} where:%
\Mpar \MStartEqua \MMath{\forall  i,j \leqslant  2n, {\text{h}^{(n)}_{{1,ij}} \mathrel{\mathop:}=  \delta _{i2} \delta _{j1} - \delta _{i1} \delta _{j2}}}.%
\MStopEqua \Mpar Using the analogue of {\seqBBBbty} for \MMath{\widetilde{\text{Sp}}^{(n)}}, together with \bseqHHHabg{theorem 3.32}, we thus get, for any \MMath{w \in  \mathds{Z} \approx  \pi _{1} \big( \text{Sp}^{(n)} \big) \subset  \widetilde{\text{Sp}}^{(n)}}:%
\Mpar \MStartEqua \MMath{\varphi (w) = \varphi  \circ  \exp_{{\widetilde{\text{Sp}}^{(n)}}} \big( 2\pi w \,  \text{h}^{(n)}_{1} \big) = \exp_{G} \big( 2i\pi  \,  w \, \hat{\text{h}}^{(n)}_{1} \big)},%
\MStopEqua \Mpar and {\seqBBBbub} then implies:%
\hypertarget{PARbuc}{}\Mpar \MStartEqua \MMath{\tilde{\mathtt{T} }^{(n)}_{\mathfrak{bos} } \circ  \varphi (w) = \exp \big( 2i\pi  \,  w \, \hat{\text{h}}^{(n)}_{1} \big) = \exp \big( - 2i\pi  \,  w \,  \mathtt{a} ^{+}_{1} \mathtt{a} _{1} - i\pi  \,  w \,  \mathds{1}  \big) = (-1)^{w} \,  \mathds{1} }.%
\NumeroteEqua{B.5}{5}\MStopEqua \Mpar Hence, \MMath{\forall  w \in  2\mathds{Z} \approx  \tilde{p }^{{(n),-1}} \left\langle  \mathds{1}  \right\rangle , \tilde{\mathtt{T} }^{(n)}_{\mathfrak{bos} } \circ  \varphi (w) = \mathds{1} }, so there exists a unitary representation \MMath{\mathtt{T} ^{(n)}_{\mathfrak{bos} }} of \MMath{\text{Mp}^{(n)}} on \MMath{\mathcal{F} ^{(n)}_{\mathfrak{bos} }}, satisfying \MMath{\tilde{\mathtt{T} }^{(n)}_{\mathfrak{bos} } \circ  \varphi  = \mathtt{T} ^{(n)}_{\mathfrak{bos} } \circ  \tilde{p }^{(n)}}.
In particular (using \bseqHHHabg{theorem 3.32} for \MMath{\varphi } and \MMath{\tilde{p }^{(n)}} together with {\seqBBBbub}), we have:%
\hypertarget{PARbue}{}\Mpar \MStartEqua \MMath{\forall  \text{h} \in  \mathfrak{sp} ^{(n)},\forall  t \in  \mathds{R}, {\mathtt{T} ^{(n)}_{\mathfrak{bos} } \big( \exp_{{\text{Mp}^{(n)}}} (t\, \text{h}) \big) = \exp \big(it\, \hat{\text{h}}\big)}}.%
\NumeroteEqua{B.5}{6}\MStopEqua \Mpar Since \MMath{i \,  \hat{\text{h}} \in  \mathfrak{g}}, \MMath{\hat{\text{h}}} is essentially self-adjoint, so {\seqBBBbug} implies {\seqBBBbuh} \bseqHHHaaq{theorem VIII.7}.%
\Mpar \vspace{\Sssaut}\hspace{\Alinea}Reciprocally, if \MMath{\mathtt{T} ^{{(n)\prime}}_{\mathfrak{bos} }} is a unitary representation of \MMath{\text{Mp}^{(n)}} satisfying {\seqBBBbuh}, then, for any \MMath{\text{h} \in  \mathfrak{sp} ^{(n)}}, \MMath{t \mapsto  \mathtt{T} ^{{(n)\prime}}_{\mathfrak{bos} } \big( \exp_{{\text{Mp}^{(n)}}} (t\, \text{h}) \big)} is a strongly continuous one-parameter unitary group (thanks to \MMath{\mathcal{D} ^{(n)}_{\mathfrak{bos} }} being dense in \MMath{\mathcal{F} ^{(n)}_{\mathfrak{bos} }} and \MMath{\mathtt{T} ^{{(n)\prime}}_{\mathfrak{bos} }(\mu )} being unitary for all \MMath{\mu  \in  \text{Mp}^{(n)}}), so by Stone's theorem \bseqHHHaaq{theorem VIII.8}, there exists a self-adjoint operators \MMath{\hat{\text{h}}'} on \MMath{\mathcal{F} ^{(n)}_{\mathfrak{bos} }} such that \MMath{\forall  t \in  \mathds{R}, {\mathtt{T} ^{{(n)\prime}}_{\mathfrak{bos} } \big( \exp_{{\text{Mp}^{(n)}}} (t\, \text{h}) \big) = \exp \big(it\, \hat{\text{h}}'\big)}} and {\seqBBBbuh} implies \MMath{\forall  \psi  \in  \mathcal{D} ^{(n)}_{\mathfrak{bos} }, {\hat{\text{h}}' \psi  = \hat{\text{h}} \psi }}. Thus, we get \MMath{\forall  \text{h} \in  \mathfrak{sp} ^{(n)}, {\mathtt{T} ^{(n)}_{\mathfrak{bos} } \big( \exp_{{\text{Mp}^{(n)}}} (\text{h}) \big) = \mathtt{T} ^{{(n)\prime}}_{\mathfrak{bos} } \big( \exp_{{\text{Mp}^{(n)}}} (\text{h}) \big)}}, so \MMath{\mathtt{T} ^{(n)}_{\mathfrak{bos} }} and \MMath{\mathtt{T} ^{{(n)\prime}}_{\mathfrak{bos} }} coincide on a neighborhood of {$\mathds{1}$} in \MMath{\text{Mp}^{(n)}}. Hence, \MMath{\text{Mp}^{(n)}} being connected for any \MMath{n>0}, they are identical, ie.~the unitary representation \MMath{\mathtt{T} ^{(n)}_{\mathfrak{bos} }} is uniquely characterized by {\seqBBBbuh}.%
\Mpar \vspace{\Sssaut}\hspace{\Alinea}Finally, using the Baker–Campbell–Hausdorff formula in \MMath{G} together with {\seqBBBbuj}, we have, for any \MMath{\text{h} \in  \mathfrak{sp} ^{(n)}} and any \MMath{\mathbf{x}  \in  \mathds{R}^{2n}}:%
\Mpar \MStartEqua \MMath{\exp \big(i\, \hat{\text{h}}\big) \exp \big(i\, \hat{\mathbf{x} }\big) \exp \big(-i\, \hat{\text{h}}\big)
 = \exp \left( e^{{\text{ad}_{{i\, \hat{\text{h}}}}}} \big( i\, \hat{\mathbf{x} } \big) \right)
 = \exp \left( i \,  \hat{\mathbf{x}  }' \right)},%
\MStopEqua \Mpar where:%
\Mpar \MStartEqua \MMath{\mathbf{x}  ' = e^{\text{h}} \,  \mathbf{x}   = p ^{(n)} \left( \exp_{{\text{Mp}^{(n)}}}(\text{h}) \right) \,  \mathbf{x} }.%
\MStopEqua \Mpar Thus, {\seqBBBbuo} holds for \MMath{\mu } in a neighborhood of {$\mathds{1}$} in \MMath{\text{Mp}^{(n)}}, and therefore for all \MMath{\mu  \in  \text{Mp}^{(n)}}.%
\MendOfProof \hypertarget{SECbup}{}\MsectionC{bup}{B.1.2}{Characteristic Function of a Density Matrix}%
\vspace{\PsectionC}\Mpar \MstartPhyMode\hspace{\Alinea}As is well known, the Fock representation routinely used to describe the quantum harmonic oscillator is unitarily equivalent to the position (aka.~Schrödinger) one (which, from the point of view of geometric quantization can be seen as arising from a \emph{real} polarization \bseqHHHaae{section 9.3}). In fact, the existence of such a unitary equivalence, just like the existence of a projective unitary representation of the symplectic group, is guaranteed by the Stone–von Neumann theorem fulfilled by finite-dimensional linear systems.%
\MleavePhyMode \hypertarget{PARbur}{}\Mpar \Mproposition{B.6}Let \MMath{\mathcal{E} ^{(n)} \mathrel{\mathop:}=  L_{2}\left(\mathds{R} \rightarrow  \mathds{C}\right)^{{\otimes n}} \approx  L_{2}\left(\mathds{R}^{n} \rightarrow  \mathds{C}\right)} equipped with its canonical complex Hilbert space structure. There exists a Hilbert space isomorphism \MMath{R  : \mathcal{F} ^{(n)}_{\mathfrak{bos} } \rightarrow  \mathcal{E} ^{(n)}} such that:%
\hypertarget{PARbus}{}\Mpar \MList{1}\MMath{\forall  \left(N_{1},\dots ,N_{n}\right) \in  \mathds{N}^{n}, {R  \left| N_{1},\dots ,N_{n} \right\rangle _{\mathfrak{bos} } = \psi _{{N_{1}}} \otimes  \psi _{{N_{2}}} \otimes  \dots  \otimes  \psi _{{N_{n}}}}}, where, for any \MMath{N \in  \mathds{N}}:%
\Mpar \MStartEqua \MMath{\forall  q \in  \mathds{R}, {\psi _{N}(q) \mathrel{\mathop:}=  \frac{\pi ^{{-\nicefrac{1}{4}}}}{\sqrt{{2^{N} N!}}} e^{{-\frac{q^{2}}{2}}} H _{N}(q)}},%
\MStopEqua \Mpar and \MMath{H _{N}} denotes the \MMath{N}-th Hermite polynomial:%
\Mpar \MStartEqua \MMath{\forall  q \in  \mathds{R}, {H _{N}(q) \mathrel{\mathop:}=  \left(-1\right)^{N} e^{{q^{2}}} \frac{d^{N}}{dq^{N}} e^{{-q^{2}}}}};%
\MStopEqua \MStopList \hypertarget{PARbuw}{}\Mpar \MList{2}for any \MMath{\mathbf{x}  \in  \mathds{R}^{2n}}, \MMath{R \,  \exp(i\, \hat{\mathbf{x} })\, R ^{-1} = T (\mathbf{x} _{1},\mathbf{x} _{2}) \otimes  T (\mathbf{x} _{3},\mathbf{x} _{4}) \otimes  \dots  \otimes  T (\mathbf{x} _{2n-1},\mathbf{x} _{2n})}, where, for any \MMath{u,v \in  \mathds{R}} and any \MMath{\psi  \in  L_{2}\left(\mathds{R} \rightarrow  \mathds{C}\right)}:%
\Mpar \MStartEqua \MMath{\forall  q \in  \mathds{R}, {\left[T (u,v) \psi \right](q) \mathrel{\mathop:}=  \exp \left( i\, u\, q - i\frac{u\, v}{2} \right) \psi \left( q - v \right)}}.%
\MStopEqua \MStopList \Mpar \Mproof That {\seqBBBbuz} defines a Hilbert space isomorphism \MMath{\mathcal{F} ^{(n)}_{\mathfrak{bos} } \rightarrow  \mathcal{E} ^{(n)}} follows from \MMath{\big( \left| N_{1},\dots ,N_{n} \right\rangle _{\mathfrak{bos} } \big)_{{\left( N_{1},\dots ,N_{n} \right) \in  \mathds{N}^{n}}}} being an orthonormal basis of \MMath{\mathcal{F} ^{(n)}_{\mathfrak{bos} }} and from the orthogonality and completeness properties of the Hermite polynomials.%
\Mpar \vspace{\Sssaut}\hspace{\Alinea}For any \MMath{u,v \in  \mathds{R}}, \MMath{T (u,v)} is a unitary transformation of \MMath{\mathcal{E} ^{(1)}} and, for any \MMath{\tau ,\tau ' \in  \mathds{R}}, we have:%
\Mpar \MStartEqua \MMath{T (\tau u,\tau v) T (\tau 'u,\tau 'v) = T \big((\tau +\tau ')u,(\tau +\tau ')v\big)}.%
\MStopEqua \Mpar Moreover, \MMath{ \tau  \mapsto  T (\tau u,\tau v)} is strongly continuous (the proof is similar to {\seqBBBbvc}).
Then, by Stone's theorem \bseqHHHaaq{theorem VIII.8}, \MMath{T (u,v) = \exp(i\, X (u,v))}, with the densely defined, essentially self-adjoint operator \MMath{X (u,v)} given by:%
\Mpar \MStartEqua \MMath{\forall  \psi  \in  \mathcal{D},\,  \forall  q \in  \mathds{R}, {\left[X (u,v) \psi \right](q) \mathrel{\mathop:}=  -i \left.\frac{d}{d\tau } \left[T (\tau u,\tau v) \psi \right]\right|_{\tau =0} (q) = u\, q\, \psi (q) +i\, v \frac{d}{dq} \psi (q)}},%
\MStopEqua \Mpar where \MMath{\mathcal{D}} denotes the space of complex-valued, smooth, rapidly decreasing functions on \MMath{\mathds{R}}.
For any \MMath{N \in  \mathds{R}}, we have:%
\Mpar \MStartEqua \MMath{\forall  q \in  \mathds{R}, {\left[X (u,v) \psi _{N}\right](q) = (u+i\, v)\, q \psi _{N}(q) -i\, v\,  \sqrt{{2 (N+1)}} \psi _{N+1}(q)}}.%
\MStopEqua \Mpar Since \MMath{X (u,v)} is a symmetric operator for any \MMath{u,v} and \MMath{\left(\psi _{N}\right)_{N\in \mathds{N}}} is an orthonormal basis of \MMath{L_{2}(\mathds{R} \rightarrow  \mathds{C})}, it follows that:%
\Mpar \MStartEqua \MMath{\forall  q \in  \mathds{R}, {q\, \psi _{N}(q) = \sqrt{{\frac{N}{2}}} \psi _{N-1}(q) + \sqrt{{\frac{N+1}{2}}} \psi _{N+1}(q)}},%
\MStopEqua \Mpar hence, for any \MMath{\mathbf{x}  \in  \mathds{R}^{2n}} and any \MMath{\psi  \in  \mathcal{D} ^{(n)}_{\mathfrak{bos} }}:%
\Mpar \MStartEqua \MMath{R  \,  \hat{\mathbf{x} } \psi  = \sum_{p=1}^{n} \mathds{1} ^{(1)} \otimes  \dots  \otimes  X ^{(p)}(\mathbf{x} _{2p-1},\, \mathbf{x} _{2p}) \otimes  \dots  \otimes  \mathds{1} ^{(n)} \,  R  \psi }.%
\MStopEqua \Mpar Therefore {\seqBBBbvl} holds.%
\MendOfProof \Mpar \MstartPhyMode\hspace{\Alinea}{\seqCCCbvn} uses two auxiliary results concerning \emph{finite-dimensional} Fock representations, viz.~that these representations are \emph{weakly continuous}, and that a density matrix defined over such a Fock space is \emph{entirely} specified once the expectation values of all exponentiated linear observables are known.
Both results can be deduced from the properties of the so-called Wigner-Weyl transform {\bseqJJJabn}, which maps density matrices on the Fock (or Schrödinger) representation to \emph{quasi-probability} distributions on the phase space.
The quasi-probability distribution of {$\rho$} is defined as the Fourier transform of its characteristic (aka.~moment-generating) function \MMath{\mathbf{x}  \mapsto  \mkop{Tr}  \big( \rho \, \exp(i\hat{\mathbf{x} }) \big)} and, as argued in {\seqBBBbvo}, the latter turns out to be more useful in the context of projective state spaces (for partial traces can be performed very easily, by simply taking the restriction of the characteristic function to a symplectic subspace).
In terms of the characteristic function, the Wigner-Weyl correspondence states that there is a \emph{bijective} mapping between the space of density matrices on \MMath{\mathcal{F} ^{(n)}_{\mathfrak{bos} }} and a certain space of continuous functions on the phase space {$V$} (or, more precisely, on the dual thereof, which is identified with {$V$} as mentioned in {\seqBBBamq}; see again {\seqBBBbvo}).%
\MleavePhyMode \hypertarget{PARbvp}{}\Mpar \Mproposition{B.7}Let \MMath{\rho ,\rho '} be density matrices on \MMath{\mathcal{F} ^{(n)}_{\mathfrak{bos} }} (ie.~trace-class, semi-definite positive operators of unit trace) such that:%
\Mpar \MStartEqua \MMath{\forall  \mathbf{x}  \in  \mathds{R}^{2n}, {\mkop{Tr}  \big( \rho \, \exp(i\hat{\mathbf{x} }) \big) = \mkop{Tr}  \big( \rho '\, \exp(i\hat{\mathbf{x} }) \big)}}.%
\MStopEqua \Mpar Then, \MMath{\rho  = \rho '}.%
\Mpar \vspace{\Sssaut}\hspace{\Alinea}Moreover, for any density matrix {$\rho$} on \MMath{\mathcal{F} ^{(n)}_{\mathfrak{bos} }}, the function \MMath{\mathds{R}^{2n} \rightarrow  \mathds{C},\,  \mathbf{x}  \mapsto  \mkop{Tr}  \big( \rho \, \exp(i\hat{\mathbf{x} }) \big)} is continuous.%
\Mpar \Mproof {$\rho$} being entirely specified by its characteristic function \MMath{\mathbf{x}  \mapsto  \mkop{Tr}  \big( \rho \, \exp(i\hat{\mathbf{x} }) \big)} follows from the invertibility of the Wigner transform (see {\seqBBBbvu}) together with the just proven equivalence of the Fock representation with the Schrödinger one.%
\Mpar \vspace{\Sssaut}\hspace{\Alinea}Similarly, the continuity of the characteristic function follows from the Schrödinger representation being weakly continuous (see {\bseqJJJabo} or {\seqBBBbvv}).%
\MendOfProof \Mpar \MstartPhyMode\hspace{\Alinea}A related result, that can be easily deduced from the previous one, is that the Fock representation is cyclic: by acting on the vacuum state with every exponentiated linear observables, we span a dense subset of \MMath{\mathcal{F} ^{(n)}_{\mathfrak{bos} }}.
This allows to completely characterize a unitary transformation on \MMath{\mathcal{F} ^{(n)}_{\mathfrak{bos} }} by giving its action on the vacuum state together with its commutator with arbitrary linear observables.
Note that it in particular means that the metaplectic representation derived above is entirely fixed (up to phase factors), since the vacuum state is determined (up to a phase) by the complex structure: it is the unique state annihilated by the quantization of those (complex-valued) linear observables which are holomorphic with respect to the chosen complex structure.%
\MleavePhyMode \hypertarget{PARbvx}{}\Mpar \Mproposition{B.8}The vector subspace \MMath{\mkop{Span} \left\{ \exp \big(i \,  \hat{\mathbf{x} }) \left| 0,\dots ,0 \right\rangle _{\mathfrak{bos} } \middlewithspace| \mathbf{x}  \in  \mathds{R}^{2n} \right\}} is dense in \MMath{\mathcal{F} ^{(n)}_{\mathfrak{bos} }}.%
\Mpar \Mproof Reasoning by contradiction, suppose that there exists \MMath{\psi  \in  \mathcal{F} ^{(n)}_{\mathfrak{bos} }} with \MMath{\left\langle  \psi  \middlewithspace| \psi  \right\rangle _{\mathfrak{bos} } = 1} such that:%
\Mpar \MStartEqua \MMath{\forall  \mathbf{x}  \in  \mathds{R}^{2n}, {\left\langle  \psi  \middlewithspace| \exp(i\,  \hat{\mathbf{x} }) \middlewithspace| 0,\dots ,0 \right\rangle _{\mathfrak{bos} } = 0}},%
\MStopEqua \Mpar and apply {\seqBBBavt} to:%
\Mpar \MStartEqua \MMath{\rho  \mathrel{\mathop:}=  \frac{\left|\psi \right\rangle _{\mathfrak{bos} } \!\!\left\langle \psi \right|_{\mathfrak{bos} } + \left|0,\dots ,0\right\rangle _{\mathfrak{bos} } \!\!\left\langle 0,\dots ,0\right|_{\mathfrak{bos} }}{2}}%
\MStopEqua \Mpar versus:%
\Mpar \MStartEqua \MMath{\rho ' \mathrel{\mathop:}=  \frac{\left|\psi \right\rangle _{\mathfrak{bos} } + \left|0,\dots ,0\right\rangle _{\mathfrak{bos} }}{\sqrt{2}} \frac{\left\langle \psi \right|_{\mathfrak{bos} } + \left\langle 0,\dots ,0\right|_{\mathfrak{bos} }}{\sqrt{2}}}.%
\MSStopEqua \MendOfProof \hypertarget{SECbwe}{}\MsectionC{bwe}{B.1.3}{Coarse-graining}%
\vspace{\PsectionC}\Mpar \MstartPhyMode\hspace{\Alinea}We now turn to coarse-graining, namely the possibility, given a symplectic frame \MMath{\left(\mathbf{e} _{1},\dots ,\mathbf{e} _{2n},\mathbf{e} _{2n+1},\dots ,\mathbf{e} _{2m}\right)}, to write a tensor product decomposition of its associated Fock space, such that the first tensor product factor carries a representation of the first \MMath{n} \dofs (spanned by \MMath{\left(\mathbf{e} _{1},\dots ,\mathbf{e} _{2n}\right)}), while the second factor collects the last \MMath{m-n} \dofs (spanned by \MMath{\left(\mathbf{e} _{1},\dots ,\mathbf{e} _{2n}\right)}).
Note that, thanks to {\seqBBBald}, it is enough to consider coarse-grainings which respect the frame in this way (see {\seqBBBacm}).%
\Mpar \vspace{\Saut}\hspace{\Alinea}Importantly, the thus-obtained tensor product decomposition is \emph{compatible} with the unitary action of the metaplectic group introduced in {\seqBBBaaq}:
metaplectic transformation acting only on the first \MMath{2n} vectors of the frame, resp.~only on the last \MMath{2m-2n}, are represented by unitary operators acting only on the first tensor product factor, resp.~only on the second. This is the key property that allows us to prove, respectively the composition {\seqDDDadb} and unambiguity of arrows {\seqDDDada} needed for projective systems (see {\seqBBBbwg}).%
\MleavePhyMode \hypertarget{PARbwh}{}\Mpar \Mdefinition{B.9}For any \MMath{m \geqslant  n \geqslant  0}, we define an isomorphism of Hilbert spaces \MMath{\Gamma ^{(m,n)}_{\mathfrak{bos} } : \mathcal{F} ^{(m)}_{\mathfrak{bos} } \rightarrow  \mathcal{F} ^{(n)}_{\mathfrak{bos} } \otimes  \mathcal{F} ^{(m-n)}_{\mathfrak{bos} }} by:%
\hypertarget{PARbwi}{}\Mpar \MStartEqua \MMath{\forall  \left( N_{1},\dots ,N_{m} \right) \in  \mathds{N}^{m}, {\Gamma ^{(m,n)}_{\mathfrak{bos} } \left| N_{1},\dots ,N_{m} \right\rangle _{\mathfrak{bos} } \mathrel{\mathop:}=  \left| N_{1},\dots ,N_{n} \right\rangle _{\mathfrak{bos} } \otimes  \left| N_{n+1},\dots ,N_{m} \right\rangle _{\mathfrak{bos} }}}.%
\NumeroteEqua{B.9}{1}\MStopEqua \hypertarget{PARbwj}{}\Mpar \Mproposition{B.10}Let \MMath{m \geqslant  n \geqslant  0}. For any \MMath{\mathbf{x}  \in  \mathds{R}^{2n}}, we have:%
\hypertarget{PARbwk}{}\Mpar \MStartEqua \MMath{\big( \hat{\mathbf{x} } \otimes  \mathds{1}  \big) \circ  \Gamma ^{(m,n)}_{\mathfrak{bos} } = \Gamma ^{(m,n)}_{\mathfrak{bos} } \circ  \hat{\mathbf{y} }},%
\NumeroteEqua{B.10}{1}\MStopEqua \Mpar where:%
\Mpar \MStartEqua \MMath{\forall  i \leqslant  2m, {\mathbf{y} _{i} = \alternative{\mathbf{x} _{i}}{\text{{ if }} i \leqslant  2n}{0}{\text{{ otherwise}}}}}.%
\MStopEqua \Mpar \vspace{\Sssaut}\hspace{\Alinea}For any \MMath{\mu  \in  \text{Mp}^{(n)}}, we have:%
\hypertarget{PARbwo}{}\Mpar \MStartEqua \MMath{\big( \mathtt{T} ^{(n)}_{\mathfrak{bos} }(\mu ) \otimes  \mathds{1}  \big) \circ  \Gamma ^{(m,n)}_{\mathfrak{bos} } = \Gamma ^{(m,n)}_{\mathfrak{bos} } \circ  \mathtt{T} ^{(m)}_{\mathfrak{bos} } \big( \ell _{m\leftarrow n}(\mu ) \big)},%
\NumeroteEqua{B.10}{2}\MStopEqua \Mpar with \MMath{\ell _{m\leftarrow n} : \text{Mp}^{(n)} \rightarrow  \text{Mp}^{(m)}} from {\seqBBBaky}.%
\Mpar \Mproof For any \MMath{p \leqslant  n}, we have:%
\hypertarget{PARbwr}{}\Mpar \MStartEqua \MMath{\Gamma ^{(m,n)}_{\mathfrak{bos} } \circ  \mathtt{a} _{p} = \left( \mathtt{a} _{p} \otimes  \mathds{1}  \right) \circ  \Gamma ^{(m,n)}_{\mathfrak{bos} } \& \Gamma ^{(m,n)}_{\mathfrak{bos} } \circ  \mathtt{a} ^{+}_{p} = \left( \mathtt{a} ^{+}_{p} \otimes  \mathds{1}  \right) \circ  \Gamma ^{(m,n)}_{\mathfrak{bos} }}%
\NumeroteEqua{B.10}{3}\MStopEqua \Mpar (with tensor products of densely-defined operators defined as in \bseqHHHaaq{section VIII.10}).
Thus, {\seqBBBauq} follows from {\seqBBBaba}, and similarly, for any \MMath{\text{h} \in  \mathfrak{sp} ^{(n)}}:%
\Mpar \MStartEqua \MMath{\big( \hat{\text{h}} \otimes  \mathds{1}  \big) \circ  \Gamma ^{(m,n)}_{\mathfrak{bos} } = \Gamma ^{(m,n)}_{\mathfrak{bos} } \circ  \hat{\text{h}}'},%
\MStopEqua \Mpar where:%
\Mpar \MStartEqua \MMath{\forall  i,j \leqslant  2m, {\text{h}'_{ij} = \alternative{\text{h}_{ij}}{\text{{ if }} i,j \leqslant  2n}{0}{\text{{ otherwise}}}} = \left[ d \iota _{m\leftarrow n} \right]_{\mathds{1} }(\text{h}) = \left[ d \ell _{m\leftarrow n} \right]_{\mathds{1} }(\text{h})}.%
\MStopEqua \Mpar Using {\seqBBBbug} and \bseqHHHabg{theorem 3.32}, we get, for any \MMath{n > 0}:%
\Mpar \MStartEqua \MMath{\forall  \text{h} \in  \mathfrak{sp} ^{(n)}, {\left(\mathtt{T} ^{(n)}_{\mathfrak{bos} } \big( \exp_{{\text{Mp}^{(n)}}} (\text{h}) \big) \otimes  \mathds{1} \right) \circ  \Gamma ^{(m,n)}_{\mathfrak{bos} } = \Gamma ^{(m,n)}_{\mathfrak{bos} } \circ  \mathtt{T} ^{(m)}_{\mathfrak{bos} } \big( \ell _{m\leftarrow n} \circ  \exp_{{\text{Mp}^{(n)}}} (\text{h}) \big)}}.%
\MStopEqua \Mpar Hence, {\seqBBBaml} holds for \MMath{\mu } in a neighborhood of {$\mathds{1}$} in \MMath{\text{Mp}^{(n)}}, and therefore for all \MMath{\mu  \in  \text{Mp}^{(n)}}.%
\Mpar \vspace{\Sssaut}\hspace{\Alinea}For \MMath{n=0}, {\seqBBBaml} follows from the definition of \MMath{\ell _{m\leftarrow 0}} {\seqDDDbpu} together with \MMath{\forall  m \geqslant  0, {\mathtt{T} _{\mathfrak{bos} }^{(m)}(\mathds{1} ^{-}) = -\text{id}_{{\mathcal{F} ^{(m)}_{\mathfrak{bos} }}}}} (for \MMath{m>0}, this is implied by {\seqBBBbxa}).%
\MendOfProof \hypertarget{PARbxb}{}\Mpar \Mproposition{B.11}Let {$V$} be a symplectic vector space {\seqDDDakg}. Let \MMath{m \geqslant  n}, \MMath{\left(\mathbf{e} _{1},\dots ,\mathbf{e} _{2m}\right)} be a symplectic family in {$V$} {\seqDDDajy}, \MMath{\left(\mathbf{e} '_{2n+1},\dots ,\mathbf{e} '_{2m}\right) \in  V ^{{2(m-n)}}} and \MMath{\mu  \in  \text{Mp}^{(m)}} such that:%
\Mpar \MStartEqua \MMath{\mu  \rhd  \left(\mathbf{e} _{1},\dots ,\mathbf{e} _{2m}\right) = \left(\mathbf{e} _{1},\dots ,\mathbf{e} _{2n},\mathbf{e} '_{2n+1},\dots ,\mathbf{e} '_{2m}\right)}.%
\MStopEqua \Mpar Then, there exists a unitary transformation \MMath{\Phi  : \mathcal{F} ^{(m-n)}_{\mathfrak{bos} } \rightarrow  \mathcal{F} ^{(m-n)}_{\mathfrak{bos} }} such that:%
\Mpar \MStartEqua \MMath{\Gamma ^{(m,n)}_{\mathfrak{bos} } \circ  \mathtt{T} ^{(m)}_{\mathfrak{bos} } \big( \mu  \big) = \big( \mathds{1}  \otimes  \Phi  \big) \circ  \Gamma ^{(m,n)}_{\mathfrak{bos} }}.%
\MStopEqua \Mpar \Mproof We define \MMath{\Pi ^{(m)} \in  \text{Sp}^{(m)}} by:%
\Mpar \MStartEqua \MMath{\Pi ^{(m)}_{ij} \mathrel{\mathop:}=  \sum_{p=1}^{m} \delta _{{i,2p-1}}\delta _{{j,2(m-p)+1}} + \delta _{{i,2p}}\delta _{{j,2(m-p)+2}}},%
\MStopEqua \Mpar and \MMath{\text{q}^{(m)} \in  \mathfrak{sp} ^{(m)}} by:%
\Mpar \MStartEqua \MMath{\text{q}^{(m)}_{ij} \mathrel{\mathop:}=  \frac{1}{2} \sum_{{p,q=1}}^{m} \big( \delta _{{p,m+1-q}} - \delta _{pq} \big) \big( \delta _{{i,2p-1}} \delta _{{j,2q}} - \delta _{{i,2p}} \delta _{{j,2q-1}} \big)}.%
\MStopEqua \Mpar Using \MMath{\big( \text{q}^{(m)} \big)^{2} = \frac{1}{2} \big( \Pi ^{(m)} - \mathds{1}  \big)} and \MMath{\text{q}^{(m)} \,  \Pi ^{(m)} = - \text{q}^{(m)}}, we get:%
\Mpar \MStartEqua \MMath{\exp_{{\text{Sp}^{(m)}}} \big( \pi  \text{q}^{(m)} \big) = \Pi ^{(m)}}.%
\MStopEqua \Mpar Thus, defining \MMath{\tilde{\mu } \mathrel{\mathop:}=  \exp_{{\text{Mp}^{(m)}}} \big( \pi  \text{q}^{(m)} \big) \,  \mu  \,  \exp_{{\text{Mp}^{(m)}}} \big( -\pi  \text{q}^{(m)} \big)}, \bseqHHHabg{theorem 3.32} together with the definition of {$\rhd$} yields:%
\Mpar \MStartEqua \MMath{\tilde{\mu } \rhd  \left(\mathbf{e} _{2m-1},\mathbf{e} _{2m},\dots ,\mathbf{e} _{1},\mathbf{e} _{2}\right) = \left(\mathbf{e} '_{2m-1},\mathbf{e} '_{2m},\dots ,\mathbf{e} '_{2n+1},\mathbf{e} '_{2n+2},\mathbf{e} _{2n-1},\mathbf{e} _{2n},\dots ,\mathbf{e} _{1},\mathbf{e} _{2}\right)}.%
\MStopEqua \Mpar Now, \MMath{\left(\mathbf{e} _{1},\dots ,\mathbf{e} _{2m}\right)} is a symplectic family, hence so is \MMath{\left(\mathbf{e} _{1},\dots ,\mathbf{e} _{2n},\mathbf{e} '_{2n+1},\dots ,\mathbf{e} '_{2m}\right)}, and therefore there exists \MMath{\sigma  \in  \text{Sp}^{(m-n)}} such that \MMath{p ^{(m)} \big( \tilde{\mu } \big) = \iota _{{m \leftarrow  (m-n)}}(\sigma )}.
Hence, \MMath{\tilde{\mu } \in  p ^{{(m),-1}} \left\langle  \iota _{{m \leftarrow  (m-n)}}(\sigma ) \right\rangle  = \ell _{{m \leftarrow  (m-n)}} \left\langle  p ^{{(n),-1}} \left\langle  \sigma  \right\rangle  \right\rangle } (where we have used that the lift \MMath{\ell _{{m \leftarrow  (m-n)}}} of \MMath{\iota _{{m \leftarrow  (m-n)}}} is injective and that both \MMath{p ^{(n)}} and \MMath{p ^{(m)}} are double covers).
In particular, \MMath{\tilde{\mu }} is in the image of \MMath{\ell _{{m \leftarrow  (m-n)}}}, so by {\seqBBBbxo} there exists a unitary transformation \MMath{\tilde{\Phi } : \mathcal{F} ^{(m-n)}_{\mathfrak{bos} } \rightarrow  \mathcal{F} ^{(m-n)}_{\mathfrak{bos} }} such that:%
\Mpar \MStartEqua \MMath{\Gamma ^{(m,m-n)}_{\mathfrak{bos} } \circ  \mathtt{T} ^{(m)}_{\mathfrak{bos} } \big( \tilde{\mu } \big) = \big( \tilde{\Phi } \otimes  \mathds{1}  \big) \circ  \Gamma ^{(m,m-n)}_{\mathfrak{bos} }}.%
\MStopEqua \Mpar \vspace{\Sssaut}\hspace{\Alinea}Next, using {\seqBBBbug}, we get:%
\Mpar \MStartEqua \MMath{\mathtt{T} ^{(m)}_{\mathfrak{bos} } \big( \tilde{\mu } \big) = \exp \big( i\pi  \hat{\text{q}}^{(m)}_{\mathfrak{bos} } \big) \,  \mathtt{T} ^{(m)}_{\mathfrak{bos} } \big( \mu  \big)\, \exp \big( -i\pi  \hat{\text{q}}^{(m)}_{\mathfrak{bos} } \big)},%
\MStopEqua \Mpar with:%
\Mpar \MStartEqua \MMath{\hat{\text{q}}^{(m)}_{\mathfrak{bos} } = \sum_{p=1}^{m} \left( \frac{\mathtt{a} ^{+}_{p} - \mathtt{a} ^{+}_{m+1-p}}{2} \right) \left( \frac{\mathtt{a} _{m+1-p} - \mathtt{a} _{p}}{2} \right) - \frac{1}{2} \lfloor \nicefrac{m}{2} \rfloor \mathds{1} },%
\MStopEqua \Mpar where \MMath{\lfloor \, \cdot \,  \rfloor} denotes the floor.
Observing that \MMath{\exp \big( i\pi  \hat{\text{q}}^{(m)}_{\mathfrak{bos} } \big) \left| 0,\dots ,0 \right\rangle _{\mathfrak{bos} } = e^{{-\nicefrac{i\pi }{2} \lfloor \nicefrac{m}{2} \rfloor}} \left| 0,\dots ,0 \right\rangle _{\mathfrak{bos} }}, and, by {\seqBBBbuo}:%
\hypertarget{PARbxv}{}\Mpar \MStartEqua \MMath{\forall  \mathbf{x}  \in  \mathds{R}^{2m}, {\exp \big( i\pi  \hat{\text{q}}^{(m)}_{\mathfrak{bos} } \big) \,  \exp\big(i \,  \hat{\mathbf{x} }\big) \,  \exp \big( -i\pi  \hat{\text{q}}^{(m)}_{\mathfrak{bos} } \big) = \exp\big(i \,  \widehat{\Pi ^{(m)} \,  \mathbf{x} }\big) = \widehat{\Pi }^{(m)}_{\mathfrak{bos} } \,  \exp\big(i \,  \hat{\mathbf{x} }\big) \,  \widehat{\Pi }^{(m)}_{\mathfrak{bos} }}},%
\NumeroteEqua{B.11}{1}\MStopEqua \Mpar where:%
\Mpar \MStartEqua \MMath{\forall  \left( N_{1},\dots ,N_{m} \right) \in  \mathds{N}^{m}, {\widehat{\Pi }^{(m)}_{\mathfrak{bos} } \left| N_{1},\dots ,N_{m} \right\rangle _{\mathfrak{bos} } \mathrel{\mathop:}=  \left| N_{m},\dots ,N_{1} \right\rangle _{\mathfrak{bos} }}},%
\MStopEqua \Mpar we get \MMath{\exp \big( i\pi  \hat{\text{q}}^{(m)}_{\mathfrak{bos} } \big) = e^{{-\nicefrac{i\pi }{2} \lfloor \nicefrac{m}{2} \rfloor}} \widehat{\Pi }^{(m)}_{\mathfrak{bos} }} (thanks to {\seqBBBbxz}).
Thus,%
\Mpar \MStartEqua \MMath{\mathtt{T} ^{(m)}_{\mathfrak{bos} } \big( \mu  \big) = \widehat{\Pi }^{(m)}_{\mathfrak{bos} } \,  \Gamma ^{{(m,m-n),-1}}_{\mathfrak{bos} } \circ  \big( \tilde{\Phi } \otimes  \mathds{1}  \big) \circ  \Gamma ^{(m,m-n)}_{\mathfrak{bos} } \,  \widehat{\Pi }^{(m)}_{\mathfrak{bos} } = \Gamma ^{{(m,n),-1}}_{\mathfrak{bos} } \circ  \Big( \mathds{1}  \otimes  \big[ \widehat{\Pi }^{(m-n)}_{\mathfrak{bos} } \,  \tilde{\Phi } \,  \widehat{\Pi }^{(m-n)}_{\mathfrak{bos} } \big] \Big) \circ  \Gamma ^{(m,n)}_{\mathfrak{bos} }},%
\MStopEqua \Mpar which yields the desired result.%
\MendOfProof \hypertarget{SECbyc}{}\MsectionC{byc}{B.1.4}{Infinite-dimensional Fock Representation}%
\vspace{\PsectionC}\Mpar \MstartPhyMode\hspace{\Alinea}Fock spaces can also be built over infinite dimensional vector spaces, but as stressed in {\seqBBBako}, this requires the choice of a \emph{polarization}, in the form of a complex structure {\seqDDDaqg}.
This complex structure, together with the symplectic structure, turns the infinite dimensional classical phase space into a (complex) Hilbert space, allowing it to play the role of the 1-particle Hilbert space on which the Fock space is modeled.%
\MleavePhyMode \hypertarget{PARbye}{}\Mpar \Mdefinition{B.12}Let \MMath{V ,\Omega } be a symplectic vector space and let {$I$} be a compatible complex structure on {$V$} {\seqDDDaqg}. Let \MMath{\overline{V }_{{\!(I )}}} denote the completion of the corresponding complex pre-Hilbert space, and \MMath{V ^{*}_{{\!(I )}}} its dual.\footnote{This is necessary in order to match the conventions adopted in the treatment of Fock spaces on finite-dimensional symplectic vector spaces. Otherwise {\seqBBBbyf} below would not come out right.}
We define the bosonic Fock space \MMath{\mathcal{F} ^{(V ,I )}_{\mathfrak{bos} }} over \MMath{V ,\Omega ,I } as:%
\Mpar \MStartEqua \MMath{\mathcal{F} ^{(V ,I )}_{\mathfrak{bos} } \mathrel{\mathop:}=  \overline{\bigoplus_{N\geqslant 0} \left( V ^{*}_{{\!(I )}} \right)^{{\otimes N,\text{{sym}}}}}}.%
\MStopEqua \Mpar \vspace{\Sssaut}\hspace{\Alinea}Let \MMath{\left(\mathbf{b} _{i}\right)_{i\in I}} be an orthonormal basis of \MMath{\overline{V }_{{\!(I )}}}, and denote by \MMath{\left(\mathbf{b} ^{*}_{i}\right)_{i\in I}} its dual basis. For any \MMath{\left(N_{i}\right)_{i\in I} \in  \mathds{N}^{I}} such that \MMath{\sum_{i\in I} N_{i} =\mathrel{\mathop:}  N < \infty }, let \MMath{i_{1},\dots ,i_{n} \in  I} such that \MMath{\left\{i_{1},\dots ,i_{n}\right\} = \left\{i\in I \middlewithspace| N_{i} \neq 0\right\}}, and define:%
\Mpar \MStartEqua \MMath{\left| \left(N_{i}\right)_{i\in I;\, }\left(\mathbf{b} _{i}\right)_{i\in I} \right\rangle _{\mathfrak{bos} } \mathrel{\mathop:}=  \frac{1}{\sqrt{{N!\, \prod_{i\in I} N_{i} !}}} \sum_{{\varepsilon  \in  S_{N}}} \hat{\varepsilon } \big( \underbrace{\mathbf{b} ^{*}_{{i_{1}}} \otimes  \dots  \otimes  \mathbf{b} ^{*}_{{i_{1}}}}_{{N_{{i_{1}}}}} \otimes  \dots  \otimes  \underbrace{\mathbf{b} ^{*}_{{i_{n}}} \otimes  \dots  \otimes  \mathbf{b} ^{*}_{{i_{n}}}}_{{N_{{i_{n}}}}} \big) \in  \left( V ^{*}_{{\!(I )}} \right)^{{\otimes N,\text{{sym}}}}}.%
\MStopEqua \Mpar \MMath{\left( \left| \left(N_{i}\right)_{i\in I;\, }\left(\mathbf{b} _{i}\right)_{i\in I} \right\rangle _{\mathfrak{bos} } \right)_{{\left(N_{i}\right)_{i\in I} \in  \mathds{N}^{I}, \sum_{i\in I} N_{i} < \infty }}} is an orthonormal basis of \MMath{\mathcal{F} ^{(V ,I )}_{\mathfrak{bos} }}.%
\Mpar \MstartPhyMode\hspace{\Alinea}In analogy to {\seqBBBama}, we can extract, from an infinite-dimensional Fock representation, partial theories associated to finite symplectic families, provided these families are compatible with the complex structure {$I$} underlying the considered Fock representation (ie.~they arise from {$I$}-orthonormal families). This is the tool we will use to construct the embedding of the infinite-dimensional Fock state space in the projective one {\seqDDDafp}.%
\MleavePhyMode \hypertarget{PARbyl}{}\Mpar \Mproposition{B.13}Let \MMath{\left(\mathbf{b} _{1},\dots ,\mathbf{b} _{n}\right)} be a \emph{finite} \emph{orthonormal} family in \MMath{\overline{V }_{{\!(I )}}}. Denotes by {$W$} the orthogonal complement of \MMath{\mkop{Span} _{\mathds{C}}\left\{\mathbf{b} _{1},\dots ,\mathbf{b} _{n}\right\}} in \MMath{\overline{V }_{{\!(I )}}} and by {$J$} its complex structure ({$W$} is a Hilbert space, hence a symplectic vector space equipped with a compatible complex structure, as shown in {\seqBBBbym}). There exists a unique Hilbert space isomorphism \MMath{\Gamma ^{(\mathbf{b} _{1},\dots ,\mathbf{b} _{n} ;V ,I )}_{\mathfrak{bos} } : \mathcal{F} ^{(V ,I )}_{\mathfrak{bos} } \rightarrow  \mathcal{F} ^{(n)}_{\mathfrak{bos} } \otimes  \mathcal{F} ^{(W ,J )}_{\mathfrak{bos} }} such that, for any orthonormal basis \MMath{\left(\mathbf{b} _{j}\right)_{j\in J}} of {$W$}:%
\hypertarget{PARbyn}{}\Mpar \MStartEqua \MMath{\forall  \left(N_{i}\right)_{i\in I} \in  \mathds{N}^{I} \big/ {\textstyle \sum_{i\in I} N_{i} < \infty },\\\hspace*{1.5cm} {\Gamma ^{(\mathbf{b} _{1},\dots ,\mathbf{b} _{n} ;V ,I )}_{\mathfrak{bos} } \left| \left(N_{i}\right)_{i\in I;\, }\left(\mathbf{b} _{i}\right)_{i\in I} \right\rangle _{\mathfrak{bos} } = \left| N_{1},\dots ,N_{n} \right\rangle _{\mathfrak{bos} } \otimes  \left| \left(N_{j}\right)_{j\in J;\, }\left(\mathbf{b} _{j}\right)_{j\in J} \right\rangle _{\mathfrak{bos} }}},%
\NumeroteEqua{B.13}{1}\MStopEqua \Mpar where \MMath{I \mathrel{\mathop:}=  \left\{1,\dots ,n\right\} \sqcup J}.%
\Mpar \Mproof Since \MMath{\left( \left| \left(N_{i}\right)_{i\in I;\, }\left(\mathbf{b} _{i}\right)_{i\in I} \right\rangle _{\mathfrak{bos} } \right)_{{\left(N_{i}\right)_{i\in I} \in  \mathds{N}^{I}, \sum_{i\in I} N_{i} < \infty }}} is an orthonormal basis of \MMath{\mathcal{F} ^{(V ,I )}_{\mathfrak{bos} }}, \MMath{\big( \left| N_{1},\dots ,N_{n} \right\rangle _{\mathfrak{bos} } \big)_{{\left( N_{1},\dots ,N_{n} \right) \in  \mathds{N}^{n}}}} of \MMath{\mathcal{F} ^{(n)}_{\mathfrak{bos} }}, and \MMath{\left( \left| \left(N_{j}\right)_{j\in J;\, }\left(\mathbf{b} _{j}\right)_{j\in J} \right\rangle _{\mathfrak{bos} } \right)_{{\left(N_{j}\right)_{j\in J} \in  \mathds{N}^{J}, \sum_{j\in J} N_{j} < \infty }}} of \MMath{\mathcal{F} ^{(W ,J )}_{\mathfrak{bos} }}, {\seqBBBbyq} uniquely defines a Hilbert space isomorphism \MMath{\mathcal{F} ^{(V ,I )}_{\mathfrak{bos} } \rightarrow  \mathcal{F} ^{(n)}_{\mathfrak{bos} } \otimes  \mathcal{F} ^{(W ,J )}_{\mathfrak{bos} }}.%
\Mpar \vspace{\Sssaut}\hspace{\Alinea}What is left to prove is that this definition is independent of the choice of the orthonormal basis \MMath{\left(\mathbf{b} _{j}\right)_{j\in J}} of {$W$}.
Let \MMath{\left(\mathbf{b} '_{j}\right)_{j\in J}} be another orthonormal basis of {$W$} and let \MMath{\left(\mathbf{b} '_{i}\right)_{1\leqslant i\leqslant n} \mathrel{\mathop:}=  \left(\mathbf{b} _{i}\right)_{1\leqslant i\leqslant n}}.
Using the definition of \MMath{\left| \left(N_{i}\right)_{i\in I;\, }\left(\mathbf{b} _{i}\right)_{i\in I} \right\rangle _{\mathfrak{bos} }}, together with the fact that \MMath{\forall  i \leqslant  n,\, \forall j\in J, {\left\langle  \mathbf{b} _{i} \middlewithspace| \mathbf{b} '_{j} \right\rangle _{I } = 0}}, we get, for any \MMath{\left(N_{i}\right)_{i\in I},\,  \left(N'_{i}\right)_{i\in I} \in  \mathds{N}^{I}} such that \MMath{\sum_{i\in I} N_{i},\,  \sum_{i\in I} N'_{i} < \infty }:%
\Mpar \MStartEqua \MMath{\left\langle  \left(N_{i}\right)_{i\in I;\, }\left(\mathbf{b} _{i}\right)_{i\in I} \middlewithspace| \left(N'_{i}\right)_{i\in I;\, }\left(\mathbf{b} '_{i}\right)_{i\in I} \right\rangle _{\mathfrak{bos} } = \delta _{{N_{1} N'_{1}}} \dots  \delta _{{N_{n} N'_{n}}} \,  \left\langle  \left(N_{j}\right)_{j\in J;\, }\left(\mathbf{b} _{j}\right)_{j\in J} \middlewithspace| \left(N'_{j}\right)_{j\in J;\, }\left(\mathbf{b} '_{j}\right)_{j\in J} \right\rangle _{\mathfrak{bos} }}.%
\MStopEqua \Mpar Hence, inserting:%
\Mpar \MStartEqua \MMath{\left| \left(N'_{i}\right)_{i\in I;\, }\left(\mathbf{b} '_{i}\right)_{i\in I} \right\rangle _{\mathfrak{bos} } = \sum_{{\deuxlignes{0pt}{\left(N_{i}\right)_{i\in I} \in  \mathds{N}^{I}}{\textstyle \sum_{i\in I} N_{i} < \infty }}} \left\langle  \left(N_{i}\right)_{i\in I;\, }\left(\mathbf{b} _{i}\right)_{i\in I} \middlewithspace| \left(N'_{i}\right)_{i\in I;\, }\left(\mathbf{b} '_{i}\right)_{i\in I} \right\rangle _{\mathfrak{bos} } \,  \left| \left(N_{i}\right)_{i\in I;\, }\left(\mathbf{b} _{i}\right)_{i\in I} \right\rangle _{\mathfrak{bos} }}%
\MStopEqua \Mpar in {\seqBBBbyq} confirms that the same equation holds with respect to the orthonormal basis \MMath{\left(\mathbf{b} '_{j}\right)_{j\in J}}.%
\MendOfProof \Mpar \MstartPhyMode\hspace{\Alinea}The extraction of finite partial theories from the infinite dimensional Fock space defined in {\seqBBBbyw} is compatible with the coarse-graining of such finite theories as introduced in {\seqBBBaki}.
It is also compatible with the finite dimensional metaplectic representation (from {\seqBBBaaq}) in the following sense:%
\Mpar \vspace{\Saut}\hspace{\Alinea}While \MMath{\mathcal{F} ^{(V ,I )}_{\mathfrak{bos} }} does not support a representation of the full, infinite dimensional symplectic group on {$V$} (see {\seqBBBabr} for more details on this), it does support a representation of its \emph{unitary subgroup}, ie.~the group of linear bijections of {$V$} that preserve \emph{both} the symplectic structure {$\Omega$} and the complex structure {$I$}, or, equivalently, that preserve the complex scalar product \MMath{\left\langle  \, \cdot \,  \middlewithspace| \, \cdot \,  \right\rangle _{I }} (see {\seqBBBbyx}, as well as the discussion preceding {\seqBBBboy}).
Indeed, such bijections correspond to a change of orthonormal basis in the 1-particle Hilbert space, which maps transparently into a change of basis in the Fock space.%
\Mpar \vspace{\Saut}\hspace{\Alinea}Considering now a unitary transformation of \MMath{V ,\Omega ,I } that only affects a \emph{finite-dimensional} subspace \MMath{F  = \mkop{Span} _{\mathds{C}} \left\{\mathbf{b} _{1},\dots ,\mathbf{b} _{n}\right\}} (leaving the {$I$}-orthogonal complement of {$F$} untouched), its action on \MMath{\mathcal{F} ^{(V ,I )}_{\mathfrak{bos} }} will decompose over the tensor product factorization from {\seqBBBbyw}, reducing, \emph{up to a phase}, to the action of the corresponding finite dimensional unitary transformation on \MMath{\mathcal{F} ^{(n)}_{\mathfrak{bos} }}.%
\MleavePhyMode \hypertarget{PARbyy}{}\Mpar \Mproposition{B.14}Let \MMath{\left(\mathbf{b} _{1},\dots ,\mathbf{b} _{n},\mathbf{b} _{n+1},\dots ,\mathbf{b} _{m}\right)} be a finite orthonormal family in \MMath{\overline{V }_{{\!(I )}}}. Denote by \MMath{W _{1},\,  J _{1}} the orthogonal complement of \MMath{\mkop{Span} _{\mathds{C}}\left\{\mathbf{b} _{1},\dots ,\mathbf{b} _{n}\right\}} and by \MMath{W _{2},\,  J _{2}} the orthogonal complement of \MMath{\mkop{Span} _{\mathds{C}}\left\{\mathbf{b} _{1},\dots ,\mathbf{b} _{m}\right\}}. Then, we have:%
\hypertarget{PARbyz}{}\Mpar \MStartEqua \MMath{\left( \Gamma ^{(m,n)}_{\mathfrak{bos} } \otimes  \text{id}_{{\mathcal{F} ^{(W _{2},J _{2})}_{\mathfrak{bos} }}} \right) \circ  \Gamma ^{(\mathbf{b} _{1},\dots ,\mathbf{b} _{m} ;V ,I )}_{\mathfrak{bos} } = \left( \text{id}_{{\mathcal{F} ^{(n)}_{\mathfrak{bos} }}} \otimes  \Gamma ^{(\mathbf{b} _{n+1},\dots ,\mathbf{b} _{m} ;W _{1},J _{1})}_{\mathfrak{bos} } \right) \circ  \Gamma ^{(\mathbf{b} _{1},\dots ,\mathbf{b} _{n} ;V ,I )}_{\mathfrak{bos} }}%
\NumeroteEqua{B.14}{1}\MStopEqua \Mpar \vspace{\Sssaut}\hspace{\Alinea}Let \MMath{\left(\mathbf{b} _{1},\dots ,\mathbf{b} _{n}\right),\,  \left(\mathbf{b} '_{1},\dots ,\mathbf{b} '_{n}\right)} be two orthonormal families in \MMath{\overline{V }_{{\!(I )}}} with \MMath{\mkop{Span} _{\mathds{C}}\left\{\mathbf{b} _{1},\dots ,\mathbf{b} _{n}\right\} = \mkop{Span} _{\mathds{C}}\left\{\mathbf{b} '_{1},\dots ,\mathbf{b} '_{n}\right\}}, and denotes by \MMath{W ,\,  J } the orthogonal complement of \MMath{\mkop{Span} _{\mathds{C}}\left\{\mathbf{b} _{1},\dots ,\mathbf{b} _{n}\right\}}. Then, we have:%
\hypertarget{PARbzb}{}\Mpar \MStartEqua \MMath{\Gamma ^{(\mathbf{b} '_{1},\dots ,\mathbf{b} '_{n} ;V ,I )}_{\mathfrak{bos} } = \left( e^{{i\phi _{\mu }}} \,  \mathtt{T} ^{(n)}_{\mathfrak{bos} }(\mu ) \otimes  \text{id}_{{\mathcal{F} ^{(W ,J )}_{\mathfrak{bos} }}}\right) \circ  \Gamma ^{(\mathbf{b} _{1},\dots ,\mathbf{b} _{n} ;V ,I )}_{\mathfrak{bos} }},%
\NumeroteEqua{B.14}{2}\MStopEqua \Mpar where {$\mu$} is some element of \MMath{\text{Mp}^{(n)}} such that \MMath{\left(\mathbf{b} '_{1},I \mathbf{b} '_{1},\dots ,\mathbf{b} '_{n},I \mathbf{b} '_{n}\right) = \mu  \rhd  \left(\mathbf{b} _{1},I \mathbf{b} _{1},\dots ,\mathbf{b} _{n},I \mathbf{b} _{n}\right)}, and \MMath{\phi _{\mu } \in  \mathds{R}}.%
\Mpar \Mproof {\seqCCCaux} can be checked from the definitions of \MMath{\Gamma ^{(m,n)}_{\mathfrak{bos} }} {\seqDDDbze} and \MMath{\Gamma ^{(\mathbf{b} _{1},\dots ,\mathbf{b} _{n} ;V ,I )}_{\mathfrak{bos} }} {\seqDDDbyq}, using some orthonormal basis \MMath{\left(\mathbf{b} _{j}\right)_{j\in J}} of \MMath{W _{2}} and using \MMath{\left(\mathbf{b} _{j}\right)_{{j\in  \left\{n+1,\dots ,m\right\} \sqcup J}}} as an orthonormal basis of \MMath{W _{1}} (taking advantage of {\seqBBBbyq} being valid with respect to any basis of the orthogonal complement as was stressed in {\seqBBBbzf}).%
\Mpar \vspace{\Sssaut}\hspace{\Alinea}Let \MMath{\left(\mathbf{b} _{1},\dots ,\mathbf{b} _{n}\right),\,  \left(\mathbf{b} '_{1},\dots ,\mathbf{b} '_{n}\right)} be two orthonormal families in \MMath{\overline{V }_{{\!(I )}}} with \MMath{\mkop{Span} _{\mathds{C}}\left\{\mathbf{b} _{1},\dots ,\mathbf{b} _{n}\right\} = \mkop{Span} _{\mathds{C}}\left\{\mathbf{b} '_{1},\dots ,\mathbf{b} '_{n}\right\}}, and denotes by \MMath{W ,\,  J } the orthogonal complement of \MMath{\mkop{Span} _{\mathds{C}}\left\{\mathbf{b} _{1},\dots ,\mathbf{b} _{n}\right\}}. 
Let u{} be the \emph{unitary} \MMath{n \times  n} matrix given by \MMath{\text{u}_{ij} \mathrel{\mathop:}=  \left\langle  \mathbf{b} ^{*}_{i} \middlewithspace| \mathbf{b} ^{{\prime *}}_{j} \right\rangle _{I } = \left\langle  \mathbf{b} '_{j} \middlewithspace| \mathbf{b} _{i} \right\rangle _{I }}.
Let \MMath{\left(\mathbf{b} _{j}\right)_{j\in J}} be an orthonormal basis of {$W$}, \MMath{\left(\mathbf{b} '_{j}\right)_{j\in J} \mathrel{\mathop:}=  \left(\mathbf{b} _{j}\right)_{j\in J}}, and \MMath{I \mathrel{\mathop:}=  \left\{1,\dots ,n\right\} \sqcup J}.
For any \MMath{\left(N_{i}\right)_{i\in I},\,  \left(N'_{i}\right)_{i\in I} \in  \mathds{N}^{I}} such that \MMath{\sum_{i\in I} N_{i},\,  \sum_{i\in I} N'_{i} < \infty }, we have:%
\hypertarget{PARbzg}{}\Mpar \MStartEqua \MMath{\left\langle  \left(N_{i}\right)_{i\in I;\, }\left(\mathbf{b} _{i}\right)_{i\in I} \middlewithspace| \left(N'_{i}\right)_{i\in I;\, }\left(\mathbf{b} '_{i}\right)_{i\in I} \right\rangle _{\mathfrak{bos} } =\\\hspace*{1.5cm}
 = \left\{ \renewcommand{\arraystretch}{1.2}\begin{array}{cl}
   \left(\prod_{j\in J} \delta _{{N_{j} N'_{j}}}\right) \,  \left\langle  N_{1},\dots ,N_{n} \middlewithspace| \text{u}^{\otimes N} \middlewithspace| N'_{1},\dots ,N'_{n} \right\rangle _{\mathfrak{bos} } \hspace{0.25cm} &\hspace{0.25cm}  \text{{ if }} \sum\limits_{i=1}^{n} N_{i} = \sum\limits_{i=1}^{n} N'_{i} =\mathrel{\mathop:}  N\\
   0 \hspace{0.25cm} &\hspace{0.25cm}  \text{{ else}}
   \end{array}\renewcommand{\arraystretch}{1} \right.}.%
\NumeroteEqua{B.14}{3}\MStopEqua \Mpar Note that \MMath{\text{u}^{\otimes N}} commutes with \MMath{\hat{\varepsilon }} for any \MMath{\varepsilon  \in  S_{N}}, hence it stabilizes the subspace \MMath{\left( \mathds{C}^{n} \right)^{{\otimes N,\text{{sym}}}}} of \MMath{\left( \mathds{C}^{n} \right)^{\otimes N}}.
Accordingly, we define a unitary automorphism \MMath{\hat{\text{u}}} on \MMath{\mathcal{F} ^{(n)}_{\mathfrak{bos} }} by:%
\Mpar \MStartEqua \MMath{\hat{\text{u}} \mathrel{\mathop:}=  \bigoplus_{N\geqslant 0} \text{u}^{\otimes N}}.%
\MStopEqua \Mpar Inserting {\seqBBBbzk} in {\seqBBBbyq} then yields:%
\Mpar \MStartEqua \MMath{\Gamma ^{(\mathbf{b} _{1},\dots ,\mathbf{b} _{n} ;V ,I )}_{\mathfrak{bos} } = \left(\hat{\text{u}} \otimes  \text{id}_{{\mathcal{F} ^{(W ,J )}_{\mathfrak{bos} }}}\right) \circ  \Gamma ^{(\mathbf{b} '_{1},\dots ,\mathbf{b} '_{n} ;V ,I )}_{\mathfrak{bos} }}.%
\MStopEqua \Mpar \vspace{\Sssaut}\hspace{\Alinea}There exists a Hermitian \MMath{n \times  n} matrix h{} such that \MMath{\text{u} = e^{{i\text{h}}}}. Thus, for any \MMath{N \in  \mathds{N}}, we have:%
\Mpar \MStartEqua \MMath{\text{u}^{\otimes N} = \exp \left(i \sum_{k=1}^{N} \mathds{1} ^{(1)} \otimes  \dots  \otimes  \text{h}^{(k)} \otimes  \dots  \otimes  \mathds{1} ^{(N)} \right)},%
\MStopEqua \Mpar and therefore:%
\Mpar \MStartEqua \MMath{\hat{\text{u}} = \exp \left(i \bigoplus_{N\geqslant 0} \sum_{k=1}^{N} \mathds{1} ^{(1)} \otimes  \dots  \otimes  \text{h}^{(k)} \otimes  \dots  \otimes  \mathds{1} ^{(N)} \right)}.%
\MStopEqua \Mpar Moreover, using the definition of \MMath{\mathtt{a} _{p}, \mathtt{a} ^{+}_{p}} {\seqDDDame}, the essentially self-adjoint operator \MMath{\bigoplus_{N\geqslant 0} \sum_{k=1}^{N} \mathds{1} ^{(1)} \otimes  \dots  \otimes  \text{h}^{(k)} \otimes  \dots  \otimes  \mathds{1} ^{(N)}} (defined on the dense domain \MMath{\mathcal{D} ^{(n)}_{\mathfrak{bos} }}) can be rewritten as:%
\Mpar \MStartEqua \MMath{\bigoplus_{N\geqslant 0} \sum_{k=1}^{N} \mathds{1} ^{(1)} \otimes  \dots  \otimes  \text{h}^{(k)} \otimes  \dots  \otimes  \mathds{1} ^{(N)} = \sum_{{i,j=1}}^{n} \text{h}_{ij} \,  \mathtt{a} ^{+}_{i} \,  \mathtt{a} _{j}}.%
\MStopEqua \Mpar \vspace{\Sssaut}\hspace{\Alinea}Next, we define a mapping \MMath{\text{M}_{n}(\mathds{C}) \rightarrow  \text{M}_{2n}(\mathds{R}),\,  \text{r} \mapsto  \text{r}^{(\mathds{R})}} by:%
\Mpar \MStartEqua \MMath{\forall  p,q \in  \left\{1,\dots ,n\right\}, {\text{r}^{(\mathds{R})}_{{2p-1,2q}} = - \text{r}^{(\mathds{R})}_{{2p,2q-1}} = \mkop{Re}(\text{r}_{pq}) \& \text{r}^{(\mathds{R})}_{{2p-1,2q-1}} = \text{r}^{(\mathds{R})}_{{2p,2q}} = - \mkop{Im}(\text{r}_{pq})}}.%
\MStopEqua \Mpar We have \MMath{\text{h}^{(\mathds{R})} \in  \mathfrak{sp} ^{(n)}} and:%
\Mpar \MStartEqua \MMath{\beta \left(\text{h}^{(\mathds{R})}\right) = i\,  \text{h} \, \&\,  \gamma \left(\text{h}^{(\mathds{R})}\right) = 0},%
\MStopEqua \Mpar so that:%
\Mpar \MStartEqua \MMath{\hat{\text{u}} = \exp \left(i\, \widehat{\text{h}^{(\mathds{R})}} - \frac{i}{2} \big( \mkop{Tr}  \text{h} \big) \,  \mathds{1} \right) = e^{{-i \phi _{\mu }}} \,  \mathtt{T} ^{(n)}_{\mathfrak{bos} } \big( \mu ^{-1} \big)}.%
\MStopEqua \Mpar where \MMath{\mu ^{-1} = \exp_{{\text{Mp}^{(n)}}}\big(\text{h}^{(\mathds{R})}\big)} and \MMath{\phi _{\mu } \mathrel{\mathop:}=  {\textstyle\frac{1}{2}} \mkop{Tr}  \text{h}}. Moreover, \MMath{i\, \text{r} \mapsto  \text{r}^{(\mathds{R})}} is an injective algebra morphism, so that \MMath{p ^{(n)}(\mu )^{-1} = \exp_{{\text{Sp}^{(n)}}}\big(\text{h}^{(\mathds{R})}\big) = (-i\, \text{u})^{(\mathds{R})}}.
Finally, by definition of u{}, we have:%
\Mpar \MStartEqua \MMath{\mathbf{b} '_{q}
 = \sum_{p=1}^{n} \text{u}^{*}_{pq} \mathbf{b} _{p}
 = \sum_{p=1}^{n} \mkop{Re}(\text{u}_{pq}) \mathbf{b} _{p} - \mkop{Im}(\text{u}_{pq}) I \, \mathbf{b} _{p}
 = \sum_{p=1}^{n} (-i\, \text{u})^{(\mathds{R})}_{{2p-1,2q-1}} \, \mathbf{b} _{p} + (-i\, \text{u})^{(\mathds{R})}_{{2p,2q-1}} \, I \, \mathbf{b} _{p}},%
\MStopEqua \Mpar as well as:%
\Mpar \MStartEqua \MMath{I \, \mathbf{b} '_{q} = \sum_{p=1}^{n} (-i\, \text{u})^{(\mathds{R})}_{{2p-1,2q}} \, \mathbf{b} _{p} + (-i\, \text{u})^{(\mathds{R})}_{{2p,2q}} \, I \, \mathbf{b} _{p}},%
\MStopEqua \Mpar so, noting that both \MMath{\left(\mathbf{b} _{1},I b_{1},\dots ,\mathbf{b} _{n},I \mathbf{b} _{n}\right)} and \MMath{\left(\mathbf{b} '_{1},I \mathbf{b} '_{1},\dots ,\mathbf{b} '_{n},I \mathbf{b} '_{n}\right)}, are finite symplectic families, we get:%
\Mpar \MStartEqua \MMath{\left(\mathbf{b} '_{1},I \mathbf{b} '_{1},\dots ,\mathbf{b} '_{n},I \mathbf{b} '_{n}\right) = \mu  \rhd  \left(\mathbf{b} _{1},I \mathbf{b} _{1},\dots ,\mathbf{b} _{n},I \mathbf{b} _{n}\right)}.%
\MStopEqua \MendOfProof \Mpar \MstartPhyMode\hspace{\Alinea}The identification of finite partial theories within the infinite-dimensional Fock space afforded by {\seqBBBbyw} allows to lift the linear observables from the finite dimensional case {\seqDDDaba} to the infinite dimensional one.
Thanks to the compatibility properties laid out in {\seqBBBbdg}, the thus constructed observables on \MMath{\mathcal{F} ^{(V ,I )}_{\mathfrak{bos} }} are consistently defined, ie.~independent of the particular truncation one uses, and their commutators obey the canonical commutation relations.%
\MleavePhyMode \hypertarget{PARcaf}{}\Mpar \Mproposition{B.15}For any \MMath{\mathbf{v}  \in  V }, there exists a unique unitary automorphism \MMath{\mathcal{O} ^{\mathfrak{bos} }_{(I )}(\mathbf{v} )} of \MMath{\mathcal{F} ^{(V ,I )}_{\mathfrak{bos} }} such that, for any finite orthonormal family \MMath{\left(\mathbf{b} _{1},\dots ,\mathbf{b} _{n}\right)} in \MMath{\overline{V }_{{\!(I )}}} with \MMath{\mathbf{v}  \in  \mkop{Span} _{\mathds{C}}\left\{\mathbf{b} _{1},\dots ,\mathbf{b} _{n}\right\}}, we have:%
\hypertarget{PARcag}{}\Mpar \MStartEqua \MMath{\Gamma ^{(\mathbf{b} _{1},\dots ,\mathbf{b} _{n} ;V ,I )}_{\mathfrak{bos} } \circ  \mathcal{O} ^{\mathfrak{bos} }_{(I )}(\mathbf{v} ) = \left( \exp \left( i\, \hat{\mathbf{x} } \right) \otimes  \text{id}_{{\mathcal{F} ^{(W ,J )}_{\mathfrak{bos} }}}\right) \circ  \Gamma ^{(\mathbf{b} _{1},\dots ,\mathbf{b} _{n} ;V ,I )}_{\mathfrak{bos} }},%
\NumeroteEqua{B.15}{1}\MStopEqua \Mpar where \MMath{\mathbf{x}  \in  \mathds{R}^{2n}} is defined by \MMath{\mathbf{v}  =\mathrel{\mathop:}  \sum_{p=1}^{n} \left(\mathbf{x} _{2p-1} + i\, \mathbf{x} _{2p}\right) \mathbf{b} _{p}}.%
\Mpar \vspace{\Sssaut}\hspace{\Alinea}Moreover, for any \MMath{\mathbf{v} ,\mathbf{w}  \in  V } we have:%
\hypertarget{PARcaj}{}\Mpar \MStartEqua \MMath{\mathcal{O} ^{\mathfrak{bos} }_{(I )}(\mathbf{v} ) \,  \mathcal{O} ^{\mathfrak{bos} }_{(I )}(\mathbf{w} ) = e^{{\nicefrac{i}{2}\, \Omega (\mathbf{v} ,\mathbf{w} )}} \,  \mathcal{O} ^{\mathfrak{bos} }_{(I )}(\mathbf{v} +\mathbf{w} )}.%
\NumeroteEqua{B.15}{2}\MStopEqua \Mpar \Mproof Since \MMath{\Gamma ^{(\mathbf{b} _{1},\dots ,\mathbf{b} _{n} ;V ,I )}_{\mathfrak{bos} }} is a Hilbert space isomorphism, {\seqBBBauj} uniquely defines a unitary automorphism of \MMath{\mathcal{F} ^{(V ,I )}_{\mathfrak{bos} }}. What is left to prove is that the thus defined \MMath{\mathcal{O} ^{\mathfrak{bos} }_{(I )}(\mathbf{v} )} does not depend on the choice of the orthonormal family \MMath{\left(\mathbf{b} _{1},\dots ,\mathbf{b} _{n}\right)}.
Let \MMath{\left(\mathbf{b} _{1},\dots ,\mathbf{b} _{n}\right)} and \MMath{\left(\mathbf{b} '_{1},\dots ,\mathbf{b} '_{m}\right)} be two orthonormal families with \MMath{\mathbf{v}  \in  \mkop{Span} _{\mathds{C}}\left\{\mathbf{b} _{1},\dots ,\mathbf{b} _{n}\right\} \cap  \mkop{Span} _{\mathds{C}}\left\{\mathbf{b} '_{1},\dots ,\mathbf{b} '_{m}\right\}}, and assume that {\seqBBBauj} holds with respect to \MMath{\left(\mathbf{b} _{1},\dots ,\mathbf{b} _{n}\right)}.
We extend \MMath{\left(\mathbf{b} _{1},\dots ,\mathbf{b} _{n}\right)}, resp.~\MMath{\left(\mathbf{b} '_{1},\dots ,\mathbf{b} '_{m}\right)}, into an orthonormal basis \MMath{\left(\mathbf{b} _{1},\dots ,\mathbf{b} _{n},\mathbf{b} _{n+1},\dots ,\mathbf{b} _{l}\right)}, resp.~\MMath{\left(\mathbf{b} '_{1},\dots ,\mathbf{b} '_{m},\mathbf{b} '_{m+1},\dots ,\mathbf{b} '_{l}\right)}, of \MMath{\mkop{Span} _{\mathds{C}}\left\{\mathbf{b} _{1},\dots ,\mathbf{b} _{n},\mathbf{b} '_{1},\dots ,\mathbf{b} '_{m}\right\}}.
Using {\seqBBBaux} together with {\seqBBBauq}, and noting that:%
\Mpar \MStartEqua \MMath{\mathbf{v}  = \sum_{p=1}^{l} \left(\mathbf{y} _{2p-1} + i\, \mathbf{y} _{2p}\right) \mathbf{b} _{p}} with \MMath{\forall  i\leqslant 2l, {\mathbf{y} _{i} \mathrel{\mathop:}=  \left\{\withAS{1.2}{cl}{\mathbf{x} _{i} \, &\,  \text{{ if }} i\leqslant 2n\\0 \, &\,  \text{{ else}}}\right.}},%
\MStopEqua \Mpar we obtain that {\seqBBBauj} holds with respect to \MMath{\left(\mathbf{b} _{1},\dots ,\mathbf{b} _{l}\right)}.
Next, using {\seqBBBbyf} together with {\seqBBBbuo}, and noting that:%
\Mpar \MStartEqua \MMath{\mathbf{v}  = \sum_{p=1}^{l} \mathbf{y} _{2p-1} \,  \mathbf{b} _{p} + \mathbf{y} _{2p} \,  I \mathbf{b} _{p} = \sum_{q=1}^{l} \big[ p ^{(n)}(\mu )\, \mathbf{y}  \big]_{2q-1} \,  \mathbf{b} '_{q} + \big[ p ^{(n)}(\mu )\, \mathbf{y}  \big]_{2q} \,  I \mathbf{b} '_{q}},%
\MStopEqua \Mpar for any \MMath{\mu  \in  \text{Mp}^{(l)}} such that \MMath{\left(\mathbf{b} '_{1},I \mathbf{b} '_{1},\dots ,\mathbf{b} '_{n},I \mathbf{b} '_{n}\right) = \mu  \rhd  \left(\mathbf{b} _{1},I \mathbf{b} _{1},\dots ,\mathbf{b} _{n},I \mathbf{b} _{n}\right)}, we obtain that {\seqBBBauj} holds with respect to \MMath{\left(\mathbf{b} '_{1},\dots ,\mathbf{b} '_{l}\right)}. Using again {\seqBBBcap}, we conclude that {\seqBBBauj} holds with respect to \MMath{\left(\mathbf{b} '_{1},\dots ,\mathbf{b} '_{m}\right)}.%
\Mpar \vspace{\Sssaut}\hspace{\Alinea}Finally, {\seqBBBcar} follows from {\seqBBBanh}, noting that:%
\Mpar \MStartEqua \MMath{\Omega \left( \sum_{p=1}^{n} \mathbf{x} _{2p-1} \,  \mathbf{b} _{p} + \mathbf{x} _{2p} \,  I \mathbf{b} _{p},\,  \sum_{p=1}^{n} \mathbf{y} _{2p-1} \,  \mathbf{b} _{p} + \mathbf{y} _{2p} \,  I \mathbf{b} _{p} \right) = {}^{\text{\sc t}} \mathbf{x}  \,  \Omega ^{(n)} \,  \mathbf{y} }.
\MStopEqua \MendOfProof %
\Mnomdefichier{lin52}%
\hypertarget{SECcau}{}\MsectionC{cau}{B.1.5}{Automorphisms of the Infinite-dimensional Fock Representation}%
\vspace{\PsectionC}\Mpar \MstartPhyMode\hspace{\Alinea}We now want to examine what is left of the Stone–von Neumann theorem in the infinite dimensional case.
As emphasized in the discussion preceding {\seqBBBbdg}, symplectomorphisms that happens to preserve the complex structure are unitary transformations of the 1-particle Hilbert space and are easily implemented on the infinite dimensional Fock space.
Going a little further, if a symplectomorphism only affects the complex structure over a finite dimensional (complex) subspace {$F$} of {$V$} (ie.~it stabilizes both {$F$} and its orthogonal complement \MMath{F ^{\perp }}, and restricts to a \MMath{\mathds{C}} linear, hence unitary, transformation on \MMath{F ^{\perp }}), we can use the tensor product decomposition from {\seqBBBbyw} together with the finite dimensional metaplectic representation from {\seqBBBakb} to implement this symplectomorphism on \MMath{\mathcal{F} ^{(V ,I )}_{\mathfrak{bos} }}.%
\Mpar \vspace{\Saut}\hspace{\Alinea}So one may ask what is the largest subgroup of the infinite dimensional symplectic group of {$V$} that can be represented on the Fock space from {\seqBBBbna}.
This question has been answered by Shale \bseqHHHaag{theorem 4.1}, and we give below an alternative proof (using the characterization of Fock states from {\seqBBBatt}) of the necessary condition that must be fulfilled by a symplectomorphism for it to be implementable on \MMath{\mathcal{F} ^{(V ,I )}_{\mathfrak{bos} }}.
That this condition is also sufficient can be proved by explicitly giving the corresponding unitary representation (see the cited reference).
In any cases, the result reproduced below confirms that there exists, in the infinite dimensional case, infinitely many inequivalent Fock representations:
given any symplectomorphism \MMath{\text{m}} on {$V$}, \MMath{I ' \mathrel{\mathop:}=  \text{m}^{-1} \,  I  \,  \text{m}} is a compatible complex structure on {$V$}, but the Fock representations \MMath{\mathcal{F} ^{(V ,I )}_{\mathfrak{bos} }} and \MMath{\mathcal{F} ^{(V ,I ')}_{\mathfrak{bos} }} will not be unitarily inequivalent unless m{} is unitarily implementable on \MMath{\mathcal{F} ^{(V ,I )}_{\mathfrak{bos} }} (indeed, m{} defines trivially a unitary mapping from \MMath{\mathcal{F} ^{(V ,I )}_{\mathfrak{bos} }} to \MMath{\mathcal{F} ^{(V ,I ')}_{\mathfrak{bos} }}, so if there is a unitary equivalence that can bring us back from \MMath{\mathcal{F} ^{(V ,I ')}_{\mathfrak{bos} }} to \MMath{\mathcal{F} ^{(V ,I )}_{\mathfrak{bos} }}, this gives a unitary implementation of m{} on \MMath{\mathcal{F} ^{(V ,I )}_{\mathfrak{bos} }}).%
\MleavePhyMode \hypertarget{PARcaw}{}\Mpar \Mproposition{B.16}Let \MMath{V ,\Omega } be a symplectic vector space and let {$I$} be a compatible complex structure on {$V$}. Denote by \MMath{\left\langle \, \cdot \, \middlewithspace|\, \cdot \, \right\rangle } the complex scalar product associated to {$I$} {\seqDDDaqg} and by \MMath{\left(\, \cdot \, \middlewithspace|\, \cdot \, \right)} the real scalar product \MMath{\left(\, \cdot \, \middlewithspace|\, \cdot \, \right) \mathrel{\mathop:}=  \mkop{Re} \left\langle \, \cdot \, \middlewithspace|\, \cdot \, \right\rangle }. Denote by \MMath{\overline{V }} the \emph{real} Hilbert space obtained by completing {$V$} with respect to the norm induced by \MMath{\left(\, \cdot \, \middlewithspace|\, \cdot \, \right)}.%
\Mpar \vspace{\Sssaut}\hspace{\Alinea}Let m{} be a bijective linear symplectomorphism of {$V$} and suppose that there exists a Hilbert space automorphism \MMath{\hat{\text{m}}} of \MMath{\mathcal{F} ^{(V ,I )}_{\mathfrak{bos} }} such that:%
\hypertarget{PARcax}{}\Mpar \MStartEqua \MMath{\forall  \mathbf{v}  \in  V , {\hat{\text{m}} \,  \mathcal{O} ^{\mathfrak{bos} }_{(I )}(\mathbf{v} ) \,  \hat{\text{m}}^{-1} = \mathcal{O} ^{\mathfrak{bos} }_{(I )}(\text{m} \mathbf{v} )}}.%
\NumeroteEqua{B.16}{1}\MStopEqua \Mpar Then, there exist an orthogonal automorphism {$O$} and a symmetric Hilbert–Schmidt operator {$T$} on \MMath{\overline{V }} such that \MMath{\text{m} = \left. \left(\text{id}_{{\overline{V }}} + T \right)\, O  \right|_{V }}.%
\hypertarget{PARcaz}{}\Mpar \Mlemma{B.17}Let \MMath{\left(\mathbf{f} _{1},\dots ,\mathbf{f} _{2n}\right)} be a finite symplectic family in {$V$} {\seqDDDajy} and define the \MMath{2n\times 2n} matrix {$S$} by:%
\Mpar \MStartEqua \MMath{S _{ij} \mathrel{\mathop:}=  \left(\mathbf{f} _{i}\middlewithspace|\mathbf{f} _{j}\right)}.%
\MStopEqua \Mpar Then, \MMath{\det S  \geqslant  1}.%
\Mpar \Mproof Let \MMath{F  \mathrel{\mathop:}=  \mkop{Span} \left\{\mathbf{f} _{1},\dots ,\mathbf{f} _{2n}\right\}}. By definition of a finite symplectic family, \MMath{\mkop{dim}  F  = 2n} and \MMath{\left. \Omega  \right|_{{F }}} is non-degenerate.
Let \MMath{\left(\mathbf{e} _{1},\dots ,\mathbf{e} _{2n}\right)} be a \MMath{\left(\, \cdot \, \middlewithspace|\, \cdot \, \right)}-orthonormal basis of \MMath{F }. The \MMath{2n\times 2n} real matrix \MMath{W } defined by:%
\Mpar \MStartEqua \MMath{\forall  i,j \leqslant  2n, {W _{ij} \mathrel{\mathop:}=  \Omega (\mathbf{e} _{i},\,  \mathbf{e} _{j})}},%
\MStopEqua \Mpar is anti-symmetric and non-degenerate. Hence, \MMath{i\, W } being a non-degenerate Hermitian matrix, there exists non-zero, real, distinct eigenvalues \MMath{\nu _{1},\dots ,\nu _{k}} and mutually orthogonal, complex vector subspaces \MMath{E _{1},\dots ,E _{k}} of \MMath{\mathds{C}^{2n}} such that:%
\Mpar \MStartEqua \MMath{W  = \sum_{r=1}^{k} i\, \nu _{r} \,  \Pi _{r}},%
\MStopEqua \Mpar where \MMath{\Pi _{r}} denotes the orthogonal projection on \MMath{E _{r}}.
Since {$W$} is a \emph{real} matrix, there exists, for any \MMath{r \leqslant  k}, an \MMath{r' \leqslant  k} such that \MMath{\nu _{r'} = - \nu _{r}} (in particular, as \MMath{\nu _{r} \neq  0}, \MMath{r' \neq  r}) and \MMath{E _{r'} = E _{r}^{*}} (where \MMath{(\, )^{*}} denotes complex conjugation).
By reordering the eigenvalues, we can enforce \MMath{\nu _{1},\dots ,\nu _{l} > 0} and \MMath{\forall  r \leqslant l, {\nu _{l+r} = - \nu _{r}}}, with \MMath{l = \nicefrac{k}{2}}.
For each \MMath{r\leqslant l}, we choose a complex orthonormal basis \MMath{\left( \mathbf{z} _{{r,1}},\dots ,\mathbf{z} _{{r,d_{r}}} \right)} of \MMath{E _{r}}. \MMath{\left( \mathbf{z} ^{*}_{{r,1}},\dots ,\mathbf{z} ^{*}_{{r,d_{r}}} \right)} is then an orthonormal basis of \MMath{E _{l+r} = E _{r}^{*}}.
We define \MMath{\left( \mathbf{z} _{1},\dots ,\mathbf{z} _{n} \right) \mathrel{\mathop:}=  \left( \mathbf{z} _{{1,1}},\dots ,\mathbf{z} _{{1,d_{1}}},\dots ,\mathbf{z} _{{l,1}},\dots ,\mathbf{z} _{{l,d_{l}}} \right)}, \MMath{\left( \nu '_{1},\dots ,\nu '_{n} \right) \mathrel{\mathop:}=  \left( \nu _{1},\dots ,\nu _{1},\dots ,\nu _{l},\dots ,\nu _{l} \right)} and, for any \MMath{p \leqslant  n}:%
\Mpar \MStartEqua \MMath{\mathbf{f} '_{2p-1} \mathrel{\mathop:}=  \sqrt{{\frac{2}{\nu '_{p}}}} \sum_{i=1}^{2n} \mkop{Re}(\mathbf{z} _{p}^{i})\, \mathbf{e} _{i} \, \&\,  \mathbf{f} '_{2p} \mathrel{\mathop:}=  \sqrt{{\frac{2}{\nu '_{p}}}} \sum_{i=1}^{2n} \mkop{Im}(\mathbf{z} _{p}^{i})\, \mathbf{e} _{i}}.%
\MStopEqua \Mpar By construction, \MMath{\left(\mathbf{f} '_{1},\dots ,\mathbf{f} '_{2n}\right)} is a symplectic, \MMath{\left(\, \cdot \, \middlewithspace|\, \cdot \, \right)}-ortho\emph{gonal} family, with, for any \MMath{p \leqslant  n}, \MMath{\left|\mathbf{f} '_{2p-1}\right| = \left|\mathbf{f} '_{2p}\right| = \nicefrac{1}{\sqrt{{\nu '_{p}}}}}.
Moreover, we know from {\seqBBBaqg}, that, for any \MMath{p\leqslant n}:%
\Mpar \MStartEqua \MMath{1 = \Omega (\mathbf{f} '_{2p-1},\, \mathbf{f} _{2p}) \leqslant  \left|\mathbf{f} '_{2p-1}\right| \,  \left|\mathbf{f} '_{2p}\right| = \frac{1}{\nu '_{p}}}.%
\MStopEqua \Mpar We denote by {$\sigma$} the element of \MMath{\text{Sp}^{(n)}} such that \MMath{\sigma  \rhd  \left(\mathbf{f} _{1},\dots ,\mathbf{f} _{2n}\right) = \left(\mathbf{f} '_{1},\dots ,\mathbf{f} '_{2n}\right)}.
Then, we have:%
\Mpar \MStartEqua \MMath{S _{ij} = \sum_{k=1}^{2n} \sigma _{ki} \,  \sigma _{kj} \,  \left|\mathbf{f} '_{k}\right|^{2}},%
\MStopEqua \Mpar so \MMath{\det S  = \left(\det \sigma \right)^{2} \,  \Pi _{p=1}^{n} \frac{1}{\nu ^{{\prime 2}}_{p}} \geqslant  1}, where we have used that, by definition of \MMath{\text{Sp}^{(n)}}, \MMath{\left(\det \sigma \right)^{2} = 1}.%
\MendOfProof \hypertarget{PARcbn}{}\Mpar \Mlemma{B.18}There exists \MMath{\epsilon  > 0} such that, for any \MMath{d\geqslant 0} and any \MMath{d\times d} positive-definite matrix \MMath{M} satisfying:%
\hypertarget{PARcbo}{}\Mpar \MStartEqua \MMath{\log\det \frac{\mathds{1}  + M}{2} - \frac{\log \det M}{2} < \epsilon },%
\NumeroteEqua{B.18}{1}\MStopEqua \Mpar we have:%
\Mpar \MStartEqua \MMath{\mkop{Tr}  \left(\sqrt{M} - 1\right)^{2} < 1},%
\MStopEqua \Mpar where \MMath{\sqrt{M}} denotes the positive square-root of \MMath{M} \bseqHHHaaq{theorem VI.9}.%
\Mpar \Mproof The function:%
\Mpar \MStartEqua \MMath{\definitionFonction{F}{\left]0,\, +\infty \right[}{\mathds{R}}{\nu }{\log \frac{1+\nu ^{2}}{2} - \log \nu }}%
\MStopEqua \Mpar is strictly decreasing on \MMath{\left]0,\, 1\right[} and strictly increasing on \MMath{\left]1,\, +\infty \right[}. At \MMath{\nu =1}, we have \MMath{F(1) = 0}, \MMath{F'(1) = 0} and \MMath{F''(1) = 1}.
Hence, there exists \MMath{\epsilon _{1} > 0} such that:%
\Mpar \MStartEqua \MMath{\forall  \nu  \in  \left]1-\epsilon _{1},\, 1+\epsilon _{1}\right[, {\frac{(\nu -1)^{2}}{3} \leqslant  F(\nu )}}.%
\MStopEqua \Mpar Moreover, defining \MMath{\epsilon _{2} \mathrel{\mathop:}=  \min\big(F(1-\epsilon _{1}),\,  F(1+\epsilon _{1})\big) > 0}, we have:%
\Mpar \MStartEqua \MMath{\forall  \nu  \in  \left]0,\, +\infty \right[, {F(\nu ) < \epsilon _{2} \Rightarrow  \left|\nu -1\right| < \epsilon _{1}}}.%
\MStopEqua \Mpar We define \MMath{\epsilon  \mathrel{\mathop:}=  \min\left( \epsilon _{2},\,  \nicefrac{1}{3} \right)}.%
\Mpar \vspace{\Sssaut}\hspace{\Alinea}Let \MMath{d\geqslant 0} and let \MMath{M} be a positive-definite matrix satisfying {\seqBBBcca}. Let \MMath{\left(\nu _{1},\dots ,\nu _{d}\right)} be the eigenvalues of \MMath{\sqrt{M}} (with multiplicities). For any \MMath{k\leqslant d}, \MMath{\nu _{k} > 0}, and {\seqBBBcca} can be rewritten:%
\Mpar \MStartEqua \MMath{\sum_{k=1}^{d} F(\nu _{k}) < \epsilon }.%
\MStopEqua \Mpar Since \MMath{F(\nu ) \geqslant  F(1) = 0} for any \MMath{\nu >0}, we have, \MMath{\forall  k\leqslant d, {F(\nu _{k}) < \epsilon  \leqslant  \epsilon _{2}}}, so \MMath{\forall  k\leqslant d, {\left|\nu _{k} - 1\right| < \epsilon _{1}}}. Thus, we get:%
\Mpar \MStartEqua \MMath{\mkop{Tr}  \left(\sqrt{M} - \mathds{1} \right)^{2} = \sum_{k=1}^{d} (\nu _{k} - 1)^{2} \leqslant  \sum_{k=1}^{d} 3\, F(\nu _{k}) < 3\, \epsilon  \leqslant  1}.%
\MStopEqua \MendOfProof \Mpar \MproofProposition{B.16}\italique{Characterization of a Fock state.} Using eg.~the Schrödinger representation {\seqDDDavj}, one can show {\seqDDDccf}, for any \MMath{n\geqslant 0} and any \MMath{\psi ,\psi ' \in  \mathcal{F} ^{(n)}_{\mathfrak{bos} }}:%
\Mpar \MStartEqua \MMath{\int_{{\mathds{R}^{2n}}} \frac{d^{{\scriptscriptstyle(2n)}}\mathbf{x} }{(2\pi )^{n}} \hspace{0.25cm}  \exp \left( - \frac{{}^{\text{\sc t}} \mathbf{x}  \,  \mathbf{x} }{4} \right) \,  \left\langle  \psi ' \middlewithspace| \exp(i\, \hat{\mathbf{x} }) \middlewithspace| \psi  \right\rangle _{\mathfrak{bos} } = \left\langle  \psi ' \middlewithspace| 0,\dots ,0 \right\rangle _{\mathfrak{bos} } \,  \left\langle  0,\dots ,0 \middlewithspace| \psi  \right\rangle _{\mathfrak{bos} }}.%
\MStopEqua \Mpar Let \MMath{\lambda _{1},\lambda _{2} \in  \mathcal{L} _{I }} with \MMath{\lambda _{1} \leqslant  \lambda _{2}} (\MMath{\mathcal{L} _{I }} was defined in {\seqBBBaao}) and let \MMath{\mu _{{\lambda _{2} \rightarrow  \lambda _{1}}}} be chosen as in {\seqBBBafp}.
For any density matrix \MMath{\tilde{\rho }} on \MMath{\mathcal{H} _{{\lambda _{2} \rightarrow  \lambda _{1}}} \mathrel{\mathop:}=  \mathcal{H} ^{\mathfrak{bos} }_{{\mu _{{\lambda _{2} \rightarrow  \lambda _{1}}}}} \approx  \mathcal{F} ^{(m-n)}_{\mathfrak{bos} }} (with \MMath{m \mathrel{\mathop:}=  d _{{\lambda _{2}}}} and \MMath{n \mathrel{\mathop:}=  d _{{\lambda _{1}}}}), we have:%
\Mpar \MStartEqua \MMath{\int_{{F }} \frac{d\mu _{F }(\mathbf{v} )}{(2\pi )^{m-n}} \hspace{0.25cm}  \exp \left( - \frac{\left(\mathbf{v} \middlewithspace|\mathbf{v} \right)}{4} \right) \,  \mkop{Tr}  \left( \tilde{\rho } \,  \mathcal{O} ^{\mathfrak{bos} }_{{\lambda _{2} \rightarrow  \lambda _{1}}}(\mathbf{v} ) \right) = \mkop{Tr}  \big( \tilde{\rho } \,  \left| 0,\dots ,0 \middlewithspace\rangle \!\!\middlewithspace\langle  0,\dots ,0 \right|_{\mathfrak{bos} } \big)},%
\MStopEqua \Mpar where \MMath{F  \mathrel{\mathop:}=  V _{{\lambda _{1}}}^{\perp } \cap  V _{{\lambda _{2}}}}, the Lebesgue measure \MMath{\mu _{F }} on \MMath{F } is normalized with respect to\MMath{\left(\, \cdot \, \middlewithspace|\, \cdot \, \right)}, and, for any \MMath{\mathbf{v}  \in  F }, the unitary operator \MMath{\mathcal{O} ^{\mathfrak{bos} }_{{\lambda _{2} \rightarrow  \lambda _{1}}}(\mathbf{v} )} on \MMath{\mathcal{H} _{{\lambda _{2} \rightarrow  \lambda _{1}}}} is defined as:%
\Mpar \MStartEqua \MMath{\mathcal{O} ^{\mathfrak{bos} }_{{\lambda _{2} \rightarrow  \lambda _{1}}}(\mathbf{v} ) \mathrel{\mathop:}=  \exp(i\, \hat{\mathbf{x} })},%
\MStopEqua \Mpar with \MMath{\mathbf{x}  = (\mathbf{x} _{2n+1},\dots ,\mathbf{x} _{2m}) \in  \mathds{R}^{{2(m-n)}}} such that \MMath{\mathbf{v}  =\mathrel{\mathop:}  \sum_{p=n+1}^{m} \left(\mathbf{x} _{2p-1} + i\, \mathbf{x} _{2p}\right) \,  \mathbf{b} _{p}} and \MMath{\left(\mathbf{b} _{1},\dots ,\mathbf{b} _{m}\right)} the orthonormal family such that \MMath{\mu _{{\lambda _{2} \rightarrow  \lambda _{1}}} \rhd  \lambda _{2} =\mathrel{\mathop:}  \left(\mathbf{b} _{1},I \mathbf{b} _{1},\dots ,\mathbf{b} _{m},I \mathbf{b} _{m}\right)}.%
\Mpar \vspace{\Sssaut}\hspace{\Alinea}Let \MMath{\rho } be a density matrix on \MMath{\mathcal{H} ^{\mathfrak{bos} }_{{\lambda _{2}}}} and let \MMath{\tilde{\rho }} be the partial trace on \MMath{\mathcal{H} ^{\mathfrak{bos} }_{{\lambda _{1}}}} of \MMath{\Phi _{{\lambda _{2} \rightarrow  \lambda _{1}}} \,  \rho  \,  \Phi _{{\lambda _{2} \rightarrow  \lambda _{1}}}^{-1}}, with \MMath{\Phi _{{\lambda _{2} \rightarrow  \lambda _{1}}} \mathrel{\mathop:}=  \text{Swap}_{{\lambda _{2} \rightarrow  \lambda _{1}}} \circ  \Phi ^{\mathfrak{bos} }_{{\mu _{{\lambda _{2} \rightarrow  \lambda _{1}}}}}: \mathcal{H} ^{\mathfrak{bos} }_{{\lambda _{2}}} \rightarrow  \mathcal{H} _{{\lambda _{2} \rightarrow  \lambda _{1}}} \otimes  \mathcal{H} ^{\mathfrak{bos} }_{{\lambda _{1}}}} {\seqDDDabd}. By definition of the partial trace, we have, for any bounded operator \MMath{A} on \MMath{\mathcal{H} _{{\lambda _{2} \rightarrow  \lambda _{1}}}}:%
\Mpar \MStartEqua \MMath{\mkop{Tr}  \big( \rho  \,  \Phi _{{\lambda _{2} \rightarrow  \lambda _{1}}}^{-1} \,  \left(A \otimes  \mathds{1} ^{\mathfrak{bos} }_{{\lambda _{1}}}\right) \,  \Phi _{{\lambda _{2} \rightarrow  \lambda _{1}}} \big) = \mkop{Tr}  \left( \tilde{\rho } \,  A \right)}.%
\MStopEqua \Mpar Hence, we get:%
\Mpar \MStartEqua \MMath{\mkop{Tr}  \left( \rho  \,  \Theta _{{\lambda _{2} |\lambda _{1}}} \right) = \int_{{F }} \frac{d\mu _{F }(\mathbf{v} )}{(2\pi )^{m-n}} \hspace{0.25cm}  \exp \left( - \frac{\left(\mathbf{v} \middlewithspace|\mathbf{v} \right)}{4} \right) \,  \mkop{Tr}  \Big( \rho  \,  \Phi _{{\lambda _{2} \rightarrow  \lambda _{1}}}^{-1} \,  \left( \mathcal{O} ^{\mathfrak{bos} }_{{\lambda _{2} \rightarrow  \lambda _{1}}}(\mathbf{v} ) \otimes  \mathds{1} ^{\mathfrak{bos} }_{{\lambda _{1}}}\right) \,  \Phi _{{\lambda _{2} \rightarrow  \lambda _{1}}} \Big)},%
\MStopEqua \Mpar where we have used the definition of \MMath{\Theta _{{\lambda _{2} | \lambda _{1}}}} and \MMath{\zeta _{{\lambda _{2} \rightarrow  \lambda _{1}}}} from {\seqBBBafp}.
Moreover, combining {\seqBBBaut} with {\seqBBBccr}, we have, for any \MMath{\mathbf{v}  \in  F }:%
\Mpar \MStartEqua \MMath{\Phi _{{\lambda _{2} \rightarrow  \lambda _{1}}}^{-1} \,  \left( \mathcal{O} ^{\mathfrak{bos} }_{{\lambda _{2} \rightarrow  \lambda _{1}}}(\mathbf{v} ) \otimes  \mathds{1} ^{\mathfrak{bos} }_{{\lambda _{1}}}\right) \,  \Phi _{{\lambda _{2} \rightarrow  \lambda _{1}}} = \mathcal{O} ^{\mathfrak{bos} }_{{\lambda _{2}}}(\mathbf{v} )}.%
\MStopEqua \Mpar \vspace{\Sssaut}\hspace{\Alinea}Let \MMath{\rho _{I }} be a density matrix on \MMath{\mathcal{F} ^{(V ,I )}_{\mathfrak{bos} }}. From {\seqBBBatt}, we have:%
\Mpar \MStartEqua \MMath{\sup_{{\lambda _{1} \in  \mathcal{L} _{I }}} \inf_{{\deuxlignes{-1pt}{\lambda _{2} \in  \mathcal{L} _{I }}{\lambda _{1} \leqslant  \lambda _{2}}}} \mkop{Tr}  \left( \rho _{{\lambda _{2}}} \,  \Theta _{{\lambda _{2} | \lambda _{1}}}\right) = 1},%
\MStopEqua \Mpar where \MMath{\rho  = \left(\rho _{\lambda }\right)_{{\lambda  \in  \mathcal{L} ^{\mathfrak{bos} }_{}}} \mathrel{\mathop:}=  \Sigma _{I }(\rho )}. Using the previous expression, together with {\seqBBBauc} yields, for any \MMath{\lambda _{1},\lambda _{2} \in  \mathcal{L} _{I }} with \MMath{\lambda _{1} \leqslant  \lambda _{2}}:%
\Mpar \MStartEqua \MMath{\mkop{Tr}  \left( \rho _{{\lambda _{2}}} \,  \Theta _{{\lambda _{2} |\lambda _{1}}} \right) = \int_{{F \mathrel{\mathop:}= V _{{\lambda _{1}}}^{\perp } \cap  V _{{\lambda _{2}}}}} \! \frac{d\mu _{F }(\mathbf{v} )}{\sqrt{{2\pi }}^{{\mkop{dim}  F }}} \hspace{0.25cm}  \exp \left( - \frac{\left(\mathbf{v} \middlewithspace|\mathbf{v} \right)}{4} \right) \,  \mkop{Tr}  \big( \rho _{(I )} \,  \mathcal{O} ^{\mathfrak{bos} }_{(I )}(\mathbf{v} ) \big)},%
\MStopEqua \Mpar so we arrive at the characterization:%
\hypertarget{PARccy}{}\Mpar \MStartEqua \MMath{\sup_{{F _{1} \in  \tilde{\mathcal{L} }_{I }}} \inf_{{\deuxlignes{-1pt}{F _{2} \in  \tilde{\mathcal{L} }_{I }}{F _{1} \subseteq F _{2}}}} \int_{{F \mathrel{\mathop:}= F _{1}^{\perp } \cap  F _{2}}} \! \frac{d\mu _{F }(\mathbf{v} )}{\sqrt{{2\pi }}^{{\mkop{dim}  F }}} \hspace{0.25cm}  \exp \left( - \frac{\left(\mathbf{v} \middlewithspace|\mathbf{v} \right)}{4} \right) \,  \mkop{Tr}  \big( \rho _{(I )} \,  \mathcal{O} ^{\mathfrak{bos} }_{(I )}(\mathbf{v} ) \big) = 1},%
\NumeroteEqua{B.16}{2}\MStopEqua \Mpar where \MMath{\tilde{\mathcal{L} }_{I }} denotes the set of \emph{finite-dimensional}, \emph{complex} (with respect to {$I$}) vector subspaces of {$V$}, and we have used {\seqBBBalc}.%
\Mpar \vspace{\Ssaut}\italique{Condition on m{}.} We define a density matrix \MMath{\rho _{\text{m}}} on \MMath{\mathcal{F} ^{(V ,I )}_{\mathfrak{bos} }} by:%
\Mpar \MStartEqua \MMath{\hat{\text{m}}^{-1} \,  \left| \left(0\right)_{i\in I;\, }\left(\mathbf{b} _{i}\right)_{i\in I} \right\rangle _{{\!\mathfrak{bos} }} \! \left\langle  \left(0\right)_{i\in I;\, }\left(\mathbf{b} _{i}\right)_{i\in I} \right|_{\mathfrak{bos} } \,  \hat{\text{m}}},%
\MStopEqua \Mpar with \MMath{\left(\mathbf{b} _{i}\right)_{i\in I}} some orthonormal basis of \MMath{\overline{V }_{{\!(I )}}} (see {\seqBBBcdd}). Note that \MMath{\rho _{\text{m}}} does \emph{not} depends on the choice of \MMath{\left(\mathbf{b} _{i}\right)_{i\in I}}, since, for any other orthonormal basis \MMath{\left(\mathbf{b} '_{i}\right)_{i\in I}} of \MMath{\overline{V }_{{\!(I )}}}, \MMath{\left| \left(0\right)_{i\in I;\, }\left(\mathbf{b} _{i}\right)_{i\in I} \right\rangle _{\mathfrak{bos} } = \left| \left(0\right)_{i\in I;\, }\left(\mathbf{b} '_{i}\right)_{i\in I} \right\rangle _{\mathfrak{bos} }}.
For any \MMath{\mathbf{v}  \in  V }, we have:%
\Mpar \MStartEqua \MMath{\mkop{Tr}  \big( \rho _{\text{m}} \,  \mathcal{O} ^{\mathfrak{bos} }_{(I )}(\mathbf{v} )\big) = \left\langle  \left(0\right)_{i\in I;\, }\left(\mathbf{b} _{i}\right)_{i\in I} \middlewithspace| \,  \mathcal{O} ^{\mathfrak{bos} }_{(I )}(\text{m} \mathbf{v} ) \,  \middlewithspace| \left(0\right)_{i\in I;\, }\left(\mathbf{b} _{i}\right)_{i\in I} \right\rangle _{\mathfrak{bos} } = \left\langle  0 \middlewithspace| \,  \exp\left( i\, \frac{\left|\text{m} \mathbf{v} \right|}{\sqrt{2}}\,  \left(\mathtt{a}  + \mathtt{a} ^{+}\right) \right) \,  \middlewithspace| 0 \right\rangle _{\mathfrak{bos} }},%
\MStopEqua \Mpar where the first equality follows from {\seqBBBcdg} and the second is obtained by applying {\seqBBBauj} to the orthonormal family \MMath{\left(\mathbf{b} \right)} where \MMath{\mathbf{b}  \mathrel{\mathop:}=  \nicefrac{\text{m} \mathbf{v} }{\left|\text{m} \mathbf{v} \right|}} and \MMath{|\text{m} \mathbf{v} | \mathrel{\mathop:}=  \sqrt{{\left(\text{m} \mathbf{v} \middlewithspace|\text{m} \mathbf{v} \right)}}}. Computing it eg.~in the Schrödinger representation {\seqDDDavj} gives:%
\Mpar \MStartEqua \MMath{\mkop{Tr}  \big( \rho _{\text{m}} \,  \mathcal{O} ^{\mathfrak{bos} }_{(I )}(\mathbf{v} )\big) = \exp \left( - \frac{\left(\text{m} \mathbf{v} \middlewithspace|\text{m} \mathbf{v} \right)}{4} \right)}.%
\MStopEqua \Mpar Thus, performing the Gaussian integration, {\seqBBBcdj} becomes:%
\Mpar \MStartEqua \MMath{\sup_{{F _{1} \in  \tilde{\mathcal{L} }_{I }}} \inf_{{\deuxlignes{-1pt}{F _{2} \in  \tilde{\mathcal{L} }_{I }}{F _{1} \subseteq F _{2}}}} \left(\det N_{{F _{1}^{\perp } \cap  F _{2}}}\right)^{{-\nicefrac{1}{2}}} = 1},%
\MStopEqua \Mpar where, for any finite dimensional subspace {$F$} of {$V$}, we have defined the positive-definite linear operators on {$F$}:%
\Mpar \MStartEqua \MMath{N_{F } \mathrel{\mathop:}=  \frac{\text{id}_{F } + M_{F}}{2} \, \&\,  M_{F} \mathrel{\mathop:}=  \Pi _{F } \,  \text{m}^{\dag} \,  \text{m} \,  \Pi _{F }},%
\MStopEqua \Mpar with \MMath{\Pi _{F }} the \MMath{\left(\, \cdot \, \middlewithspace|\, \cdot \, \right)}-orthogonal projection on {$F$}, and the adjoint \MMath{\text{m}^{\dag}} being taken with respect to the \emph{real} scalar product \MMath{\left(\, \cdot \, \middlewithspace|\, \cdot \, \right)} (indeed, m{} is \emph{not} assumed to be \MMath{\mathds{C}}-linear).
Since the function \MMath{\left]0,\, \infty \right[ \rightarrow  \left]0,\, \infty \right[,\,  x \mapsto  x^{{-\nicefrac{1}{2}}}} is strictly decreasing, this condition can be rewritten as:%
\hypertarget{PARcdo}{}\Mpar \MStartEqua \MMath{\inf_{{F _{1} \in  \tilde{\mathcal{L} }_{I }}} \sup_{{\deuxlignes{-1pt}{F _{2} \in  \tilde{\mathcal{L} }_{I }}{F _{1} \subseteq F _{2}}}} \det N_{{F _{1}^{\perp } \cap  F _{2}}} = 1}.%
\NumeroteEqua{B.16}{3}\MStopEqua \Mpar \vspace{\Ssaut}\italique{Condition on \MMath{\left|\overline{\text{m}}\right| - \text{id}_{V }}.} Let \MMath{\epsilon  > 0} be as in {\seqBBBcdq}. Since \MMath{e^{\epsilon } > 1}, there exists \MMath{F _{1} \in  \tilde{\mathcal{L} }_{I }} such that:%
\Mpar \MStartEqua \MMath{\forall  F _{2} \in  \tilde{\mathcal{L} }_{I } \big/ F _{1} \subseteq F _{2}, {\log\det N_{{F _{1}^{\perp } \cap  F _{2}}} < \epsilon }}.%
\MStopEqua \Mpar Let \MMath{F _{2} \in  \tilde{\mathcal{L} }_{I }} such that \MMath{F _{1} \subseteq F _{2}}, and let \MMath{\left(\mathbf{b} _{1},\dots ,\mathbf{b} _{n}\right)} be a \MMath{\left\langle \, \cdot \, \middlewithspace|\, \cdot \, \right\rangle }-orthonormal basis of \MMath{F  \mathrel{\mathop:}=  F _{1}^{\perp } \cap  F _{2}}. Let \MMath{\left(\mathbf{e} _{1},\dots ,\mathbf{e} _{2n}\right) \mathrel{\mathop:}=  \left(\mathbf{b} _{1},I \mathbf{b} _{1},\dots ,\mathbf{b} _{n},I \mathbf{b} _{n}\right)}, which is both a \MMath{\left(\, \cdot \, \middlewithspace|\, \cdot \, \right)}-orthonormal basis of \MMath{F } and a finite symplectic family (as follows from {\seqBBBaqg}).
Let \MMath{\left(\mathbf{f} _{1},\dots ,\mathbf{f} _{2n}\right) \mathrel{\mathop:}=  \left(\text{m} \mathbf{e} _{1},\dots ,\text{m} \mathbf{e} _{2n}\right)}. Since m{} is a symplectomorphism, \MMath{\left(\mathbf{f} _{1},\dots ,\mathbf{f} _{2n}\right)} is a finite symplectic family.
Thus, defining the \MMath{2n\times 2n} matrix \MMath{M} by:%
\Mpar \MStartEqua \MMath{\forall i,j\leqslant 2n, {M_{ij} \mathrel{\mathop:}=  \left(e_{i} \middlewithspace| M_{F } \,  e_{j}\right) = \left(\mathbf{f} _{i} \middlewithspace| \mathbf{f} _{j}\right)}},%
\MStopEqua \Mpar {\seqBBBcdv} implies that \MMath{\det M_{F } = \det M \geqslant  1}.
Hence, we have:%
\Mpar \MStartEqua \MMath{\log\det \frac{\text{id}_{F } + M_{F}}{2} - \frac{\log\det M_{F }}{2} < \epsilon },%
\MStopEqua \Mpar so {\seqBBBcdq} implies \MMath{\mkop{Tr}  \left(\sqrt{{M_{F }}} - \text{id}_{F }\right)^{2} < 1}.
In particular, it implies \MMath{\left\| M_{F } \right\| < 4}, so \MMath{\left\| \left.\text{m}\right|_{{F _{2}}} \right\| < 2 + \left\|\left.\text{m}\right|_{{F _{1}}}\right\|}.%
\Mpar \vspace{\Sssaut}\hspace{\Alinea}Now, for any \MMath{\mathbf{v}  \in  V }, there exists \MMath{F  \in  \tilde{\mathcal{L} }_{I }} such that \MMath{\mathbf{v}  \in  F }, so \MMath{\left|\text{m} \mathbf{v} \right| \leqslant  \left\|\left.\text{m}\right|_{{F \oplus F _{1}}}\right\| \,  \left|\mathbf{v} \right| < \left( 2 + \left\|\left.\text{m}\right|_{{F _{1}}}\right\| \right) \,  \left|\mathbf{v} \right|}. Hence, m{} is a bounded operator on {$V$}, and therefore, there exists a bounded operator \MMath{\overline{\text{m}}} on \MMath{\overline{V }} such that \MMath{\text{m} = \left. \overline{\text{m}} \right|_{V }}.
\MMath{\overline{\text{m}}^{\dag} \,  \overline{\text{m}}} is then a bounded, positive, symmetric operator on \MMath{\overline{V }}, so there exists a projection-valued measure on \MMath{\left[0,\, \left\|\text{m}\right\|\right]} such that \bseqHHHabp{Remark 20.18}:%
\Mpar \MStartEqua \MMath{\overline{\text{m}}^{\dag} \,  \overline{\text{m}} = \int_{0}^{\left\|\text{m}\right\|} d\Pi (\nu ) \,  \nu }.%
\MStopEqua \Mpar For any \MMath{\delta  \in  \left]0,\, 1\right[}, we define the spectral projectors \MMath{\Pi _{{\delta ,-}} \mathrel{\mathop:}=  \int_{0}^{{(1-\delta )^{2}}} d\Pi (\nu )}, resp.~\MMath{\Pi _{{\delta ,+}} \mathrel{\mathop:}=  \int_{{(1+\delta )^{2}}}^{\left\|\text{m}\right\|} d\Pi (\nu )}, as well as the subspaces \MMath{W _{{\delta ,\pm }} \mathrel{\mathop:}=  \Pi _{{\delta ,\pm }}\left\langle \overline{V }\right\rangle }.%
\Mpar \vspace{\Sssaut}\hspace{\Alinea}Let \MMath{\left(\mathbf{g} _{1},\dots ,\mathbf{g} _{k}\right)} be a \MMath{\left(\, \cdot \, \middlewithspace|\, \cdot \, \right)}-orthonormal family in \MMath{W _{{\delta ,\pm }}}.
Using {\seqBBBcec}, together with the density of {$V$} in \MMath{\overline{V }}, there exists a \MMath{\left(\, \cdot \, \middlewithspace|\, \cdot \, \right)} orthonormal family \MMath{\left(\mathbf{g} '_{1},\dots ,\mathbf{g} '_{k}\right)} in {$V$}, satisfying:%
\Mpar \MStartEqua \MMath{\forall  i\leqslant k, {\left|\mathbf{g} '_{i} - \mathbf{g} _{i}\right| < \frac{\nicefrac{\delta }{2}}{(k+1)\, (\left\|\text{m}\right\|+1)}}}.%
\MStopEqua \Mpar For any \MMath{\mathbf{w}  \in  W  \mathrel{\mathop:}=  \mkop{Span} \left\{\mathbf{g} '_{1},\dots ,\mathbf{g} '_{k}\right\}}, we have:%
\Mpar \MStartEqua \MMath{\left|\text{m} \,  \mathbf{w} \right| = \left|\sum_{i=1}^{k} \mathbf{w} ^{i} \,  \text{m} \,  \mathbf{g} '_{i} \right| \geqslant  \left|\sum_{i=1}^{k} \mathbf{w} ^{i} \,  \overline{\text{m}} \,  \mathbf{g} _{i} \right| - \sum_{i=1}^{k} \left|\mathbf{w} \right| \,  \left\|\text{m}\right\| \,  \left|\mathbf{g} _{i} - \mathbf{g} '_{i}\right| \geqslant  \left(1+\nicefrac{\delta }{2}\right) \,  \left|\mathbf{w} \right|},%
\MStopEqua \Mpar resp:%
\Mpar \MStartEqua \MMath{\left|\text{m} \,  \mathbf{w} \right| = \left|\sum_{i=1}^{k} \mathbf{w} ^{i} \,  \text{m} \,  \mathbf{g} '_{i} \right| \leqslant  \left|\sum_{i=1}^{k} \mathbf{w} ^{i} \,  \overline{\text{m}} \,  \mathbf{g} _{i} \right| + \sum_{i=1}^{k} \left|\mathbf{w} \right| \,  \left\|\text{m}\right\| \,  \left|\mathbf{g} _{i} - \mathbf{g} '_{i}\right| \leqslant  \left(1-\nicefrac{\delta }{2}\right) \,  \left|\mathbf{w} \right|}.%
\MStopEqua \Mpar Next, \MMath{F  \mathrel{\mathop:}=  \mkop{Span} \left\{\mathbf{g} '_{1},I \mathbf{g} '_{1},\dots ,\mathbf{g} '_{k},I \mathbf{g} '_{k}\right\}} is a finite-dimensional, complex (with respect to {$I$}) vector subspace of {$V$}, hence \MMath{F  \in  \tilde{\mathcal{L} }_{I }}, and defining \MMath{F _{2} \mathrel{\mathop:}=  F _{1} \oplus  F  \supseteq W }, we have \MMath{F _{2} \in  \tilde{\mathcal{L} }_{I }} with \MMath{F _{1} \subseteq F _{2}}. Let \MMath{\left(\mathbf{g} ''_{1},\dots ,\mathbf{g} ''_{2n}\right)} be an eigenbasis of the positive operator \MMath{M_{{F _{1}^{\perp } \cap  F _{2}}}}, with eigenvalues \MMath{\left(\nu _{1},\dots ,\nu _{2n}\right)}, and let \MMath{W  ' \mathrel{\mathop:}=  F _{1} \oplus  \mkop{Span} \left\{ \mathbf{g} ''_{i} \middlewithspace| \nu _{i} \geqslant  \left(1+\nicefrac{\delta }{2}\right)^{2} \right\}}, resp.~\MMath{F _{1} \oplus  \mkop{Span} \left\{ \mathbf{g} ''_{i} \middlewithspace| \nu _{i} \leqslant  \left(1-\nicefrac{\delta }{2}\right)^{2} \right\} \subseteq F _{2}}.
For any \MMath{\mathbf{w}  \in  W ^{{\prime \perp }} \cap  F _{2}} with \MMath{\mathbf{w}  \neq  0}, we have:%
\Mpar \MStartEqua \MMath{\left|\text{m} \,  \mathbf{w} \right| < \left(1 + \nicefrac{\delta }{2}\right) \,  \left|\mathbf{w} \right|}, resp.~\MMath{> \left(1 - \nicefrac{\delta }{2}\right) \,  \left|\mathbf{w} \right|}.%
\MStopEqua \Mpar Hence, \MMath{W  \cap  W ^{{\prime \perp }} = \left\{0\right\}}, so the orthogonal projection on \MMath{W  '} induces an injection \MMath{W  \rightarrow  W  '}, which requires \MMath{k = \mkop{dim}  W  \leqslant  \mkop{dim}  W  ' = \mkop{dim}  F _{1} + N^{(F _{2})}_{{\delta ,\pm }}}, where \MMath{N^{(F _{2})}_{{\delta ,+}} \mathrel{\mathop:}=  \# \left\{ \nu _{i} \geqslant  \left(1+\nicefrac{\delta }{2}\right)^{2} \right\}}, resp.~\MMath{N^{(F _{2})}_{{\delta ,-}} \mathrel{\mathop:}=  \# \left\{ \nu _{i} \leqslant  \left(1-\nicefrac{\delta }{2}\right)^{2} \right\}}.%
\Mpar \vspace{\Sssaut}\hspace{\Alinea}On the other hand, \MMath{\mkop{Tr}  \left(\sqrt{{M_{{F _{1}^{\perp } \cap  F _{2}}}}} - \text{id}_{{F _{1}^{\perp } \cap  F _{2}}}\right)^{2} = \sum_{i=1}^{2n} \left(\sqrt{{\nu _{i}}} - 1\right)^{2} < 1}, so \MMath{N^{(F _{2})}_{{\delta _{\pm }}} \leqslant  \frac{4}{\delta ^{2}}}. Therefore, \MMath{k \leqslant  \mkop{dim}  F _{1} + \frac{4}{\delta ^{2}}}.
In particular, this implies \MMath{\mkop{dim}  W _{{\delta ,\pm }} < \infty } for any \MMath{\delta  \in  \left]0,\, 1\right[}, so the eigenvalues of \MMath{\overline{\text{m}}^{\dag} \,  \overline{\text{m}}} are discrete, with \MMath{1} as unique accumulation point.
Moreover, the previous reasoning in fact shows that, for any \MMath{\delta  \in  \left]0,\, 1\right[}, there exists \MMath{F _{{\delta ,\pm }} \in  \tilde{\mathcal{L} }_{I }}, with \MMath{F _{1} \subseteq F _{{\delta _{\pm }}}} such that, for any \MMath{F _{2} \in  \tilde{\mathcal{L} }_{I }} with \MMath{F _{{\delta _{\pm }}} \subseteq F _{2}}, we have \MMath{\mkop{dim}  W _{{\delta ,\pm }} \leqslant  \mkop{dim}  F _{1} + N^{(F _{2})}_{{\delta ,\pm }}}.%
\Mpar \vspace{\Sssaut}\hspace{\Alinea}Let \MMath{\nu _{1} \geqslant  \dots  \geqslant  \nu _{k} > 1} be the \MMath{k}-th highest eigenvalues of \MMath{\overline{\text{m}}^{\dag} \,  \overline{\text{m}}}, and, for any \MMath{i\leqslant k}, let \MMath{\delta _{i} \mathrel{\mathop:}=  \min\big( \sqrt{{\nu _{i}}} - 1, \nicefrac{1}{2} \big) \in  \left]0,\, 1\right[}.
Since \MMath{\mathcal{L} _{I }} is directed {\seqDDDafp}, there exists \MMath{F _{2} \in  \tilde{\mathcal{L} }_{I }}, with \MMath{F _{1} \subseteq F _{2}}, such that, \MMath{\forall  i \leqslant  k, {N^{(F _{2})}_{{\delta _{i},+}} \geqslant  \mkop{dim}  W _{{\delta _{i},+}} - \mkop{dim}  F _{1} \geqslant  i - \mkop{dim}  F _{1}}}.
So, we get:%
\Mpar \MStartEqua \MMath{\sum_{{i=\mkop{dim} _{{F _{1}}} + 1}}^{k} \frac{\delta _{i}^{2}}{4} \leqslant  \mkop{Tr}  \left(\sqrt{{M_{{F _{1}^{\perp } \cap  F _{2}}}}} - \text{id}_{{F _{1}^{\perp } \cap  F _{2}}}\right)^{2} < 1}.%
\MStopEqua \Mpar Since this holds for any \MMath{k \leqslant  \mkop{dim}  \left\{ \int_{{1^{(+)}}}^{{\left\|\text{m}\right\|}} d\Pi (\nu ) \,  \mathbf{v}  \middlewithspace| \mathbf{v}  \in  V  \right\}}, we conclude:%
\Mpar \MStartEqua \MMath{\sum_{{i>\mkop{dim} _{{F _{1}}}}} \min\left( \left(\sqrt{{\nu _{i}}} - 1\right)^{2}, \nicefrac{1}{4} \right) \leqslant  4},%
\MStopEqua \Mpar where \MMath{\left\|\text{m}\right\| \geqslant  \nu _{1} \geqslant  \dots  \geqslant  \nu _{i} \geqslant  \dots  > 1} are the eigenvalues of \MMath{\overline{\text{m}}^{\dag} \,  \overline{\text{m}}} strictly greater than \MMath{1}.
Similarly, we have:%
\Mpar \MStartEqua \MMath{\sum_{{j>\mkop{dim} _{{F _{1}}}}} \left(1 - \sqrt{{\nu _{-j}}}\right)^{2} \leqslant  4},%
\MStopEqua \Mpar where \MMath{0 < \nu _{-1} \leqslant  \dots  \leqslant  \nu _{-j} \leqslant  \dots  < 1} are the eigenvalues of \MMath{\overline{\text{m}}^{\dag} \,  \overline{\text{m}}} strictly smaller than \MMath{1}.%
\Mpar \vspace{\Sssaut}\hspace{\Alinea}Defining \MMath{\tilde{T } \mathrel{\mathop:}=  \sqrt{{\overline{\text{m}}^{\dag} \,  \overline{\text{m}}}} - \text{id}_{{\overline{V }}}} (the square-root being defined by spectral resolution), we have \MMath{\mkop{Tr}  \tilde{T }^{2} < \infty }, ie.~\MMath{\tilde{T }} is Hilbert–Schmidt. Finally, we can define \MMath{O  \mathrel{\mathop:}=  \overline{\text{m}} \,  \big(\text{id}_{{\overline{V }}} + \tilde{T }\big)^{-1}} and \MMath{T  \mathrel{\mathop:}=  O  \,  \tilde{T } \,  O ^{\dag}}.
\MendOfProof %
\Mnomdefichier{lin53}%
\hypertarget{SECceu}{}\MsectionB{ceu}{B.2}{Fermionic Fock Spaces}%
\vspace{\PsectionB}\Mpar \MstartPhyMode\hspace{\Alinea}Repeating the previous section for the fermionic case, the main difference is that we need to carefully keep track of the additional sign factors caused by the anti-commutation of the fermionic annihilation and creation operators.
On the other hand, fermionic Fock spaces have the advantage of being finite-dimensional (at least when considering only finitely many classical \dofs), which simplifies some proofs.%
\MleavePhyMode \hypertarget{SECcew}{}\MsectionC{cew}{B.2.1}{Finite-dimensional Case -- Spin Representation}%
\vspace{\PsectionC}\hypertarget{PARcex}{}\Mpar \Mdefinition{B.19}For any \MMath{N \geqslant  0} and any Hilbert space {$\mathcal{H}$}, we define the Hilbert space \MMath{\mathcal{H} ^{{\otimes N,\text{{alt}}}}} by:%
\Mpar \MStartEqua \MMath{\mathcal{H} ^{{\otimes N,\text{{alt}}}} \mathrel{\mathop:}=  \left\{ \Psi  \in  \mathcal{H} ^{{\otimes N}} \middlewithspace| \forall  \varepsilon  \in  S_{N}, \hat{\varepsilon } \Psi  = \text{sig}(\varepsilon ) \,  \Psi  \right\}},%
\MStopEqua \Mpar where, for any permutation \MMath{\varepsilon  \in  S_{N}}, the unitary operator \MMath{\hat{\varepsilon }} has been defined in {\seqBBBcfa} and \MMath{\text{sig}(\varepsilon ) = \pm  1} denotes the signature of {$\varepsilon$}. By convention, \MMath{\mathcal{H} ^{{\otimes 0,\text{{alt}}}} \approx  \mathds{C}}.%
\Mpar \vspace{\Sssaut}\hspace{\Alinea}For any \MMath{n \geqslant  0}, we define the fermionic Fock space over \MMath{n} states as:%
\Mpar \MStartEqua \MMath{\mathcal{F} ^{(n)}_{\mathfrak{ferm} } \mathrel{\mathop:}=  \bigoplus_{N=0}^{n} \left( \mathds{C}^{n} \right)^{{\otimes N,\text{{alt}}}}}.%
\MStopEqua \Mpar For any \MMath{\left( N_{1},\dots ,N_{n} \right) \in  \left\{0,1\right\}^{n}}, we define:%
\Mpar \MStartEqua \MMath{\left| N_{1},\dots ,N_{n} \right\rangle _{\mathfrak{ferm} } \mathrel{\mathop:}=  \frac{1}{\sqrt{{N!}}} \sum_{{\varepsilon  \in  S_{N}}} \text{sig}(\varepsilon ) \,  \hat{\varepsilon } \big( \mathbf{b} _{1}^{{N_{1}}} \otimes  \dots  \otimes  \mathbf{b} _{n}^{{N_{n}}} \big) \in  \mathcal{F} ^{(n)}_{\mathfrak{ferm} }}%
\MStopEqua \Mpar where \MMath{N \mathrel{\mathop:}=  N_{1} + \dots  + N_{n}} (recall that \MMath{\left( \mathbf{b} _{1},\dots ,\mathbf{b} _{n} \right)} denotes the canonical basis of \MMath{\mathds{C}^{n}}). \MMath{\big( \left| N_{1},\dots ,N_{n} \right\rangle _{\mathfrak{ferm} } \big)_{{\left( N_{1},\dots ,N_{n} \right) \in  \left\{0,1\right\}^{n}}}} is an orthonormal basis of \MMath{\mathcal{F} ^{(n)}_{\mathfrak{ferm} }}.
In particular, \MMath{\mathcal{F} ^{(n)}_{\mathfrak{ferm} }} is a complex finite dimensional Hilbert space (of dimension \MMath{2^{n}}).%
\Mpar \vspace{\Sssaut}\hspace{\Alinea}For any \MMath{p \leqslant  n}, we define the linear operators \MMath{\mathtt{a} _{p}, \mathtt{a} ^{+}_{p} : \mathcal{F} ^{(n)}_{\mathfrak{ferm} } \rightarrow  \mathcal{F} ^{(n)}_{\mathfrak{ferm} }} by:%
\hypertarget{PARcfh}{}\Mpar \MStartEqua \MMath{\forall  \left( N_{1},\dots ,N_{n} \right) \in  \left\{0,1\right\}^{n}, {\mathtt{a} _{p} \left| N_{1},\dots ,N_{n} \right\rangle _{\mathfrak{ferm} } \mathrel{\mathop:}=  (-1)^{{N_{1} + \dots  + N_{p-1}}} \,  N_{p} \left| N_{1},\dots ,N_{p} -1,\dots ,N_{n} \right\rangle _{\mathfrak{ferm} }}\\
\hphantom{\forall  \left( N_{1},\dots ,N_{n} \right) \in  \left\{0,1\right\}^{n}, }\llap{$\&$}{\mathtt{a} ^{+}_{p} \left| N_{1},\dots ,N_{n} \right\rangle _{\mathfrak{ferm} } \mathrel{\mathop:}=  (-1)^{{N_{1} + \dots  + N_{p-1}}} \,  (1-N_{p}) \left| N_{1},\dots ,N_{p} +1,\dots ,N_{n} \right\rangle _{\mathfrak{ferm} }}}.%
\NumeroteEqua{B.19}{1}\MStopEqua \Mpar \MstartPhyMode\hspace{\Alinea}In analogy to the bosonic case {\seqDDDaba}, we give the quantization of the linear and quadratic observables.
The former satisfy anti-commutation relations, that are the counterpart of the symmetric scalar product at the classical level.
The latter, being quadratic in the ladder operators, satisfy commutation relations, and implement the generators of the orthogonal group, which will give rise to the spin representation described below.%
\Mpar \vspace{\Saut}\hspace{\Alinea}Like at the classical level {\seqDDDcfj}, the sign differences with respect to the bosonic case affect the anti-\MMath{\mathds{C}}-linear part of h{}: recall that those generators that are \MMath{\mathds{C}}-linear are shared by the symplectic and orthogonal Lie algebra (they belongs to the Lie algebra of the unitary group, on which they overlap).
Note that the extra phase in \MMath{\hat{\text{h}}} (discussed before {\seqBBBcfk}) is also affected: although this phase is expressed in terms of the \MMath{\mathds{C}}-linear part of h{}, we stressed above that its presence is only necessary to ensure that the commutators of the anti-\MMath{\mathds{C}}-linear generators comes out right.%
\MleavePhyMode \hypertarget{PARcfl}{}\Mpar \Mdefinition{B.20}For any \MMath{\mathbf{x}  \in  \mathds{R}^{2n}}, we define a linear operator \MMath{\hat{\mathbf{x} }} on \MMath{\mathcal{F} ^{(n)}_{\mathfrak{ferm} }} by:%
\Mpar \MStartEqua \MMath{\hat{\mathbf{x} } \mathrel{\mathop:}=  \sum_{p=1}^{n} \frac{\mathbf{x} _{2p-1} + i \,  \mathbf{x} _{2p}}{\sqrt{2}} \mathtt{a} _{p} + \frac{\mathbf{x} _{2p-1} - i \,  \mathbf{x} _{2p}}{\sqrt{2}} \mathtt{a} ^{+}_{p}}.%
\MStopEqua \Mpar \vspace{\Sssaut}\hspace{\Alinea}For any \MMath{\text{h} \in  \mathfrak{so} ^{(n)}}, we define a linear operator \MMath{\hat{\text{h}}} on \MMath{\mathcal{F} ^{(n)}_{\mathfrak{ferm} }} by:%
\Mpar \MStartEqua \MMath{\hat{\text{h}} \mathrel{\mathop:}=  \sum_{{p,q=1}}^{n} \left[ - i \,  \beta _{pq}(\text{h}) \,  \mathtt{a} ^{+}_{p} \mathtt{a} _{q} - \frac{i}{2} \left( \gamma _{pq}(\text{h}) \,  \mathtt{a} _{p} \mathtt{a} _{q} \heavyplus  \gamma ^{*}_{pq}(\text{h}) \,  \mathtt{a} ^{+}_{p} \mathtt{a} ^{+}_{q} \right) \right] \heavyplus  \frac{i}{2} \big( \mkop{Tr}  \beta (\text{h}) \big) \,  \mathds{1} }%
\MStopEqua \Mpar (with \MMath{\beta (\text{h}), \gamma (\text{h})} from {\seqBBBcfq}).%
\hypertarget{PARcfr}{}\Mpar \Mproposition{B.21}For any \MMath{\mathbf{x}  \in  \mathds{R}^{2n}}, resp.\ any \MMath{\text{h} \in  \mathfrak{so} ^{(n)}}, \MMath{\hat{\mathbf{x} }}, resp.\ \MMath{\hat{\text{h}}}, is a symmetric (hence self-adjoint) operator.%
\Mpar \vspace{\Sssaut}\hspace{\Alinea}Moreover, for any \MMath{\mathbf{x} ,\mathbf{x} ' \in  \mathds{R}^{2n}} and any \MMath{\text{h},\text{h}' \in  \mathfrak{so} ^{(n)}}, we have the following (anti-)commutators:%
\hypertarget{PARcfs}{}\Mpar \MStartEqua \MMath{\left[ \hat{\mathbf{x} }, \hat{\mathbf{x} '} \right]_{+} = \left({}^{\text{\sc t}} \mathbf{x}  \,  \mathbf{x}  '\right) \mathds{1} ,\hspace{0.25cm} 
\left[ \hat{\text{h}}, \hat{\mathbf{x} } \right] = - i \,  \widehat{\text{h} \,  \mathbf{x}  } \&
\left[ \hat{\text{h}}, \hat{\text{h}'} \right] = - i \,  \widehat{[\text{h}, \text{h}']}},%
\NumeroteEqua{B.21}{1}\MStopEqua \Mpar where \MMath{\left[\, \cdot \, ,\, \cdot \, \right]_{+}} denotes the anti-commutator.%
\Mpar \Mproof For any \MMath{p \leqslant  n}, \MMath{\mathtt{a} ^{+}_{p}} is the adjoint of \MMath{\mathtt{a} _{p}}, hence, for any \MMath{\mathbf{x}  \in  \mathds{R}^{2n}}, \MMath{\hat{\mathbf{x} }} is symmetric, and, for any \MMath{\text{h} \in  \mathfrak{so} ^{(n)}}, \MMath{\hat{\text{h}}} is symmetric (using \abbrevProposition{\seqBBBcfv}). Since \MMath{\mathcal{F} ^{(n)}_{\mathfrak{ferm} }} is finite dimensional, these operators are self-adjoint.%
\Mpar \vspace{\Sssaut}\hspace{\Alinea}For any \MMath{p,q \leqslant  n}, \MMath{\left[ \mathtt{a} _{p}, \mathtt{a} ^{+}_{q} \right]_{+} = \delta _{pq} \,  \mathds{1} } and \MMath{\left[ \mathtt{a} _{p}, \mathtt{a} _{q} \right]_{+} = \left[ \mathtt{a} ^{+}_{p}, \mathtt{a} ^{+}_{q} \right]_{+} = 0}. Now, using that, for any operators \MMath{A,B,C,D} on \MMath{\mathcal{F} ^{(n)}_{\mathfrak{ferm} }}, we have:%
\Mpar \MStartEqua \MMath{\left[ AB,\,  C \right] = A \left[ B,\,  C \right]_{+} - \left[ A,\,  C \right]_{+} B
\& \left[ AB,\,  CD \right] = \left[ AB,\,  C \right] D + C \left[ AB,\,  D \right]},%
\MStopEqua \Mpar we get, for any  \MMath{u, u' \in  \mathds{C}^{n}} and any \MMath{\beta ,\beta ',\gamma ,\gamma ' \in  \text{M}_{n}(\mathds{C})}:%
\Mpar \MStartEqua \MMath{\left[ X(u), X^{+}(u') \right]_{+} = \left\langle  u', u \right\rangle  \,  \mathds{1} \\[3pt]
\left[ B(\beta ), X(u) \right] = - X \left({}^{\text{\sc t}} \beta  u\right) \&
\left[ B(\beta ), X^{+}(u) \right] = X^{+} \left(\beta ^{*} u\right)\\[3pt]
\left[ A(\gamma ), X^{+}(u) \right] = X \big( \left(\gamma  \heavyminus  {}^{\text{\sc t}} \gamma \right) u^{*} \big) \&
\left[ A^{+}(\gamma ), X(u) \right] = - X^{+} \big( \left(\gamma  \heavyminus  {}^{\text{\sc t}} \gamma \right) u^{*} \big)\\[3pt]
\left[ B(\beta ), B(\beta ') \right] = B \big( \left[ \beta , \beta ' \right] \big)\\[3pt]
\left[ B(\beta ), A(\gamma ') \right] = - A \left( \gamma ' \beta  + {}^{\text{\sc t}} \beta  \gamma ' \right) = - A \big( \left(\gamma ' \heavyminus  {}^{\text{\sc t}} \gamma '\right) \beta  \big)\\[3pt]
\left[ B(\beta ), A^{+}(\gamma ') \right] = A^{+} \left( \beta ^{*} \gamma ' + \gamma ' \beta ^{\dag} \right) = A^{+} \big( \left( \gamma ' \heavyminus  {}^{\text{\sc t}} \gamma ' \right) \beta ^{\dag} \big)\\[3pt]
\left[ A(\gamma ), A^{+}(\gamma ') \right] = \heavyminus  \frac{1}{2} \mkop{Tr}  \big( \left(\gamma ^{{\prime*}} \heavyminus  {}^{\text{\sc t}} \gamma ^{{\prime*}}\right) \left(\gamma  \heavyminus  {}^{\text{\sc t}} \gamma \right) \big) + B \big( (\gamma ^{{\prime*}} \heavyminus  {}^{\text{\sc t}} \gamma ^{{\prime*}}) (\gamma  \heavyminus  {}^{\text{\sc t}} \gamma ) \big)}%
\MStopEqua \Mpar where \MMath{X(u)}, \MMath{X^{+}(u)}, \MMath{A(\gamma )}, \MMath{B(\beta )} were defined in {\seqBBBbsp}, and \MMath{A^{+}(\gamma )} is redefined as (note the sign difference!):%
\Mpar \MStartEqua \MMath{A^{+}(\gamma ) \mathrel{\mathop:}=  \heavyminus  \sum_{{p,q=1}}^{n} \gamma ^{*}_{pq} \,  \mathtt{a} ^{+}_{p} \mathtt{a} ^{+}_{q}}.%
\MStopEqua \Mpar The desired commutators then follow from {\seqBBBcfj} like in {\seqBBBbsp}.%
\MendOfProof \Mpar \MstartPhyMode\hspace{\Alinea}Exponentiating the representation of the Lie algebra into a (projective) representation of the orthogonal group is easier than deriving the metaplectic representation in the bosonic case, because we are here only dealing with \emph{bounded} symmetric operators (as stressed at the beginning of the present section, we are working on a finite-dimensional Hilbert space).%
\MleavePhyMode \hypertarget{PARcgd}{}\Mpar \Mproposition{B.22}For any \MMath{n>0}, there exists a unique unitary representation \MMath{\mathtt{T} ^{(n)}_{\mathfrak{ferm} }} of the spin group \MMath{\text{Spin}^{(n)}} {\seqDDDalo} on \MMath{\mathcal{F} ^{(n)}_{\mathfrak{ferm} }} such that:%
\hypertarget{PARcge}{}\Mpar \MStartEqua \MMath{\forall  \text{h} \in  \mathfrak{so} ^{(n)}, \forall  t \in  \mathds{R}, {\mathtt{T} ^{(n)}_{\mathfrak{ferm} } \big( \exp_{{\text{Spin}^{(n)}}}(t\, \text{h}) \big) = \exp \big( it \,  \hat{\text{h}} \big)}}%
\NumeroteEqua{B.22}{1}\MStopEqua \Mpar (where \MMath{\exp_{{\text{Spin}^{(n)}}}} denotes the exponential mapping \MMath{\mathfrak{so} ^{(n)} \rightarrow  \text{Spin}^{(n)}} and \MMath{\exp} the exponential of operators), and it satisfies:%
\hypertarget{PARcgg}{}\Mpar \MStartEqua \MMath{\forall  \mu  \in  \text{Spin}^{(n)}, \forall  \mathbf{x}  \in  \mathds{R}^{2n}, {\mathtt{T} ^{(n)}_{\mathfrak{ferm} } \big( \mu  \big) \,  \hat{\mathbf{x} } \,  \mathtt{T} ^{(n)}_{\mathfrak{ferm} } \big( \mu ^{-1} \big) = \widehat{p ^{(n)}(\mu ) \,  \mathbf{x} }}}%
\NumeroteEqua{B.22}{2}\MStopEqua \Mpar (with the covering map \MMath{p ^{(n)} : \text{Spin}^{(n)} \rightarrow  \text{SO}^{(n)}} from {\seqBBBalo}).%
\Mpar \vspace{\Sssaut}\hspace{\Alinea}For \MMath{n=0}, we \emph{define} \MMath{\mathtt{T} ^{(0)}_{\mathfrak{ferm} }} by:%
\Mpar \MStartEqua \MMath{\mathtt{T} _{\mathfrak{ferm} }^{(0)}(\mathds{1} ) = \text{id}_{{\mathcal{F} ^{(0)}_{\mathfrak{ferm} }}} \, \&\,  \mathtt{T} _{\mathfrak{ferm} }^{(0)}(\mathds{1} ^{-}) = -\text{id}_{{\mathcal{F} ^{(0)}_{\mathfrak{ferm} }}}}.%
\MStopEqua \Mpar \Mproof From {\seqBBBant}, \MMath{\text{h} \mapsto  i\hat{\text{h}}} is a Lie algebra morphism, hence by \bseqHHHabg{theorems 3.27 and 3.32}, there exists unitary representation \MMath{\tilde{\mathtt{T} }^{(n)}_{\mathfrak{ferm} }} of the universal cover \MMath{\widetilde{\text{SO}}^{(n)}} of \MMath{\text{SO}^{(n)}} such that:%
\Mpar \MStartEqua \MMath{\forall  \text{h} \in  \mathfrak{so} ^{(n)}, \forall  t \in  \mathds{R}, {\tilde{\mathtt{T} }^{(n)}_{\mathfrak{ferm} } \big( \exp_{{\widetilde{\text{SO}}^{(n)}}}(t\, \text{h}) \big) = \exp \big( it \,  \hat{\text{h}} \big)}}.%
\MStopEqua \Mpar When \MMath{n=1}, \MMath{\widetilde{\text{SO}}^{(n)} \neq  \text{Spin}^{(n)}}, but using the expression for a representative of \MMath{1 \in  \pi _{1} \big( \text{SO}^{(1)} \big)} given in {\seqBBBcgn}, and proceeding like in {\seqBBBbhz}, there exists a map \MMath{\mathtt{T} ^{(n)}_{\mathfrak{ferm} }} from \MMath{\text{Spin}^{(n)}} into the space of unitary operators on \MMath{\mathcal{F} ^{(n)}_{\mathfrak{ferm} }} such that \MMath{\tilde{\mathtt{T} }^{(n)}_{\mathfrak{ferm} } = \mathtt{T} ^{(n)}_{\mathfrak{ferm} } \circ  \tilde{p }^{(n)}} (where \MMath{\tilde{p }^{(n)}} denotes the covering map \MMath{\widetilde{\text{SO}}^{(n)} \rightarrow  \text{Spin}^{(n)}}).
Since \MMath{\tilde{p }^{(n)}} is a group homomorphism and a covering map, \MMath{\mathtt{T} ^{(n)}_{\mathfrak{ferm} }} is a unitary representation of \MMath{\text{Spin}^{(n)}} and it satisfies {\seqBBBcgo}.
The uniqueness is guaranteed because {\seqBBBcgo} completely specifies \MMath{\mathtt{T} ^{(n)}_{\mathfrak{ferm} }} in a neighborhood of the group unit and \MMath{\text{Spin}^{(n)}} is connected for any \MMath{n>0}.%
\Mpar \vspace{\Sssaut}\hspace{\Alinea}Finally, {\seqBBBcgq} can be obtained in a neighborhood of the group unit, using the commutators computed in {\seqBBBant} together with the definition of the exponential of a bounded operator, and then extended by connectedness.
In the \MMath{n=0} case, it can be checked directly.%
\MendOfProof \hypertarget{SECcgr}{}\MsectionC{cgr}{B.2.2}{Moments of a Density Matrix}%
\vspace{\PsectionC}\Mpar \MstartPhyMode\hspace{\Alinea}In the bosonic case, we have shown {\seqDDDavt} that density matrices can be characterized by their moment-generating function \MMath{\mathbf{x}  \mapsto  \mkop{Tr}  \big( \rho \, \exp(i\hat{\mathbf{x} }) \big)}. Since exponentiation is not needed in the fermionic case (the linear observables are \emph{bounded} operators), we can directly characterize a density matrix by its moments (aka.~\MMath{n}-points functions).%
\MleavePhyMode \hypertarget{PARcgt}{}\Mpar \Mproposition{B.23}Let \MMath{\rho ,\rho '} be density matrices on \MMath{\mathcal{F} ^{(n)}_{\mathfrak{ferm} }} (ie.~semi-definite positive operators of unit trace; note that, since \MMath{\mathcal{F} ^{(n)}_{\mathfrak{ferm} }} is finite-dimensional, any operator is trace-class) such that:%
\Mpar \MStartEqua \MMath{\forall k \geqslant  0,\,  \forall  \mathbf{x} _{1},\dots ,\mathbf{x} _{k} \in  \mathds{R}^{2n}, {\mkop{Tr}  \big( \rho \, \hat{\mathbf{x} }_{1} \dots  \hat{\mathbf{x} }_{k} \big) = \mkop{Tr}  \big( \rho '\, \hat{\mathbf{x} }_{1} \dots  \hat{\mathbf{x} }_{k} \big)}}.%
\MStopEqua \Mpar Then, \MMath{\rho  = \rho '}.%
\Mpar \vspace{\Sssaut}\hspace{\Alinea}Moreover, for any density matrix {$\rho$} on \MMath{\mathcal{F} ^{(n)}_{\mathfrak{ferm} }}, and any \MMath{k\geqslant 0}, the function \MMath{\left(\mathds{R}^{2n}\right)^{k} \rightarrow  \mathds{C},\,  \mathbf{x} _{1},\dots ,\mathbf{x} _{k} \mapsto  \mkop{Tr}  \big( \rho \, \hat{\mathbf{x} }_{1} \dots  \hat{\mathbf{x} }_{k} \big)} is continuous.%
\Mpar \Mproof Using the definition of \MMath{\mathtt{a} _{p},\, \mathtt{a} ^{+}_{p}} {\seqDDDanj}, we have:%
\Mpar \MStartEqua \MMath{\forall  \left( N_{1},\dots ,N_{n} \right) \in  \left\{0,1\right\}^{n}, {\mathtt{a} _{p} \mathtt{a} ^{+}_{p} \left| N_{1},\dots ,N_{n} \right\rangle _{\mathfrak{ferm} } = (1 - N_{p}) \,  \left| N_{1},\dots ,N_{n} \right\rangle _{\mathfrak{ferm} }}},%
\MStopEqua \Mpar hence:%
\Mpar \MStartEqua \MMath{\left| 0,\dots ,0 \middlewithspace\rangle \!\!\middlewithspace\langle  0,\dots ,0 \right|_{\mathfrak{ferm} } = \mathtt{a} _{1} \,  \mathtt{a} ^{+}_{1} \dots  \mathtt{a} _{n} \,  \mathtt{a} ^{+}_{n} = \left(\frac{1}{2} + i\,  \hat{\mathbf{x} }^{(1)} \,  \hat{\mathbf{x} }^{(2)}\right) \dots  \left(\frac{1}{2} + i\,  \hat{\mathbf{x} }^{(2n-1)} \,  \hat{\mathbf{x} }^{(2n)}\right)},%
\MStopEqua \Mpar where \MMath{\forall  i,j\leqslant 2n, {\mathbf{x} ^{(j)}_{i} \mathrel{\mathop:}=  \delta _{ij}}}. Moreover, we also have:%
\Mpar \MStartEqua \MMath{\mathcal{F} ^{(n)}_{\mathfrak{ferm} } = \mkop{Span}  \left\{ \hat{\mathbf{x} }_{1} \dots  \hat{\mathbf{x} }_{k} \,  \left|0,\dots ,0\right\rangle  \middlewithspace| k\geqslant 0,\,  \mathbf{x} _{1},\dots ,\mathbf{x} _{k} \in  \mathds{R}^{2n} \right\}}.%
\MStopEqua \Mpar Thus, an operator {$\rho$} on \MMath{\mathcal{F} ^{(n)}_{\mathfrak{ferm} }} is entirely characterized by the value of:%
\Mpar \MStartEqua \MMath{\left\langle  0,\dots 0 \right| \,  \hat{\mathbf{x} }_{k} \dots  \hat{\mathbf{x} }_{1} \,  \rho  \,  \hat{\mathbf{x} }'_{1} \dots  \hat{\mathbf{x} }'_{k'} \left| 0,\dots ,0 \right\rangle _{\mathfrak{ferm} } = \mkop{Tr}  \big( \rho \hspace{0.25cm} \hat{\mathbf{x} }'_{1} \dots  \hat{\mathbf{x} }'_{k'} \,  \left| 0,\dots ,0 \middlewithspace\rangle \!\!\middlewithspace\langle  0,\dots ,0 \right|_{\mathfrak{ferm} } \,  \hat{\mathbf{x} }_{k} \dots  \hat{\mathbf{x} }_{1} \big)},%
\MStopEqua \Mpar for any \MMath{k,k' \geqslant  0} and any \MMath{\mathbf{x} _{1},\dots ,\mathbf{x} _{k},\mathbf{x} '_{1},\dots ,\mathbf{x} '_{k'} \in  \mathds{R}^{2n}}, and therefore by the value of \MMath{\mkop{Tr}  \big( \rho \, \hat{\mathbf{x} }_{1} \dots  \hat{\mathbf{x} }_{k} \big)} for any \MMath{k \geqslant  0} and any \MMath{\mathbf{x} _{1},\dots ,\mathbf{x} _{k} \in  \mathds{R}^{2n}}.%
\Mpar \vspace{\Sssaut}\hspace{\Alinea}For any \MMath{\mathbf{x}  \in  \mathds{R}^{2n}}, we have \MMath{\left\|\hat{\mathbf{x} }\right\|^{2} = \left\|\hat{\mathbf{x} }^{2}\right\| = {\textstyle\frac{1}{2}}\,  \left|\mathbf{x} \right|^{2}} (where \MMath{\left|\mathbf{x} \right|^{2} \mathrel{\mathop:}=  {}^{\text{\sc t}} \mathbf{x}  \,  \mathbf{x} }). This ensures that the function \MMath{\mathbf{x}  \mapsto  \hat{\mathbf{x} }} is strongly continuous, so that, for any operator {$\rho$}, and any \MMath{k\geqslant 0}, the function \MMath{\mathbf{x} _{1},\dots ,\mathbf{x} _{k} \mapsto  \mkop{Tr}  \big( \rho \, \hat{\mathbf{x} }_{1} \dots  \hat{\mathbf{x} }_{k} \big)} is continuous.%
\MendOfProof \hypertarget{SECchh}{}\MsectionC{chh}{B.2.3}{Coarse-graining}%
\vspace{\PsectionC}\Mpar \MstartPhyMode\hspace{\Alinea}Coarse-graining is not as straightforward as it was in the bosonic case, due to the sign prefactors entering the definition of the ladder operators {\seqDDDanj}.
In particular, operators acting on the last \MMath{m-n} modes may pick up a sign prefactor dependent on the occupation numbers in the first \MMath{n} modes. Of course, we could have set up the conventions differently, and/or introduced additional signs in the definition of the tensor product decomposition below {\seqDDDchj} to compensate for the ones in {\seqBBBanj}. But this would merely shift the problem to the operators acting on the first \MMath{n} modes.
Our particular choice of convention has the merit that projective fermionic linear observables can be treated just like the bosonic ones (see {\seqBBBchk}).%
\Mpar \vspace{\Saut}\hspace{\Alinea}This is where it is important to have restricted ourselves to orientation preserving automorphisms (ie.~\MMath{\text{SO}^{(n)}} rather than \MMath{\text{O}^{(n)}}, with double cover \MMath{\text{Spin}^{(n)}} rather than \MMath{\text{Pin}^{(n)}}, see {\seqBBBchl}, as well as \bseqHHHabq{section 4.3}). For orientation reversing transformations, {\seqBBBchm} would \emph{not} hold, and no choice of sign conventions would allow to enforce \emph{at the same time} both the composition and unambiguity properties for arrows ({\seqBBBael} guaranteed by {\seqBBBchn} respectively).%
\MleavePhyMode \hypertarget{PARcho}{}\Mpar \Mdefinition{B.24}For any \MMath{m \geqslant  n \geqslant  0}, we define an isomorphism of Hilbert spaces \MMath{\Gamma ^{(m,n)}_{\mathfrak{ferm} } : \mathcal{F} ^{(m)}_{\mathfrak{ferm} } \rightarrow  \mathcal{F} ^{(n)}_{\mathfrak{ferm} } \otimes  \mathcal{F} ^{(m-n)}_{\mathfrak{ferm} }} by:%
\hypertarget{PARchp}{}\Mpar \MStartEqua \MMath{\forall  \left( N_{1},\dots ,N_{m} \right) \in  \mathds{N}^{m}, {\Gamma ^{(m,n)}_{\mathfrak{ferm} } \left| N_{1},\dots ,N_{m} \right\rangle _{\mathfrak{ferm} } \mathrel{\mathop:}=  \left| N_{1},\dots ,N_{n} \right\rangle _{\mathfrak{ferm} } \otimes  \left| N_{n+1},\dots ,N_{m} \right\rangle _{\mathfrak{ferm} }}}.%
\NumeroteEqua{B.24}{1}\MStopEqua \hypertarget{PARchq}{}\Mpar \Mproposition{B.25}Let \MMath{m \geqslant  n \geqslant  0}. For any \MMath{\mathbf{x}  \in  \mathds{R}^{2n}}, we have:%
\hypertarget{PARchr}{}\Mpar \MStartEqua \MMath{\big( \hat{\mathbf{x} } \otimes  \mathds{1}  \big) \circ  \Gamma ^{(m,n)}_{\mathfrak{ferm} } = \Gamma ^{(m,n)}_{\mathfrak{ferm} } \circ  \hat{\mathbf{y} }},%
\NumeroteEqua{B.25}{1}\MStopEqua \Mpar where:%
\Mpar \MStartEqua \MMath{\forall  i \leqslant  2m, {\mathbf{y} _{i} = \alternative{\mathbf{x} _{i}}{\text{{ if }} i \leqslant  2n}{0}{\text{{ otherwise}}}}}.%
\MStopEqua \Mpar \vspace{\Sssaut}\hspace{\Alinea}For any \MMath{\mu  \in  \text{Mp}^{(n)}}, we have:%
\hypertarget{PARchv}{}\Mpar \MStartEqua \MMath{\big( \mathtt{T} ^{(n)}_{\mathfrak{ferm} }(\mu ) \otimes  \mathds{1}  \big) \circ  \Gamma ^{(m,n)}_{\mathfrak{ferm} } = \Gamma ^{(m,n)}_{\mathfrak{ferm} } \circ  \mathtt{T} ^{(m)}_{\mathfrak{ferm} } \big( \ell _{m\leftarrow n}(\mu ) \big)},%
\NumeroteEqua{B.25}{2}\MStopEqua \Mpar with \MMath{\ell _{m\leftarrow n} : \text{Spin}^{(n)} \rightarrow  \text{Spin}^{(m)}} from {\seqBBBalq}.%
\Mpar \Mproof The proof is the same as in the bosonic case ({\seqBBBbxo}, except for the complications due to unbounded operators in the latter).%
\Mpar \vspace{\Ssaut}\italique{Note.} The signs pre-factors in {\seqBBBanj} have been chosen so that the analogue of {\seqBBBchy} holds: this is what allows the bosonic proof to work unchanged.%
\MendOfProof \hypertarget{PARchz}{}\Mpar \Mproposition{B.26}Let \MMath{V} be an even or infinite-dimensional real pre-Hilbert space (like in {\seqBBBabl}). Let \MMath{m \geqslant  n}, \MMath{\left(\mathbf{e} _{1},\dots ,\mathbf{e} _{2m}\right)} be an orthonormal family in \MMath{V }, \MMath{\left(\mathbf{e} '_{2n+1},\dots ,\mathbf{e} '_{2m}\right) \in  V ^{{2(m-n)}}} and \MMath{\mu  \in  \text{Mp}^{(m)}} such that:%
\Mpar \MStartEqua \MMath{\mu  \rhd  \left(\mathbf{e} _{1},\dots ,\mathbf{e} _{2m}\right) = \left(\mathbf{e} _{1},\dots ,\mathbf{e} _{2n},\mathbf{e} '_{2n+1},\dots ,\mathbf{e} '_{2m}\right)}.%
\MStopEqua \Mpar Then, there exists a unitary transformation \MMath{\Phi  : \mathcal{F} ^{(m-n)}_{\mathfrak{ferm} } \rightarrow  \mathcal{F} ^{(m-n)}_{\mathfrak{ferm} }} such that:%
\Mpar \MStartEqua \MMath{\Gamma ^{(m,n)}_{\mathfrak{ferm} } \circ  \mathtt{T} ^{(m)}_{\mathfrak{ferm} } \big( \mu  \big) = \big( \mathds{1}  \otimes  \Phi  \big) \circ  \Gamma ^{(m,n)}_{\mathfrak{ferm} }}.%
\MStopEqua \hypertarget{PARcid}{}\Mpar \Mlemma{B.27}We define the operator \MMath{\mathtt{S} ^{(n)}} on \MMath{\mathcal{F} ^{(n)}_{\mathfrak{ferm} }} by:%
\Mpar \MStartEqua \MMath{\forall  \left( N_{1},\dots ,N_{n} \right) \in  \mathds{N}^{n}, {\mathtt{S} ^{(n)} \left| N_{1},\dots ,N_{n} \right\rangle _{\mathfrak{ferm} } \mathrel{\mathop:}=  (-1)^{{N_{1} + \dots  + N_{n}}} \left| N_{1},\dots ,N_{n} \right\rangle _{\mathfrak{ferm} }}}.%
\MStopEqua \Mpar For any \MMath{\mu  \in  \text{Spin}^{(n)}}, we have:%
\hypertarget{PARcig}{}\Mpar \MStartEqua \MMath{\mathtt{T} ^{(n)}_{\mathfrak{ferm} } \big( \mu  \big) \,  \mathtt{S} ^{(n)} = \mathtt{S} ^{(n)} \,  \mathtt{T} ^{(n)}_{\mathfrak{ferm} } \big( \mu  \big)}.%
\NumeroteEqua{B.27}{1}\MStopEqua \Mpar \Mproof Let \MMath{\text{h} \in  \mathfrak{so} ^{(n)}}. Since \MMath{\hat{\text{h}}} only contains terms either quadratic in the operators \MMath{\mathtt{a} _{p}, \mathtt{a} ^{+}_{p}} or proportional to the identity, \MMath{\hat{\text{h}}} commutes with \MMath{\mathtt{S} ^{(n)}}. Hence, for any \MMath{n>0}, {\seqBBBcgo} ensures that {\seqBBBcii} holds in a neighborhood of the identity, and therefore, by connectedness, for any \MMath{\mu  \in  \text{Spin}^{(n)}}. For \MMath{n=0}, \MMath{\mathtt{S} ^{(0)} = \text{id}_{{\mathcal{F} ^{(0)}_{\mathfrak{ferm} }}}}, so {\seqBBBcii} can be checked directly.%
\MendOfProof \Mpar \MproofProposition{B.26}The matrix \MMath{\Pi ^{(m)}}, resp.~\MMath{\text{q}^{(m)}}, defined in {\seqBBBcik}, is in \MMath{\text{Sp}^{(m)} \cap  \text{SO}^{(n)}}, resp.~\MMath{\mathfrak{sp} ^{(m)} \cap  \mathfrak{so} ^{(n)}}. Thus, observing that \MMath{\left(\mathbf{e} _{1},\dots ,\mathbf{e} _{2m}\right)}  and \MMath{\left(\mathbf{e} _{1},\dots ,\mathbf{e} _{2n},\mathbf{e} '_{2n+1},\dots ,\mathbf{e} '_{2m}\right)} are orthonormal families with the \emph{same} orientation and proceeding like in {\seqBBBcik}, there exists \MMath{\tilde{\mu } \in  \text{Spin}^{(m-n)}} such that:%
\Mpar \MStartEqua \MMath{\exp_{{\text{Spin}^{(m)}}} \big( \pi  \text{q}^{(m)} \big) \,  \mu  \,  \exp_{{\text{Spin}^{(m)}}} \big( -\pi  \text{q}^{(m)} \big) = \ell _{{m \leftarrow  (m-n)}}(\tilde{\mu })}.%
\MStopEqua \Mpar Let \MMath{\tilde{\Phi } \mathrel{\mathop:}=  \mathtt{T} ^{(m-n)}_{\mathfrak{ferm} } \big( \tilde{\mu } \big)}. Using {\seqBBBcgo} and {\seqBBBamm}, we get:%
\Mpar \MStartEqua \MMath{\exp \big( i\pi  \,  \hat{\text{q}}^{(m)}_{\mathfrak{ferm} } \big) \,  \mathtt{T} ^{(m)}_{\mathfrak{ferm} } \big( \mu  \big) \,  \exp \big( -i\pi  \,  \hat{\text{q}}^{(m)}_{\mathfrak{ferm} } \big) = \Gamma ^{{(m,m-n),-1}}_{\mathfrak{ferm} } \circ  \big( \tilde{\Phi } \otimes  \mathds{1}  \big) \circ  \Gamma ^{(m,m-n)}_{\mathfrak{ferm} }},%
\MStopEqua \Mpar with:%
\Mpar \MStartEqua \MMath{\hat{\text{q}}^{(m)}_{\mathfrak{ferm} } = \sum_{p=1}^{m} \left( \frac{\mathtt{a} ^{+}_{p} - \mathtt{a} ^{+}_{m+1-p}}{2} \right) \left( \frac{\mathtt{a} _{m+1-p} - \mathtt{a} _{p}}{2} \right) + \frac{1}{2} \lfloor \nicefrac{m}{2} \rfloor \mathds{1} }%
\MStopEqua \Mpar (the sign difference with respect to {\seqBBBcik} comes from the sign difference in {\seqBBBcir}).
From:%
\Mpar \MStartEqua \MMath{\exp \big( i\pi  \,  \hat{\text{q}}^{(m)}_{\mathfrak{ferm} } \big) \,  \mathtt{a} ^{+}_{q} \,  \exp \big( -i\pi  \,  \hat{\text{q}}^{(m)}_{\mathfrak{ferm} } \big) = \mathtt{a} ^{+}_{m+1-q}
 = \widehat{\Pi }^{(m)}_{\mathfrak{ferm} \, } \mathtt{a} ^{+}_{q} \,  \widehat{\Pi }^{{(m),-1}}_{\mathfrak{ferm} }},%
\MStopEqua \Mpar where:%
\Mpar \MStartEqua \MMath{\forall  \left( N_{1},\dots ,N_{m} \right) \in  \left\{0,1\right\}^{m}, {\widehat{\Pi }^{(m)}_{\mathfrak{ferm} } \left| N_{1},\dots ,N_{m} \right\rangle _{\mathfrak{ferm} } \mathrel{\mathop:}=  \left(-1\right)^{{\sum_{i<j} N_{i} N_{j}}} \left| N_{m},\dots ,N_{1} \right\rangle _{\mathfrak{ferm} }}},%
\MStopEqua \Mpar we have \MMath{\exp \big( i\pi  \,  \hat{\text{q}}^{(m)}_{\mathfrak{ferm} } \big) = e^{{\nicefrac{i\pi }{2} \lfloor \nicefrac{m}{2} \rfloor}} \,  \widehat{\Pi }^{(m)}_{\mathfrak{ferm} }} (for the vectors of the form \MMath{\mathtt{a} ^{+}_{{q_{1}}} \dots  \mathtt{a} ^{+}_{{q_{N}}} \left| 0, \dots , 0 \right\rangle _{\mathfrak{ferm} }} generate \MMath{\mathcal{F} ^{(m)}_{\mathfrak{ferm} }}).
Using:%
\Mpar \MStartEqua \MMath{\sum_{1\leqslant i<j\leqslant m} N_{i} N_{j} = \sum_{1\leqslant i<j\leqslant n} N_{i} N_{j} + \left( \sum_{1\leqslant i\leqslant n} N_{i} \right) \left( \sum_{n<j\leqslant m} N_{j} \right) + \sum_{n<i<j\leqslant m} N_{i} N_{j}},%
\MStopEqua \Mpar we get:%
\Mpar \MStartEqua \MMath{\forall  \left( N_{1},\dots ,N_{m} \right) \in  \left\{0,1\right\}^{m}, \Gamma ^{(m,m-n)}_{\mathfrak{ferm} } \circ  \widehat{\Pi }^{(m)}_{\mathfrak{ferm} } \left| N_{1},\dots ,N_{m} \right\rangle _{\mathfrak{ferm} } =\\\hspace*{1.2cm}= \left(-1\right)^{{\left( \sum_{1\leqslant i\leqslant n} N_{i} \right) \left( \sum_{n<j\leqslant m} N_{j} \right)}} \left( \widehat{\Pi }^{(m-n)}_{\mathfrak{ferm} } \left| N_{n+1},\dots ,N_{m} \right\rangle _{\mathfrak{ferm} } \right) \otimes  \left( \widehat{\Pi }^{(n)}_{\mathfrak{ferm} } \left| N_{1},\dots ,N_{n} \right\rangle _{\mathfrak{ferm} } \right)},%
\MStopEqua \Mpar and therefore:%
\Mpar \MStartEqua \MMath{\forall  \left( N_{1},\dots ,N_{n} \right) \in  \left\{0,1\right\}^{n}, \forall  \left| \psi  \right\rangle  \in  \mathcal{F} ^{(m-n)}_{\mathfrak{ferm} }, \Gamma ^{(m,m-n)}_{\mathfrak{ferm} } \circ  \widehat{\Pi }^{(m)}_{\mathfrak{ferm} } \circ  \Gamma ^{{(m,n),-1}}_{\mathfrak{ferm} } \Big( \left| N_{1},\dots ,N_{n} \right\rangle _{\mathfrak{ferm} } \otimes  \left| \psi  \right\rangle  \Big) =\\\hspace*{1.2cm}= \left( \left(\mathtt{S} ^{(m-n)}\right)^{{\left( \sum_{1\leqslant i\leqslant n} N_{i} \right)}} \widehat{\Pi }^{(m-n)}_{\mathfrak{ferm} } \left| \psi  \right\rangle _{\mathfrak{ferm} } \right) \otimes  \left( \widehat{\Pi }^{(n)}_{\mathfrak{ferm} } \left| N_{1},\dots ,N_{n} \right\rangle _{\mathfrak{ferm} } \right)}.%
\MStopEqua \Mpar Putting everything together, we obtain:%
\Mpar \MStartEqua \MMath{\forall  \left( N_{1},\dots ,N_{n} \right) \in  \left\{0,1\right\}^{n}, \forall  \left| \psi  \right\rangle  \in  \mathcal{F} ^{(m-n)}_{\mathfrak{ferm} }, \Gamma ^{(m,n)}_{\mathfrak{ferm} } \circ  \mathtt{T} ^{(m)}_{\mathfrak{ferm} } \big( \mu  \big) \circ  \Gamma ^{{(m,n),-1}} \Big( \left| N_{1},\dots ,N_{n} \right\rangle _{\mathfrak{ferm} } \otimes  \left| \psi  \right\rangle  \Big) =\\
\hspace*{1.2cm}= \left| N_{1},\dots ,N_{n} \right\rangle _{\mathfrak{ferm} } \otimes  \left( \widehat{\Pi }^{(m-n)}_{\mathfrak{ferm} } \left(\mathtt{S} ^{(m-n)}\right)^{{\left( \sum_{1\leqslant i\leqslant n} N_{i} \right)}} \tilde{\Phi } \left(\mathtt{S} ^{(m-n)}\right)^{{\left( \sum_{1\leqslant i\leqslant n} N_{i} \right)}} \widehat{\Pi }^{(m-n)}_{\mathfrak{ferm} } \left| \psi  \right\rangle  \right)\\
\hspace*{1.2cm}= \left| N_{1},\dots ,N_{n} \right\rangle _{\mathfrak{ferm} } \otimes  \left( \widehat{\Pi }^{(m-n)}_{\mathfrak{ferm} } \tilde{\Phi } \widehat{\Pi }^{(m-n)}_{\mathfrak{ferm} } \left| \psi  \right\rangle  \right)},%
\MStopEqua \Mpar where the second equality comes from {\seqBBBchm}. Setting \MMath{\Phi  \mathrel{\mathop:}=  \widehat{\Pi }^{(m-n)}_{\mathfrak{ferm} } \tilde{\Phi } \widehat{\Pi }^{(m-n)}_{\mathfrak{ferm} }} yields the desired result.%
\MendOfProof \hypertarget{SECcje}{}\MsectionC{cje}{B.2.4}{Infinite-dimensional Fock Representation}%
\vspace{\PsectionC}\Mpar \MstartPhyMode\hspace{\Alinea}Building Fock spaces over infinitely many fermionic \dofs works in a way very similar to the bosonic case {\seqDDDbna}, except that, due to the anti-commutation of the ladder operators, we need an \emph{ordered} orthonormal basis of the 1-particle Hilbert space to seed an orthonormal basis of the corresponding Fock space.
This does not represent any restriction, since, given some {$I$}-orthonormal basis of {$V$}, we can always order it (any set can be equipped with a total order), and, of course, the fermionic Fock space itself does \emph{not} depend on the choice of ordering: different orderings simply corresponds to different orthonormal basis of \MMath{\mathcal{F} ^{(V ,I )}_{\mathfrak{ferm} }}, and are related by a straightforward unitary transformation.%
\MleavePhyMode \hypertarget{PARcjg}{}\Mpar \Mdefinition{B.28}Let \MMath{V ,\left(\, \cdot \, \middlewithspace|\, \cdot \, \right)} be an infinite-dimensional real pre-Hilbert space and let {$I$} be a compatible complex structure on {$V$} {\seqDDDaqh}. Let \MMath{\overline{V }_{{\!(I )}}} denote the completion of the corresponding complex pre-Hilbert space, and \MMath{V ^{*}_{{\!(I )}}} its dual.
We define the fermionic Fock space \MMath{\mathcal{F} ^{(V ,I )}_{\mathfrak{ferm} }} over \MMath{V ,\left(\, \cdot \, \middlewithspace|\, \cdot \, \right),I } as:%
\Mpar \MStartEqua \MMath{\mathcal{F} ^{(V ,I )}_{\mathfrak{ferm} } \mathrel{\mathop:}=  \overline{\bigoplus_{N\geqslant 0} \left( V ^{*}_{{\!(I )}} \right)^{{\otimes N,\text{{alt}}}}}}.%
\MStopEqua \Mpar \vspace{\Sssaut}\hspace{\Alinea}Let \MMath{\left(\mathbf{b} _{i}\right)_{i\in I}} be an orthonormal basis of \MMath{\overline{V }_{{\!(I )}}}, with \MMath{I} a \emph{totally} ordered set\footnote{By the well-ordering theorem, aka.~Zermelo's theorem, any set can be equipped with a total order.}, and denote by \MMath{\left(\mathbf{b} ^{*}_{i}\right)_{i\in I}} its dual basis. For any \MMath{\left(N_{i}\right)_{i\in I} \in  \left\{0,1\right\}^{I}} such that \MMath{\sum_{i\in I} N_{i} =\mathrel{\mathop:}  N < \infty }, let \MMath{i_{1} < \dots  < i_{N}} such that \MMath{\left\{i_{1},\dots ,i_{N}\right\} = \left\{ i \in  I \middlewithspace| N_{i} \neq  0\right\}} and define:%
\Mpar \MStartEqua \MMath{\left| \left(N_{i}\right)_{i\in I;\, }\left(\mathbf{b} _{i}\right)_{i\in I} \right\rangle _{\mathfrak{ferm} } \mathrel{\mathop:}=  \frac{1}{\sqrt{{N!}}} \sum_{{\varepsilon  \in  S_{N}}} \text{sig}(\varepsilon ) \,  \hat{\varepsilon } \big( \mathbf{b} ^{*}_{{i_{1}}} \otimes  \dots  \otimes  \mathbf{b} ^{*}_{{i_{N}}} \big) \in  \left( V ^{*}_{{\!(I )}} \right)^{{\otimes N,\text{{alt}}}}}.%
\MStopEqua \Mpar \MMath{\left( \left| \left(N_{i}\right)_{i\in I;\, }\left(\mathbf{b} _{i}\right)_{i\in I} \right\rangle _{\mathfrak{ferm} } \right)_{{\left(N_{i}\right)_{i\in I} \in  \left\{0,1\right\}^{I}, \sum_{i\in I} N_{i} < \infty }}} is an orthonormal basis of \MMath{\mathcal{F} ^{(V ,I )}_{\mathfrak{ferm} }}.%
\hypertarget{PARcjl}{}\Mpar \Mproposition{B.29}Let \MMath{\left(\mathbf{b} _{1},\dots ,\mathbf{b} _{n}\right)} be a \emph{finite} \emph{orthonormal} family in \MMath{\overline{V }_{{\!(I )}}}. Denotes by {$W$} the orthogonal complement of \MMath{\mkop{Span} _{\mathds{C}}\left\{\mathbf{b} _{1},\dots ,\mathbf{b} _{n}\right\}} in \MMath{\overline{V }_{{\!(I )}}} and by {$J$} its complex structure. There exists a unique Hilbert space isomorphism \MMath{\Gamma ^{(\mathbf{b} _{1},\dots ,\mathbf{b} _{n} ;V ,I )}_{\mathfrak{ferm} } : \mathcal{F} ^{(V ,I )}_{\mathfrak{ferm} } \rightarrow  \mathcal{F} ^{(n)}_{\mathfrak{ferm} } \otimes  \mathcal{F} ^{(W ,J )}_{\mathfrak{ferm} }} such that, for any orthonormal basis \MMath{\left(\mathbf{b} _{j}\right)_{j\in J}} of {$W$} (with \MMath{J} a totally ordered set):%
\hypertarget{PARcjm}{}\Mpar \MStartEqua \MMath{\forall  \left(N_{i}\right)_{i\in I} \in  \left\{0,1\right\}^{I} \big/ {\textstyle \sum_{i\in I} N_{i} < \infty },\\\hspace*{1.5cm} {\Gamma ^{(\mathbf{b} _{1},\dots ,\mathbf{b} _{n} ;V ,I )}_{\mathfrak{ferm} } \left| \left(N_{i}\right)_{i\in I;\, }\left(\mathbf{b} _{i}\right)_{i\in I} \right\rangle _{\mathfrak{ferm} } = \left| N_{1},\dots ,N_{n} \right\rangle _{\mathfrak{ferm} } \otimes  \left| \left(N_{j}\right)_{j\in J;\, }\left(\mathbf{b} _{j}\right)_{j\in J} \right\rangle _{\mathfrak{ferm} }}},%
\NumeroteEqua{B.29}{1}\MStopEqua \Mpar where \MMath{I \mathrel{\mathop:}=  \left\{1,\dots ,n\right\} \sqcup J} (the total orders on \MMath{\left\{1,\dots ,n\right\}} and \MMath{J} being extended so that \MMath{\forall  i \in  \left\{1,\dots ,n\right\}, \forall  j \in  J, {i < j}}).%
\Mpar \Mproof The proof is similar to the one of {\seqBBBbyw}.%
\MendOfProof \hypertarget{PARcjp}{}\Mpar \Mproposition{B.30}Let \MMath{\left(\mathbf{b} _{1},\dots ,\mathbf{b} _{n},\mathbf{b} _{n+1},\dots ,\mathbf{b} _{m}\right)} be a finite orthonormal family in \MMath{\overline{V }_{{\!(I )}}}. Denote by \MMath{W _{1},\,  J _{1}} the orthogonal complement of \MMath{\mkop{Span} _{\mathds{C}}\left\{\mathbf{b} _{1},\dots ,\mathbf{b} _{n}\right\}} and by \MMath{W _{2},\,  J _{2}} the orthogonal complement of \MMath{\mkop{Span} _{\mathds{C}}\left\{\mathbf{b} _{1},\dots ,\mathbf{b} _{m}\right\}}. Then, we have:%
\hypertarget{PARcjq}{}\Mpar \MStartEqua \MMath{\left( \Gamma ^{(m,n)}_{\mathfrak{ferm} } \otimes  \text{id}_{{\mathcal{F} ^{(W _{2},J _{2})}_{\mathfrak{ferm} }}} \right) \circ  \Gamma ^{(\mathbf{b} _{1},\dots ,\mathbf{b} _{m} ;V ,I )}_{\mathfrak{ferm} } = \left( \text{id}_{{\mathcal{F} ^{(n)}_{\mathfrak{ferm} }}} \otimes  \Gamma ^{(\mathbf{b} _{n+1},\dots ,\mathbf{b} _{m} ;W _{1},J _{1})}_{\mathfrak{ferm} } \right) \circ  \Gamma ^{(\mathbf{b} _{1},\dots ,\mathbf{b} _{n} ;V ,I )}_{\mathfrak{ferm} }}%
\NumeroteEqua{B.30}{1}\MStopEqua \Mpar \vspace{\Sssaut}\hspace{\Alinea}Let \MMath{\left(\mathbf{b} _{1},\dots ,\mathbf{b} _{n}\right),\,  \left(\mathbf{b} '_{1},\dots ,\mathbf{b} '_{n}\right)} be two orthonormal families in \MMath{\overline{V }_{{\!(I )}}} with \MMath{\mkop{Span} _{\mathds{C}}\left\{\mathbf{b} _{1},\dots ,\mathbf{b} _{n}\right\} = \mkop{Span} _{\mathds{C}}\left\{\mathbf{b} '_{1},\dots ,\mathbf{b} '_{n}\right\}}, and denotes by \MMath{W ,\,  J } the orthogonal complement of \MMath{\mkop{Span} _{\mathds{C}}\left\{\mathbf{b} _{1},\dots ,\mathbf{b} _{n}\right\}}. Then, we have:%
\hypertarget{PARcjs}{}\Mpar \MStartEqua \MMath{\Gamma ^{(\mathbf{b} '_{1},\dots ,\mathbf{b} '_{n} ;V ,I )}_{\mathfrak{ferm} } = \left( e^{{i\phi _{\mu }}} \,  \mathtt{T} ^{(n)}_{\mathfrak{ferm} }(\mu ) \otimes  \text{id}_{{\mathcal{F} ^{(W ,J )}_{\mathfrak{ferm} }}}\right) \circ  \Gamma ^{(\mathbf{b} _{1},\dots ,\mathbf{b} _{n} ;V ,I )}_{\mathfrak{ferm} }},%
\NumeroteEqua{B.30}{2}\MStopEqua \Mpar where {$\mu$} is some element of \MMath{\text{Spin}^{(n)}} such that \MMath{\left(\mathbf{b} '_{1},I \mathbf{b} '_{1},\dots ,\mathbf{b} '_{n},I \mathbf{b} '_{n}\right) = \mu  \rhd  \left(\mathbf{b} _{1},I \mathbf{b} _{1},\dots ,\mathbf{b} _{n},I \mathbf{b} _{n}\right)}, and \MMath{\phi _{\mu } \in  \mathds{R}}.%
\Mpar \Mproof The proof is similar to the one of {\seqBBBbdg}.
In particular, paying attention to the sign factors in {\seqBBBamf}, the following equality holds on \MMath{\mathcal{F} ^{(n)}_{\mathfrak{ferm} }} for any \MMath{n\times n} complex matrix h{}:%
\Mpar \MStartEqua \MMath{\bigoplus_{N=0}^{n} \sum_{k=1}^{N} \mathds{1} ^{(1)} \otimes  \dots  \otimes  \text{h}^{(k)} \otimes  \dots  \otimes  \mathds{1} ^{(N)} = \sum_{{i,j=1}}^{n} \text{h}_{ij} \,  \mathtt{a} ^{+}_{i} \,  \mathtt{a} _{j}}.%
\MStopEqua \Mpar Also, note that a \MMath{\mathds{C}}-linear bijection on a finite-dimensional complex vector space is always an \emph{orientation-preserving} transformation of the underlying \MMath{\mathds{R}}-vector space (as \MMath{\text{GL}_{n}(\mathds{C})} is connected), which confirms that \MMath{\left(\mathbf{b} '_{1},I \mathbf{b} '_{1},\dots ,\mathbf{b} '_{n},I \mathbf{b} '_{n}\right)} and \MMath{\left(\mathbf{b} _{1},I \mathbf{b} _{1},\dots ,\mathbf{b} _{n},I \mathbf{b} _{n}\right)} have the same orientation.%
\MendOfProof \hypertarget{PARcjx}{}\Mpar \Mproposition{B.31}For any \MMath{\mathbf{v}  \in  V }, there exists a unique bounded self-adjoint operator \MMath{\mathcal{O} ^{\mathfrak{ferm} }_{(I )}(\mathbf{v} )} on \MMath{\mathcal{F} ^{(V ,I )}_{\mathfrak{ferm} }} such that, for any finite orthonormal family \MMath{\left(\mathbf{b} _{1},\dots ,\mathbf{b} _{n}\right)} in \MMath{\overline{V }_{{\!(I )}}} with \MMath{\mathbf{v}  \in  \mkop{Span} _{\mathds{C}}\left\{\mathbf{b} _{1},\dots ,\mathbf{b} _{n}\right\}}, we have:%
\hypertarget{PARcjy}{}\Mpar \MStartEqua \MMath{\Gamma ^{(\mathbf{b} _{1},\dots ,\mathbf{b} _{n} ;V ,I )}_{\mathfrak{ferm} } \circ  \mathcal{O} ^{\mathfrak{ferm} }_{(I )}(\mathbf{v} ) = \left(\hat{\mathbf{x} } \otimes  \text{id}_{{\mathcal{F} ^{(W ,J )}_{\mathfrak{ferm} }}}\right) \circ  \Gamma ^{(\mathbf{b} _{1},\dots ,\mathbf{b} _{n} ;V ,I )}_{\mathfrak{ferm} }},%
\NumeroteEqua{B.31}{1}\MStopEqua \Mpar where \MMath{\mathbf{x}  \in  \mathds{R}^{2n}} is defined by \MMath{\mathbf{v}  =\mathrel{\mathop:}  \sum_{p=1}^{n} \left(\mathbf{x} _{2p-1} + i\, \mathbf{x} _{2p}\right) \mathbf{b} _{p}}.%
\Mpar \vspace{\Sssaut}\hspace{\Alinea}Moreover, for any \MMath{\mathbf{v} ,\mathbf{w}  \in  V } we have:%
\hypertarget{PARckb}{}\Mpar \MStartEqua \MMath{\left[ \mathcal{O} ^{\mathfrak{ferm} }_{(I )}(\mathbf{v} ),\,  \mathcal{O} ^{\mathfrak{ferm} }_{(I )}(\mathbf{w} ) \right]_{+} = \left(\mathbf{v} \middlewithspace|\mathbf{w} \right) \,  \text{id}_{{\mathcal{F} ^{(V ,I )}_{\mathfrak{ferm} }}}}.%
\NumeroteEqua{B.31}{2}\MStopEqua \Mpar \Mproof The proof is similar to the one of {\seqBBBckd} using {\seqBBBcke} in place of {\seqBBBckf} respectively.
\MendOfProof %
\Mfin%
\bibliography{refs}%
\end{document}